\pdfoutput=1 
\documentclass[cernpreprint,english]{na61doc}
\usepackage[T1]{fontenc}

\usepackage{lineno}
\usepackage{mathptmx} 
\usepackage{enumerate}
\usepackage{mathtools} 
\usepackage{amssymb}
\usepackage{placeins}
\usepackage{appendix}
\usepackage[utf8]{inputenc}
\usepackage{chngcntr}
\usepackage{microtype}
\usepackage{lineno}
\usepackage{color}
\usepackage{epstopdf}
\usepackage{colortbl}
\usepackage{tabularx}
\usepackage{multirow}

\usepackage{subcaption}
\usepackage{booktabs}
\usepackage{url}
\usepackage{cite}
\usepackage{caption}

\usepackage{hyperref}

\usepackage{svg}
\usepackage{bbding}
\usepackage{tikz}
\usetikzlibrary{patterns}
\usetikzlibrary{plotmarks}
\usepackage{pgfplots}
\usepackage{ulem}

\usepackage{array}
\newcolumntype{g}{>{\global\let\currentrowstyle\relax}}
\newcolumntype{^}{>{\currentrowstyle}}

\newcommand{\eV}{\ensuremath{\mbox{e\kern-0.1em V}}\xspace}
\newcommand{\GeV}{\ensuremath{\mbox{Ge\kern-0.1em V}}\xspace}
\newcommand{\MeV}{\ensuremath{\mbox{Me\kern-0.1em V}}\xspace}
\newcommand{\GeVc}{\ensuremath{\mbox{Ge\kern-0.1em V}\!/\!c}\xspace}
\newcommand{\GeVcc}{\ensuremath{\mbox{Ge\kern-0.1em V}\!/\!c^2}\xspace}
\newcommand{\AGeV}{\ensuremath{A\,\mbox{Ge\kern-0.1em V}}\xspace}
\newcommand{\AGeVc}{\ensuremath{A\,\mbox{Ge\kern-0.1em V}\!/\!c}\xspace}
\newcommand{\MeVc}{\ensuremath{\mbox{Me\kern-0.1em V}/c}\xspace}

\newcommand{\dd}{\ensuremath{{\textrm d}}\xspace}
\newcommand{\dedx}{\ensuremath{\dd E/\dd x}\xspace}
\newcommand{\tofdedx}{\ensuremath{\textup{\emph{tof}}-\dd E/\dd x}\xspace}
\newcommand{\tof}{\ensuremath{\textup{\emph{tof}}}\xspace}
\newcommand{\y}{\ensuremath{y}\xspace}
\newcommand{\p}{\ensuremath{p}\xspace}
\newcommand{\pt}{\ensuremath{p_{T}}\xspace}

\newcommand{\snn}{\ensuremath{\sqrt{s_{NN}}}\xspace}

\newcommand{\Urqmd}{{\scshape U}r{\scshape qmd}\xspace}

\newcommand{\Geant}{{\scshape Geant}\xspace}
\newcommand{\GeantThree}{{\scshape Geant3}\xspace}

\newcommand{\Epos}{{\scshape Epos}\xspace}
\newcommand{\EposLong}{{\scshape Epos1.99}\xspace}

\newcommand{\Phsd}{{\scshape Phsd}\xspace}
\newcommand{\Smash}{{\scshape Smash}\xspace}
\newcommand{\Smes}{{\scshape Smes}\xspace}

\newcommand{\NASixtyOne}{NA61\slash SHINE\xspace}
\newcommand{\CernVM}{\textsc{Cern\-\kern-0.05emVM}\xspace}
\newcommand{\pp}{\mbox{\textit{p}+\textit{p}}\xspace}

\newcommand{\dEdx}{$\text{d}E/\text{d}x$ }
\newcommand{\eqdEdx}{\text{d}E/\text{d}x}

\definecolor{tab_blue}{HTML}{1f77b4}
\definecolor{tab_orange}{HTML}{ff7f0e}
\definecolor{tab_green}{HTML}{2ca02c}
\definecolor{tab_red}{HTML}{d62728}
\definecolor{tab_violet}{HTML}{9467bd}
\definecolor{tab_brown}{HTML}{8c564b}
\definecolor{tab_pink}{HTML}{e377c2}
\definecolor{tab_gray}{HTML}{7f7f7f}
\definecolor{tab_yellow}{HTML}{bcbd22}
\definecolor{tab_cyan}{HTML}{17becf}

\definecolor{darkred}{rgb}{0.5,0,0}
\definecolor{darkblue}{rgb}{0,0,0.5}
\definecolor{firebrick}{rgb}{0.75,0.125,0.125}
\definecolor{darkgreen}{rgb}{0,0.5,0}
\definecolor{kPink+2}{RGB}{204,102,153}
\definecolor{kOrange+8}{RGB}{255,102,51}
\definecolor{kGreen+2}{RGB}{0,153,0}
\definecolor{kCyan+2}{RGB}{0,153,153}
\definecolor{kBlue+2}{RGB}{0,0,153}
\definecolor{kRed+1}{RGB}{204,0,0}
\definecolor{kBlue}{RGB}{0,0,204}
\definecolor{kBlue-9}{RGB}{153,153,255}
\definecolor{kGreen}{RGB}{0,153,0}
\definecolor{kRed}{RGB}{204,0,0}
\definecolor{kCyan}{RGB}{51,204,204}
\definecolor{kMagenta}{RGB}{153,0,153}
\definecolor{kPink}{RGB}{204,0,102}
\definecolor{kGray}{RGB}{204,204,204}
\definecolor{kBlack}{RGB}{0,0,0}

\definecolor{kRed+3}{RGB}{102,0,0}
\definecolor{kRed+2}{RGB}{153,0,0}
\definecolor{kRed-4}{RGB}{255,51,51}
\definecolor{kRed-7}{RGB}{255,102,102}
\definecolor{kRed-9}{RGB}{255,153,153}
\definecolor{redShading}{RGB}{229,127,127}
\definecolor{kBlue+1}{RGB}{0,0,204}

\newcommand{\dashedLine}{\tikz[baseline=-0.5ex]\draw [thick,dashed] (0,0) -- (0.5,0);}

\newcommand{\solidLine}{\tikz[baseline=-0.5ex]\draw [thick] (0,0) -- (0.5,0);}

\ShineTitle{
Measurements of $\pi^\pm$, $K^\pm$, $p$ and $\bar{p}$ spectra in $^{40}$Ar+$^{45}$Sc collisions at 13$A$ to 150$A$~\GeVc
}

\PreprintIdNumber{CERN-EP-2023-179}

\ShineAbstract{ 
The \NASixtyOne experiment at the CERN Super Proton Synchrotron studies the onset
of deconfinement in strongly interacting matter through a beam energy scan of particle production in collisions of nuclei of varied sizes. This paper presents results on inclusive double-differential spectra, transverse momentum and rapidity distributions and mean multiplicities of $\pi^\pm$, $K^\pm$, \textit{p} and $\bar{p}$ produced in $^{40}$Ar+$^{45}$Sc collisions at beam momenta of 13$A$, 19$A$, 30$A$, 40$A$, 75$A$ and 150$A$~\GeVc. The analysis uses the 10\% most central collisions, where the observed forward energy defines centrality.
The energy dependence of the $K^\pm$/$\pi^\pm$ ratios as well as of inverse slope parameters of the $K^\pm$ transverse mass distributions are placed in between those found in inelastic $p$+$p$ and central Pb+Pb collisions. 
The results obtained here establish a system-size dependence of hadron production properties that so far cannot be explained either within statistical or dynamical models.
}
\begin{document}

\maketitle

\pagebreak

\small
\tableofcontents
\normalsize

\pagebreak

\newpage 
{\Large The \NASixtyOne Collaboration}
\bigskip

\noindent
\mbox{H.\;Adhikary\href{https://orcid.org/0000-0002-5746-1268}{\includegraphics[height=1.7ex]{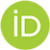}}\textsuperscript{\,13}},
\mbox{P.\;Adrich\href{https://orcid.org/0000-0002-7019-5451}{\includegraphics[height=1.7ex]{figures/orcid-logo.png}}\textsuperscript{\,15}},
\mbox{K.K.\;Allison\href{https://orcid.org/0000-0002-3494-9383}{\includegraphics[height=1.7ex]{figures/orcid-logo.png}}\textsuperscript{\,26}},
\mbox{N.\;Amin\href{https://orcid.org/0009-0004-7572-3817}{\includegraphics[height=1.7ex]{figures/orcid-logo.png}}\textsuperscript{\,5}},
\mbox{E.V.\;Andronov\href{https://orcid.org/0000-0003-0437-9292}{\includegraphics[height=1.7ex]{figures/orcid-logo.png}}\textsuperscript{\,22}},
\mbox{T.\;Anti\'ci\'c\href{https://orcid.org/0000-0002-6606-0191}{\includegraphics[height=1.7ex]{figures/orcid-logo.png}}\textsuperscript{\,3}},
\mbox{I.-C.\;Arsene\href{https://orcid.org/0000-0003-2316-9565}{\includegraphics[height=1.7ex]{figures/orcid-logo.png}}\textsuperscript{\,12}},
\mbox{M.\;Bajda\href{https://orcid.org/0009-0005-8859-1099}{\includegraphics[height=1.7ex]{figures/orcid-logo.png}}\textsuperscript{\,16}},
\mbox{Y.\;Balkova\href{https://orcid.org/0000-0002-6957-573X}{\includegraphics[height=1.7ex]{figures/orcid-logo.png}}\textsuperscript{\,18}},
\mbox{M.\;Baszczyk\href{https://orcid.org/0000-0002-2595-0104}{\includegraphics[height=1.7ex]{figures/orcid-logo.png}}\textsuperscript{\,17}},
\mbox{D.\;Battaglia\href{https://orcid.org/0000-0002-5283-0992}{\includegraphics[height=1.7ex]{figures/orcid-logo.png}}\textsuperscript{\,25}},
\mbox{A.\;Bazgir\href{https://orcid.org/0000-0003-0358-0576}{\includegraphics[height=1.7ex]{figures/orcid-logo.png}}\textsuperscript{\,13}},
\mbox{S.\;Bhosale\href{https://orcid.org/0000-0001-5709-4747}{\includegraphics[height=1.7ex]{figures/orcid-logo.png}}\textsuperscript{\,14}},
\mbox{M.\;Bielewicz\href{https://orcid.org/0000-0001-8267-4874}{\includegraphics[height=1.7ex]{figures/orcid-logo.png}}\textsuperscript{\,15}},
\mbox{A.\;Blondel\href{https://orcid.org/0000-0002-1597-8859}{\includegraphics[height=1.7ex]{figures/orcid-logo.png}}\textsuperscript{\,4}},
\mbox{M.\;Bogomilov\href{https://orcid.org/0000-0001-7738-2041}{\includegraphics[height=1.7ex]{figures/orcid-logo.png}}\textsuperscript{\,2}},
\mbox{Y.\;Bondar\href{https://orcid.org/0000-0003-2773-9668}{\includegraphics[height=1.7ex]{figures/orcid-logo.png}}\textsuperscript{\,13}},
\mbox{N.\;Bostan\href{https://orcid.org/0000-0002-1129-4345}{\includegraphics[height=1.7ex]{figures/orcid-logo.png}}\textsuperscript{\,25}},
\mbox{A.\;Brandin\textsuperscript{\,22}},
\mbox{W.\;Bryli\'nski\href{https://orcid.org/0000-0002-3457-6601}{\includegraphics[height=1.7ex]{figures/orcid-logo.png}}\textsuperscript{\,21}},
\mbox{J.\;Brzychczyk\href{https://orcid.org/0000-0001-5320-6748}{\includegraphics[height=1.7ex]{figures/orcid-logo.png}}\textsuperscript{\,16}},
\mbox{M.\;Buryakov\href{https://orcid.org/0009-0008-2394-4967}{\includegraphics[height=1.7ex]{figures/orcid-logo.png}}\textsuperscript{\,22}},
\mbox{A.F.\;Camino\textsuperscript{\,28}},
\mbox{M.\;\'Cirkovi\'c\href{https://orcid.org/0000-0002-4420-9688}{\includegraphics[height=1.7ex]{figures/orcid-logo.png}}\textsuperscript{\,23}},
\mbox{M.\;Csan\'ad\href{https://orcid.org/0000-0002-3154-6925}{\includegraphics[height=1.7ex]{figures/orcid-logo.png}}\textsuperscript{\,8}},
\mbox{J.\;Cybowska\href{https://orcid.org/0000-0003-2568-3664}{\includegraphics[height=1.7ex]{figures/orcid-logo.png}}\textsuperscript{\,21}},
\mbox{T.\;Czopowicz\href{https://orcid.org/0000-0003-1908-2977}{\includegraphics[height=1.7ex]{figures/orcid-logo.png}}\textsuperscript{\,13}},
\mbox{C.\;Dalmazzone\href{https://orcid.org/0000-0001-6945-5845}{\includegraphics[height=1.7ex]{figures/orcid-logo.png}}\textsuperscript{\,4}},
\mbox{N.\;Davis\href{https://orcid.org/0000-0003-3047-6854}{\includegraphics[height=1.7ex]{figures/orcid-logo.png}}\textsuperscript{\,14}},
\mbox{A.\;Dmitriev\href{https://orcid.org/0000-0001-7853-0173}{\includegraphics[height=1.7ex]{figures/orcid-logo.png}}\textsuperscript{\,22}},
\mbox{P.~von\;Doetinchem\href{https://orcid.org/0000-0002-7801-3376}{\includegraphics[height=1.7ex]{figures/orcid-logo.png}}\textsuperscript{\,27}},
\mbox{W.\;Dominik\href{https://orcid.org/0000-0001-7444-9239}{\includegraphics[height=1.7ex]{figures/orcid-logo.png}}\textsuperscript{\,19}},
\mbox{P.\;Dorosz\href{https://orcid.org/0000-0002-8884-0981}{\includegraphics[height=1.7ex]{figures/orcid-logo.png}}\textsuperscript{\,17}},
\mbox{J.\;Dumarchez\href{https://orcid.org/0000-0002-9243-4425}{\includegraphics[height=1.7ex]{figures/orcid-logo.png}}\textsuperscript{\,4}},
\mbox{R.\;Engel\href{https://orcid.org/0000-0003-2924-8889}{\includegraphics[height=1.7ex]{figures/orcid-logo.png}}\textsuperscript{\,5}},
\mbox{G.A.\;Feofilov\href{https://orcid.org/0000-0003-3700-8623}{\includegraphics[height=1.7ex]{figures/orcid-logo.png}}\textsuperscript{\,22}},
\mbox{L.\;Fields\href{https://orcid.org/0000-0001-8281-3686}{\includegraphics[height=1.7ex]{figures/orcid-logo.png}}\textsuperscript{\,25}},
\mbox{Z.\;Fodor\href{https://orcid.org/0000-0003-2519-5687}{\includegraphics[height=1.7ex]{figures/orcid-logo.png}}\textsuperscript{\,7,20}},
\mbox{M.\;Friend\href{https://orcid.org/0000-0003-4660-4670}{\includegraphics[height=1.7ex]{figures/orcid-logo.png}}\textsuperscript{\,9}},
\mbox{M.\;Ga\'zdzicki\href{https://orcid.org/0000-0002-6114-8223}{\includegraphics[height=1.7ex]{figures/orcid-logo.png}}\textsuperscript{\,13,6}},
\mbox{O.\;Golosov\href{https://orcid.org/0000-0001-6562-2925}{\includegraphics[height=1.7ex]{figures/orcid-logo.png}}\textsuperscript{\,22}},
\mbox{V.\;Golovatyuk\href{https://orcid.org/0009-0006-5201-0990}{\includegraphics[height=1.7ex]{figures/orcid-logo.png}}\textsuperscript{\,22}},
\mbox{M.\;Golubeva\href{https://orcid.org/0009-0003-4756-2449}{\includegraphics[height=1.7ex]{figures/orcid-logo.png}}\textsuperscript{\,22}},
\mbox{K.\;Grebieszkow\href{https://orcid.org/0000-0002-6754-9554}{\includegraphics[height=1.7ex]{figures/orcid-logo.png}}\textsuperscript{\,21}},
\mbox{F.\;Guber\href{https://orcid.org/0000-0001-8790-3218}{\includegraphics[height=1.7ex]{figures/orcid-logo.png}}\textsuperscript{\,22}},
\mbox{S.N.\;Igolkin\textsuperscript{\,22}},
\mbox{S.\;Ilieva\href{https://orcid.org/0000-0001-9204-2563}{\includegraphics[height=1.7ex]{figures/orcid-logo.png}}\textsuperscript{\,2}},
\mbox{A.\;Ivashkin\href{https://orcid.org/0000-0003-4595-5866}{\includegraphics[height=1.7ex]{figures/orcid-logo.png}}\textsuperscript{\,22}},
\mbox{A.\;Izvestnyy\href{https://orcid.org/0009-0009-1305-7309}{\includegraphics[height=1.7ex]{figures/orcid-logo.png}}\textsuperscript{\,22}},
\mbox{K.\;Kadija\textsuperscript{\,3}},
\mbox{N.\;Kargin\textsuperscript{\,22}},
\mbox{N.\;Karpushkin\href{https://orcid.org/0000-0001-5513-9331}{\includegraphics[height=1.7ex]{figures/orcid-logo.png}}\textsuperscript{\,22}},
\mbox{E.\;Kashirin\href{https://orcid.org/0000-0001-6062-7997}{\includegraphics[height=1.7ex]{figures/orcid-logo.png}}\textsuperscript{\,22}},
\mbox{M.\;Kie{\l}bowicz\href{https://orcid.org/0000-0002-4403-9201}{\includegraphics[height=1.7ex]{figures/orcid-logo.png}}\textsuperscript{\,14}},
\mbox{V.A.\;Kireyeu\href{https://orcid.org/0000-0002-5630-9264}{\includegraphics[height=1.7ex]{figures/orcid-logo.png}}\textsuperscript{\,22}},
\mbox{H.\;Kitagawa\textsuperscript{\,10}},
\mbox{R.\;Kolesnikov\href{https://orcid.org/0009-0006-4224-1058}{\includegraphics[height=1.7ex]{figures/orcid-logo.png}}\textsuperscript{\,22}},
\mbox{D.\;Kolev\href{https://orcid.org/0000-0002-9203-4739}{\includegraphics[height=1.7ex]{figures/orcid-logo.png}}\textsuperscript{\,2}},
\mbox{Y.\;Koshio\textsuperscript{\,10}},
\mbox{V.N.\;Kovalenko\href{https://orcid.org/0000-0001-6012-6615}{\includegraphics[height=1.7ex]{figures/orcid-logo.png}}\textsuperscript{\,22}},
\mbox{S.\;Kowalski\href{https://orcid.org/0000-0001-9888-4008}{\includegraphics[height=1.7ex]{figures/orcid-logo.png}}\textsuperscript{\,18}},
\mbox{B.\;Koz{\l}owski\href{https://orcid.org/0000-0001-8442-2320}{\includegraphics[height=1.7ex]{figures/orcid-logo.png}}\textsuperscript{\,21}},
\mbox{A.\;Krasnoperov\href{https://orcid.org/0000-0002-1425-2861}{\includegraphics[height=1.7ex]{figures/orcid-logo.png}}\textsuperscript{\,22}},
\mbox{W.\;Kucewicz\href{https://orcid.org/0000-0002-2073-711X}{\includegraphics[height=1.7ex]{figures/orcid-logo.png}}\textsuperscript{\,17}},
\mbox{M.\;Kuchowicz\href{https://orcid.org/0000-0003-3174-585X}{\includegraphics[height=1.7ex]{figures/orcid-logo.png}}\textsuperscript{\,20}},
\mbox{M.\;Kuich\href{https://orcid.org/0000-0002-6507-8699}{\includegraphics[height=1.7ex]{figures/orcid-logo.png}}\textsuperscript{\,19}},
\mbox{A.\;Kurepin\href{https://orcid.org/0000-0002-1851-4136}{\includegraphics[height=1.7ex]{figures/orcid-logo.png}}\textsuperscript{\,22}},
\mbox{A.\;L\'aszl\'o\href{https://orcid.org/0000-0003-2712-6968}{\includegraphics[height=1.7ex]{figures/orcid-logo.png}}\textsuperscript{\,7}},
\mbox{M.\;Lewicki\href{https://orcid.org/0000-0002-8972-3066}{\includegraphics[height=1.7ex]{figures/orcid-logo.png}}\textsuperscript{\,14}},
\mbox{G.\;Lykasov\href{https://orcid.org/0000-0002-1544-6959}{\includegraphics[height=1.7ex]{figures/orcid-logo.png}}\textsuperscript{\,22}},
\mbox{V.V.\;Lyubushkin\href{https://orcid.org/0000-0003-0136-233X}{\includegraphics[height=1.7ex]{figures/orcid-logo.png}}\textsuperscript{\,22}},
\mbox{M.\;Ma\'ckowiak-Paw{\l}owska\href{https://orcid.org/0000-0003-3954-6329}{\includegraphics[height=1.7ex]{figures/orcid-logo.png}}\textsuperscript{\,21}},
\mbox{Z.\;Majka\href{https://orcid.org/0000-0003-3064-6577}{\includegraphics[height=1.7ex]{figures/orcid-logo.png}}\textsuperscript{\,16}},
\mbox{A.\;Makhnev\href{https://orcid.org/0009-0002-9745-1897}{\includegraphics[height=1.7ex]{figures/orcid-logo.png}}\textsuperscript{\,22}},
\mbox{B.\;Maksiak\href{https://orcid.org/0000-0002-7950-2307}{\includegraphics[height=1.7ex]{figures/orcid-logo.png}}\textsuperscript{\,15}},
\mbox{A.I.\;Malakhov\href{https://orcid.org/0000-0001-8569-8409}{\includegraphics[height=1.7ex]{figures/orcid-logo.png}}\textsuperscript{\,22}},
\mbox{A.\;Marcinek\href{https://orcid.org/0000-0001-9922-743X}{\includegraphics[height=1.7ex]{figures/orcid-logo.png}}\textsuperscript{\,14}},
\mbox{A.D.\;Marino\href{https://orcid.org/0000-0002-1709-538X}{\includegraphics[height=1.7ex]{figures/orcid-logo.png}}\textsuperscript{\,26}},
\mbox{H.-J.\;Mathes\href{https://orcid.org/0000-0002-0680-040X}{\includegraphics[height=1.7ex]{figures/orcid-logo.png}}\textsuperscript{\,5}},
\mbox{T.\;Matulewicz\href{https://orcid.org/0000-0003-2098-1216}{\includegraphics[height=1.7ex]{figures/orcid-logo.png}}\textsuperscript{\,19}},
\mbox{V.\;Matveev\href{https://orcid.org/0000-0002-2745-5908}{\includegraphics[height=1.7ex]{figures/orcid-logo.png}}\textsuperscript{\,22}},
\mbox{G.L.\;Melkumov\href{https://orcid.org/0009-0004-2074-6755}{\includegraphics[height=1.7ex]{figures/orcid-logo.png}}\textsuperscript{\,22}},
\mbox{A.\;Merzlaya\href{https://orcid.org/0000-0002-6553-2783}{\includegraphics[height=1.7ex]{figures/orcid-logo.png}}\textsuperscript{\,12}},
\mbox{{\L}.\;Mik\href{https://orcid.org/0000-0003-2712-6861}{\includegraphics[height=1.7ex]{figures/orcid-logo.png}}\textsuperscript{\,17}},
\mbox{A.\;Morawiec\href{https://orcid.org/0009-0001-9845-4005}{\includegraphics[height=1.7ex]{figures/orcid-logo.png}}\textsuperscript{\,16}},
\mbox{S.\;Morozov\href{https://orcid.org/0000-0002-6748-7277}{\includegraphics[height=1.7ex]{figures/orcid-logo.png}}\textsuperscript{\,22}},
\mbox{Y.\;Nagai\href{https://orcid.org/0000-0002-1792-5005}{\includegraphics[height=1.7ex]{figures/orcid-logo.png}}\textsuperscript{\,8}},
\mbox{T.\;Nakadaira\href{https://orcid.org/0000-0003-4327-7598}{\includegraphics[height=1.7ex]{figures/orcid-logo.png}}\textsuperscript{\,9}},
\mbox{M.\;Naskr\k{e}t\href{https://orcid.org/0000-0002-5634-6639}{\includegraphics[height=1.7ex]{figures/orcid-logo.png}}\textsuperscript{\,20}},
\mbox{S.\;Nishimori\href{https://orcid.org/~0000-0002-1820-0938}{\includegraphics[height=1.7ex]{figures/orcid-logo.png}}\textsuperscript{\,9}},
\mbox{V.\;Ozvenchuk\href{https://orcid.org/0000-0002-7821-7109}{\includegraphics[height=1.7ex]{figures/orcid-logo.png}}\textsuperscript{\,14}},
\mbox{O.\;Panova\href{https://orcid.org/0000-0001-5039-7788}{\includegraphics[height=1.7ex]{figures/orcid-logo.png}}\textsuperscript{\,13}},
\mbox{V.\;Paolone\href{https://orcid.org/0000-0003-2162-0957}{\includegraphics[height=1.7ex]{figures/orcid-logo.png}}\textsuperscript{\,28}},
\mbox{O.\;Petukhov\href{https://orcid.org/0000-0002-8872-8324}{\includegraphics[height=1.7ex]{figures/orcid-logo.png}}\textsuperscript{\,22}},
\mbox{I.\;Pidhurskyi\href{https://orcid.org/0000-0001-9916-9436}{\includegraphics[height=1.7ex]{figures/orcid-logo.png}}\textsuperscript{\,13,6}},
\mbox{R.\;P{\l}aneta\href{https://orcid.org/0000-0001-8007-8577}{\includegraphics[height=1.7ex]{figures/orcid-logo.png}}\textsuperscript{\,16}},
\mbox{P.\;Podlaski\href{https://orcid.org/0000-0002-0232-9841}{\includegraphics[height=1.7ex]{figures/orcid-logo.png}}\textsuperscript{\,19}},
\mbox{B.A.\;Popov\href{https://orcid.org/0000-0001-5416-9301}{\includegraphics[height=1.7ex]{figures/orcid-logo.png}}\textsuperscript{\,22,4}},
\mbox{B.\;P\'orfy\href{https://orcid.org/0000-0001-5724-9737}{\includegraphics[height=1.7ex]{figures/orcid-logo.png}}\textsuperscript{\,7,8}},
\mbox{M.\;Posiada{\l}a-Zezula\href{https://orcid.org/0000-0002-5154-5348}{\includegraphics[height=1.7ex]{figures/orcid-logo.png}}\textsuperscript{\,19}},
\mbox{D.S.\;Prokhorova\href{https://orcid.org/0000-0003-3726-9196}{\includegraphics[height=1.7ex]{figures/orcid-logo.png}}\textsuperscript{\,22}},
\mbox{D.\;Pszczel\href{https://orcid.org/0000-0002-4697-6688}{\includegraphics[height=1.7ex]{figures/orcid-logo.png}}\textsuperscript{\,15}},
\mbox{S.\;Pu{\l}awski\href{https://orcid.org/0000-0003-1982-2787}{\includegraphics[height=1.7ex]{figures/orcid-logo.png}}\textsuperscript{\,18}},
\mbox{J.\;Puzovi\'c\textsuperscript{\,23}\textsuperscript{\dag}},
\mbox{R.\;Renfordt\href{https://orcid.org/0000-0002-5633-104X}{\includegraphics[height=1.7ex]{figures/orcid-logo.png}}\textsuperscript{\,18}},
\mbox{L.\;Ren\href{https://orcid.org/0000-0003-1709-7673}{\includegraphics[height=1.7ex]{figures/orcid-logo.png}}\textsuperscript{\,26}},
\mbox{V.Z.\;Reyna~Ortiz\href{https://orcid.org/0000-0002-7026-8198}{\includegraphics[height=1.7ex]{figures/orcid-logo.png}}\textsuperscript{\,13}},
\mbox{D.\;R\"ohrich\textsuperscript{\,11}},
\mbox{E.\;Rondio\href{https://orcid.org/0000-0002-2607-4820}{\includegraphics[height=1.7ex]{figures/orcid-logo.png}}\textsuperscript{\,15}},
\mbox{M.\;Roth\href{https://orcid.org/0000-0003-1281-4477}{\includegraphics[height=1.7ex]{figures/orcid-logo.png}}\textsuperscript{\,5}},
\mbox{{\L}.\;Rozp{\l}ochowski\href{https://orcid.org/0000-0003-3680-6738}{\includegraphics[height=1.7ex]{figures/orcid-logo.png}}\textsuperscript{\,14}},
\mbox{B.T.\;Rumberger\href{https://orcid.org/0000-0002-4867-945X}{\includegraphics[height=1.7ex]{figures/orcid-logo.png}}\textsuperscript{\,26}},
\mbox{M.\;Rumyantsev\href{https://orcid.org/0000-0001-8233-2030}{\includegraphics[height=1.7ex]{figures/orcid-logo.png}}\textsuperscript{\,22}},
\mbox{A.\;Rustamov\href{https://orcid.org/0000-0001-8678-6400}{\includegraphics[height=1.7ex]{figures/orcid-logo.png}}\textsuperscript{\,1,6}},
\mbox{M.\;Rybczynski\href{https://orcid.org/0000-0002-3638-3766}{\includegraphics[height=1.7ex]{figures/orcid-logo.png}}\textsuperscript{\,13}},
\mbox{A.\;Rybicki\href{https://orcid.org/0000-0003-3076-0505}{\includegraphics[height=1.7ex]{figures/orcid-logo.png}}\textsuperscript{\,14}},
\mbox{K.\;Sakashita\href{https://orcid.org/0000-0003-2602-7837}{\includegraphics[height=1.7ex]{figures/orcid-logo.png}}\textsuperscript{\,9}},
\mbox{K.\;Schmidt\href{https://orcid.org/0000-0002-0903-5790}{\includegraphics[height=1.7ex]{figures/orcid-logo.png}}\textsuperscript{\,18}},
\mbox{A.Yu.\;Seryakov\href{https://orcid.org/0000-0002-5759-5485}{\includegraphics[height=1.7ex]{figures/orcid-logo.png}}\textsuperscript{\,22}},
\mbox{P.\;Seyboth\href{https://orcid.org/0000-0002-4821-6105}{\includegraphics[height=1.7ex]{figures/orcid-logo.png}}\textsuperscript{\,13}},
\mbox{U.A.\;Shah\href{https://orcid.org/0000-0002-9315-1304}{\includegraphics[height=1.7ex]{figures/orcid-logo.png}}\textsuperscript{\,13}},
\mbox{Y.\;Shiraishi\textsuperscript{\,10}},
\mbox{A.\;Shukla\href{https://orcid.org/0000-0003-3839-7229}{\includegraphics[height=1.7ex]{figures/orcid-logo.png}}\textsuperscript{\,27}},
\mbox{M.\;S{\l}odkowski\href{https://orcid.org/0000-0003-0463-2753}{\includegraphics[height=1.7ex]{figures/orcid-logo.png}}\textsuperscript{\,21}},
\mbox{P.\;Staszel\href{https://orcid.org/0000-0003-4002-1626}{\includegraphics[height=1.7ex]{figures/orcid-logo.png}}\textsuperscript{\,16}},
\mbox{G.\;Stefanek\href{https://orcid.org/0000-0001-6656-9177}{\includegraphics[height=1.7ex]{figures/orcid-logo.png}}\textsuperscript{\,13}},
\mbox{J.\;Stepaniak\href{https://orcid.org/0000-0003-2064-9870}{\includegraphics[height=1.7ex]{figures/orcid-logo.png}}\textsuperscript{\,15}},
\mbox{M.\;Strikhanov\textsuperscript{\,22}},
\mbox{H.\;Str\"obele\textsuperscript{\,6}},
\mbox{T.\;\v{S}u\v{s}a\href{https://orcid.org/0000-0001-7430-2552}{\includegraphics[height=1.7ex]{figures/orcid-logo.png}}\textsuperscript{\,3}},
\mbox{L.\;Swiderski\href{https://orcid.org/0000-0001-5857-2085}{\includegraphics[height=1.7ex]{figures/orcid-logo.png}}\textsuperscript{\,15}},
\mbox{J.\;Szewi\'nski\href{https://orcid.org/0000-0003-2981-9303}{\includegraphics[height=1.7ex]{figures/orcid-logo.png}}\textsuperscript{\,15}},
\mbox{R.\;Szukiewicz\href{https://orcid.org/0000-0002-1291-4040}{\includegraphics[height=1.7ex]{figures/orcid-logo.png}}\textsuperscript{\,20}},
\mbox{A.\;Taranenko\href{https://orcid.org/0000-0003-1737-4474}{\includegraphics[height=1.7ex]{figures/orcid-logo.png}}\textsuperscript{\,22}},
\mbox{A.\;Tefelska\href{https://orcid.org/0000-0002-6069-4273}{\includegraphics[height=1.7ex]{figures/orcid-logo.png}}\textsuperscript{\,21}},
\mbox{D.\;Tefelski\href{https://orcid.org/0000-0003-0802-2290}{\includegraphics[height=1.7ex]{figures/orcid-logo.png}}\textsuperscript{\,21}},
\mbox{V.\;Tereshchenko\textsuperscript{\,22}},
\mbox{A.\;Toia\href{https://orcid.org/0000-0001-9567-3360}{\includegraphics[height=1.7ex]{figures/orcid-logo.png}}\textsuperscript{\,6}},
\mbox{R.\;Tsenov\href{https://orcid.org/0000-0002-1330-8640}{\includegraphics[height=1.7ex]{figures/orcid-logo.png}}\textsuperscript{\,2}},
\mbox{L.\;Turko\href{https://orcid.org/0000-0002-5474-8650}{\includegraphics[height=1.7ex]{figures/orcid-logo.png}}\textsuperscript{\,20}},
\mbox{T.S.\;Tveter\href{https://orcid.org/0009-0003-7140-8644}{\includegraphics[height=1.7ex]{figures/orcid-logo.png}}\textsuperscript{\,12}},
\mbox{M.\;Unger\href{https://orcid.org/0000-0002-7651-0272~}{\includegraphics[height=1.7ex]{figures/orcid-logo.png}}\textsuperscript{\,5}},
\mbox{M.\;Urbaniak\href{https://orcid.org/0000-0002-9768-030X}{\includegraphics[height=1.7ex]{figures/orcid-logo.png}}\textsuperscript{\,18}},
\mbox{F.F.\;Valiev\href{https://orcid.org/0000-0001-5130-5603}{\includegraphics[height=1.7ex]{figures/orcid-logo.png}}\textsuperscript{\,22}},
\mbox{D.\;Veberi\v{c}\href{https://orcid.org/0000-0003-2683-1526}{\includegraphics[height=1.7ex]{figures/orcid-logo.png}}\textsuperscript{\,5}},
\mbox{V.V.\;Vechernin\href{https://orcid.org/0000-0003-1458-8055}{\includegraphics[height=1.7ex]{figures/orcid-logo.png}}\textsuperscript{\,22}},
\mbox{V.\;Volkov\href{https://orcid.org/0000-0002-4785-7517}{\includegraphics[height=1.7ex]{figures/orcid-logo.png}}\textsuperscript{\,22}},
\mbox{A.\;Wickremasinghe\href{https://orcid.org/0000-0002-5325-0455}{\includegraphics[height=1.7ex]{figures/orcid-logo.png}}\textsuperscript{\,24}},
\mbox{K.\;W\'ojcik\href{https://orcid.org/0000-0002-8315-9281}{\includegraphics[height=1.7ex]{figures/orcid-logo.png}}\textsuperscript{\,18}},
\mbox{O.\;Wyszy\'nski\href{https://orcid.org/0000-0002-6652-0450}{\includegraphics[height=1.7ex]{figures/orcid-logo.png}}\textsuperscript{\,13}},
\mbox{A.\;Zaitsev\href{https://orcid.org/0000-0003-4711-9925}{\includegraphics[height=1.7ex]{figures/orcid-logo.png}}\textsuperscript{\,22}},
\mbox{E.D.\;Zimmerman\href{https://orcid.org/0000-0002-6394-6659}{\includegraphics[height=1.7ex]{figures/orcid-logo.png}}\textsuperscript{\,26}},
\mbox{A.\;Zviagina\href{https://orcid.org/0009-0007-5211-6493}{\includegraphics[height=1.7ex]{figures/orcid-logo.png}}\textsuperscript{\,22}}, and
\mbox{R.\;Zwaska\href{https://orcid.org/0000-0002-4889-5988}{\includegraphics[height=1.7ex]{figures/orcid-logo.png}}\textsuperscript{\,24}}
\\\rule{2cm}{.5pt}\\[-.5ex]\textit{\textsuperscript{\dag} \footnotesize deceased}


\noindent
\textsuperscript{1}~National Nuclear Research Center, Baku, Azerbaijan\\
\textsuperscript{2}~Faculty of Physics, University of Sofia, Sofia, Bulgaria\\
\textsuperscript{3}~Ru{\dj}er Bo\v{s}kovi\'c Institute, Zagreb, Croatia\\
\textsuperscript{4}~LPNHE, University of Paris VI and VII, Paris, France\\
\textsuperscript{5}~Karlsruhe Institute of Technology, Karlsruhe, Germany\\
\textsuperscript{6}~University of Frankfurt, Frankfurt, Germany\\
\textsuperscript{7}~Wigner Research Centre for Physics, Budapest, Hungary\\
\textsuperscript{8}~E\"otv\"os Lor\'and University, Budapest, Hungary\\
\textsuperscript{9}~Institute for Particle and Nuclear Studies, Tsukuba, Japan\\
\textsuperscript{10}~Okayama University, Japan\\
\textsuperscript{11}~University of Bergen, Bergen, Norway\\
\textsuperscript{12}~University of Oslo, Oslo, Norway\\
\textsuperscript{13}~Jan Kochanowski University, Kielce, Poland\\
\textsuperscript{14}~Institute of Nuclear Physics, Polish Academy of Sciences, Cracow, Poland\\
\textsuperscript{15}~National Centre for Nuclear Research, Warsaw, Poland\\
\textsuperscript{16}~Jagiellonian University, Cracow, Poland\\
\textsuperscript{17}~AGH - University of Science and Technology, Cracow, Poland\\
\textsuperscript{18}~University of Silesia, Katowice, Poland\\
\textsuperscript{19}~University of Warsaw, Warsaw, Poland\\
\textsuperscript{20}~University of Wroc{\l}aw,  Wroc{\l}aw, Poland\\
\textsuperscript{21}~Warsaw University of Technology, Warsaw, Poland\\
\textsuperscript{22}~Affiliated with an institution covered by a cooperation agreement with CERN\\
\textsuperscript{23}~University of Belgrade, Belgrade, Serbia\\
\textsuperscript{24}~Fermilab, Batavia, USA\\
\textsuperscript{25}~University of Notre Dame, Notre Dame, USA\\
\textsuperscript{26}~University of Colorado, Boulder, USA\\
\textsuperscript{27}~University of Hawaii at Manoa, Honolulu, USA\\
\textsuperscript{28}~University of Pittsburgh, Pittsburgh, USA\\


\section{Introduction}

This paper presents experimental results on inclusive spectra and mean multiplicities of
 $\pi^\pm, K^\pm, p$ and $\bar{p}$  produced in the 10\% most central $^{40}$Ar+$^{45}$Sc collisions at beam momenta of 13$A$, 19$A$, 30$A$, 40$A$, 75$A$, and 150\AGeVc (\snn = 5.12, 6.12, 7.62, 8.77, 11.9 and 16.8 GeV).
These studies form a part of the strong interactions program of \NASixtyOne~\cite{Antoniou:2006mh} at the CERN SPS investigating the properties of the onset of deconfinement and searching for the possible existence of a critical point. The program is mainly motivated by the observed rapid changes in hadron production properties in central Pb+Pb collisions at about 30\AGeVc by the NA49 experiment~\cite{Afanasiev:2002mx, Alt:2007aa}. These findings were interpreted as the onset of deconfinement; they were confirmed by the RHIC beam energy program~\cite{Adamczyk:2017iwn} and their interpretation is supported by the LHC results (see Ref.~\cite{Rustamov:2012np} and references therein). 

The goals of the \NASixtyOne strong interaction program are pursued experimentally by a two-dimensional scan in collision energy and size of colliding nuclei.  This allows us to systematically explore the phase diagram of strongly interacting matter~\cite{Antoniou:2006mh}. In particular, the analysis of the existing data within the framework of statistical models suggests that by increasing collision energy one increases the temperature and decreases the baryon chemical potential of the fireball of strongly interacting matter at kinetic freeze-out~\cite{Becattini:2005xt}, whereas by increasing the nuclear mass of the colliding nuclei the temperature decreases~\cite{NA49:2007fqa, Becattini:2005xt, Gazdzicki:2014sva, Vovchenko:2015idt}.

Within this program \NASixtyOne recorded data on \pp, Be+Be, Ar+Sc, Xe+La, and Pb+Pb collisions during 2009-2018 running. Further high-statistics measurements of Pb+Pb collisions with an upgraded detector started in 2022~\cite{PbAddendum, UpgradeAddendum22}. Comprehensive results on particle spectra and multiplicities have already been published for \pp interactions~\cite{Abgrall:2013pp_pim,Aduszkiewicz:2017sei,Aduszkiewicz:2019zsv}  and Be+Be collisions~\cite{NA61SHINE:2020ggt, NA61SHINE:2020czq} at 19$A$-150\AGeVc (20-158~\GeVc for \pp). For Ar+Sc collisions, only results on $\pi^-$ production were published up to now \cite{NA61SHINE:2021nye}.

The Ar+Sc collisions became crucial for the \NASixtyOne scan program. As the results obtained for the Be+Be system closely resemble inelastic \pp interactions, the collisions of Ar+Sc are the lightest of the studied systems for which a significant increase in the $K^+/\pi^+$ ratio was observed. The properties of measured spectra and multiplicities indicate that the Ar+Sc system is on a boundary between light (\pp, Be+Be) and heavy (Pb+Pb) systems.

The paper is organized as follows. After this introduction, the experiment is briefly presented in Sec.~\ref{sec:exp_description}. The analysis procedure, as well as statistical and systematic uncertainties, are discussed in Sec.~\ref{sec:analysis}. Section~\ref{sec:results} presents experimental results and compares them with measurements of \NASixtyOne in inelastic \pp interactions~\cite{Abgrall:2013pp_pim,Aduszkiewicz:2017sei,Aduszkiewicz:2019zsv} and central Be+Be~\cite{NA61SHINE:2020ggt,NA61SHINE:2020czq} collisions, as well as NA49's results on Pb+Pb, C+C and Si+Si reactions~\cite{Afanasiev:2002mx,Alt:2007aa}. Section~\ref{sec:discussion} discusses model predictions. A summary in Section~\ref{sec:summary} closes the paper. Additionally, Appendix \ref{sec:additional_plots}, containing plots presenting details of the analysis is included.

The following variables and definitions are used in this paper. The particle rapidity $y$ is calculated in the collision center of mass system (cms),
$y=0.5 \cdot \ln{[(E+p_{L}c)/(E-p_{L}c)]}$,
assuming proton mass, where $E$ and $p_{L}$ are the particle energy and longitudinal momentum, respectively. The transverse component of the momentum is denoted as \pt and the transverse mass $m_{T}$ is defined as
$m_{T} = \sqrt{m^2 + (c\pt)^2}$
where $m$ is the particle mass in \GeV. The momentum in the laboratory frame is denoted \p and the collision energy per nucleon pair in the center of mass by \snn.

The Ar+Sc collisions are selected by requiring a low value of the forward energy -- the energy emitted into the region populated by projectile spectators. These collisions are referred to as \textit{central} collisions and a selection of collisions based on the forward energy is called a \textit{centrality selection}. The term \textit{central} is written in italics throughout this paper to denote the specific event selection procedure based on measurements of the forward energy.

\FloatBarrier
\section{Experimental setup}
\label{sec:exp_description}

\begin{figure}[h] 
        \centering
        {\small
        \includegraphics[width = 1\linewidth]{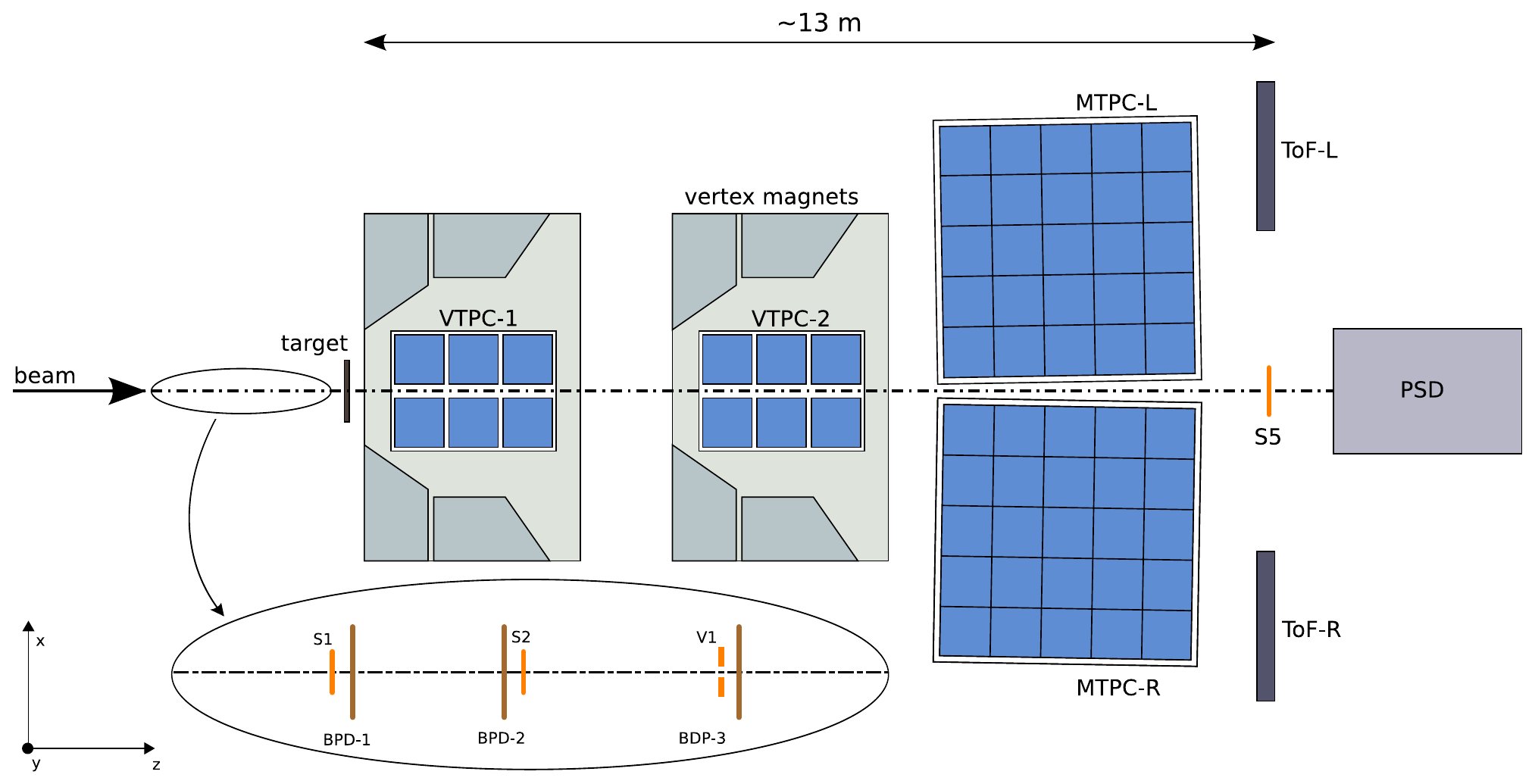}}
        \caption{The schematic layout of the \NASixtyOne experiment at the CERN SPS ~\cite{Abgrall:2014fa} showing the components used for the Ar+Sc energy scan. The beam instrumentation is sketched in the inset. The alignment of the chosen coordinate system as shown in the figure. The nominal beam direction is along the $z$-axis. The magnetic field bends charged particle trajectories in the $x$-$z$ (horizontal) plane. The drift direction in the TPCs is along the $y$ (vertical) axis.}
        \label{fig:detectorSetup}
\end{figure}

The \NASixtyOne experiment is a multi-purpose facility designed to measure particle production in
nucleus-nucleus, hadron-nucleus and \pp interactions~\cite{Abgrall:2014fa}. The detector
is situated at the CERN Super Proton Synchrotron (SPS) in the H2 beamline of the North experimental area.
A schematic diagram of the setup is shown in Fig.~\ref{fig:detectorSetup}.
The main components of the particle detection system used in the 2015 Ar+Sc data-taking campaign are four large-volume
Time Projection Chambers (TPC). Two of them, called Vertex TPCs (VTPC), are located
downstream of the target inside superconducting magnets with a maximum combined bending power of 9~Tm.
The magnetic field was scaled down in proportion to the beam momentum in order to obtain similar $y-\pt$ acceptance at all beam momenta. The main TPCs (MTPC) and two walls of pixel Time-of-Flight (ToF-L/R) detectors are placed symmetrically on either side of the beamline downstream of the magnets. The TPCs are filled with Ar:CO$_{2}$ gas mixtures in proportions 90:10 for the VTPCs and 95:5 for the MTPCs. The Projectile Spectator Detector (PSD) is positioned 20.5 m (16.7 m) downstream of the MTPCs at beam momenta of 75$A$ and 150\AGeVc (13$A$, 19$A$, 30$A$, 40\AGeVc), centered in the transverse plane on the deflected position of the beam. A degrader in the form of a 5 cm diameter brass cylinder was placed in front of the center of the PSD in order to reduce electronic saturation effects and shower leakage from the downstream side. The length of the cylinder was 10 cm except the 19\AGeVc measurements, when the length was 5 cm. No degrader was used at 13\AGeVc.

Primary beams of fully ionized ${}^{40}$Ar nuclei were extracted from the SPS accelerator at beam momenta of 13$A$, 19$A$, 30$A$, 40$A$, 75$A$ and 150\AGeVc.
Two scintillation counters, S1 and S2, provide beam trigger definition, together with a veto counter V1 with a 1~cm diameter hole, which defines the beam before the target. The S1 counter also provides the timing reference (start time for all counters). Beam particles are selected by the trigger system requiring the coincidence $\textrm{S1} \wedge\textrm{S2} \wedge\overline{\textrm{V1}}$. Individual beam particles are precisely measured by the three Beam Position Detectors (BPDs) placed upstream of the target~\cite{Abgrall:2014fa}. Collimators in the beam line were adjusted to obtain beam rates of the order of $10^4$/s during the $10.4$ s spill within a $32.4$ s accelerator super cycle.

The target was a stack of six Sc plates of 1 mm thickness and 2 x 2 cm$^2$ area placed 75 cm upstream of VTPC-1. Mass concentrations of impurities were measured at 0.3~\% resulting in an estimated increase of the produced pion multiplicity by less than 0.2~\% due to the small admixture of heavier elements~\cite{Banas:2018sak}. No correction was applied for this negligible contamination. Data were taken with target inserted (93\%) and target removed (7\%). 

Interactions in the target are selected with the trigger system by requiring an incoming Ar ion and a signal below that of beam ions from S5, a small 2~cm diameter scintillation counter placed on the beam trajectory behind the MTPCs. This minimum bias trigger selects inelastic collisions of the beam ion with the target and with matter between the target and S5. In addition, central collisions were selected by requiring an energy signal below a threshold set on the summed signal from the 16 central modules of the PSD, which measure mainly the energy of projectile spectators. The cut was set to retain only the event triggers with roughly 30\% smallest energies in the PSD, which was studied quantitatively in offline analysis. The central event trigger condition thus was $\textrm{S1} \wedge\textrm{S2} \wedge\overline{\textrm{V1}}\wedge\overline{\textrm{S5}}\wedge\overline{\textrm{PSD}}$. The statistics of recorded events are summarized in Table~\ref{tab:statbeam}.

\begin{table}[h]
	\centering
	\begin{tabular}{c||c|c|c|c}
	\parbox[][2cm][c]{2.5cm}{\centering $p_{\textrm{beam}}$ \\  {(GeV/$c$)}} & \parbox[][1.5cm][c]{2cm}{\centering \snn \\(GeV)}& \parbox[][1.5cm][c]{2cm}{\centering total recorded events} & \parbox[][2.5cm][c]{3cm}{\centering ~~central \newline trigger events \newline (good events only)} & \parbox[][1.5cm][c]{2cm}{\centering ~~0-10\% \newline most central after cuts}\\[-0.2cm]
	\hline \hline
	13$A$  & 5.12 & 3.9 $\cdot 10^6$ & 2.95 $\cdot 10^6$ & 5.22 $\cdot 10^5$\\
	19$A$  & 6.12 & 4.2 $\cdot 10^6$ & 2.97 $\cdot 10^6$ & 5.35 $\cdot 10^5$\\
	30$A$  & 7.62 & 5.2 $\cdot 10^6$ & 4.09 $\cdot 10^6$ & 9.14 $\cdot 10^5$\\
	40$A$  & 8.77 & 1.0 $\cdot 10^7$ & 6.36 $\cdot 10^6$ & 1.28 $\cdot 10^6$\\
	75$A$  & 11.94~~ & 4.8 $\cdot 10^6$ & 3.08 $\cdot 10^6$ & 1.14 $\cdot 10^6$\\
	150$A$~ & 16.83~~ & 3.7 $\cdot 10^6$ & 1.54 $\cdot 10^6$ & 5.56 $\cdot 10^5$\\
	\end{tabular}
   \caption{
Basic beam properties and the number of events recorded and used in the analysis of 10\% most \textit{central} Ar+Sc interactions.
Event selection criteria will be discussed in Sec. \ref{sec:analysis}.}
	\label{tab:statbeam}
\end{table}
\normalsize

\FloatBarrier
\section{Analysis procedure}
\label{sec:analysis}
This section starts with a brief overview of the data analysis procedure and the corrections applied to the experimental results.
It also defines to which species of particles the final results correspond.
A description of the detector calibration and the track and vertex reconstruction procedures can be found in
Ref.~\cite{Abgrall:2013pp_pim}.

The analysis procedure consists of the following steps:
\begin{enumerate}[(i)]

  \item application of event and track selection criteria,
  \item determination of raw spectra of identified charged hadrons
        using the selected events and tracks,
  \item evaluation of corrections to the raw spectra based on
        experimental data and simulations,
  \item calculation of the corrected spectra and mean multiplicities,
  \item calculation of statistical and systematic uncertainties.

\end{enumerate}

Corrections for the following biases were evaluated:
\begin{enumerate}[(a)]
 \item contribution from off-target interactions,
 \item losses of in-target interactions due to the event selection criteria,
 \item geometric acceptance,
 \item reconstruction and detector inefficiencies,
 \item losses of tracks due to track selection criteria,
 \item contribution of particles other than \emph{primary} (see below)
       charged particles produced in Ar+Sc collisions,
 \item losses of primary charged particles
       due to their decays and secondary interactions.
\end{enumerate}

Correction (a) was found to be negligible ($\mathcal{O}(10^{-4})$) and was therefore not applied.

Corrections (b)-(g) were estimated by data and simulations. MC events were generated with the \EposLong model (version CRMC 1.5.3)~\cite{Werner:2005jf}, passed through detector simulation employing the \Geant 3.21 package~\cite{Geant3} and then reconstructed using standard procedures, exactly matching the ones used in the processing of experimental data. The selection of central events in the simulation was based on the number of projectile spectator nucleons available in the \Epos model.

The final results refer to particles produced in \textit{central} Ar+Sc collisions by strong and electromagnetic processes. Such hadrons are referred to as \emph{primary} hadrons. The definition of \textit{central} collisions is given in Sec. \ref{sec:centrality}.

\subsection{\textit{Central} collisions}
\label{sec:centrality}

A short description of the procedure for defining \textit{central} collisions is given below.
For more details, see Refs.~\cite{Acharya:2020cyb,Kaptur:2017,NA61SHINE:2021nye}.

The final results presented in this paper refer to the 10\% of Ar+Sc collisions with the lowest value of the forward energy $E_\text{F}$ (\textit{central} collisions). The quantity $E_\text{F}$ is defined as the sum of energies (measured in the laboratory reference frame) of all particles produced in Ar+Sc collisions via strong and electromagnetic processes in the forward momentum region defined by the acceptance map in Ref.~\cite{PSD_acceptance}. The forward region defined by the acceptance map roughly corresponds to polar angles (angle between beam momentum and secondary particle momentum vectors in LAB frame of reference) $\theta$ smaller than $\theta_\text{max} = 0.1-0.2$ (charged particles) and $\theta_\text{max} = 0.03-0.05$ (neutral particles).  The final results on \textit{central} collisions, derived using this procedure, allow a precise comparison with model predictions without additional information about the \NASixtyOne setup and used magnetic field. Using this definition, the mean number of wounded nucleons $\langle W \rangle$ was calculated in the Wounded Nucleon Model (WNM)~\cite{Bialas:1976ed} implemented in \Epos \cite{Pierog:2018}.

\begin{figure}[!htbp]
       \centering 
       \includegraphics[width=0.245\textwidth]{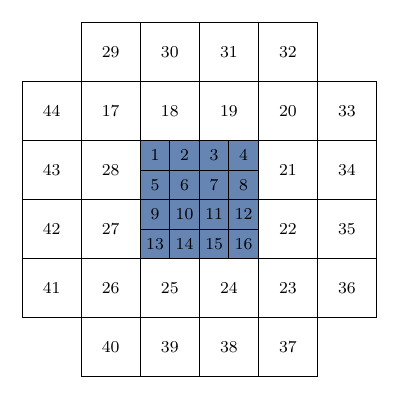}
       \includegraphics[width=0.245\textwidth]{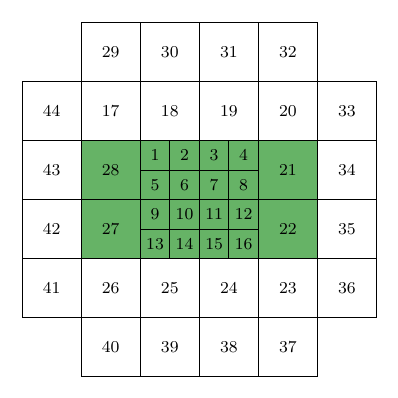}
       \includegraphics[width=0.245\textwidth]{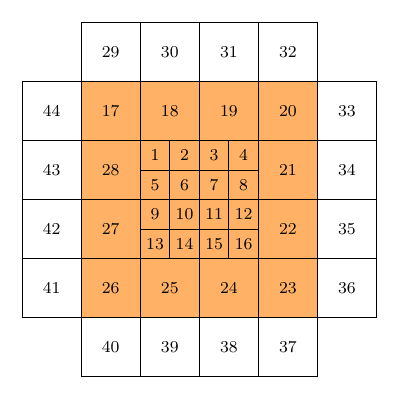}
       \includegraphics[width=0.245\textwidth]{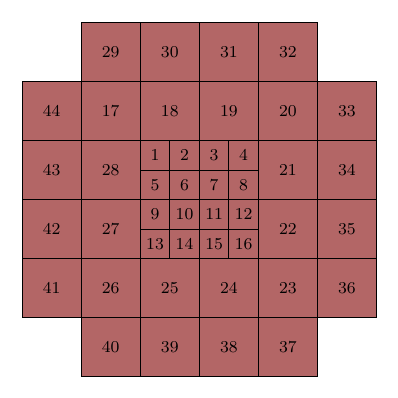}
       {	\centering
	    \begin{tabular}{p{0.5cm}p{3.5cm}p{3cm}p{4cm}p{4cm}}
	    & Central trigger & 150\AGeVc & 30$A$, 40$A$, 75\AGeVc & 13$A$, 19\AGeVc
	    \end{tabular}
	   }

       \caption{PSD modules used in the online and offline event selection. The online central trigger is derived from the energy in the central 16 modules, while the set of modules used to determine the offline PSD energy $E_\text{PSD}$ changes with beam momentum.}
\label{fig:PSDAllModuleSelections}
\end{figure}

For data analysis, the event selection was based on the 10\% of collisions with the lowest value of the energy $E_\text{PSD}$ measured by a subset of PSD modules (see Fig.~\ref{fig:PSDAllModuleSelections}) optimized for the sensitivity to projectile spectators. The acceptance in the definition of the forward energy $E_\text{F}$ corresponds closely to the acceptance of this subset of PSD modules at all energies \cite{NA61SHINE:2020ggt, NA61SHINE:2021nye}. 

\begin{figure}[!htbp]
	\centering
	\includegraphics[width=0.48\textwidth]{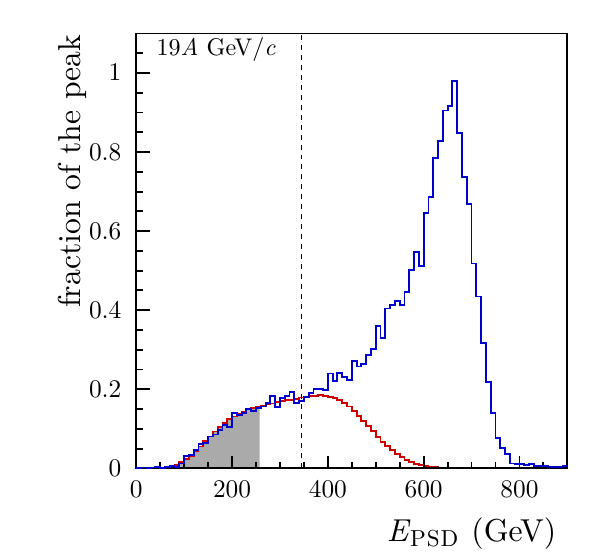}
	\includegraphics[width=0.48\textwidth]{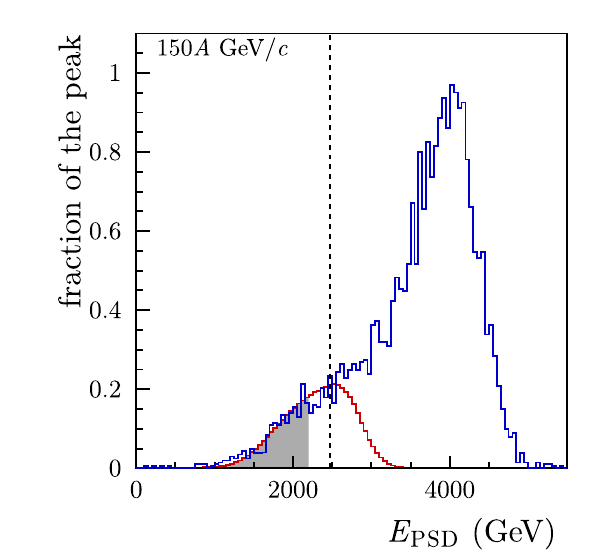}
	{\textcolor{kBlue+2}\solidLine} beam trigger \hspace{.5cm}\textcolor{kRed+1}{\solidLine} central trigger (scaled) \hspace{.5cm} \textcolor{gray}{\tiny\SquareSolid} 10\% centrality \hspace{.5cm} {\textcolor{kBlack}\dashedLine} Normalization region\\
\caption{Distributions of the energy $E_\text{PSD}$ measured by the PSD calorimeter for 19$A$ (left plot) and 150\AGeVc (right plot) beam momentum, for minimum-bias selected (blue histograms) and central trigger selected (red histograms) events. Histograms are normalized to agree in the overlap region (from the beginning of the distribution to the black dashed line). The on-line central trigger was set to accept approximately 30\% of most central inelastic events. The shaded area indicates 10\% collisions with the smallest $E_\text{PSD}$.}
\label{fig:PSDEnergy_cent}
\end{figure}

Online event selection by the central hardware trigger used a threshold on the sum of electronic signals from the 16 central modules of the PSD set to accept approx.~30\% of the inelastic interactions. Measured distributions of $E_\text{PSD}$ for minimum-bias and central trigger selected events, calculated in the offline analysis, are shown in Fig.~\ref{fig:PSDEnergy_cent} at beam momenta of 19$A$ and 150\AGeVc. The accepted region corresponding to the 10\% of most \textit{central} collisions is indicated by shading. The minimum-bias distribution was obtained using the data from the beam trigger with an offline selection of events by requiring an event vertex in the target region. A properly normalized spectrum for target-removed events was subtracted.

The forward energy $E_\text{F}$ cannot be measured directly. However, both $E_\text{F}$ and $E_\text{PSD}$ can be obtained from simulations using the \EposLong (version CRMC 1.5.3)~\cite{Werner:2005jf, Pierog:2009zt, Pierog:2018} model. A global factor $c_{\textrm{cent}}$ (listed in Table~\ref{tab:w}) was calculated as the ratio of mean negatively charged pion multiplicities obtained with the two selection procedures in the 10\% most \textit{central} events. A possible dependence of the scaling factor on rapidity and transverse momentum was neglected. The resulting factors $c_{\textrm{cent}}$ range between 1.002 and 1.005, corresponding to a correction at least an order of magnitude smaller compared to the systematic uncertainties of the measured particle multiplicities (see Sec. \ref{sec:systematic}). The correction was therefore not applied and its possible impact was neglected in the final uncertainty calculation.

\begin{figure}[!htbp]
  \centering
     \includegraphics[width=0.49\textwidth]{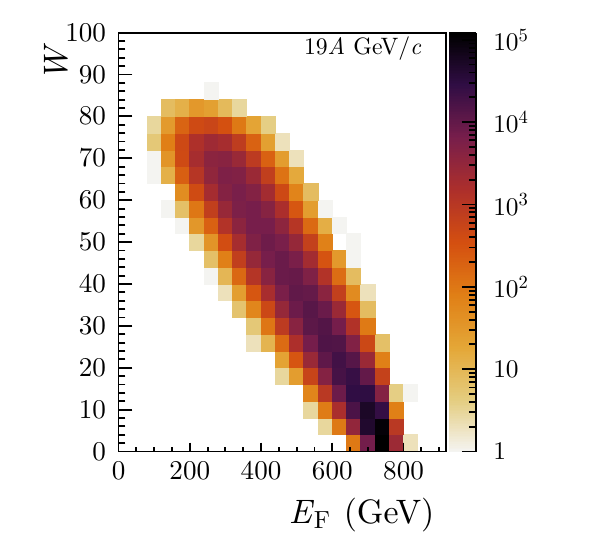}
     \includegraphics[width=0.49\textwidth]{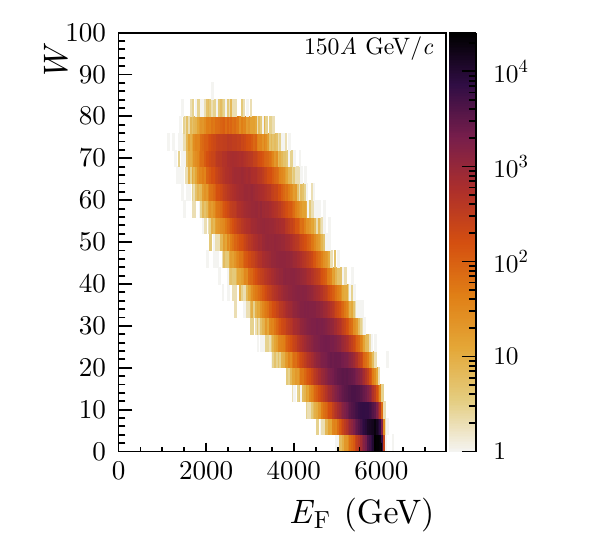}\\
  \caption{Distributions of number of wounded nucleons as a function of $E_F$ for all inelastic Ar+Sc collisions at 19$A$ and 150$A$ GeV/$c$ beam momentum calculated from \Epos \cite{Pierog:2018}.}
  \label{fig:woundedDistribution}
\end{figure}

Finally, the \Epos WNM~\cite{Pierog:2018} simulation was used to estimate the average number of wounded nucleons~$\langle W \rangle$ for the 10\% of events with the smallest number of spectator nucleons and with the smallest value of $E_\text{F}$. The average impact parameter $\langle b \rangle$ was also obtained for the latter selection. Results are listed in Table~\ref{tab:w}. Example distributions of $E_F-W$ for 19$A$ and 150\AGeVc beam momenta are shown in Fig.~\ref{fig:woundedDistribution}. These distributions are quite broad and emphasize the importance of proper simulation of the centrality selection when comparing model calculations with the experimental results.

\begin{table*}
 \centering
 \footnotesize
 \begin{tabular}{l||cccccc}
  $p_{\textrm{beam}}$ (\GeVc) & 13$A$ & 19$A$ & 30$A$ & 40$A$ & 75$A$ & 150$A$\\ 
  \hline \hline
  $\langle W \rangle$ in \Epos WNM & $61.4$ & $61.2$ & $60.9$ & $60.9$ & $60.8$ & $61.0$\\
  $\langle W \rangle$ in \Epos $E_\text{F}$ & $61.0$ & $60.8$ & $60.6$ & $60.4$ & $60.3$ & $60.6$\\
  $\langle b \rangle$ (fm) & $1.82$ & $1.95$ & $2.00$ & $2.09$ & $2.23$ & $2.08$\\
 \hline
  $c_{\textrm{cent}}$ & $1.005$ & $1.004$ & $1.002$ & $1.004$ & $1.005$ & $1.003$ \\
 \end{tabular}
 \caption{The average number of wounded nucleons $\langle W \rangle$ and the average impact parameter $\langle b\rangle$ in the 10\% most \textit{central} Ar+Sc collisions estimated from simulations using \Epos~\cite{Werner:2005jf, Pierog:2009zt, Pierog:2018}. Results from \Epos~WNM are for centrality selection using the smallest number of spectators, whereas the \Epos~$E_\text{F}$ results are obtained using the forward energy $E_\text{F}$ within the acceptance map in Ref.~\cite{PSD_acceptance}. The last line presents numerical values of the $c_\text{cent}$ factor for all the beam momenta (see text and Ref. \cite{NA61SHINE:2021nye} for more details).}
 \label{tab:w}
\end{table*}

\subsection{Event and track selection}\label{sec:cuts}

\subsubsection{Event selection}
\label{sec:event_selection}

For further analysis, Ar+Sc events were selected using the following criteria:
\begin{enumerate}[(i)]
    \item No off-time beam particle detected within a time window of $\pm$4$~\mu$s around the trigger particle and no other event trigger detected within a time window of $\pm$25$~\mu$s around the trigger particle, reducing pile-up events in the data sample.
    \item Beam particle detected in at least three planes out of four of BPD-1 and BPD-2 and in both planes of BPD-3, providing good reconstruction of the beam trajectory.
    \item A well-reconstructed interaction vertex with $z$-coordinate (fitted using the beam trajectory and TPC tracks) not farther away than 10~cm from the center of the Sc target.
    \item An upper limit on the measured energy $E_{PSD}$ selecting 10\% of all inelastic collisions.
\end{enumerate}
The event statistics after applying the selection criteria is summarized in Table~\ref{tab:statbeam}.

\subsubsection{Track selection}
\label{sec:track_selection}

To select tracks of primary charged hadrons and to reduce the contamination by particles from secondary interactions and weak decays, the following track selection criteria were applied:

\begin{enumerate}[(i)]
    \item Fitted $x$ component of particle rigidity $q \cdot \p$ should be positive. This selection minimizes the angle between the track trajectory and the TPC pad direction for the chosen magnetic field direction, reducing uncertainties of the reconstructed cluster position, energy deposition and track parameters.
    \item Total number of reconstructed points on the track should be greater than 30, ensuring good resolution of \dEdx measurement.
    \item Sum of the number of reconstructed points inside the vertex magnets (VTPC-1 and VTPC-2) should be greater than 15, which ensures good accuracy of track momentum fit.
    \item The distance between the track extrapolated to the interaction plane and the reconstructed vertex (track impact parameter) should be smaller than 4~cm in the horizontal (bending) plane and 2~cm in the vertical (drift) plane.
\end{enumerate}

In the case of \dEdx analysis, an additional criterion was used:
\begin{enumerate}[(i)]
\item track azimuthal angle $\phi$ should be within 30$^\circ$ with respect to the horizontal plane ($x$-$z$).
\end{enumerate}
Similarly, specifically for \tofdedx analysis, the following supplementary cuts were implemented:
\begin{enumerate}[(i)]
\item the extrapolated trajectory (as measured in the TPCs) reaches one of the ToF walls,
\item the last measured point on the track is located at least 70~cm upstream of the back wall of MTPCs, and its distance from the fitted track is within 4~cm,
\item measured flight time and charge for the pixel are of good quality (as in Refs.~\cite{Anticic:2011ny,Podlaski:2022}).
\end{enumerate}
\FloatBarrier
\subsection{Identification techniques}
\label{sec:identification}

\begin{figure}[!ht]
        \begin{center}
        \includegraphics[width=0.9\textwidth]{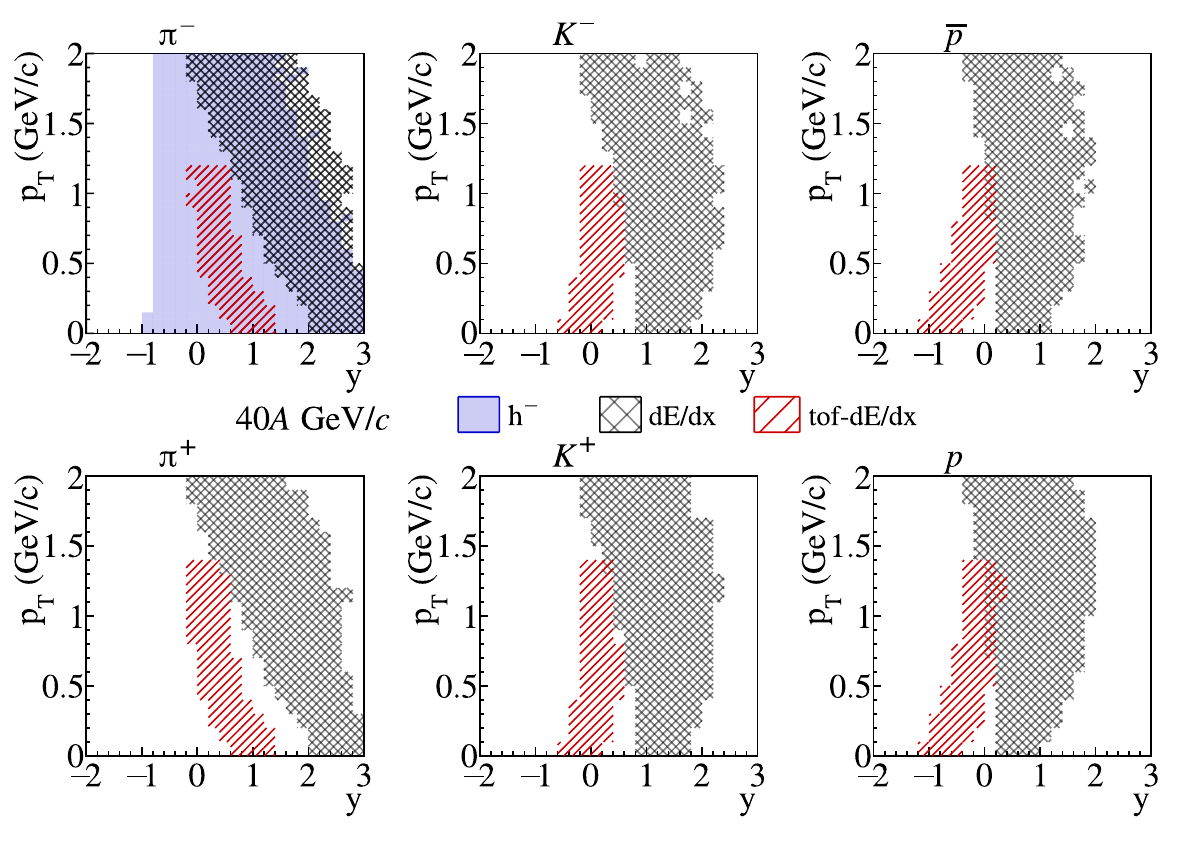}
        \end{center}
        \caption{Acceptance of the \tofdedx and \dEdx methods for identification of pions, kaons and protons in the 10\% most \textit{central} Ar+Sc interactions at 40\AGeVc. Negatively charged pion yield is also calculated using a so-called $h^-$ method (see Ref. \cite{NA61SHINE:2021nye}).
        }
        \label{fig:methodacc40}
\end{figure}

\begin{figure}[!ht]
        \begin{center}
        \includegraphics[width=0.9\textwidth]{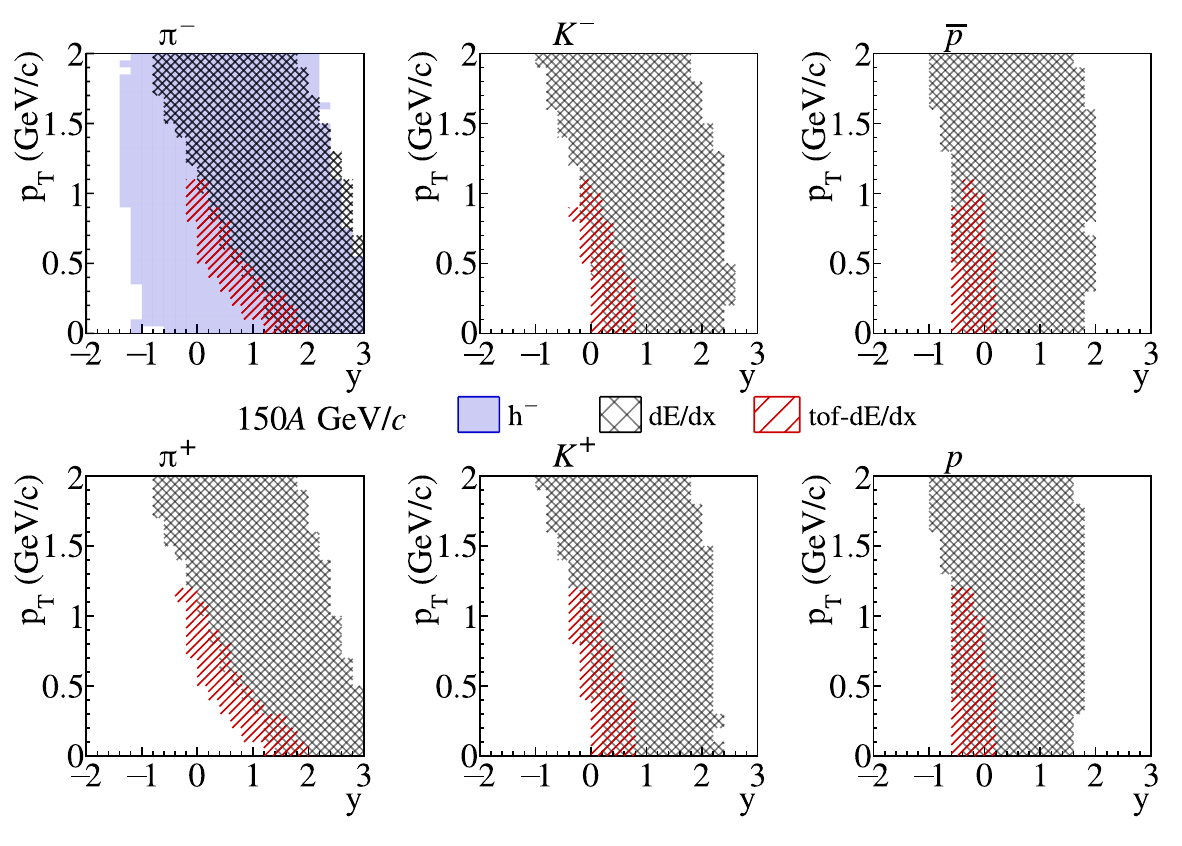}
        \end{center}
        \caption{Acceptance of the \tofdedx and \dEdx methods for identification of pions, kaons and protons in the 10\% most \textit{central} Ar+Sc interactions at 150\AGeVc. Negatively charged pion yield is also calculated using a so-called $h^-$ method \cite{NA61SHINE:2021nye}.
        }
        \label{fig:methodacc150}
\end{figure}

Charged particle identification in \NASixtyOne is based on the ionization energy loss, \dEdx, in the gas of the TPCs and the time of flight, \tof, obtained from the ToF-L and ToF-R walls. In the region of the relativistic rise of the ionization at large momenta, the measurement of \dEdx alone allows identification. At lower momenta, the \dEdx bands for different particle species overlap, and an additional measurement of \tof is used for unambiguous particle identification. These two methods allow covering most of the relevant space in rapidity and transverse momentum, in particular the mid-rapidity region of $K^+$ and $K^-$ spectra, which is an important part of the strong interaction program of \NASixtyOne. 
The acceptance of the two methods is shown in Figs.~\ref{fig:methodacc40} and~\ref{fig:methodacc150} for the 10\% most \textit{central} Ar+Sc interactions at 40$A$ and 150\AGeVc, respectively. The figures also display the $h^-$ analysis method \cite{NA61SHINE:2021nye}, which provides large-acceptance measurements of $\pi^-$ yields.
At low beam energies, the \tofdedx method extends the identification acceptance, while at the top SPS energy it overlaps with the \dEdx method.

\subsubsection{Identification based on energy loss measurement \textmd{\dEdx}}
\label{sec:dedx_id}

\begin{figure}[!ht]
        \begin{center}
       \includegraphics[width=\textwidth]{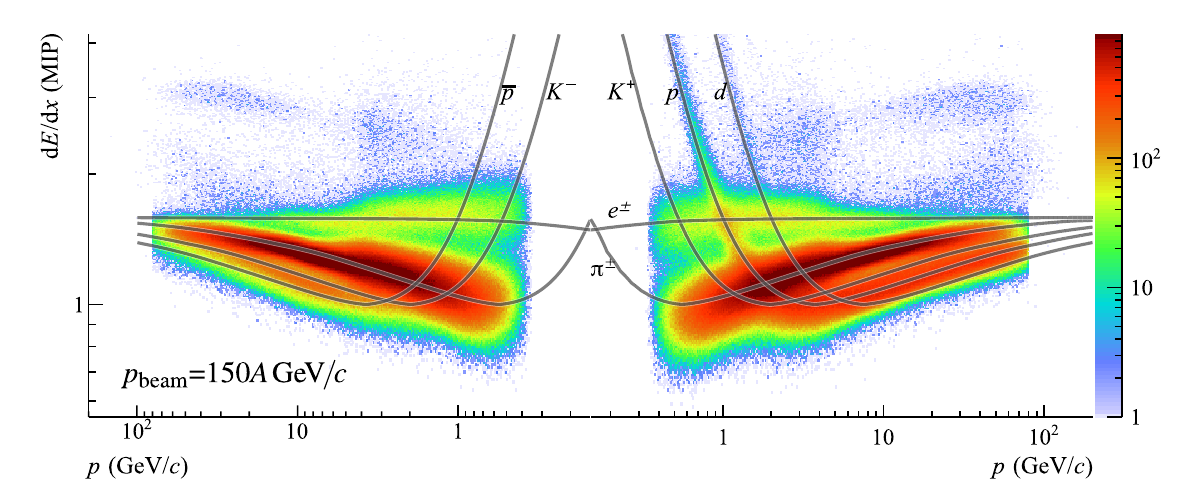}
        \end{center}
        \caption{
        Distribution of charged particles in the \dEdx -- \p~plane.
        The energy loss (in units of the minimum ionizing particle) in the TPCs is shown for different charged particles produced in Ar+Sc collisions at 150\AGeVc.
        Expectations for the dependence of the mean \dEdx on ~\p for the considered particle types are shown by the curves calculated based on the Bethe-Bloch function.
        }
        \label{fig:dedx}
\end{figure}

Time projection chambers provide measurements of energy loss \dEdx of charged particles 
in the chamber gas along their trajectories. Simultaneous measurements of \dEdx and \p~ 
allow extraction of information on particle mass. The mass assignment follows the procedure that was developed
for the analysis of \textit{p+p} reactions as described in Ref.~\cite{Aduszkiewicz:2017sei}.
Values of \dEdx are calculated as the truncated mean (smallest 50\%)
of ionization energy loss measurements along the track trajectory. As an example, \dEdx measured in
Ar+Sc interactions at 150\AGeVc~is presented in Fig.~\ref{fig:dedx}, for positively and negatively charged particles, as a function of momentum.

In the \dEdx method the contributions of $e^+, e^-, \pi^+, \pi^-, K^+, K^-, p,\,\bar{p} \textrm{~and~} d$ are obtained by fitting the \dEdx distributions in bins of laboratory momentum \p and transverse momentum \pt. The data are divided into 13 logarithmic bins in \p in the range 5--100~\GeVc and into linear bins in \pt. Thin binning in \pt is used up to $\pt=0.6$~\GeVc (bin width 0.05 GeV/\textit{c}) and wider bins are used above this value (0.1 GeV/\textit{c}). Due to the crossing of Bethe-Bloch curves at low momenta, the applicability of particle identification based solely on \dEdx measurement is limited to tracks with $\p>5$~\GeVc. Only bins with a total number of selected tracks greater than 100 were used in the further analysis.

\begin{figure}[ht]
        \begin{center}
        \includegraphics[width=0.45\textwidth]{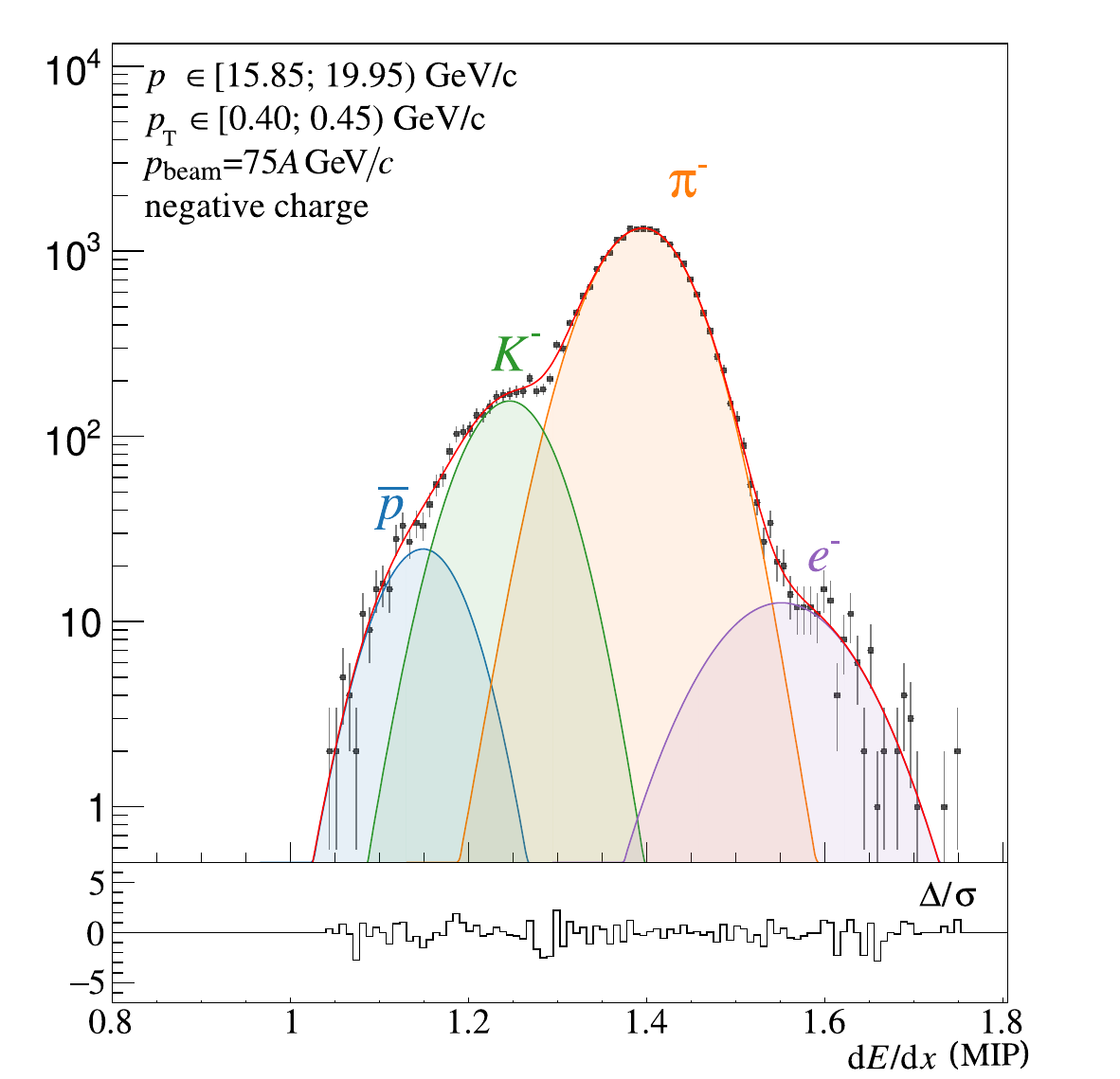}
        \includegraphics[width=0.45\textwidth]{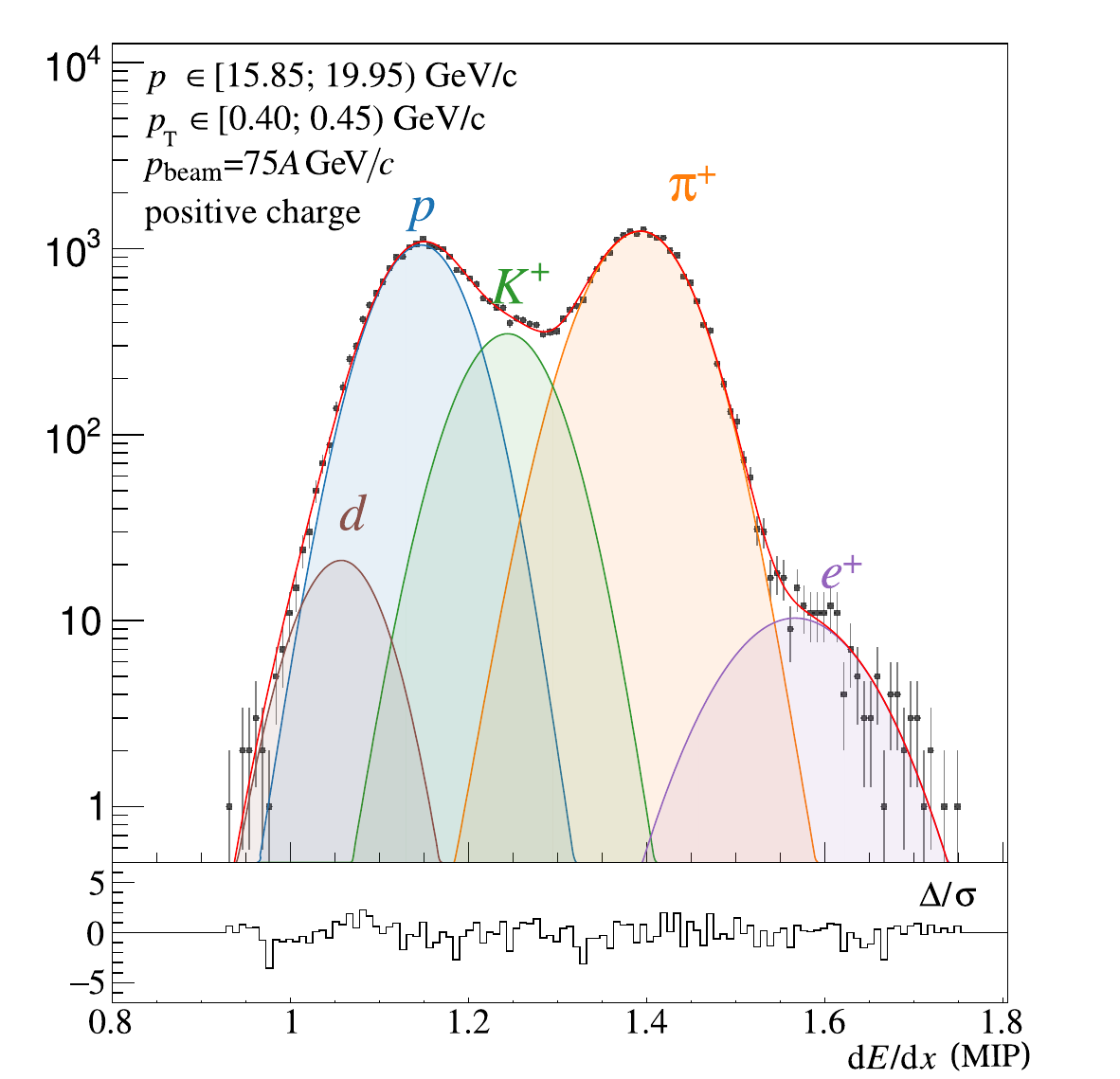}
        \end{center}
        \caption{
        The \dEdx distributions for negatively (\textit{left}) and positively (\textit{right}) charged
        particles in a selected $\p-\pt$ bin produced in \textit{central} Ar+Sc collisions at 75\AGeVc.
        The fits by a sum of contributions from different particle types are shown by solid lines.
        The corresponding residuals (the difference between the data and fit divided by the statistical uncertainty of the data) are shown in the bottom plots.
        }
        \label{fig:exfit}
\end{figure}

Due to the characteristic long tails in the distribution of charge deposited in a single ionization cluster, the mean \dEdx for a given track is calculated using 50\% of the lowest charge deposits. Such a truncation may introduce an asymmetry of the final \dEdx distribution, shift of the peak and it also affects its width.
Therefore, the \dEdx spectrum in each $p, p_T$ bin is fitted by the sum of asymmetric Gaussians with widths $\sigma_{i, l}$ depending on the particle type $i$ and the number of points $l$ measured in the TPCs (the method is based on previous work described in Refs.~\cite{vanLeeuwen:2003ru,note_MvL}): 
\begin{equation}
\rho(x) = \sum_{i=\pi^\pm,~K^\pm,p,\bar{p},d,e^\pm} N_i \frac{1}{\sum_l n_l}\sum_l \frac{n_l}{\sqrt{2\pi}\sigma_{i,l}}\exp\left[-\frac{1}{2}\left( \frac{x-x_i+\frac{2}{\sqrt{2\pi}} \delta_{l} \sigma_{i,l}}{(1 \pm \delta_{l})~\sigma_{i,l}} \right)^2\right]~,
\label{eq:model}
\end{equation}
where truncated mean energy loss \dEdx is denoted with $x$, the amplitude of the contribution of particles of type $i$ is expressed as $N_i$ and variable $n_l$ is the number of tracks with the number of points $l$ in the sample. The peak position of the \dEdx distribution for particle type $i$ is expressed as $x_i$ and the expression $2 \delta_{l} \sigma /\sqrt{2\pi}$ accounts for the drift of the peak related to the asymmetry of the distribution introduced with
the parameter $\delta_{l}=\delta_0/l$, which is taken with a negative sign if $x<x_i-\frac{2}{\sqrt{2\pi}} \delta_{l} \sigma_{i,l}$ and with a positive sign otherwise.  The width, $\sigma_{i,l}$ depends on the particle species and the track length in the following way:
\begin{equation}
\sigma_{i,l} = \frac{\sigma_0}{\sqrt{l}} \left( \frac{x_i}{x_{\pi}} \right)^{\alpha}~,
\label{eq:sigma}
\end{equation}
where $\sigma_0$ is common for all particle types and $\alpha$ is a universal constant. The details about the fitting procedure can be found in Ref. \cite{Lewicki:2020}. Examples of final fits are shown in Fig.~\ref{fig:exfit}.

\subsubsection{Identification based on time of flight and energy loss measurements
\textmd{(\tofdedx)}}
\label{sec:tof_id}
Identification of $\pi^{+}$, $\pi^{-}$, $K^{+}$, $K^{-}$, $p$ and $\bar{p}$ at low momenta (0.5-10~\GeVc) is possible when measurement of \dEdx  is combined with time-of-flight information \tof. Timing signals from the constant-fraction discriminators and signal amplitude information are recorded for each tile of the ToF-L/R walls. The coordinates of the track intersection with the front face are used to match the track to tiles with valid \tof hits. The position of the extrapolation point on the scintillator tile is used to correct the measured value of \tof for the propagation time of the light signal inside the tile. The distribution of the difference between the corrected \tof measurement and the value calculated from the extrapolated track trajectory length with the assumed mass hypothesis can be described well by a Gaussian with a standard deviation of 80 ps for ToF-R and 100 ps for ToF-L. These values represent the \tof resolution including all detector effects.

Momentum phase space is subdivided into bins of 1 \GeVc in \p~and 0.1 \GeVc~in \pt. Only bins with more than 1000 entries were used for extracting yields with the \tofdedx method.

\begin{figure}[!ht]
        \begin{center}
        \includegraphics[width=0.49\textwidth]{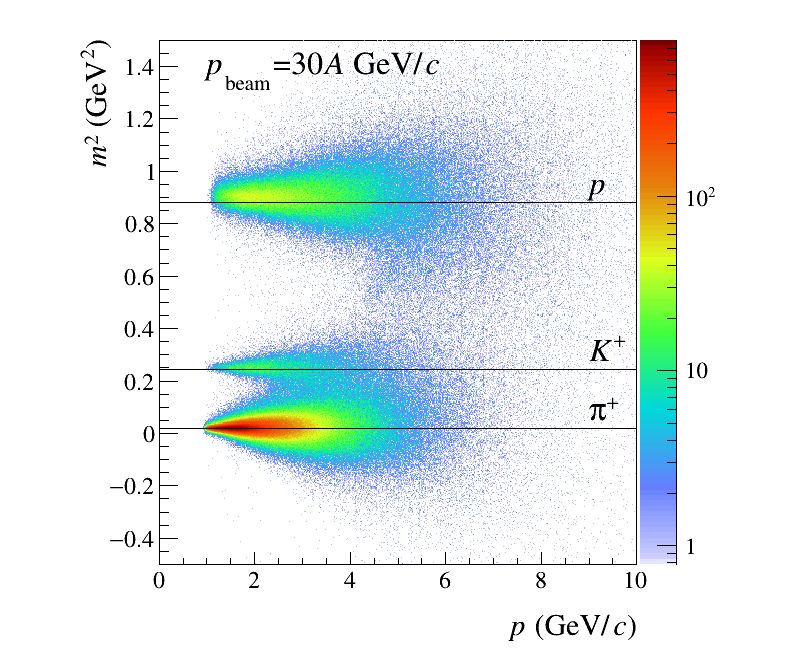}
        \includegraphics[width=0.49\textwidth]{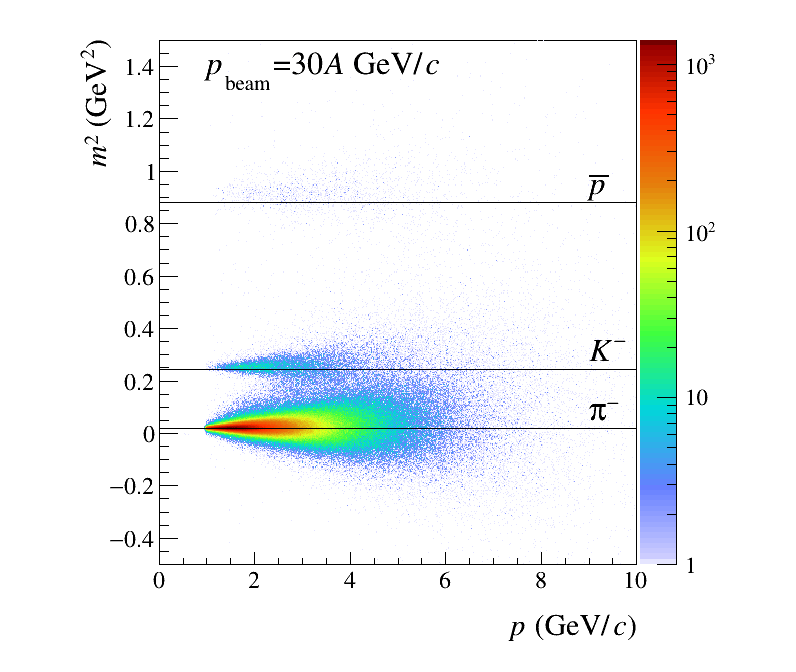}
        \end{center}
        \caption{Distribution of particles in the plane laboratory momentum and mass squared derived using time-of-flight measured by ToF-R ({\em right}) and ToF-L ({\em left}) produced in \textit{central} Ar+Sc collisions at 30\AGeVc. The lines show the expected mass squared values for different hadron species.}
        \label{fig:m2vsp}
\end{figure}
The square of the particle mass $m^2$ is obtained from \tof, from the momentum \p and from
the fitted trajectory length~$L$:
\begin{equation} 
m^2=(c\p)^2 \left(\frac{c^2~\tof^2}{L^2}-1 \right)~.
\label{eq:m2}
\end{equation}
For illustration distributions of $m^2$ versus \p~are plotted in Fig.~\ref{fig:m2vsp}
for positively ({\em left}) and negatively ({\em right}) charged hadrons produced in 10\% \textit{central} Ar+Sc interactions at 30\AGeVc.
Bands that correspond to different particle types are visible.

\begin{figure}[!ht]
           \centering
    \includegraphics[page=1, width=0.45\textwidth , trim = 50 0 50 0, clip]{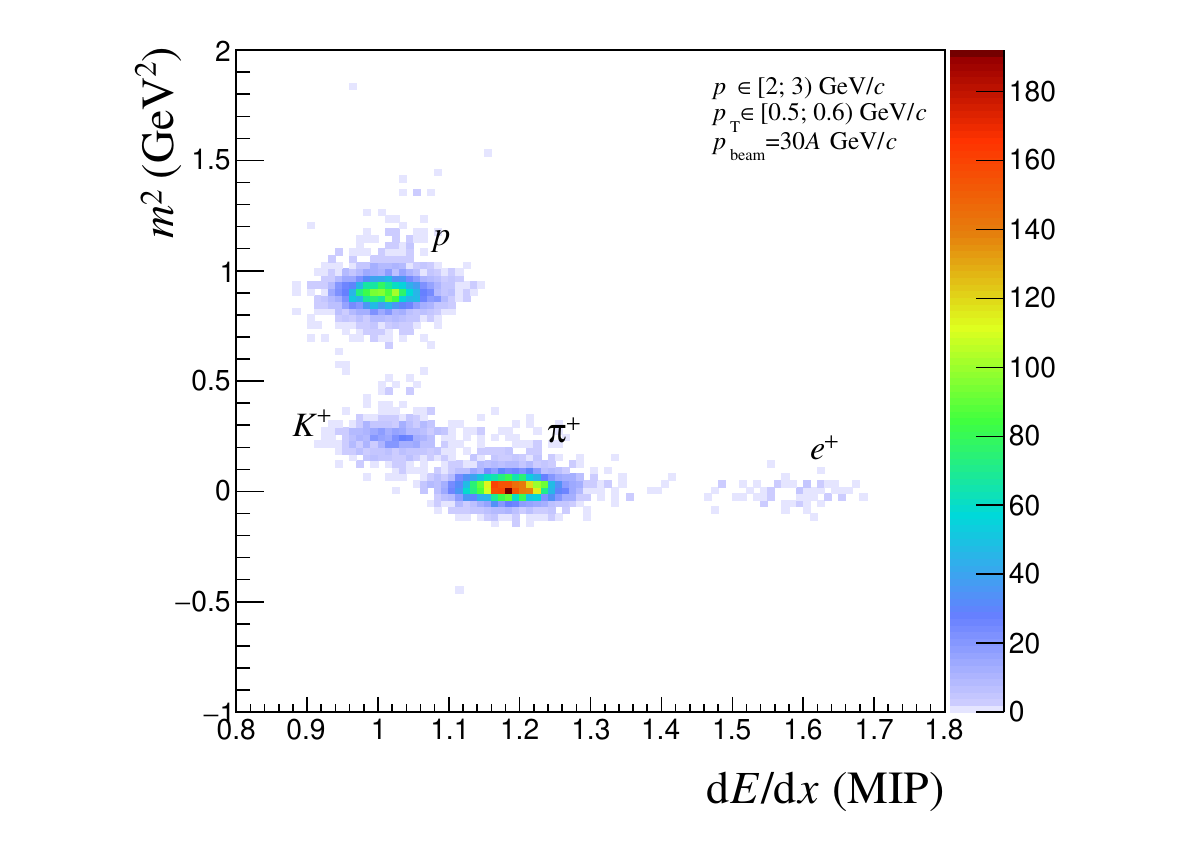}
    \includegraphics[page=2, width=0.45\textwidth , trim = 50 0 50 0, clip]{figures/analysis/m2_dedx.pdf}
    \caption{Example distributions of particles in $m^2$ and \dEdx plane in a single bin ($2~\GeVc<\p<3~\GeVc$, $0.5~\GeVc<\pt<0.6~\GeVc$) for positively (\textit{left}) and negatively (\textit{right}) charged particles measured in \textit{central} Ar+Sc collisions at 30\AGeVc. }
        \label{fig:tofdedx}
\end{figure}

Example distributions of particles in the $m^2$ -- \dEdx plane for the selected Ar+Sc interactions 
at 30\AGeVc are presented in Fig.~\ref{fig:tofdedx}. Simultaneous \dEdx and \tof measurements lead to 
improved separation between different hadron types. In this case, a simple Gaussian parametrization of 
the \dEdx distribution for a given hadron type can be used.

The \tofdedx identification method proceeds by fitting the two-dimensional distribution of particles in the \dEdx -- $m^2$ plane. Fits were performed in the momentum range from 2-10~\GeVc and transverse momentum range 0-1.5~\GeVc. Particles with total momentum less than 2~\GeVc are identified based on $m^2$ measurement alone, as different species of particles are separated enough. Contamination of electrons to pions identified in such a way is removed with a dedicated \dEdx cut (see Fig. \ref{fig:tofdedx} and Ref. \cite{Podlaski:2022}). For positively charged particles the fit function included contributions of $p$, $K^+$, $\pi^+$, and $e^+$, and for negatively charged particles the corresponding anti-particles were considered. The deuterons are not accounted for in the fits, as they are removed with a cut on measured $m^2$. The fit function for a given particle type was assumed to be the product of a Gaussian function in \dEdx and a sum of two Gaussian functions in $m^2$ (in order to properly describe the tails of the $m^2$ distributions).  In order to simplify the notation in the fit formulae, the peak positions of the \dEdx and $m^2$ Gaussians for particle type $j$ are denoted as $x_{j}$ and $y_{j}$, respectively. The fitted function reads:

\begin{equation}
\begin{split}
\rho(x,y) &= \sum _{j = p,\pi,K,e}\rho_j(x, y)\\
&= \sum _{j}\ \frac{N_j}{2\pi\ \sigma_{x}}\exp \left[ {-\frac{(x-x_j)^2}{2 \sigma^2 _{x}}} \right] \Bigg(
          \frac{f}{\sigma_{y_1}}\exp \left [{\frac{(y-y_j )^2}{2 \sigma^2 _{y_1}}} \right] 
        + \frac{(1-f)}{\sigma_{y_2}} \exp \left[ {\frac{(y-y_j)^2}{2 \sigma^2 _{y_2}}} \right] \Bigg) ,
        \end{split}
\label{eq:2dgaus}
\end{equation}
where $N_{j}$  and $f$ are amplitude parameters, $x_j$, $\sigma_{x}$ are means and width of the \dEdx
Gaussians and $y_j$, $\sigma_{y1}$, $\sigma_{y2}$ are means and widths of the $m^2$ Gaussians, respectively.
The total number of parameters in Eq.~\ref{eq:2dgaus} is 16. Imposing the constraint of normalization to
the total number of tracks $N$ in the kinematic bin 
\begin{equation}
N=\sum _{i} N_i~,
\end{equation}
the number of parameters is reduced to 15. Two additional assumptions were adopted:
\begin{enumerate}[(i)]
\item the fitted amplitudes were required to be greater than or equal to 0,
\item $\sigma_{y1} < \sigma_{y2} $ and $f > 0.7 $, the ''core'' distribution dominates the
$m^2$ fit.
\end{enumerate}

\begin{figure}[!ht]
        \begin{center}
        \includegraphics[width=0.7\textwidth]{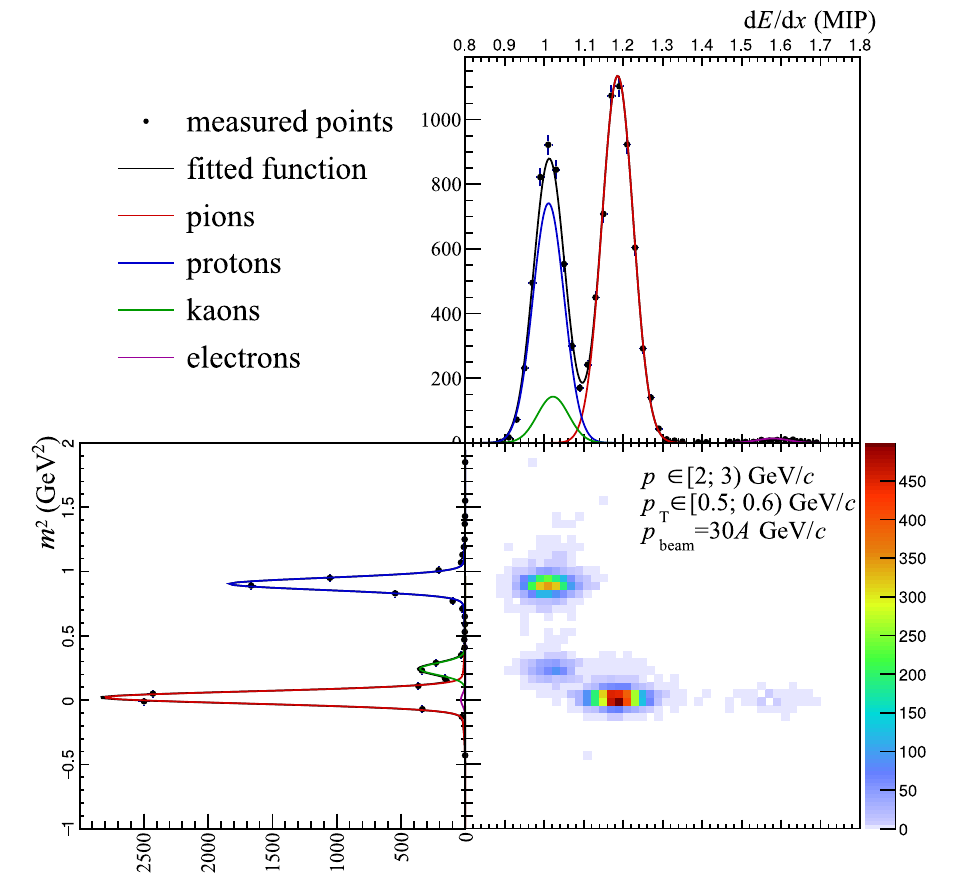}
        \end{center}
        \caption{
                 Example of the \tofdedx fit (Eq.~\protect{\ref{eq:2dgaus}}) obtained in a single bin 
                 (2 $<$ \p $<$ 3~\GeVc and 0.5 $<$ \pt $<$ 0.6~\GeVc) for positively charged particles in \textit{central} Ar+Sc collisions at 30\AGeVc.
                }
        \label{fig:exmtof}
\end{figure}

An example of the \tofdedx fit obtained in a single phase-space bin for positively charged particles 
in \textit{central} Ar+Sc collisions at 30\AGeVc is shown in Fig.~\ref{fig:exmtof}.

The \tofdedx method allows fitting the kaon yield close to mid-rapidity. This is not possible using the \dEdx method alone. Moreover, the kinematic domain in which pion and proton yields can be fitted is enlarged by the \tofdedx analysis. The results from both methods partly overlap at the highest beam momenta. In these regions, the results from both PID methods were combined using standard formulae \cite{Agashe:2014kda}.

\subsubsection{Probability method}
\label{sec:propability}

The 1D (\dEdx) and 2D (\tofdedx) models fitted to experimental distributions provide information on the contribution of individual particle species to total measured yields in bins of \p and \pt.
In order to unfold these contributions in the \dEdx method, for each particle trajectory with measured charge $q$, \p, \pt and \dEdx a probability $P_i$ of being a given species can be calculated as:

\begin{equation}
P_{i}(\p,\pt,\eqdEdx) =
    \frac{\rho_{i}^{\p,\pt}(\eqdEdx)}{\sum\limits_{i=\pi, K, p, e, d}^{} \rho_{i}^{\p,\pt}(\eqdEdx)}~,
    \label{eq:propdedx}
\end{equation}    
where $\rho_{i}^{\p,\pt}$ is the probability density according to the model with parameters fitted in a given (\p, \pt) bin calculated for \dEdx of the particle.

\begin{figure}[!ht]
        \begin{center}
        \includegraphics[width=0.9\textwidth]{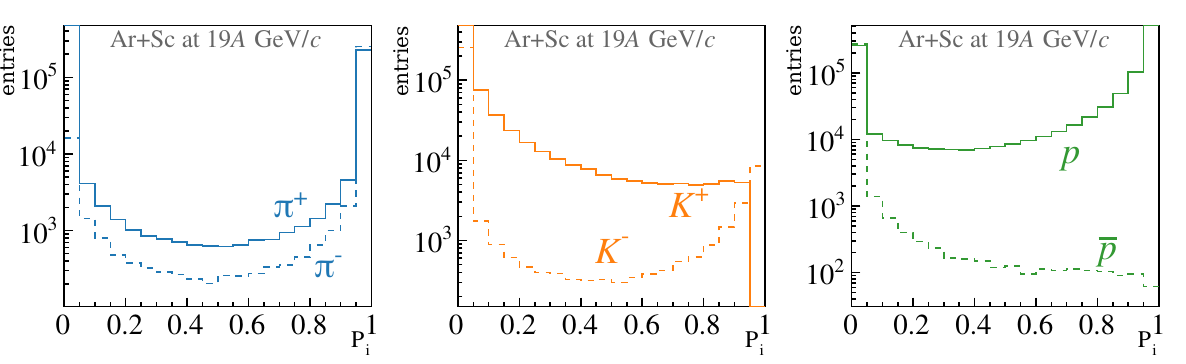}\\
        \includegraphics[width=0.9\textwidth]{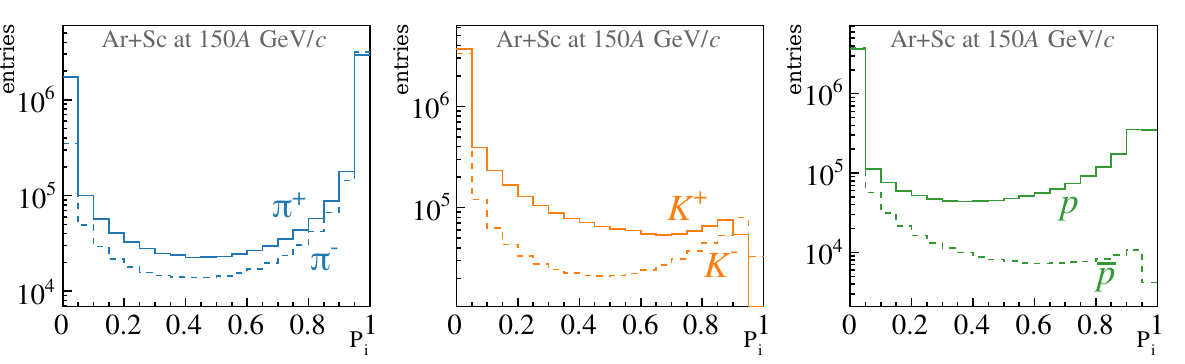}
        \end{center}
        \caption{
                 Distribution of probabilities of a track being a pion, kaon, proton for positively (solid lines)
                 and negatively (dashed lines) charged particles from \dEdx measurements in \textit{central} Ar+Sc collisions at 19$A$ (\textit{top}) and 150\AGeVc (\textit{bottom}).
                 }
        \label{fig:propdedx}
\end{figure}

Similarly, in the \tofdedx method (see Eq.~\ref{eq:2dgaus}) for \p>2~\GeVc the particle type probability is given by
\begin{equation}
P_{i}(\p,\pt,\eqdEdx,m^2) =
    \frac{\rho_{i}^{\p,\pt}(\eqdEdx, m^2)}{\sum\limits_{i=\pi, K, p, e}^{} \rho_{i}^{\p,\pt}(\eqdEdx, m^2)}~,
\label{eq:proptof}
\end{equation}

\begin{figure}[!ht]
        \begin{center}
        \includegraphics[width=0.9\textwidth]{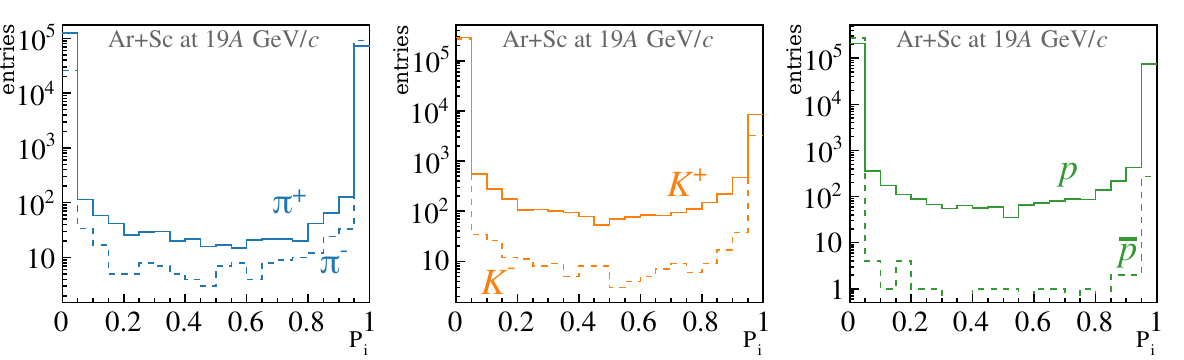}\\
        \includegraphics[width=0.9\textwidth]{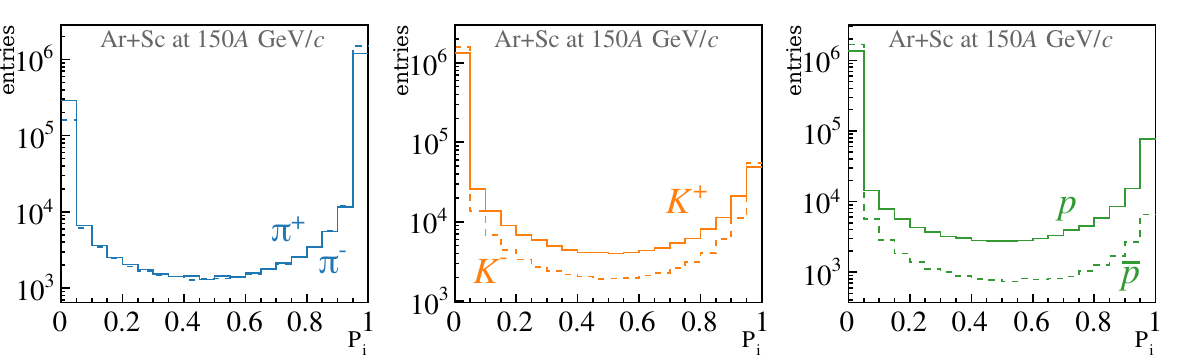}
        \end{center}
        \caption{
                Distribution of probabilities of a track being a pion, kaon, proton for positively (solid lines)
                 and negatively (dashed lines) charged particles from \tofdedx measurements in \textit{central} Ar+Sc collisions at 19$A$ (\textit{top}) and 150\AGeVc (\textit{bottom}).}
        \label{fig:proptof}
\end{figure}

In the case of low-momentum particles (\p<2~\GeVc), the assigned probability is either 0 or 1 based on the measured $m^2$. For illustration, particle type probability distributions for positively and negatively charged particles produced in \textit{central} Ar+Sc collisions at 19$A$ and 150\AGeVc are presented in Fig.~\ref{fig:propdedx} for the \dEdx fits and in Fig.~\ref{fig:proptof} for the \tofdedx fits. In the case of perfect particle type identification, the probability distributions in Figs.~\ref{fig:propdedx} and \ref{fig:proptof} will show entries at 0 and 1 only. In the case of incomplete particle identification (overlapping \dEdx or \tofdedx distributions) values between these extremes will also be populated.

The probability method allows transforming the fit results performed in
(\p, \pt) bins to results in (\y, \pt) bins.
Hence, for the probability method the mean number of identified particles in a given kinematical bin (e.g. (\y, \pt)) is given by~\cite{Rustamov:2012bx}:
\begin{equation}
n[i]^{\text{raw}}_{\eqdEdx}(\y,\pt)=\frac{1}{N_{\text{ev}}}\sum_{j=1}^{N_{\text{trk}}}P_{i}(\p,\pt,\eqdEdx)~,
\label{eq:ntrkdEdx}
\end{equation}
for the \dEdx identification method and:
\begin{equation}
n[i]^{\text{raw}}_{\eqdEdx,m^2}(\y,\pt)=\frac{1}{N_{\text{ev}}}\sum_{j=1}^{N_{\text{trk}}}P_{i}(\p,\pt,\eqdEdx,m^2)~,
\label{eq:ntrktof}
\end{equation}
for the \tofdedx procedure,
where $P_i$ is the probability of particle type $i$ given by
Eqs.~\ref{eq:propdedx} and~\ref{eq:proptof}, $j$ the summation index running over all entries $N_{\text{trk}}$ in the bin, $N_{\text{ev}}$ is the number of selected events. In the case of the \dEdx analysis, the probabilities $P_i$ were linearly interpolated in the $(\p,\pt)$ plane in order to minimize bin-edge effects.
\FloatBarrier
\subsection{Corrections and uncertainties}
\label{sec:Corrections}

In order to estimate the true number of each type of identified particle produced in Ar+Sc interactions, a set of corrections was applied to the extracted raw results.
These were obtained from a simulation of the \NASixtyOne detector followed by event reconstruction using the standard reconstruction chain. Only inelastic Ar+Sc collisions were simulated in the target material. 
The \EposLong model~\cite{Werner:2005jf} was selected to generate primary inelastic interactions as it best describes the \NASixtyOne measurements \cite{Abgrall:2013pp_pim}.
A \GeantThree-based program chain was used to track particles through the spectrometer, generate decays and secondary interactions, and simulate the detector response (for details see Ref.~\cite{Abgrall:2013pp_pim}).
Simulated events were then processed using the standard \NASixtyOne reconstruction chain. The reconstructed tracks were matched to the simulated particles based on the cluster positions of the reconstructed simulated tracks.
The selection of \textit{central} events was based on the number of forward spectators.
Corrections depend on the particle identification technique 
(i. e. \dEdx or \tofdedx).
Hadrons that were not produced in the primary interaction can amount to a significant fraction of the selected tracks, thus a special effort was undertaken to evaluate and subtract this contribution.
The correction factors were calculated in the same bins of \y and \pt as the particle spectra. The magnitude of correction factors reflects the effects of detector acceptance, track selection criteria, and reconstruction efficiency.
The generated \Epos events are referred to as ``MCgen'' and the label ``MCsel'' is given to the events with simulated detector response and reconstructed using standard \NASixtyOne chain with event and track selection criteria matching the ones used in the analysis of the experimental data.

\subsubsection{Corrections of the spectra}
The total correction for biasing effects listed in Sec.~\ref{sec:analysis} items (b) - (g) (influence of item (a) on the final result was found to be negligible) was calculated in the following way:
\begin{equation}
n[i]^{\textrm{corrected}} = \left(n[i]^{\textrm{raw data}}-n[i]^{\textrm{MCsel decay}}\right) \cdot \frac{n[i]^{\textrm{MCgen}}}{n[i]^{\textrm{MCsel primary}}}~,
\label{eq:ndedx}
\end{equation}
where, $n[i]$ stands for the per-event yield in the bin $i$ of the $y-\pt$ histogram of a given particle type, specifically:
\begin{center}
\begin{tabular}{ll}
	$n[i]^{\textrm{corrected}}$ & stands for the corrected yield,\\
	$n[i]^{\textrm{raw data}}$ & stands for the raw data yield,\\
	$n[i]^{\textrm{MCsel decay}}$ & is the contribution of feed-down from weak decays in MCsel,\\
	$n[i]^{\textrm{MCsel primary}}$ & is the contribution of primary particles in MCsel,\\
	$n[i]^{\textrm{MCgen}}$ & is a pure MC simulated yield.\\
\end{tabular}
\end{center}
The correction of spectra due to contamination by weak decays $(n[i]^{\textrm{MCsel decay}})$ is weakly correlated with the primary hadron yields, therefore this contribution is accounted for in an additive way (later referred to as $c_{\text{add}}$). The combined geometrical and efficiency correction is applied as the quotient in the second term of the Equation \ref{eq:ndedx} of the numbers of reconstructed primary tracks and all simulated tracks in a given momentum space bin (later referred to as $c_{\text{mult}}$).

The corrections for the spectra obtained with the \tofdedx PID method account additionally for the ToF tile efficiency $\epsilon_{\text{pixel}}(\p,\pt)$. It was calculated from the measured data as the probability of observing a valid reconstructed ToF hit if there exists an extrapolated TPC track that intersects with a given ToF tile. The ToF hit was considered valid if the signal satisfied the quality criteria given in Ref.~\cite{Anticic:2011ny}.

The ToF pixel efficiency factor $\epsilon_{\text{pixel}}(\p,\pt)$ was used in the MC simulation by weighting each reconstructed MC track passing all event and track selection cuts by the efficiency factor of the corresponding ToF tile.
Then, the number of selected MC tracks $n[i]^{\text{MCsel primary}}$ originating from primary particles becomes the sum of weights of those tiles which contribute to bin $i$:
\begin{equation}
n[i]^{\text{MCsel primary}} = \sum_{j=1}^{N_{\text{trk}}}
\epsilon_{\text{pixel}}^{j}(\p,\pt)~.
\label{eq:mcseltof}
\end{equation}
$n[i]^{\text{MCsel decay}}$ is defined in the same manner for particles originating from weak decays. Only hits in working tiles, with efficiency higher than 50\%, were taken into account in the identification and correction procedures.

The uncertainty of the multiplicative part of the correction was calculated assuming that the ``MCsel primary'' sample is a subset of the ``MCgen'' sample and thus has a binomial distribution. The uncertainty of the $c_\text{mult}$ ratio is thus given by:\\
\begin{equation}
    \sigma_{\text{stat}}\left(c_{\text{mult}}\right) = \frac{n[i]^{\textrm{MCgen}}}{n[i]^{\textrm{MCsel primary}}} \sqrt{\frac{N[i]^{\textrm{MCgen}}-N[i]^{\textrm{MCsel primary}}}{N[i]^{\textrm{MCgen}} \cdot N[i]^{\textrm{MCsel primary}}}}~,
\label{eq:sigmacdedx}
\end{equation}
where $N[i]$ is the number of tracks in bin $i$ (not normalized with the number of events, unlike $n[i]$). Absolute values of correction factors in phase space bins weakly depend on the original shapes of $y$-$\pt$ distributions provided by the model, due to the fact that the core part of the correction is calculated as a ratio (Eq. \ref{eq:ndedx}) in small $y$-$\pt$ bins.

The statistical uncertainty of the additive weak-decay feed-down correction ($c_{\text{add}}$) is added to the statistical uncertainty as a quadratic average.

\subsubsection{Tuning of feed-down in MC corrections}
\label{sec:feeddown}

\begin{figure}[h]
	\centering
	\includegraphics[width=0.7\linewidth]{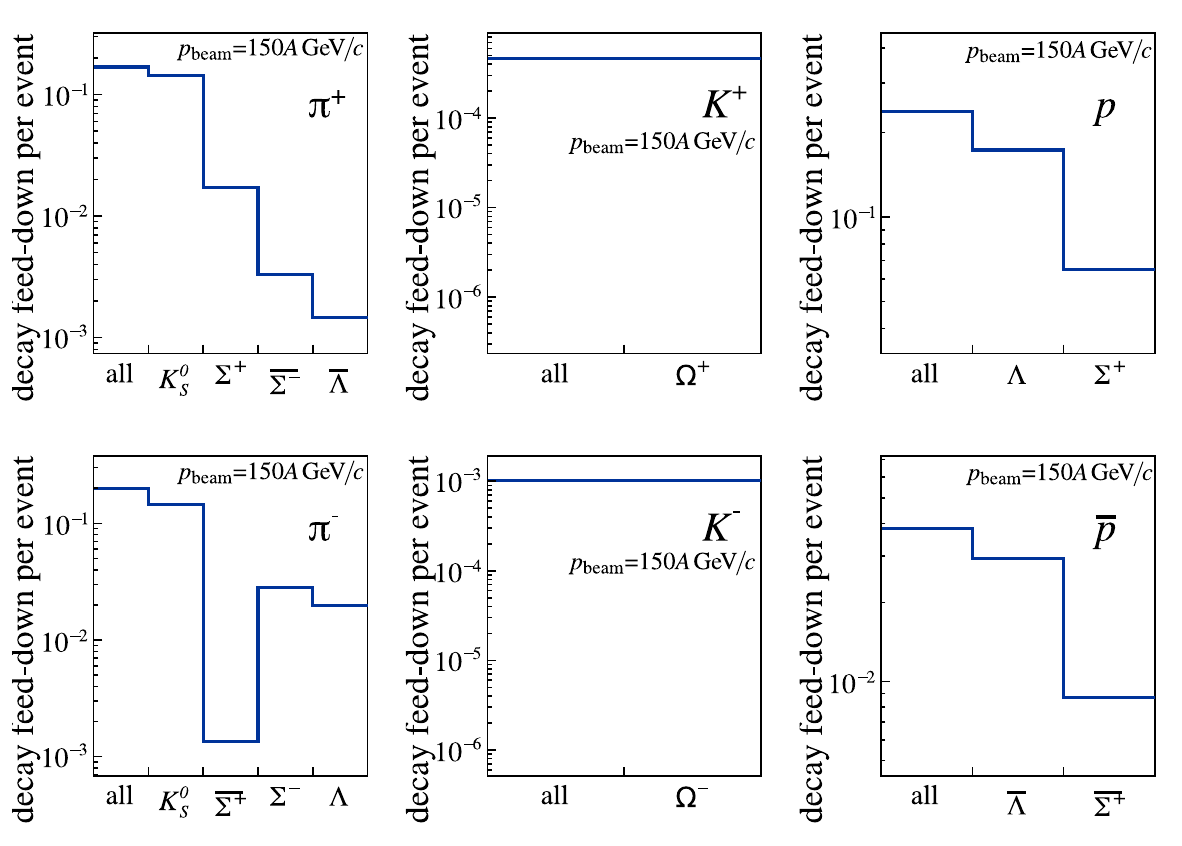}
	\caption{Contribution per event of reconstructed secondary tracks in the fiducial volume of the detector originating from weak decays and erroneously identified as products of primary interaction (based on \Epos at 150\AGeVc).}
	\label{fig:decayparents}
\end{figure}

\begin{figure}[h]
	\centering
	\includegraphics[width=0.9\linewidth]{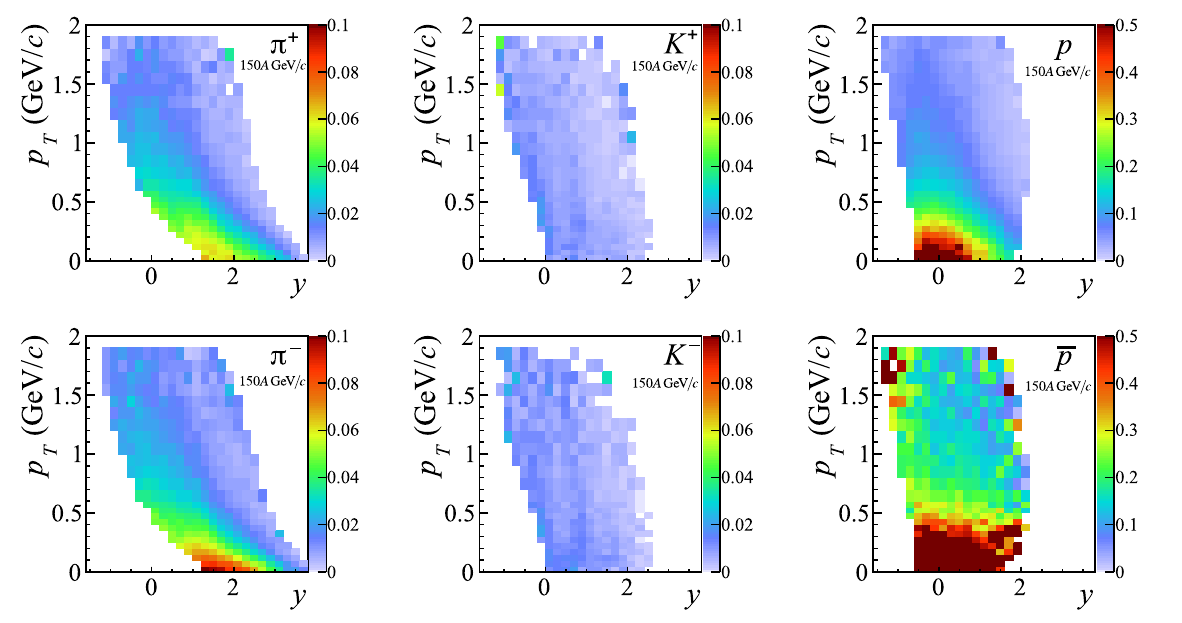}
	\caption{An example of the relative contribution of weak decay products in spectra of hadrons identified with \dEdx method at 150\AGeVc plotted in $y-\pt$ plane. The corrections are based on the \Epos model with data-derived tuning (see text for details). The feed-down for $\pi^-$ and $\pi^+$ is typically on the level of 1\%-3\%, reaching up to 10\% in the case of low-$p_T$ $\pi^-$. Secondary kaons can only originate from sparsely produced $\Omega^+$ and $\Omega^-$decays, thus the correction is well below 1\%. Spectra of protons and anti-protons are heavily influenced by the decays of $\Lambda$ and $\Sigma^+$ ($\bar{\Lambda}$ and $\bar{\Sigma^+}$ for anti-protons) -- in these cases, feed-down correction can reach over 50\% at low $p_T$. Similar numbers hold true for all collision energies.}
	\label{fig:decaycorrections}
\end{figure}

Another source of contamination of experimental results are the secondary particles originating from weak decays, that are reconstructed as primary ones. Figure \ref{fig:decayparents} shows the contribution of decay products originating from different decay parents and Fig. \ref{fig:decaycorrections} shows the relative cumulative contribution of weak-decay feed-down to the measured particle spectra. While the yield of weak-decay products is negligibly small in the case of kaons ($K^+$ and $K^-$) it is significant in the case of pions ($\pi^+$ and $\pi^-$) and (anti-)protons which are contaminated by the decay products from $K_S^0$ and (anti-)hyperons.
The \Epos model used in the MC simulation does not reproduce the properties of strangeness enhancement in nucleus-nucleus collisions, thus the yields of strange mesons and strange baryons are typically underestimated.
A procedure for tuning the contribution of weak decays was developed to improve the precision of the calculated corrections. It is based on data-derived quantities: mean multiplicities of particles estimated in measured data are compared with the ones extracted from MC simulation. Thus, using the preliminary results on charged kaon multiplicities from this analysis it is possible to construct an auto-tuning factor for yields of $K^0_S$:
\begin{equation}
K^0_{S~\textrm{(EPOS tuned)}} = K^0_{S~\textrm{(EPOS)}} \times \frac{\langle K^+ \rangle_{data}+\langle K^- \rangle_{data}}{\langle K^+ \rangle_{\textrm{(EPOS)}}+\langle K^- \rangle_{\textrm{(EPOS)}}}~.
\end{equation}
In the absence of measurements of strange (anti-)baryons in Ar+Sc collisions, the best effort was made to estimate their yields using existing data. At the SPS collision energies mean multiplicities of $\Lambda$ baryons are well approximated by the following relation:
\begin{equation}
\langle \Lambda \rangle = \alpha \cdot ( \langle K^+ \rangle - \langle K^- \rangle )~,
\end{equation}
where $\alpha$ is usually close to unity. A relevant parametrization of $\alpha$ was extracted from NA49's Pb+Pb data \cite{na49comp} and used to get an approximate estimate of $\Lambda$ yield in Ar+Sc at each collision energy.
Scaling of the yields of $\Lambda$ and $\Sigma^\pm$ are calculated as:
\begin{equation}
\Lambda_\text{(EPOS~tuned)} = \Lambda_\text{(EPOS)} \times \frac{\alpha \cdot ( \langle K^+ \rangle - \langle K^- \rangle )}{\langle \Lambda \rangle_\text{(EPOS)}}~.
\end{equation}
Obtained tuning factors are presented in Table \ref{tab:tuning_factors}, showing also the uncertainties assigned to these corrections. Furthermore, the yields of other strange and multi-strange baryons were tuned with the same factors as $\Lambda$ and $\Sigma^{\pm}$. The imperfect description of rapidity and transverse momentum dependence in the \Epos model is not accounted for in the presented calculation, hence the large values of assigned uncertainties.
In Figs. \ref{fig:systematic_dedx} and \ref{fig:systematic_tof} the total uncertainty introduced by the contribution of secondary particles is denoted with grey lines.
\begin{table}[h]
\centering
\begin{tabular}{l||llllll|l}
	&&&&&&& assigned\\
	$p_{\textrm{beam}}$ (\GeVc) & 13$A$ & 19$A$ & 30$A$ & 40$A$ & 75$A$ & 150$A$ & uncertainty\\ \hline \hline
	$K^0_S$ tuning factor & 1.339 & 1.339 & 1.297 & 1.371 & 1.332 & 1.307 & 5\%\\
	$\alpha$ & 0.89 & 0.89 & 1.00 & 1.08 & 1.12 & 0.95 & 5\%\\
	$\Lambda$, $\Sigma^+$, $\Sigma^-$ tuning factor & 1.582 & 1.582 & 1.562 & 1.634 & 1.522 & 1.507 & 10\%\\
\end{tabular}
	\caption{Multiplicative factors used for tuning the feed-down contribution in the \Epos MC simulation of \textit{central} Ar+Sc collisions.}
	\label{tab:tuning_factors}
\end{table}

\color{black}

\subsection{Corrected spectra}

\newcommand{\wdth}{0.12\textwidth}
\begin{figure}[h]

	\begin{minipage}{0.05\linewidth}
	\footnotesize
	~\\[0.6cm]
	\rotatebox[origin=c]{90}{\textbf{13\textit{A} GeV/\textit{c}}} \\[0.6cm]
	\rotatebox[origin=c]{90}{\textbf{19\textit{A} GeV/\textit{c}}} \\[0.6cm]
	\rotatebox[origin=c]{90}{\textbf{30\textit{A} GeV/\textit{c}}} \\[0.6cm]
	\rotatebox[origin=c]{90}{\textbf{40\textit{A} GeV/\textit{c}}} \\[0.6cm]
	\rotatebox[origin=c]{90}{\textbf{75\textit{A} GeV/\textit{c}}} \\[0.6cm]
	\rotatebox[origin=c]{90}{\textbf{150\textit{A} GeV/\textit{c}}}\\[0.6cm]
	\end{minipage}
	\begin{minipage}{0.96\linewidth}
	\hspace*{1.2cm} $\pi^+$
	\hspace*{1.8cm} $\pi^-$
	\hspace*{1.8cm} $K^+$ 
	\hspace*{1.8cm} $K^-$
	\hspace*{1.8cm} $p$
	\hspace*{1.8cm} $\bar{p}$\\
	\includegraphics[width=\linewidth]{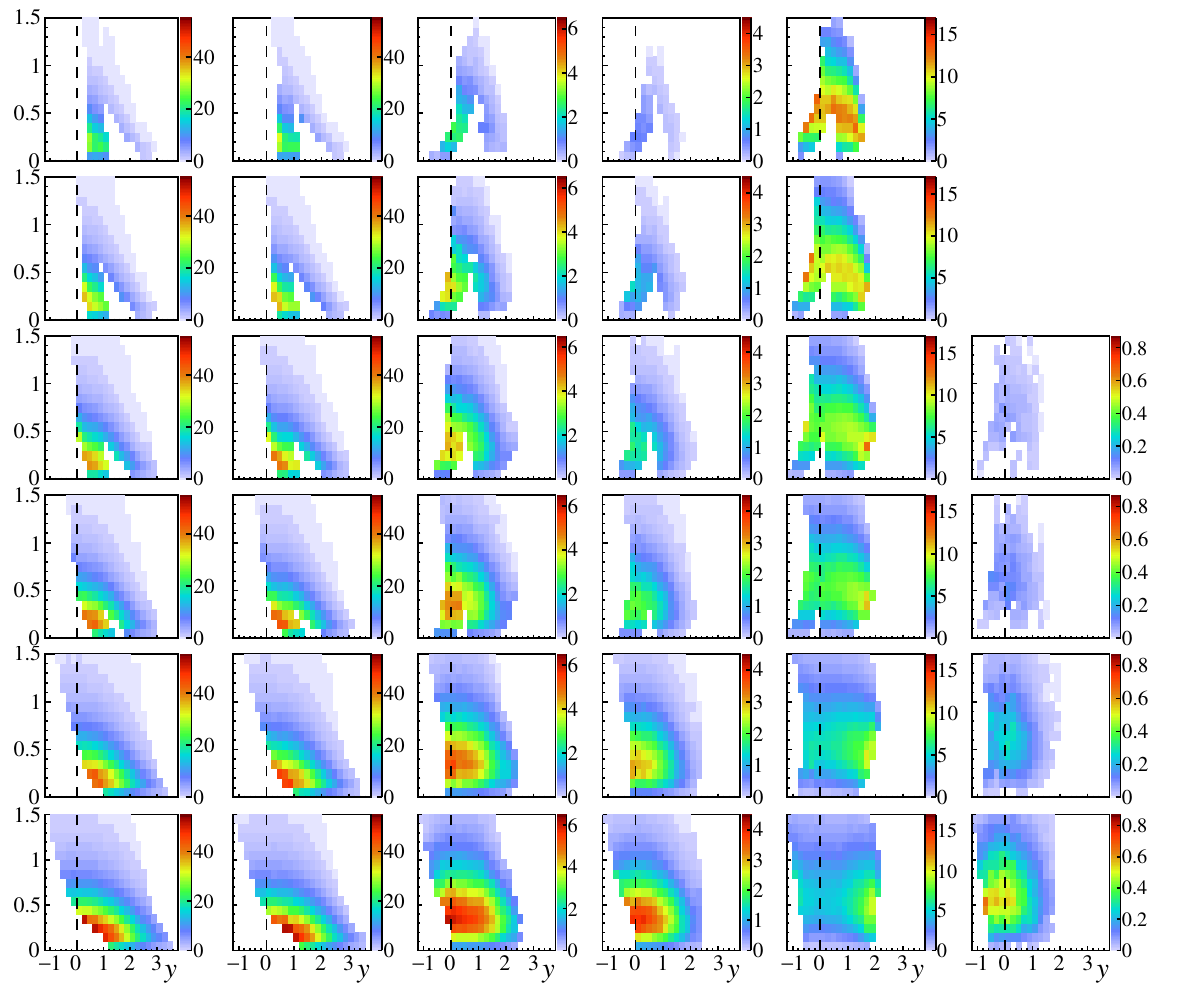}
	\end{minipage}~
	\hspace*{-0.99\linewidth}
	\begin{minipage}{0.13\linewidth}
	\scriptsize
	~\\[0.6cm]
	\rotatebox[origin=c]{90}{\pt (\GeVc)}\\[0.78cm]
	\rotatebox[origin=c]{90}{\pt (\GeVc)}\\[0.78cm]
	\rotatebox[origin=c]{90}{\pt (\GeVc)}\\[0.78cm]
	\rotatebox[origin=c]{90}{\pt (\GeVc)}\\[0.78cm]
	\rotatebox[origin=c]{90}{\pt (\GeVc)}\\[0.78cm]
	\rotatebox[origin=c]{90}{\pt (\GeVc)}\\[0.78cm]
	\end{minipage}

	\caption{Two-dimensional distributions (\y vs. $p_{T}$) of double differential yields (Eq. \ref{eq:double_diff})
	         of $\pi^{-}$, $\pi^{+}$, $K^{-}$, $K^{+}$, $p$ and $\bar{p}$ produced in the 10\% most \textit{central} Ar+Sc interactions at 13$A$, 19$A$, 30$A$, 40$A$, 75$A$ and 150\AGeVc.}
	\label{fig:final2D}
\end{figure}

The final spectra of different types of hadrons produced in Ar+Sc collisions are defined as:
\begin{equation}
\frac{\text{d}^{2}n}{\text{d}y~\text{d}p_T} =  \frac{1}{\Delta\y\cdot\Delta\pt} n[i]^{\textrm{corrected}}~,
\label{eq:double_diff}
\end{equation}
where $\Delta\y$ and $\Delta\pt$ are the bin sizes and $n[i]^{\text{corrected}}$ represents the mean multiplicity of given particle type in the $i$-th bin in \y and \pt obtained with either \dEdx or \tofdedx identification method, as introduced in Eq. \ref{eq:ndedx}.

The resulting two-dimensional distributions $\frac{\text{d}^{2}n}{\text{d}y~\text{d}p_T}$ of $\pi^{-}, \pi^{+}, K^{-}, K^{+}, 
p$ and $\bar{p}$ produced in the 10\% most \textit{central} Ar+Sc collisions at different SPS energies are presented in Fig.~\ref{fig:final2D}.

\subsubsection{Statistical uncertainties}

Statistical uncertainties of multiplicities calculated in the \tofdedx method were derived under the assumption of Poissonian statistics in a single bin and no correlation between the bins. The resulting uncertainty in (\p,\pt) bin is given by:

\begin{equation}
\sigma^{2}_{stat}(n[i]^{\text{raw}}) = \frac{1}{N_{\text{ev}}} \sum_{j=1}^{N_{\text{trk}}} {P_{i}^j}(\p,\pt,\eqdEdx,m^2)^2~.
\label{eq:sigmaraw}
\end{equation}

An alternative method, bootstrapping, was used to calculate statistical uncertainties in the case of the \dEdx identification technique. One hundred bootstrap samples were generated through random sampling with replacement, performed on the level of events. Each bootstrap sample is injected into the procedure of particle identification and calculation of $y$-\pt spectra. The errors are then estimated as standard deviations of yields at all bootstrap samples. It was verified that the number of bootstrap samples was large enough and that the distribution of yields resembles the normal distribution, allowing to assign the standard deviation as the statistical error. It was found that bootstrapping and weighted variances (Eq. \ref{eq:sigmaraw}) methods yield similar values of uncertainty.

The contribution to statistical uncertainties from the MC correction factors (discussed in detail in Sec. \ref{sec:Corrections}) is propagated into final uncertainties using the standard procedure.

\subsubsection{Systematic uncertainties}
\label{sec:systematic}
\begin{figure}[h]
	\centering
	\hspace*{-0.5cm}
	\includegraphics[width=0.8\linewidth]{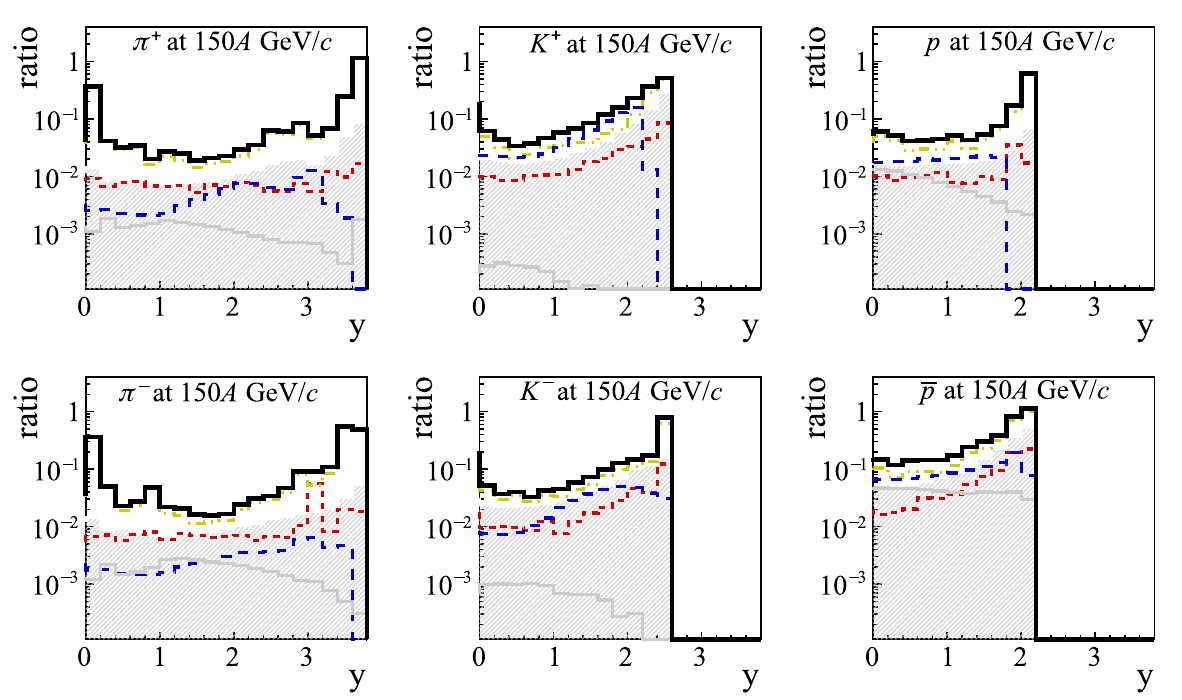}~
	\includegraphics[width=0.33\linewidth]{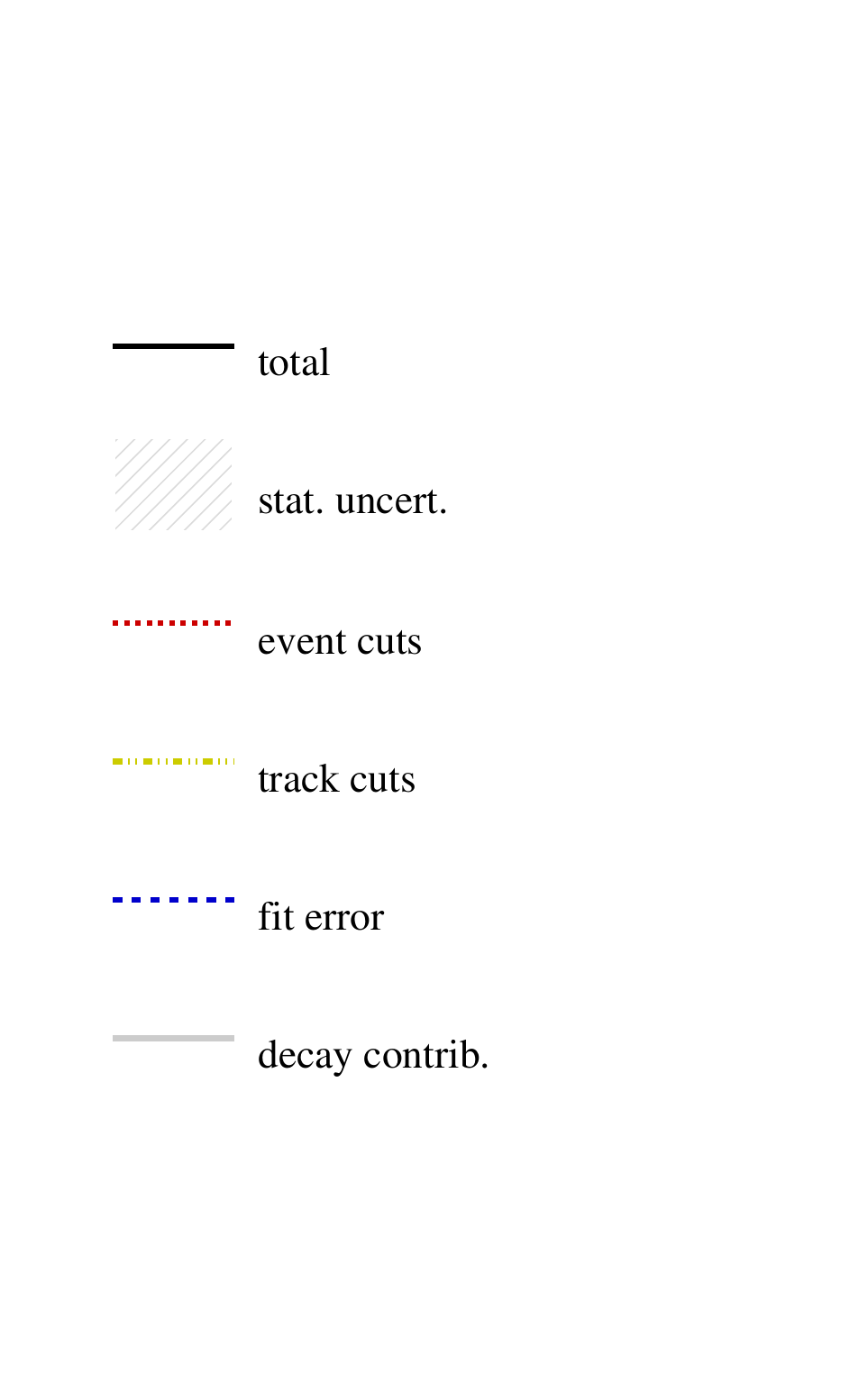}
	\caption{Systematic uncertainty relative to the measured yield of double-differential distributions obtained with \dEdx PID method, integrated in \pt, shown for each identified species in dependence on rapidity $y$ at $p_{\textrm{beam}}$=150\AGeVc. Different contributions to the total uncertainty are plotted, along with statistical error (shaded area). Large fluctuations seen at low-rapidity uncertainties of pion spectra are due to narrow acceptance in \pt.}
	\label{fig:systematic_dedx}
\end{figure}

\begin{figure}[h]
	\centering
	\hspace*{-0.5cm}
	\includegraphics[width=0.8\linewidth]{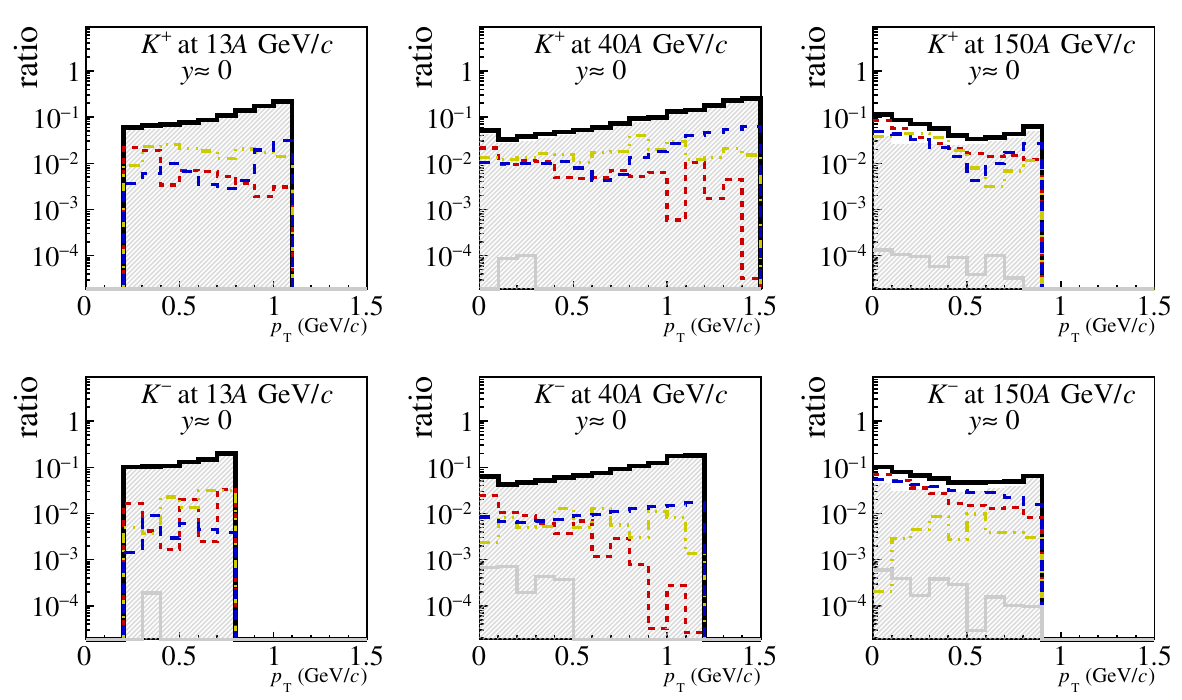}~
	\includegraphics[width=0.33\linewidth]{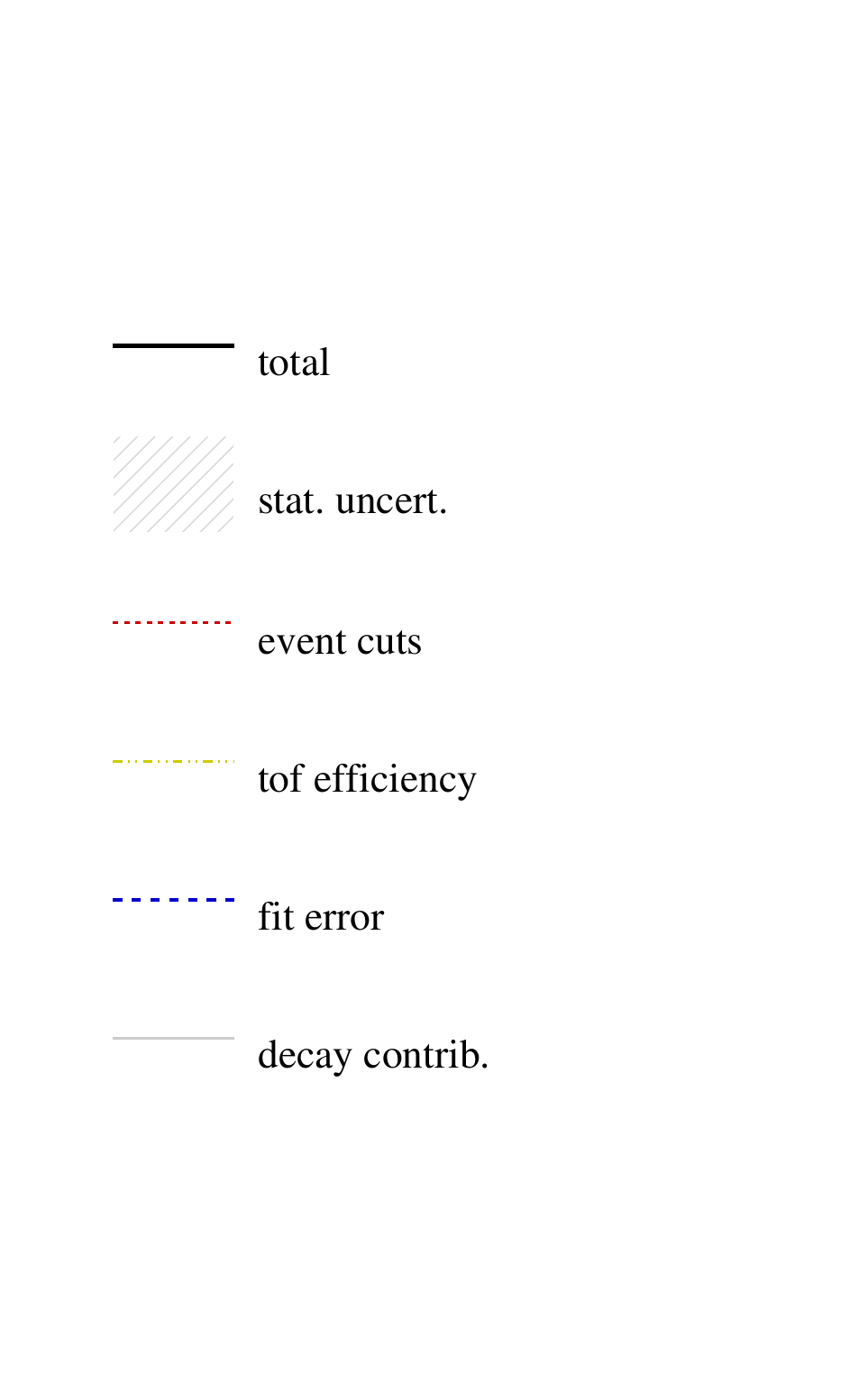}
	\caption{Systematic uncertainty relative to the measured yield of double-differential distributions in $y$ and \pt for charged kaons, obtained with \textit{tof-}\dEdx PID method, shown in dependence on transverse momentum \pt at mid-rapidity for \textit{central} Ar+Sc collisions at $p_{\textrm{beam}}$=13$A$, 40$A$ and 150\AGeVc. Different contributions to the total uncertainty are plotted, along with statistical error (shaded area).}
	\label{fig:systematic_tof}
\end{figure}

The following sources of systematic uncertainties were considered in this study:
\begin{enumerate}[(I)]
	\item \textbf{Particle identification} methods utilized in this analysis provide measurements of particle yields through the fits of multi-parameter models. In order to increase stability, some of the parameters need to be fixed. Moreover, it may happen that the fitted variable reaches the imposed limit. Such cases may lead to biases in the estimation of particle yields and therefore were carefully studied.
	\begin{enumerate}[(a)]
		\item \dEdx method\\
    In the $dE/dx$ method the fits of peak positions of kaons and protons were found to have the largest influence on particle yields and their ratios, while also having a relatively high variance, in particular in sparsely populated bins. The strategy used in this study (described in Sec.~\ref{sec:dedx_id}) involved fixing these parameters at pre-fitted values and assuming their independence of transverse momentum. The differences between prefits and results of bin-by-bin fits were studied and the spread within a single momentum bin was found at approx.~0.2\%.
    Therefore in order to determine a potential bias introduced by fixing relative peak positions, they are varied by $\pm$0.1\%.
    Contribution to the biases from other fit parameters was found negligible.
		\item \textit{tof-}\dEdx\\
		Systematic uncertainties were estimated by shifting the mean ($x_j$ and $y_j$) of the two-dimensional Gaussians (Eq. \ref{eq:2dgaus}) fitted to the $m^2$-\dEdx distributions by $\pm 1\%$, which corresponds to typical uncertainty of the fitted parameters. Additional systematic uncertainty arises for the \tofdedx method from the quality requirements on the signals registered in the ToF pixels. In order to estimate this uncertainty the nominal signal selection thresholds were varied by $\pm 10\%$
	\end{enumerate}
	
	\item \textbf{Event selection} criteria based on any measurements downstream of the target may also introduce bias in the results. Uncertainties due to this were estimated through an independent variation of criteria listed below:
	\begin{enumerate}[(a)]
		\item Removal of events with off-time particles -- the time window in which no off-time beam particle is allowed was varied by $\pm$2 $\mu$s with respect to the default value of 4 $\mu$s.
		\item Fitted main vertex position -- the range of allowed main vertex $z$-coordinate was varied by $\pm$5 cm at both ends.
	\end{enumerate}
	\item \textbf{Track selection}:
	
	The contribution to systematic uncertainty from track selection criteria was estimated by varying the following parameters:
	\begin{enumerate}[(a)]
		\item The required minimum of the total number of clusters was varied by $+$5 and $-$5 points.
		\item Similarly, the minimum number of clusters in VTPCs was varied by $\pm$5 points. Note that both of the cuts on the number of points affect the acceptance of the \dEdx PID method as well, which was also taken into account.
		\item The influence of the selection of azimuthal angle was investigated in the case of \dEdx\hspace*{-1pt}-only PID by comparing the results obtained for $|\phi|<30^\circ$ (default value), $|\phi|<20^\circ$ and $|\phi|<40^\circ$.
	\end{enumerate}
	\item \textbf{Feed-down correction}:
	
	Uncertainties of weak decays feed-down correction were accounted for as described in Sec. \ref{sec:feeddown}.
\end{enumerate}
The maximum difference of the particle yields in each bin of $y$ and \pt obtained under each varied criterium was assigned as the partial contribution to the systematic uncertainty. The final systematic uncertainty was taken as the square root of squares of its partial contributions. The relative contribution of each of the listed sources to the systematic uncertainties of the final spectra of identified particles is shown in Figs. \ref{fig:systematic_dedx} (\dEdx method) and \ref{fig:systematic_tof} (\textit{tof-}\dEdx method). The total uncertainty is typically 3--10\% for charged pions, charged kaons, and protons, while it exceeds 10\% in the case of anti-protons. The relative total uncertainties tend to increase at lower collision energies.

\FloatBarrier
\section{Results}
\label{sec:results}

Figure~\ref{fig:final2D} displays two dimensional distributions $\text{d}^{2}n / (\text{d}y\,\text{d}p_\text{T})$ of $\pi^{-}$, $\pi^{+}$, $K^{-}$, $K^{+}$, $p$ and $\bar{p}$ produced in 10\% most \textit{central} Ar+Sc collisions at beam momenta of 13$A$, 19$A$, 30$A$, 40$A$, 75$A$ and 150\AGeVc.
The spectra obtained using \dEdx and \tofdedx PID methods were combined to ensure a maximal momentum space coverage. Bins were removed from the final spectrum in the case of insufficient bin entries for the identification methods used in the analysis or if the yield uncertainty, either statistical or systematic, exceeded 80\%. The gaps in the acceptance grow with decreasing collision energies, however, reliable measurement of key properties of charged hadron production is still possible even at the lowest beam momentum. In $y$-$p_\text{T}$ bins where both \tofdedx and \dEdx measurements exist, a weighted average is calculated using standard formulae \cite{Agashe:2014kda}. For rapidity bins where there is an overlap, a comparison of \tofdedx and \dEdx results is provided in Appendix \ref{sec:additional_plots}.

The transverse momentum spectra of identified hadrons are extrapolated to account for the missing acceptance. Extrapolation of $p_\text{T}$ spectra allows for an accurate calculation of rapidity distribution, which in turn is also extrapolated into regions of missing measurements to calculate mean multiplicities.
Only the experimental results up to $p_\text{T}<$1.5 GeV/\textit{c} are considered since the contribution of misidentified particles becomes large at higher values of \pt, which results in a higher systematic uncertainty. The contribution of the extrapolation towards high $p_\text{T}$ ($>$1.5 GeV/\textit{c}) is typically of the order of 1\%. At small $p_\text{T}$ ($\approx~0-0.1~\GeVc$) the extrapolation or interpolation (when a gap between \tofdedx and \dEdx data exists) ranges from 0 to 40\%.
The extrapolation methods and their applicability differ for each of the studied particle species and thus are described separately in Secs. \ref{sec:pions}, \ref{sec:kaons} and \ref{sec:protons}. A complete set of plots depicting transverse momentum spectra of all particles in rapidity slices, together with fitted functions is available in Appendix \ref{sec:additional_plots}.

Subsequently, Sec. \ref{sec:discussion} reviews presented measurements in terms of collision energy and system size dependence, including also a comparison with relevant models. Presented results are then discussed in the context of the onset of deconfinement and an emerging phenomenon of the onset of QGP fireball.

\subsection{Charged pions}
\label{sec:pions}
\subsubsection{Transverse momentum spectra}
The measured double differential charged pion spectra in rapidity and transverse momentum at 13\textit{A}--150\AGeVc beam momenta are presented in Fig. \ref{fig:final2D}.

\begin{figure}[h]
\centering
\includegraphics[width=0.7\linewidth]{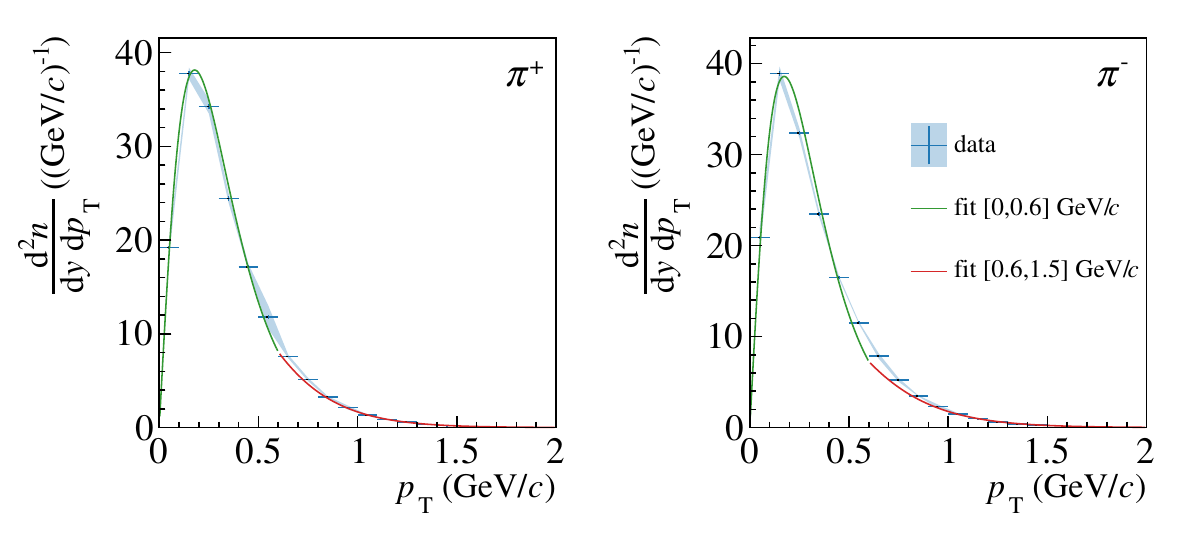}
\includegraphics[width=0.7\linewidth]{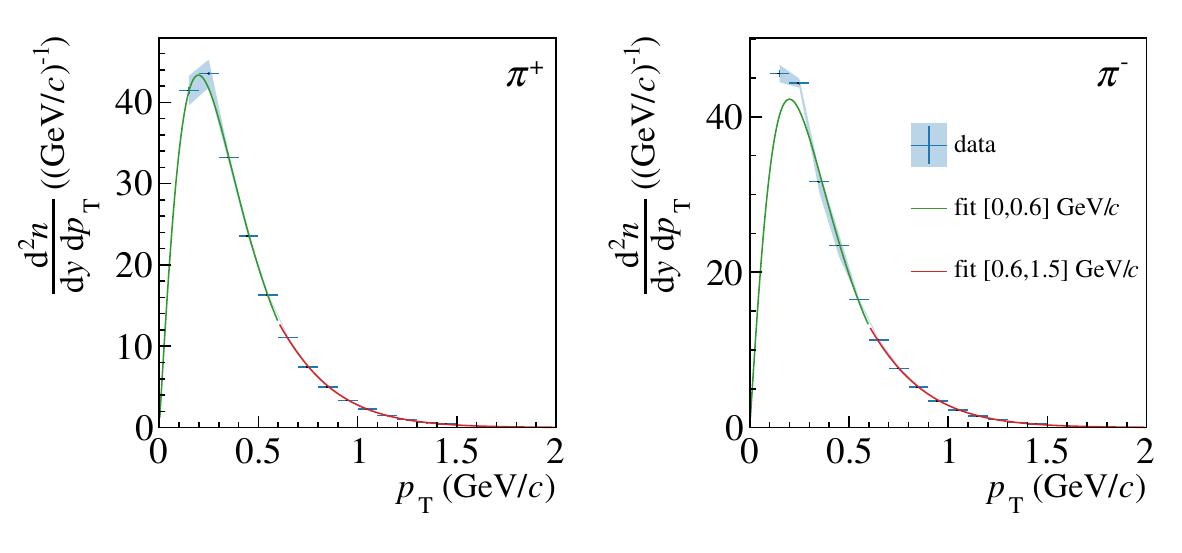}
\caption{Example fits to charged pion, $\pi^+$ (left) and $\pi^-$ (right), transverse momentum spectra ($p_{\text{beam}}=30$\AGeVc and $75$\AGeVc at $0.6<y<0.8$, \textit{top} and \textit{bottom} panels respecitvely). Exponential fits (Eq \ref{eq:pion_pT}) are performed in two regions separately: $p_T \in$ [0.0, 0.6] and $p_T \in$ [0.6, 1.5] \GeVc. The fitted functions are used to extrapolate the yields beyond $p_T=1.5$ \GeVc and interpolate the yields in case a gap in acceptance appears due to different coverage of PID methods. A full set of transverse momentum spectra with corresponding fits is presented in Figures \ref{fig:rap_slices_pi_pos} and \ref{fig:rap_slices_pi_neg} in Appendix \ref{sec:additional_plots}. In rare cases, the yield is extrapolated also in the region of \pt$<0.1$ GeV/$c$. The vertical bars represent statistical uncertainties and the shaded bands stand for the systematic uncertainties.}
\label{fig:pTfits_pions}
\end{figure}

In order to account for the regions outside \dEdx and \tofdedx PID acceptance, the $p_T$ distributions in each bin of rapidity were fitted independently in two separate $p_T$ intervals: [0.0,0.6] and [0.6,1.5] GeV/\textit{c}. Such a procedure was employed due to the influence of radial flow and a large contribution from resonance decays, which is difficult to model reliably. Dividing the $p_T$ range into two intervals allows for an accurate interpolation as well as the extrapolation of the \pt spectra. A fit to combined data points from both PID methods is performed using the following formula:
\begin{equation}
f(p_T) = \frac{A\cdot p_T}{T (m_\pi+T)} \exp\left(\frac{m_\pi-m_T}{T}\right),
\label{eq:pion_pT}
\end{equation}
where $T$ is the inverse slope parameter, $m_\pi$ and $m_T$ denote pion's rest and transverse masses respectively and $A$ is a normalization factor, $T$ and $A$ are fit parameters. Additionally, it is required that the fitted function is continuous between the intervals. Example fit results are shown in Fig. \ref{fig:pTfits_pions}. The inverse slope parameter $T$ is decreasing from mid-rapidity towards higher values of rapidity.

\subsubsection{Rapidity spectra}
\label{sec:pions_rapidity_spec}
\begin{figure}[h]
\centering
\includegraphics[width=0.5\linewidth]{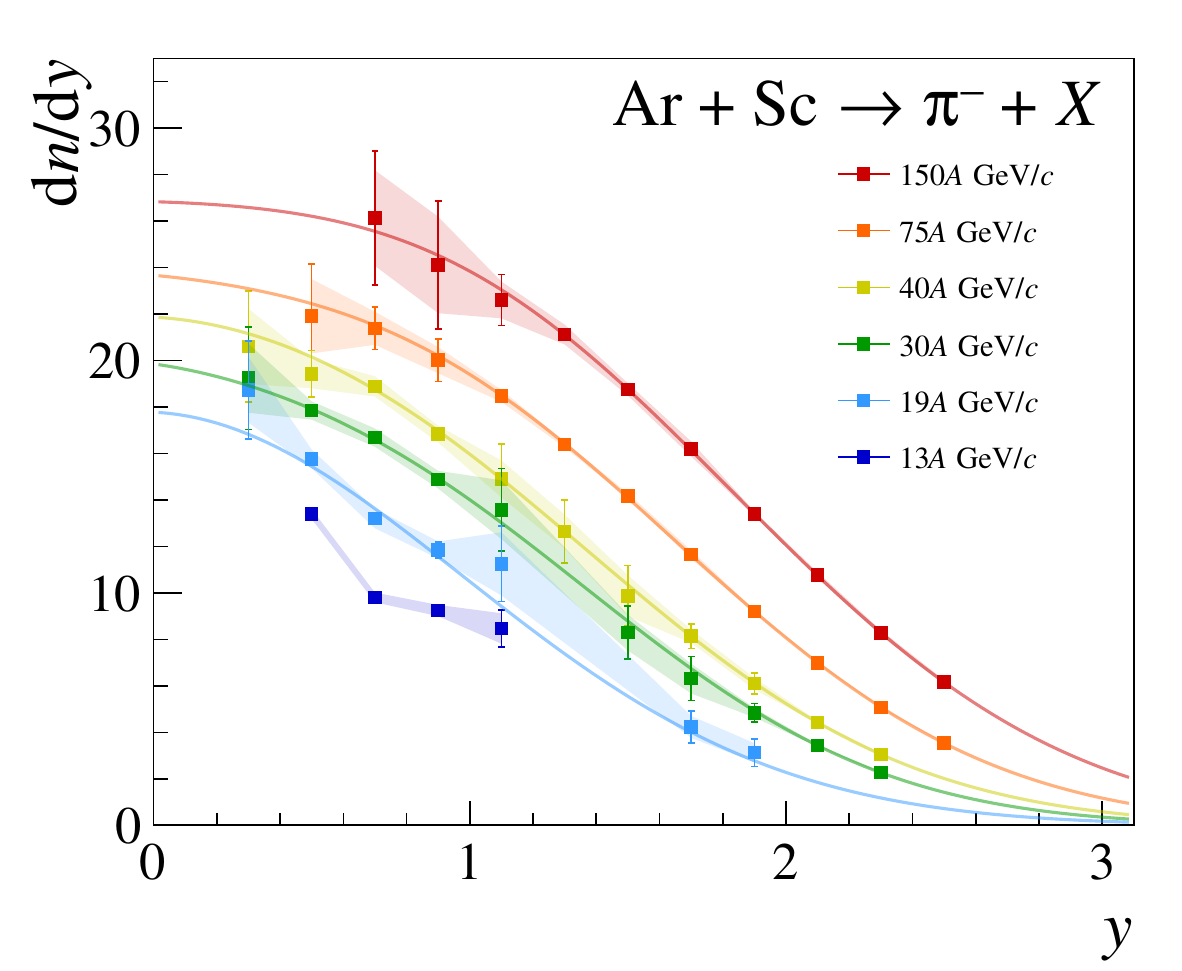}~
\includegraphics[width=0.5\linewidth]{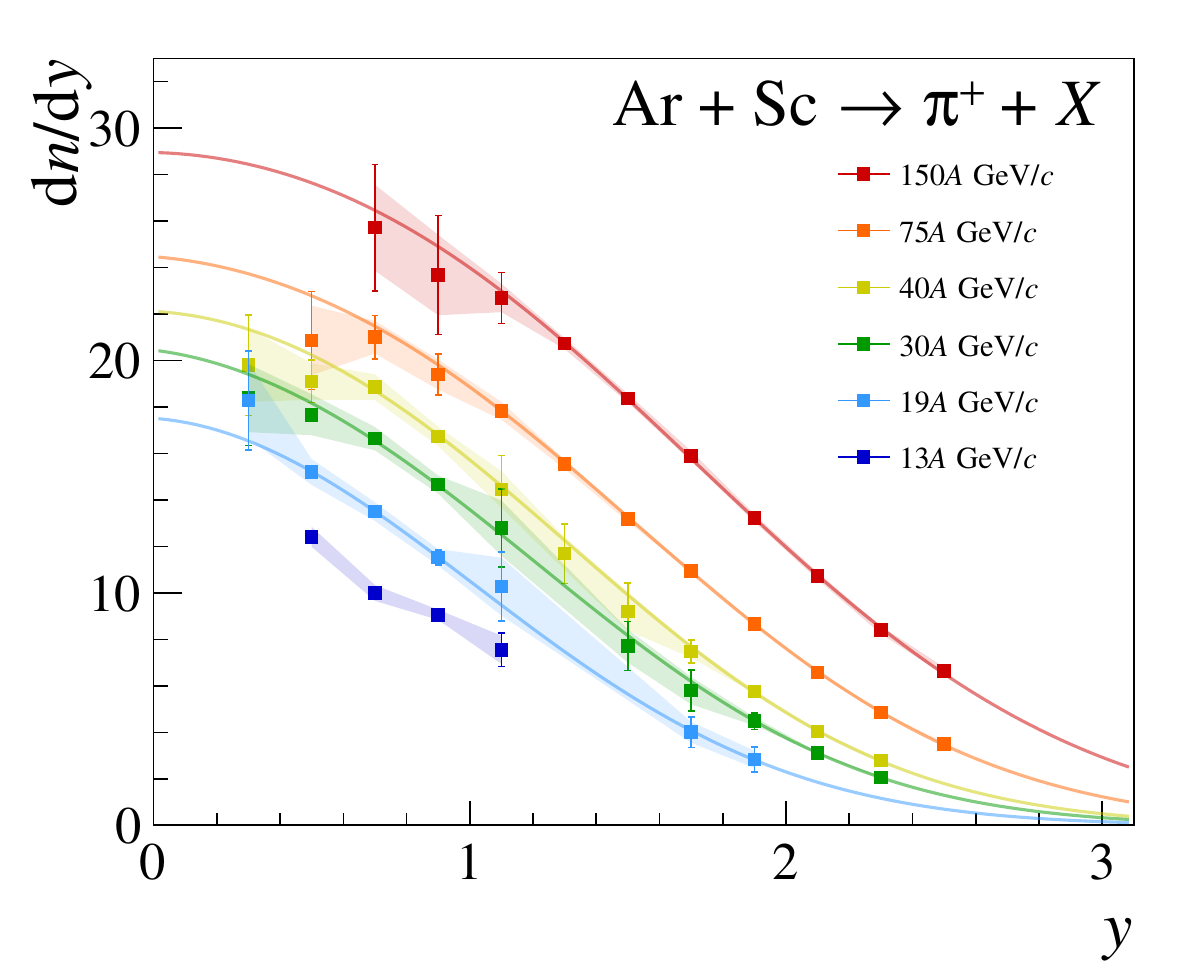}
\caption{Rapidity spectra of charged pions ($\pi^+$ and $\pi^-$) measured with \dEdx and tof-\dEdx methods in  10\% most \textit{central} Ar+Sc collisions. The line is a sum of two Gaussians, equidistant from mid-rapidity with differing amplitudes (Eq. \ref{eq:pion_yfit}). The statistical uncertainties are shown with error bars (if not visible, they do not exceed the size of the markers) and the systematic uncertainties are shown as shaded bands.
}
\label{fig:yspectra_pions}
\end{figure}

The $\text{d}n/\text{d}y$ yields are obtained by integration of the $\text{d}^2n/\text{d}y\,\text{d}p_T$ spectra and the addition of the integral of the fitted functions in the regions of missing acceptance.
An additional contribution to the systematic uncertainty of 25\% of the extrapolated yield is added to account for a possible bias due to model selection. Figure \ref{fig:yspectra_pions} displays the resulting $\text{d}n/\text{d}y$ distributions for all collision energies. The one-dimensional rapidity spectra are fitted with double-Gaussians with means equidistant from mid-rapidity:

\begin{equation}
f(y) = \frac{A_0 ~ A_\text{rel}}{\sigma_0\sqrt{2\pi}}~ \exp\left(-\frac{(y-y_0)^2}{2\sigma_0^2}\right) ~~+~~ \frac{A_0}{\sigma_0\sqrt{2\pi}}~ \exp\left(-\frac{(y+y_0)^2}{2\sigma_0^2}\right),
\label{eq:pion_yfit}
\end{equation}
where $A_0$ is the amplitude, $A_\text{rel}$ is a parameter reflecting the asymmetry between forward and backward rapidity hemispheres, $\sigma_0$ is the width of individual peaks and $y_0$ stands for the displacement of contributing distributions from mid-rapidity. Within such parametrization, RMS width of the obtained rapidity distributions, $y_\text{RMS}$, can be calculated as follows:

\begin{equation}
    y_\text{RMS}=\sqrt{\sigma_0^2+\frac{4A_\text{rel}}{(A_\text{rel}+1)^2}\cdot y_0^2},
    \label{eq:rapidiry_rms}
\end{equation}
which in the case of symmetrical rapidity distribution ($A_\text{rel}=1$) reduces to $y_\text{RMS}=\sqrt{\sigma_0^2+ y_0^2}$.
The measured data covers only the region of positive rapidity, thus the parameter $A_\text{rel}$ is not fitted, but instead taken from the published results of complementary analyses, with the $h^-$ method, described in detail in Ref. \cite{NA61SHINE:2021nye}. The asymmetry was found between 3\% at 150\AGeVc and 17\% at 13\AGeVc towards the backward hemisphere.
The fit quality is satisfactory for beam momenta from 19\AGeVc to 150\AGeVc. Good agreement of the measured data with the fitted model allows to extrapolate the spectra into the unmeasured region and thus calculation of mean multiplicities of charged pions. In the case of 13\AGeVc, the extrapolation was not performed, as the measurement covers too small acceptance region. Moreover, the spectrum of negatively charged pions at 13\AGeVc shows an unphysical shape and does not agree with the results from the $h^-$ method, thus suggesting a possible bias, not fully accounted for in the estimation of measurement uncertainties.
Table \ref{tab:pion_mult} displays the mean multiplicity as the integral of the fitted function for beam momenta 19$A$--150\AGeVc. The statistical uncertainty is calculated as the integral error due to parameter uncertainties and their covariance matrix. The fit accounts for both statistical and systematic uncertainties of measured data, however, an additional systematic uncertainty of 5\% is imposed on the data based on the largest discrepancies observed when comparing the fit integrals with parameters fixed at values taken from $h^-$ method \cite{NA61SHINE:2021nye} against the unconstrained fit.
\begin{table}[h]
\small
\centering 
\begin{tabular}{c||c|c}
\parbox{1.1cm}{\,\,$p_{\textrm{beam}}$ (GeV/\textit{c})}& $\langle \pi^+ \rangle$ &  $\langle \pi^- \rangle$ \\[0.3cm]
\hline \hline
19 & 45.4 $\pm$ 2.4 $\pm$ 1.8 & 45.6 $\pm$ 1.9 $\pm$ 1.8 \\ 
30 & 58.2 $\pm$ 0.8 $\pm$ 2.2 & 60.0 $\pm$ 0.8 $\pm$ 2.2 \\ 
40 & 64.8 $\pm$ 0.8 $\pm$ 2.3 & 66.5 $\pm$ 0.9 $\pm$ 2.3 \\ 
75 & 81.7 $\pm$ 0.7 $\pm$ 2.8 & 83.5 $\pm$ 0.6 $\pm$ 2.8 \\ 
150 & 107.8 $\pm$ 1.2 $\pm$ 3.8 & 104.9 $\pm$ 1.6 $\pm$ 3.8 \\ 
\end{tabular}
\caption{Mean multiplicities of $\pi^+$ and $\pi^-$ mesons calculated as the integrals of the fits (Eq. \ref{eq:pion_yfit}) to the measured identified pion spectra in 10\% most \textit{central} Ar+Sc collisions at 19$A$-30\AGeVc. The values are provided with statistical ($\sigma_\text{stat}$) and systematic ($\sigma_\text{sys}$) uncertainties in the form: $\langle \pi \rangle \pm \sigma_\text{stat} \pm \sigma_\text{sys}$.}
\label{tab:pion_mult}
\end{table}

\begin{figure}[h]
\centering
\includegraphics[width=0.4\linewidth, page=4]{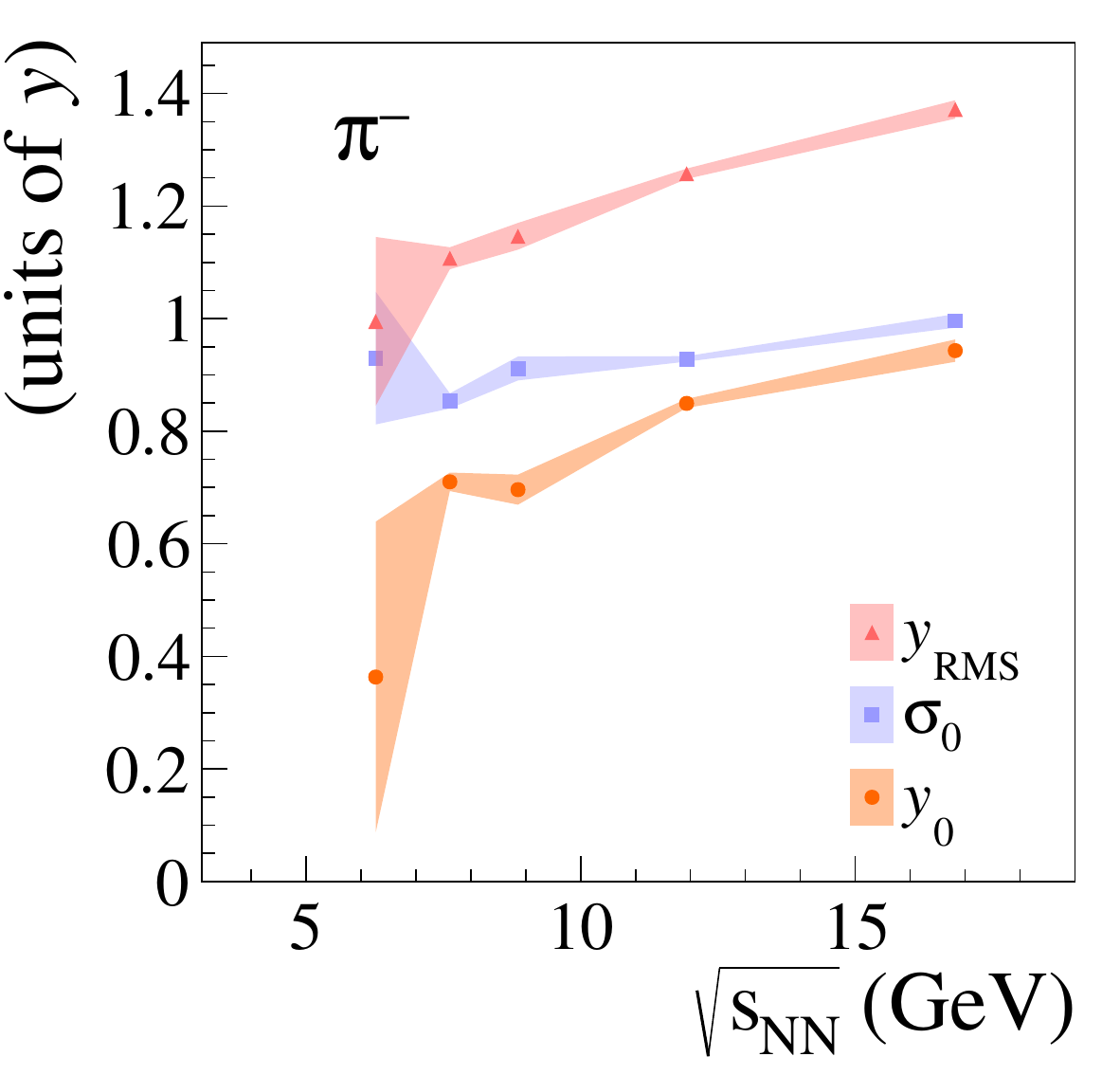}~
\includegraphics[width=0.4\linewidth, page=3]{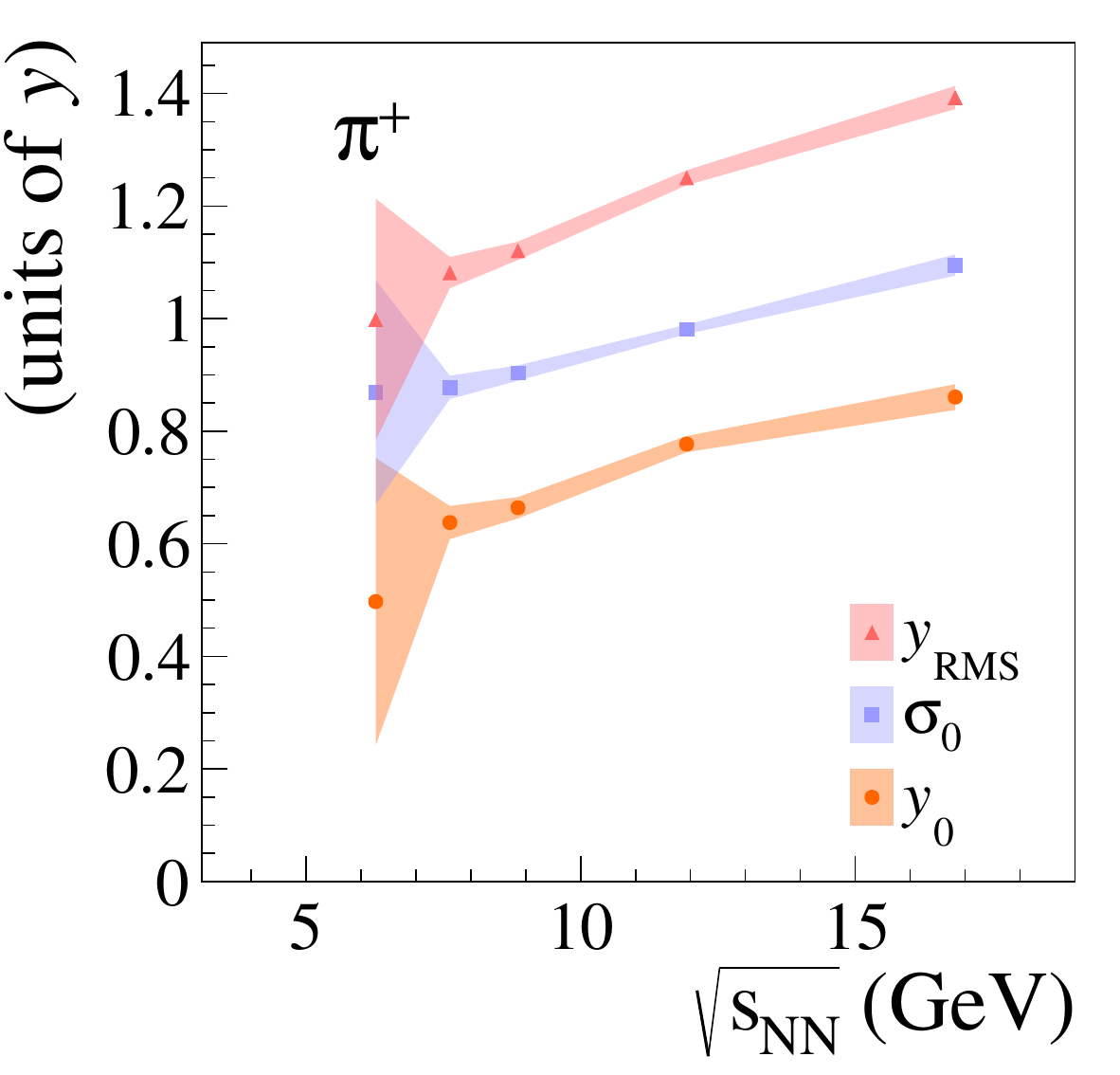}\\
\caption{The parameters of double-Gaussian fit (Eqs. \ref{eq:pion_yfit} and \ref{eq:rapidiry_rms}) to charged pion rapidity spectra in 10\% most \textit{central} Ar+Sc collisions. The shaded bands illustrate the statistical uncertainties of the fit.
}
\label{fig:yspectra_pions_fit}
\end{figure}

The parameters of the function fitted to charged pion rapidity distribution are shown in dependence on collision energy in Fig. \ref{fig:yspectra_pions_fit}. Both the width $\sigma$ and the distance between the peaks increase with increasing collision energy. Comparison of $y_\text{RMS}$ with other particle species is presented in Fig. \ref{fig:yspectra_y_rms_comparison}.

\begin{figure}[h]
\centering
\includegraphics[width=0.4\linewidth]{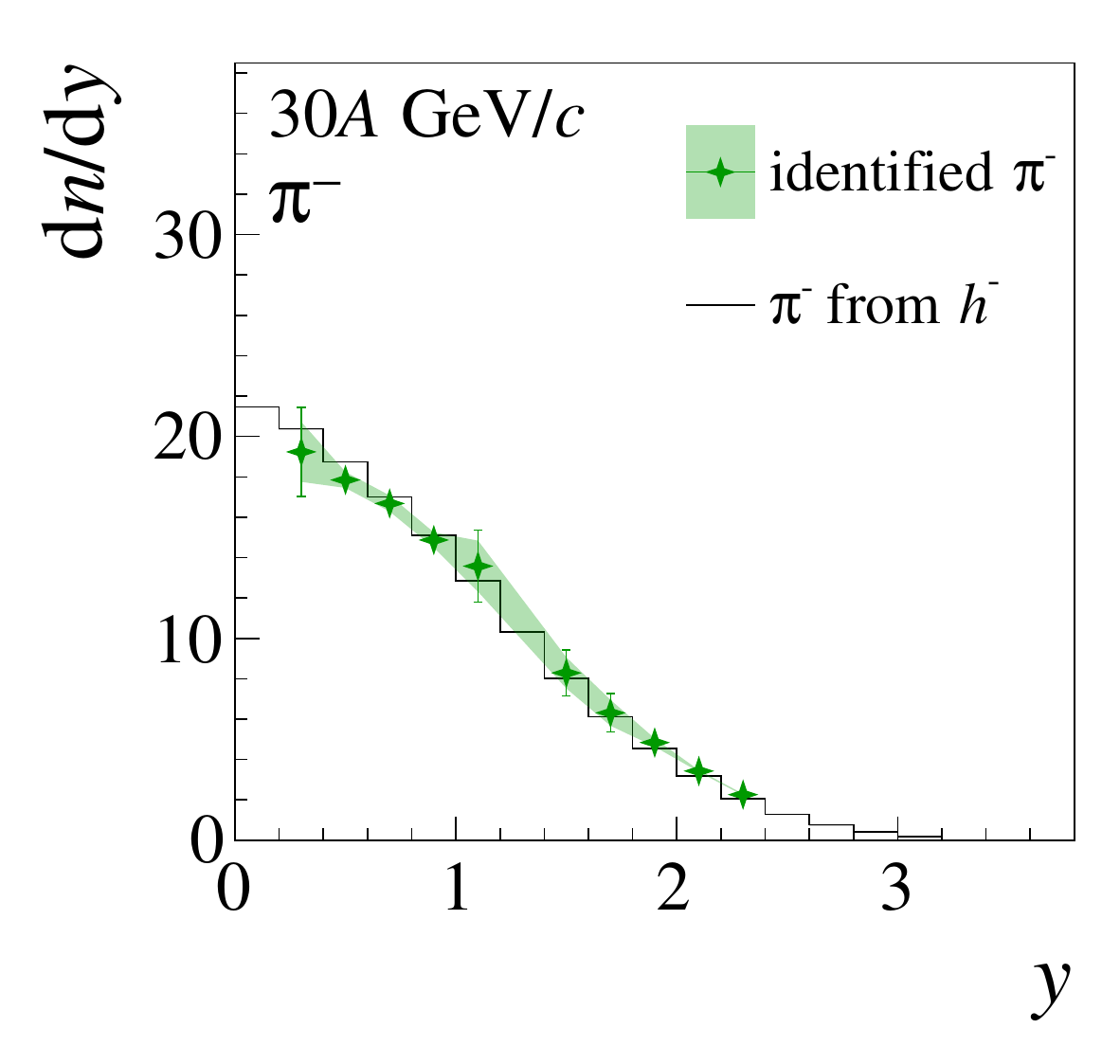}~
\includegraphics[width=0.4\linewidth]{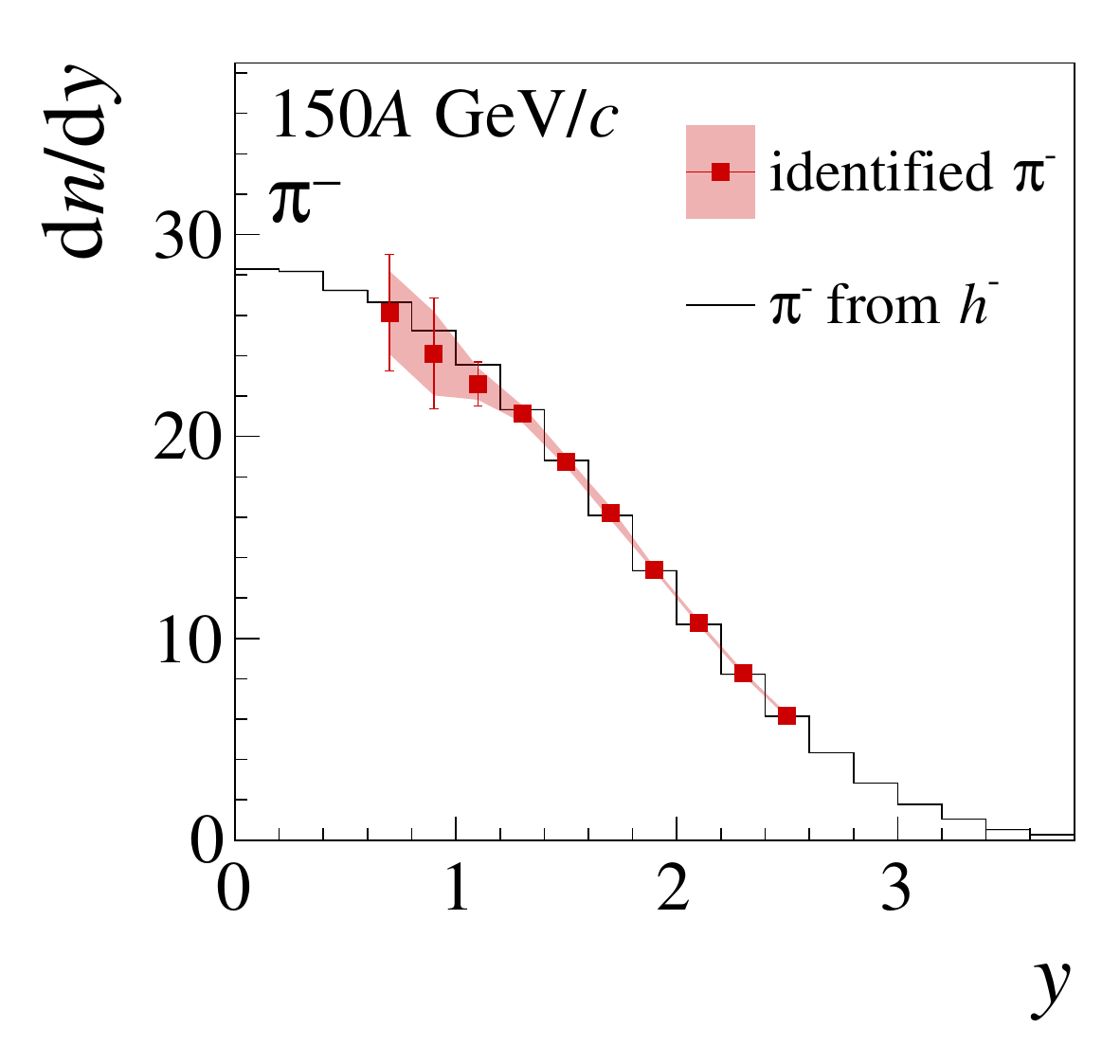}
\caption{Comparison of rapidity spectra of negatively charged pions measured with \dEdx and tof-\dEdx methods (colored markers) against the results obtained with the $h^-$ method \cite{NA61SHINE:2021nye} (black line). Both results consider 10\% most \textit{central} Ar+Sc collisions, and the comparison is displayed for beam momenta of 30$A$ and 150 \AGeVc. The statistical uncertainties do not exceed the size of the markers and the systematic uncertainties are shown as shaded bands.
}
\label{fig:pions_hminus_comparison}
\end{figure}

Finally, rapidity spectra of negatively charged pions were compared with the results of $h^-$ analysis \cite{NA61SHINE:2021nye} for all beam momenta. Comparisons for 30$A$ and 150\AGeVc are shown in Fig. \ref{fig:pions_hminus_comparison}. The results from the $h^-$ method were obtained for the 0-5\% centrality interval and to match the results of this study were scaled by the ratio of numbers of wounded nucleons $N_W^{\text{0-10\%}}/N_W^{\text{0-5\%}}$. The results from the two different analysis methods agree within calculated uncertainties.

\FloatBarrier

\subsection{Charged kaons}
\label{sec:kaons}

\subsubsection{Transverse momentum spectra}

\begin{figure}[h]
\centering
\includegraphics[width=0.7\linewidth]{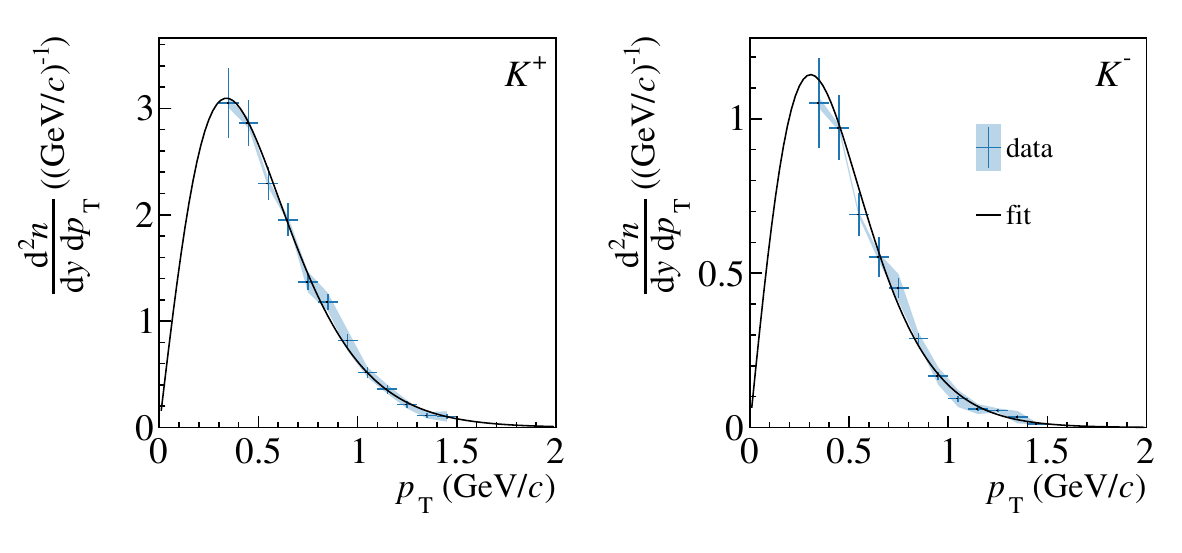}
\includegraphics[width=0.7\linewidth]{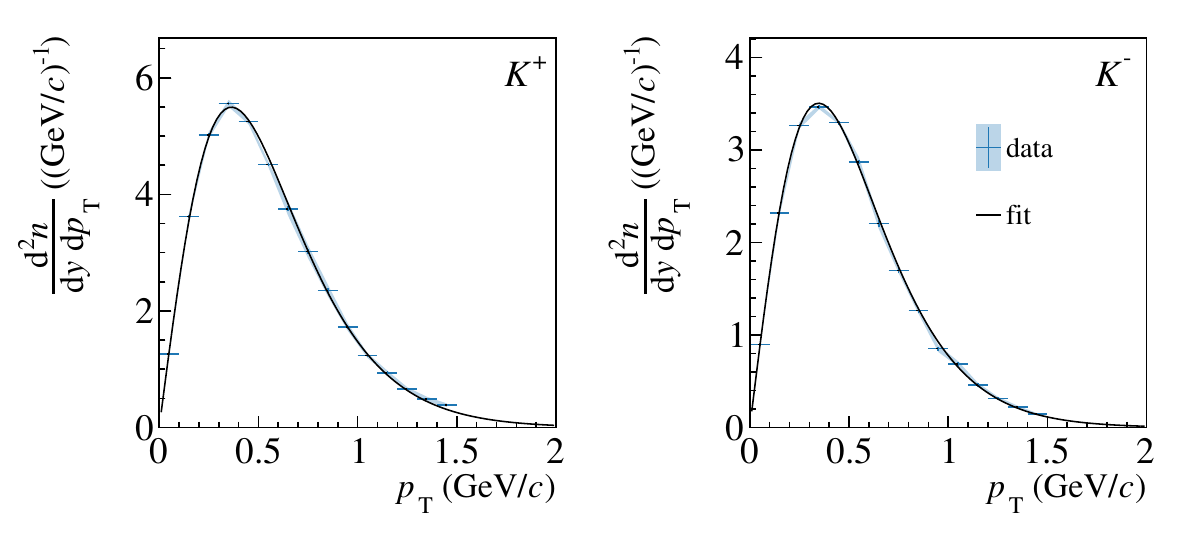}
\caption{Example fits to charged kaon transverse momentum spectra ($p_{\text{beam}}=19$\AGeVc at $0.4<y<0.6$ and $p_{\text{beam}}=150$\AGeVc at $0.6<y<0.8$, \textit{top} and \textit{bottom} panels respectively) obtained for 10\% most \textit{central} Ar+Sc collisions. 
Exponential fits are performed in the range of $\pt \in [0.0,1.5]$. The fitted functions are used to extrapolate the yields beyond $p_T=1.5$ GeV/$c$ and interpolate the yields in case a gap in acceptance appears due to different coverage of PID methods. A full set of transverse momentum spectra with corresponding fits is presented in Figures \ref{fig:rap_slices_k_pos} and \ref{fig:rap_slices_k_neg} in Appendix \ref{sec:additional_plots}. The vertical bars represent statistical uncertainties and the shaded bands stand for the systematic uncertainties.
}
\label{fig:pTfits_kaons}
\end{figure}
The acceptance for charged kaons, identified with \dEdx and \tofdedx PID methods at 13$A$--150\AGeVc, covers the region close to mid-rapidity as well as a large part of the spectrum in forward rapidity.

In order to obtain $dn/dy$ yields, the data is extrapolated in $p_\text{T}$ to account for unmeasured regions at high values of $p_\text{T}$ and in rare cases at low $p_\text{T}$ as well.
Kaon spectra are well approximated by single-interval exponential fits. The effects of collective flow bend the spectra up (or down) in the case of lighter
(or more massive) particles, while kaon \pt distribution remains approximately exponential. Moreover, the contribution to the spectra from products of weak decays is small.
The fit formula is analogous to the one used in the fits to the pion spectra (Eq. \ref{eq:pion_pT}).

\begin{figure}[h]
	\centering
\includegraphics[width=0.5\linewidth]{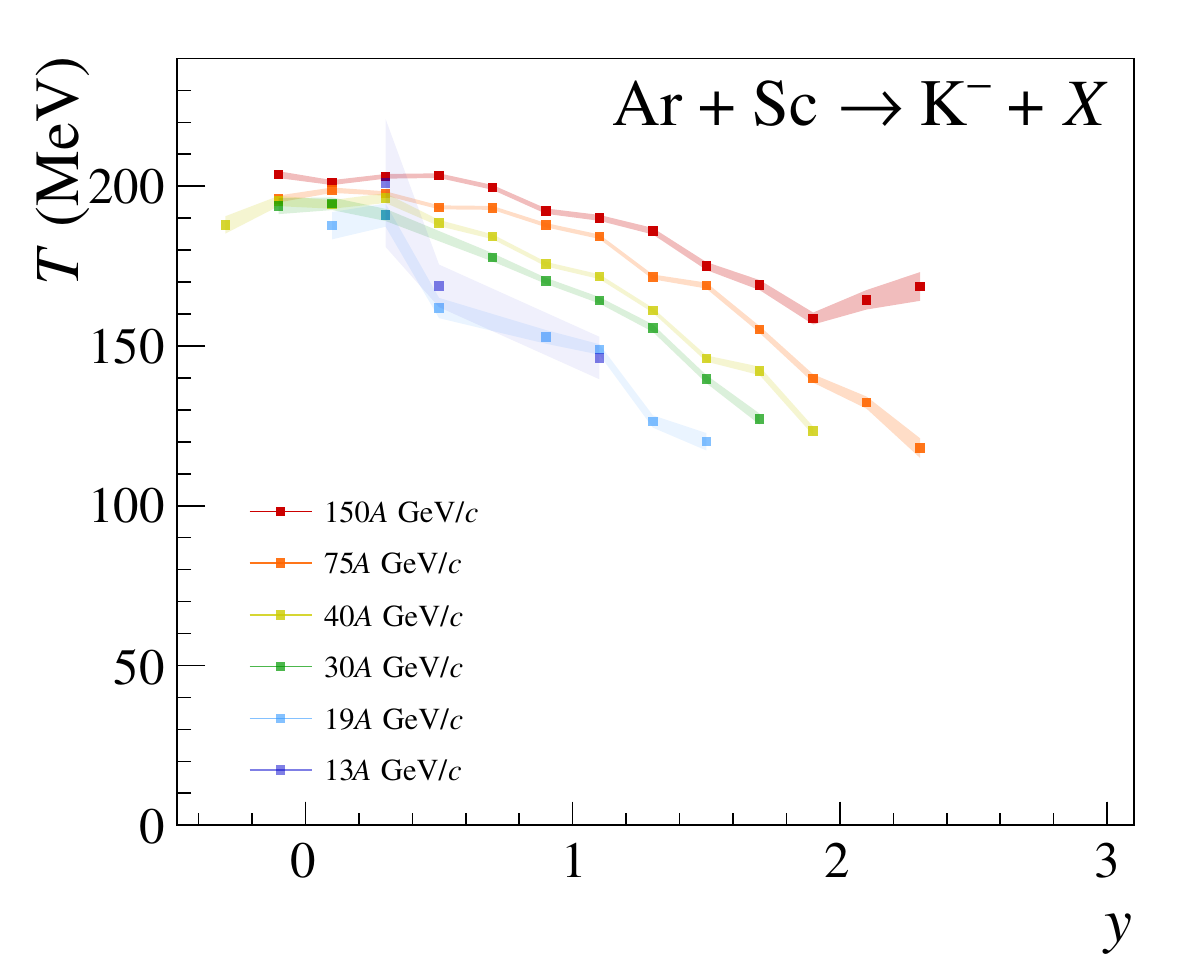}~
\includegraphics[width=0.5\linewidth]{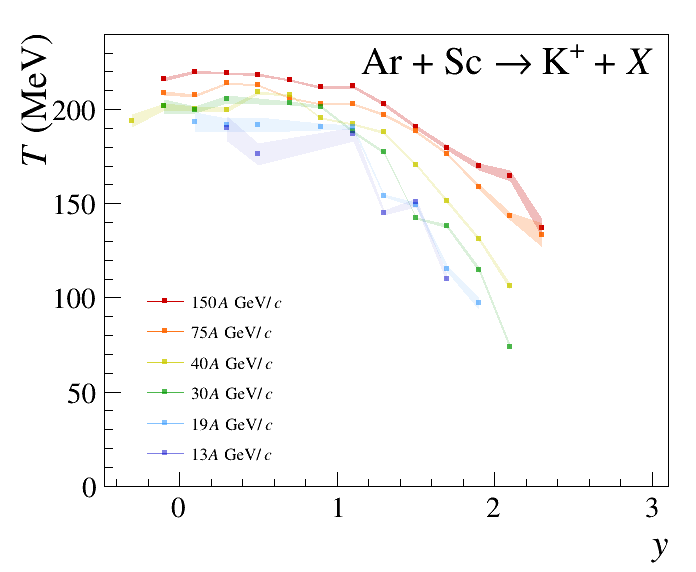}
	\caption{Rapidity dependence of the inverse slope parameter $T$ fitted to charged kaon $p_\text{T}$ distribution in 10\% most \textit{central} Ar+Sc collisions. The shaded bands illustrate the statistical uncertainties of the fit.}
	\label{fig:slope_ydep}
\end{figure}

The function is fitted in the acceptance region and its integral beyond the acceptance is added to the measured data. Example fits are shown in Fig. \ref{fig:pTfits_kaons}.
The fit of the transverse momentum spectra with Eq.~\ref{eq:pion_pT} determines the inverse slope parameter $T$. Figure~\ref{fig:slope_ydep} shows the rapidity dependence of $T$ for each measured beam momentum.
It can be observed that at 75\textit{A} and 150\AGeVc the fitted values of $T$ are approximately constant up to $y$$\approx$0.8, which agrees with observations made in the analysis of Pb+Pb \cite{note_MvL,vanLeeuwen:2003it} data.

\begin{figure}[h]
\centering
\includegraphics[page=1, width=0.49\textwidth, trim=30 10 100 0, clip]{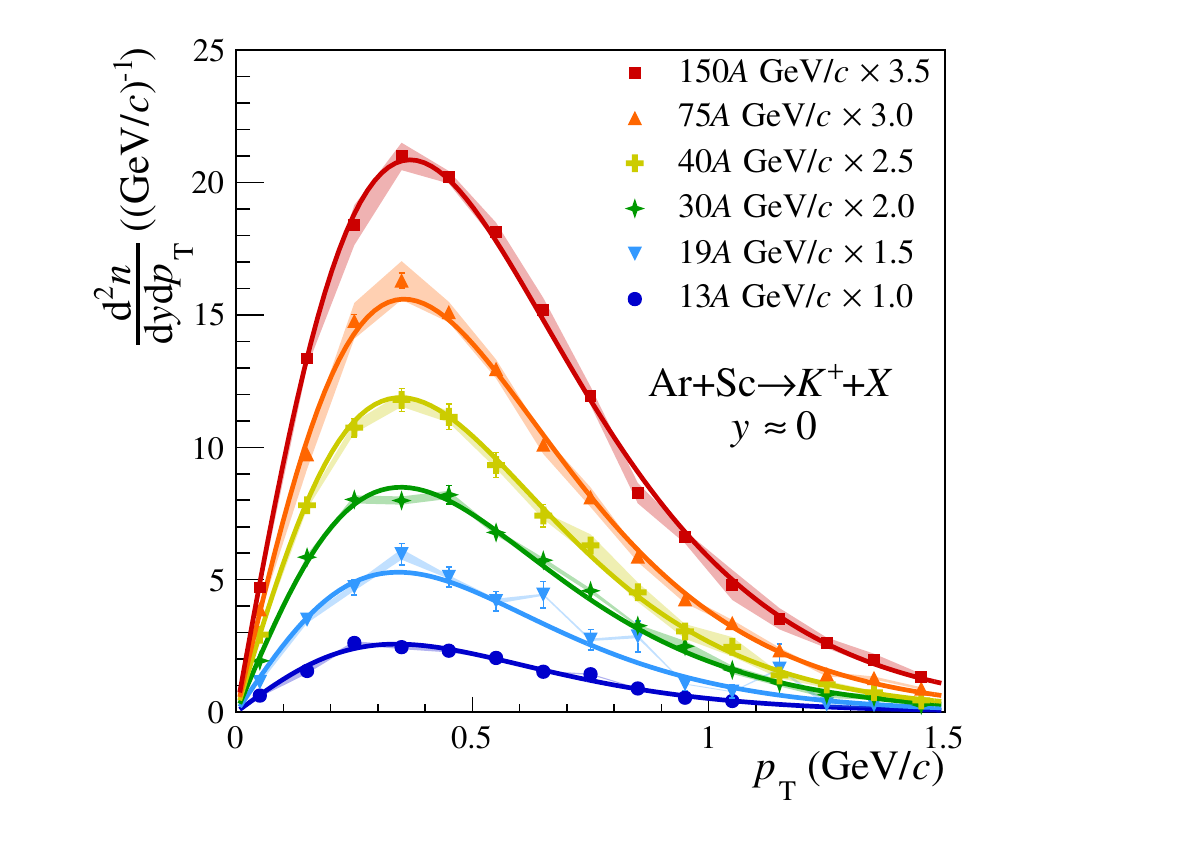}
\includegraphics[page=1, width=0.49\textwidth, trim=30 10 100 0, clip]{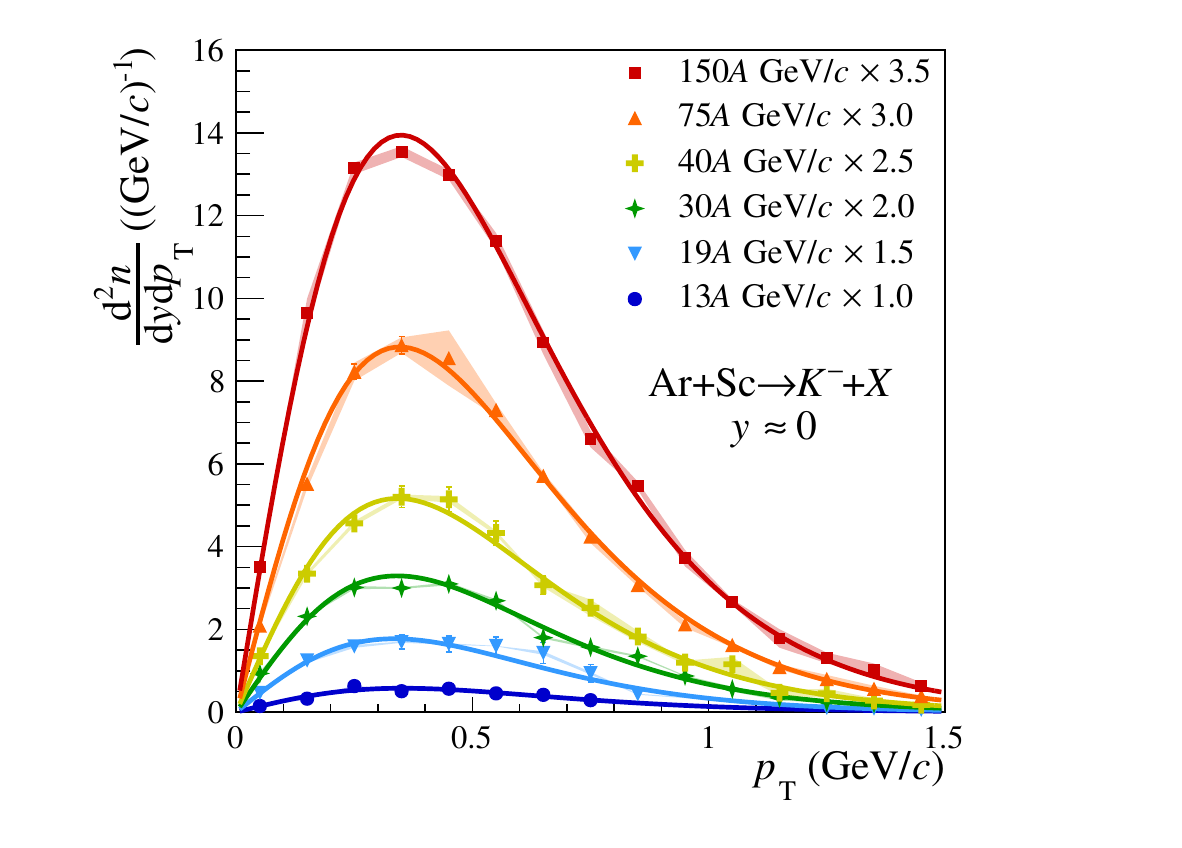}
\caption{Transverse momentum spectra of $K^+$ (\textit{left}) and $K^-$ (\textit{right}) at mid-rapidity produced in 10\% most \textit{central} Ar+Sc collisions. Error bars show statistical uncertainties, while shaded bands correspond to systematic uncertainties. The lines represent the fitted functions. The spectra were scaled for better readability.}
\label{fig:pT_midrap}
\end{figure}

The transverse momentum spectra of charged kaons at mid-rapidity for six
beam momenta are presented in Fig. \ref{fig:pT_midrap}. The mid-rapidity is defined as the range from 0.0 to 0.2, with the exception of $p_\text{beam}$=13\AGeVc, where due to limited acceptance the mid-rapidity range is taken as from 0.2 to 0.4. The symmetry of the charged kaon spectra with respect to $y=0$ was verified with available theoretical models to be within 1\%. Thus for 13\textit{A}, 19\textit{A} and 30\AGeVc the missing points in the low \pt region (see Fig. \ref{fig:final2D}) are complemented with measured reflections of the spectrum with respect to $y=0$.

The $\text{d}n/\text{d}y$ yield of charged kaons at mid-rapidity is obtained from the measured spectrum (including points reflected with respect to $y=0$) and fitted with an exponential function (Eq. \ref{eq:pion_pT}). The $\text{d}n/\text{d}y$ yield calculated with the inclusion of reflected points agrees within total uncertainties with the value obtained without reflection. The yields of $K^+$ and $K^-$ calculated with this procedure and the $K/\pi$ ratios at mid-rapidity together with their statistical and systematic uncertainties are summarized in Table \ref{tab:midrap_dndy_ktopi}. Table \ref{tab:midrap_inv_slope_mean_pt} summarizes values of the inverse slope parameter $T$ resulting from the fit of \pt spectra of charged kaons. Additionally, mean transverse momenta of $K$ mesons at mid-rapidity were calculated, based on their \pt spectra and fitted functions, and are listed in Table \ref{tab:midrap_inv_slope_mean_pt}. As the analysis presented in this paper does not offer acceptance for charged pions at mid-rapidity, negatively charged pion yields at mid-rapidity were taken from Ref. \cite{NA61SHINE:2021nye}, and scaled to $10\%$ most \textit{central} Ar+Sc events with the use of number on wounded nucleons calculated within \Epos WNM (see Sec. \ref{sec:pions_rapidity_spec}). Based on the agreement of $\pi^\pm$ mean multiplicities well within total uncertainties (Table \ref{tab:pion_mult}) and very similar $y$ distribution shapes (Fig. \ref{fig:yspectra_pions_fit}), it was assumed that the yields of charged pions at mid-rapidity are close enough for the purpose of the $K/\pi$ ratio calculation.

\begin{table}[h]
\small
{\centering
	\begin{tabular}{c||c|c|c|c}
		\parbox{1.1cm}{\,\,$p_{\textrm{beam}}$ (GeV/\textit{c})}& $\left( \frac{\text{d}n}{\text{d}y}\right)_{y\approx0} (K^+)$ & $\left( \frac{\text{d}n}{\text{d}y}\right)_{y\approx0} (K^-)$ & $K^+ /\pi^+ $ at $y$$\approx$0 & $K^-/\pi^-$ at $y$$\approx$0 
        \\[0.3cm]
		\hline \hline
13$A$  & $ 1.748 \pm 0.040 \pm 0.088$ & $ 0.411 \pm 0.014 \pm 0.032$ & $ 0.1225 \pm 0.0028 \pm 0.0093$ & $ 0.0288 \pm 0.0010 \pm 0.0034$ \\
19$A$  & $ 2.624 \pm 0.102 \pm 0.076$ & $ 0.815 \pm 0.035 \pm 0.029$ & $ 0.1449 \pm 0.0056 \pm 0.0065$ & $ 0.0450 \pm 0.0019 \pm 0.0025$ \\
30$A$  & $ 2.994 \pm 0.042 \pm 0.090$ & $ 1.109 \pm 0.021 \pm 0.025$ & $ 0.1395 \pm 0.0020 \pm 0.0064$ & $ 0.0517 \pm 0.0010 \pm 0.0018$ \\
40$A$  & $ 3.283 \pm 0.041 \pm 0.118$ & $ 1.417 \pm 0.024 \pm 0.048$ & $ 0.1438 \pm 0.0018 \pm 0.0079$ & $ 0.0621 \pm 0.0011 \pm 0.0032$ \\
75$A$  & $ 3.732 \pm 0.016 \pm 0.148$ & $ 2.029 \pm 0.012 \pm 0.069$ & $ 0.1476 \pm 0.0007 \pm 0.0090$ & $ 0.0802 \pm 0.0005 \pm 0.0042$ \\
150$A$ & $ 4.422 \pm 0.013 \pm 0.154$ & $ 2.785 \pm 0.010 \pm 0.072$ & $ 0.1563 \pm 0.0005 \pm 0.0085$ & $ 0.0984 \pm 0.0004 \pm 0.0041$ \\
	\end{tabular}\\[0.2cm]}
\normalsize
\caption{Numerical values of the $K$ meson yields ($\text{d}n/\text{d}y$) and the ratios of $K^+/\pi^+$ and $K^-/\pi^-$ yields at mid-rapidity in 10\% most \textit{central} Ar+Sc collisions. The values are provided with statistical ($\sigma_\text{stat}$) and systematic ($\sigma_\text{sys}$) uncertainties in the form: $(\cdots) \pm \sigma_\text{stat} \pm \sigma_\text{sys}$. The $\pi^+$ and $\pi^-$ yields at mid-rapidity are both taken as the $\pi^-$ yield measured with the $h^-$ method \cite{NA61SHINE:2021nye}, scaled to 10\% most \textit{central} events (see text for details).}
\label{tab:midrap_dndy_ktopi}
\end{table}

\begin{table}[h]
\small
\centering
\begin{tabular}{c||c|c|c|c}
		\parbox{1.1cm}{\,\,$p_{\textrm{beam}}$ (GeV/\textit{c})} & $T(K^+)$ (MeV) &  $T(K^-)$ (MeV) &$\langle p_T\rangle_{K^+}$ (MeV)&$\langle p_T\rangle_{K^-}$ (MeV)\\[0.3cm]
  \hline \hline
13$A$  & $ 191.9 \pm 6.4 \pm 12.8$ & $ 204   \pm 18  \pm 15 $ & $ 508 \pm 22 \pm 7 $ & $ 526 \pm 38 \pm 8 $ \\
19$A$  & $ 195.7 \pm 5.3 \pm 7.2$  & $ 187.5 \pm 4.2 \pm 3.9$ & $ 536 \pm 45 \pm 9 $ & $ 507 \pm 40 \pm 7 $ \\
30$A$  & $ 201.1 \pm 1.9 \pm 12.7$ & $ 193.3 \pm 1.9 \pm 4.0$ & $ 528 \pm 13 \pm 9 $ & $ 518 \pm 17 \pm 7 $ \\
40$A$  & $ 200.3 \pm 1.4 \pm 8.8$  & $ 194.3 \pm 1.4 \pm 2.8$ & $ 528 \pm 11 \pm 12$ & $ 518 \pm 15 \pm 12$ \\
75$A$  & $ 207.4 \pm 0.8 \pm 6.5$  & $ 198.8 \pm 0.8 \pm 2.7$ & $ 539 \pm 4  \pm 11$ & $ 525 \pm 5  \pm 12$ \\
150$A$ & $ 219.9 \pm 0.7 \pm 11.8$ & $ 201.1 \pm 0.8 \pm 6.2$ & $ 563 \pm 3  \pm 10$ & $ 527 \pm 3  \pm 8 $ \\

\end{tabular}\\[0.2cm]
\normalsize
\caption{Inverse slope parameter $T$ and mean transverse momenta of $K$ mesons at mid-rapidity in 10\% most \textit{central} Ar+Sc collisions. The values are provided with statistical ($\sigma_\text{stat}$) and systematic ($\sigma_\text{sys}$) uncertainties in the form: $(\cdots) \pm \sigma_\text{stat} \pm \sigma_\text{sys}$.}
\label{tab:midrap_inv_slope_mean_pt}
\end{table}

\subsubsection{Rapidity spectra}
\label{sec:kaon_rapidity}
\begin{figure}[h]
	\centering
	\includegraphics[width=0.3\linewidth]{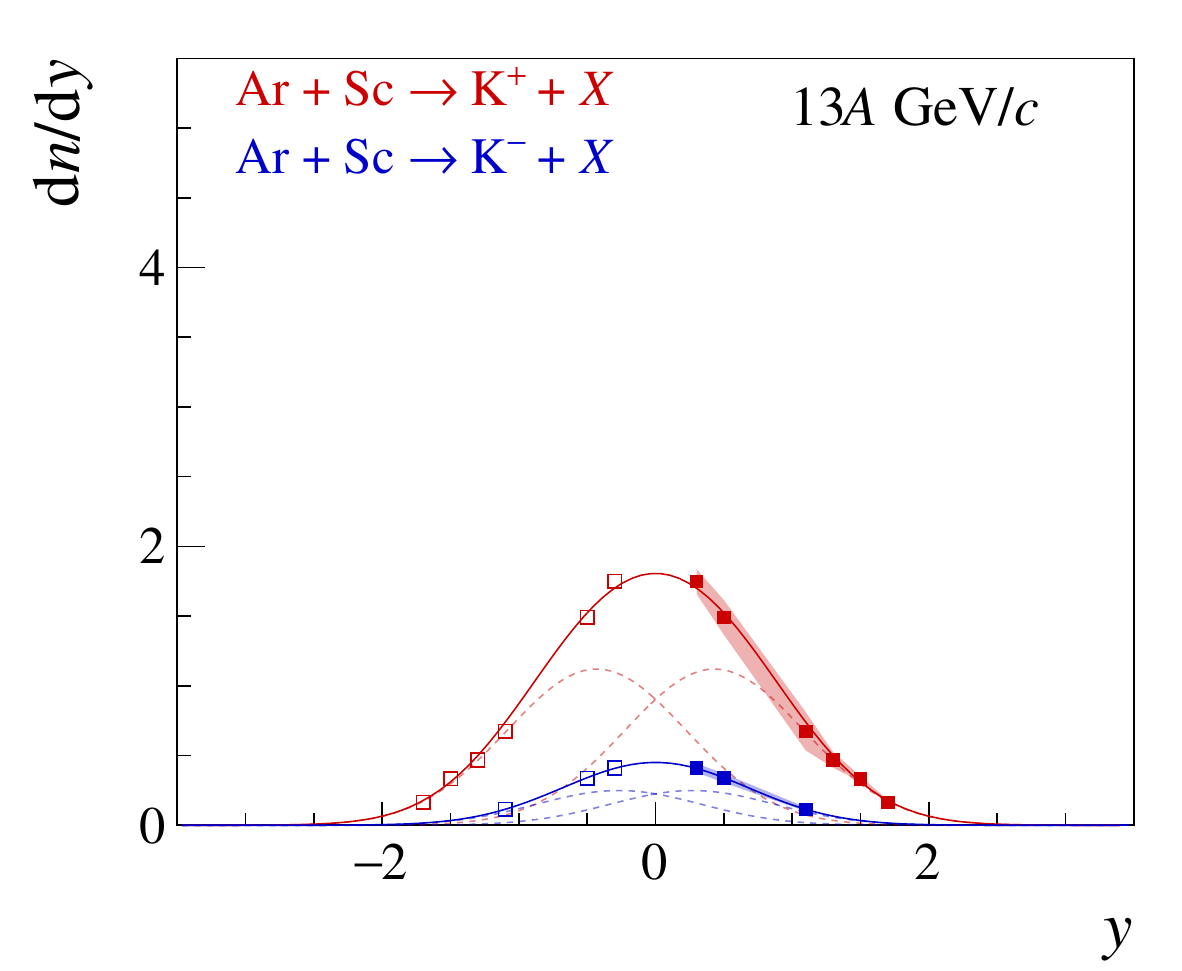}~
	\includegraphics[width=0.3\linewidth]{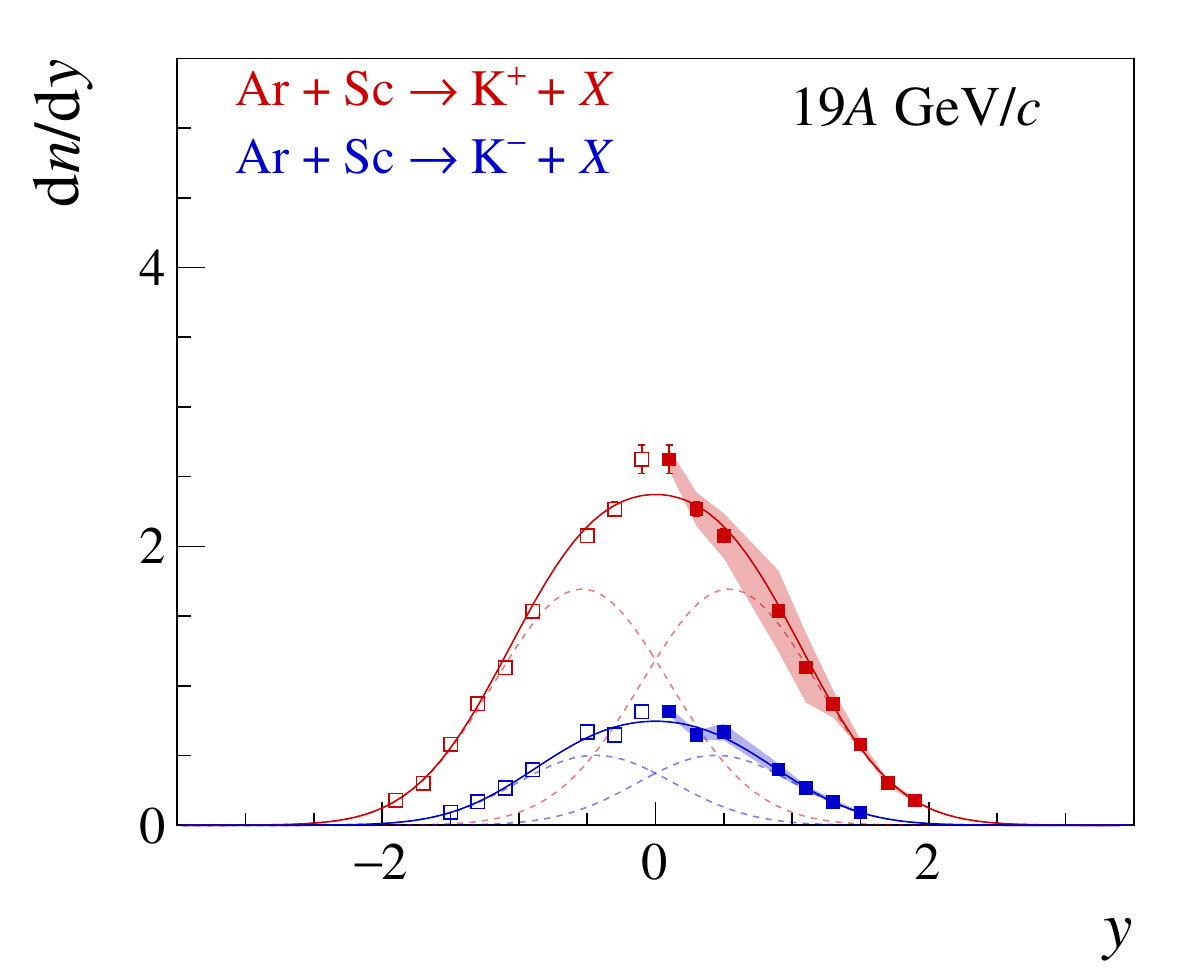}~
	\includegraphics[width=0.3\linewidth]{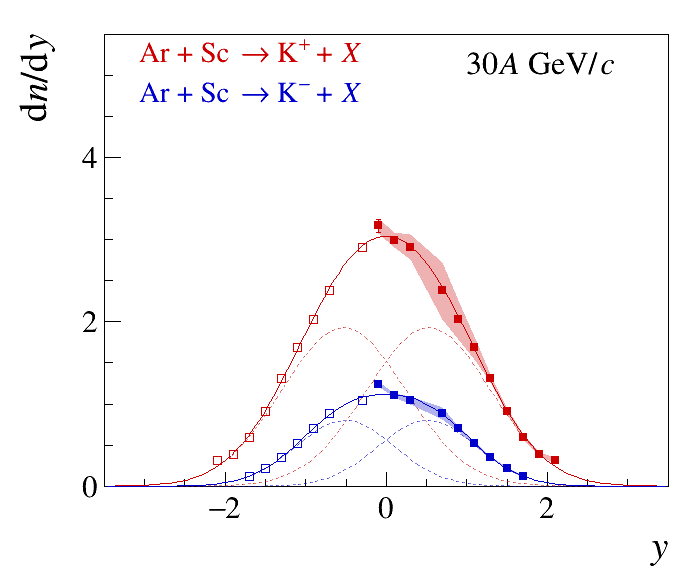}\\
	\includegraphics[width=0.3\linewidth]{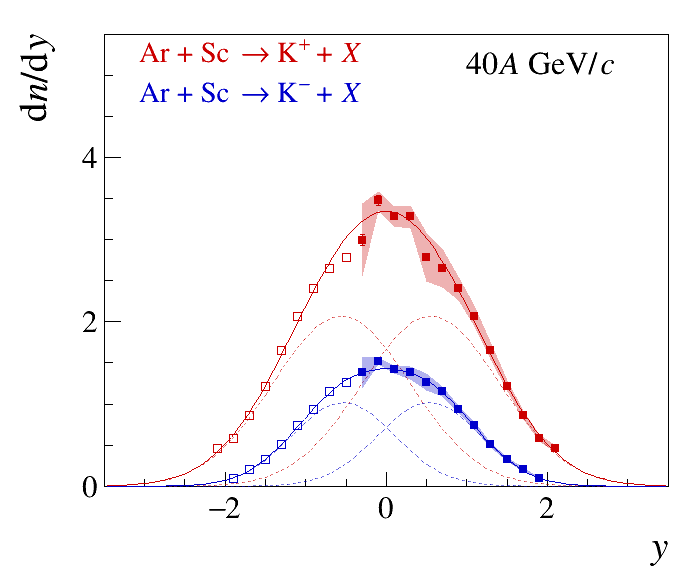}~
	\includegraphics[width=0.3\linewidth]{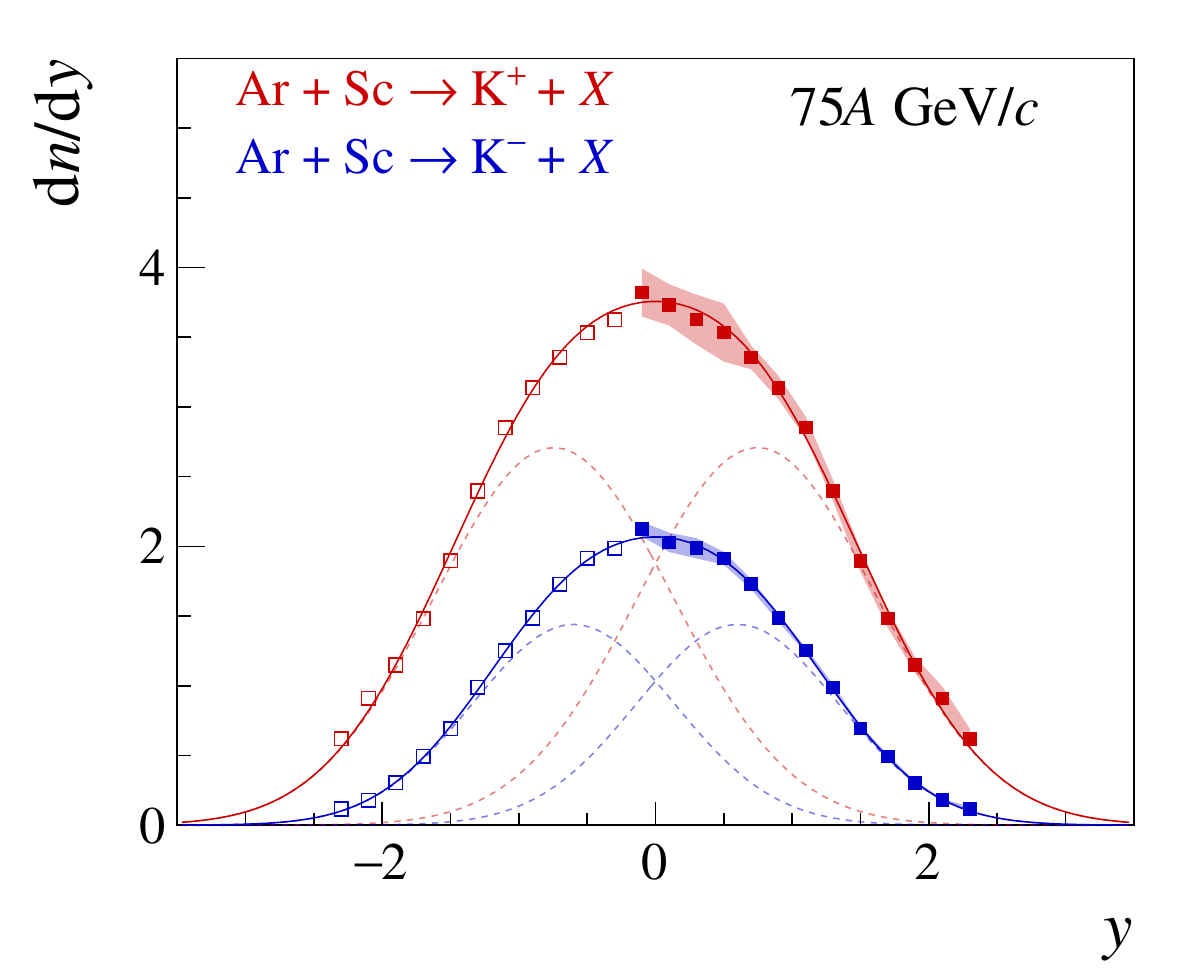}~
	\includegraphics[width=0.3\linewidth]{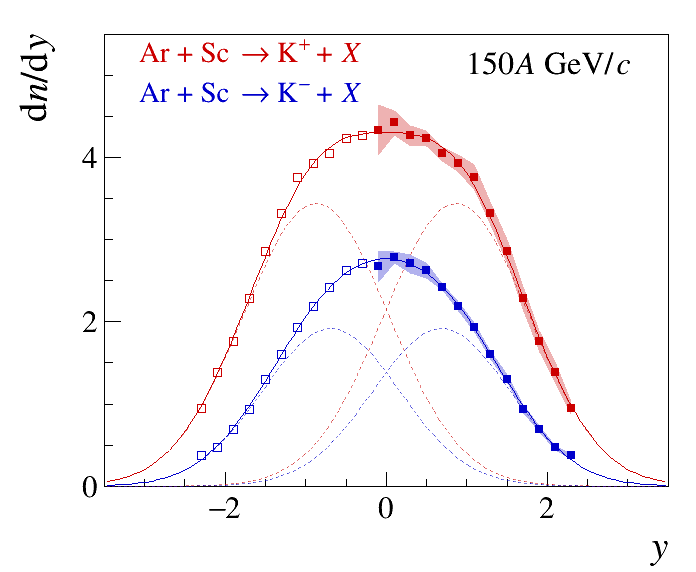}
	\caption{Rapidity distributions of $K^+$ and $K^-$ in 10\% most \textit{central} Ar+Sc collision at 13\textit{A}--150\textit{A} GeV/\textit{c} beam momenta. Data points are plotted along with double-Gaussian fits (Eq. \ref{eq:pion_yfit}). The vertical bars represent statistical uncertainties and the shaded bands stand for the systematic uncertainties. Open markers represent data reflected wrt. $y=0$.}
	\label{fig:kaonrapidity}
\end{figure}

Figure \ref{fig:kaonrapidity} shows the kaon rapidity distributions.
The spectra are fitted with the sum of two Gaussians placed symmetrically with respect to $y=0$, defined in Eq. \ref{eq:pion_yfit}. In the absence of data in the backward hemisphere, a symmetry of amplitudes in the forward and backward hemispheres is assumed ($A_\text{rel}=1$).

\begin{figure}[h]
\centering
\includegraphics[width=0.4\linewidth, page=6]{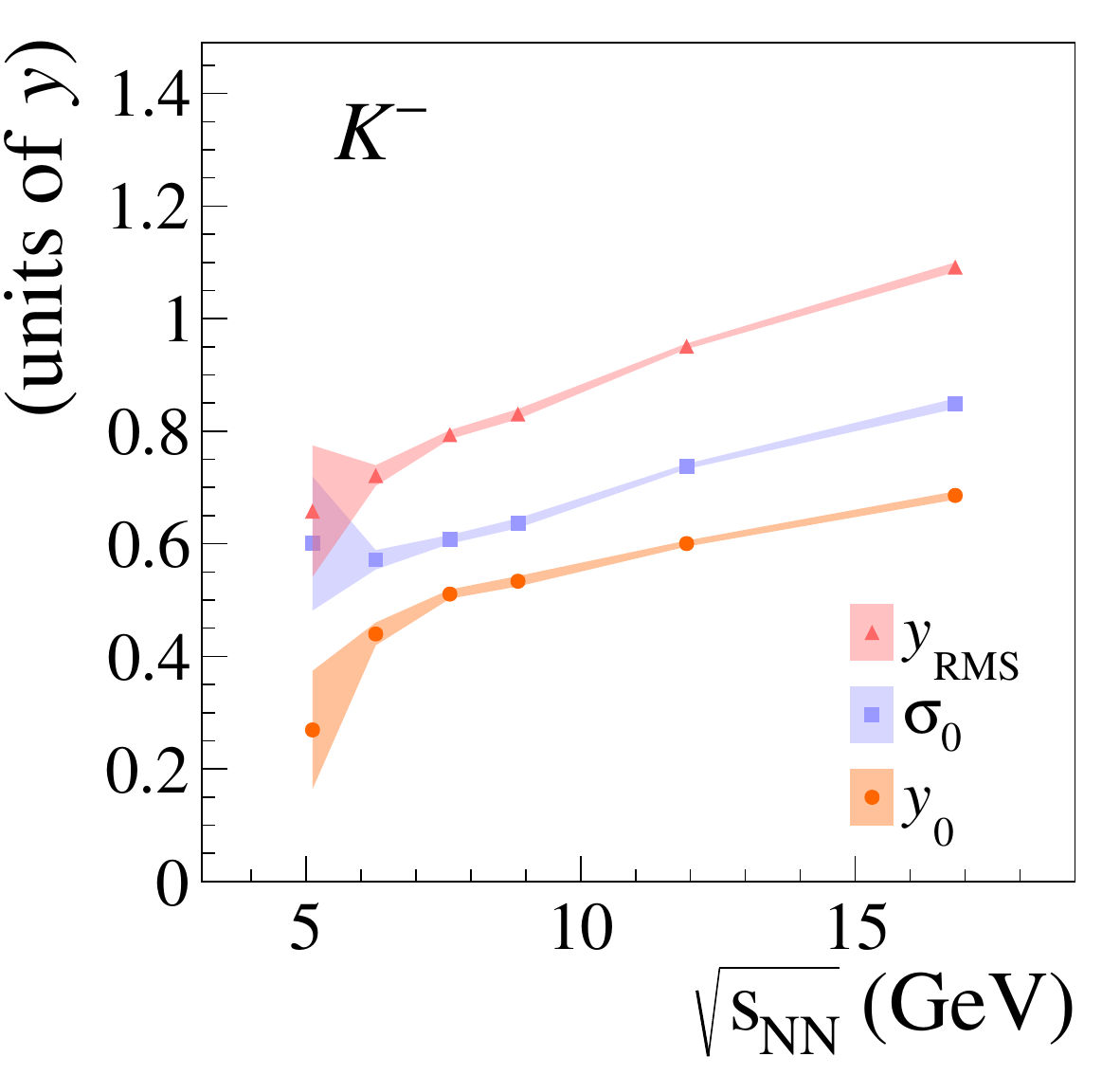}~
\includegraphics[width=0.4\linewidth, page=5]{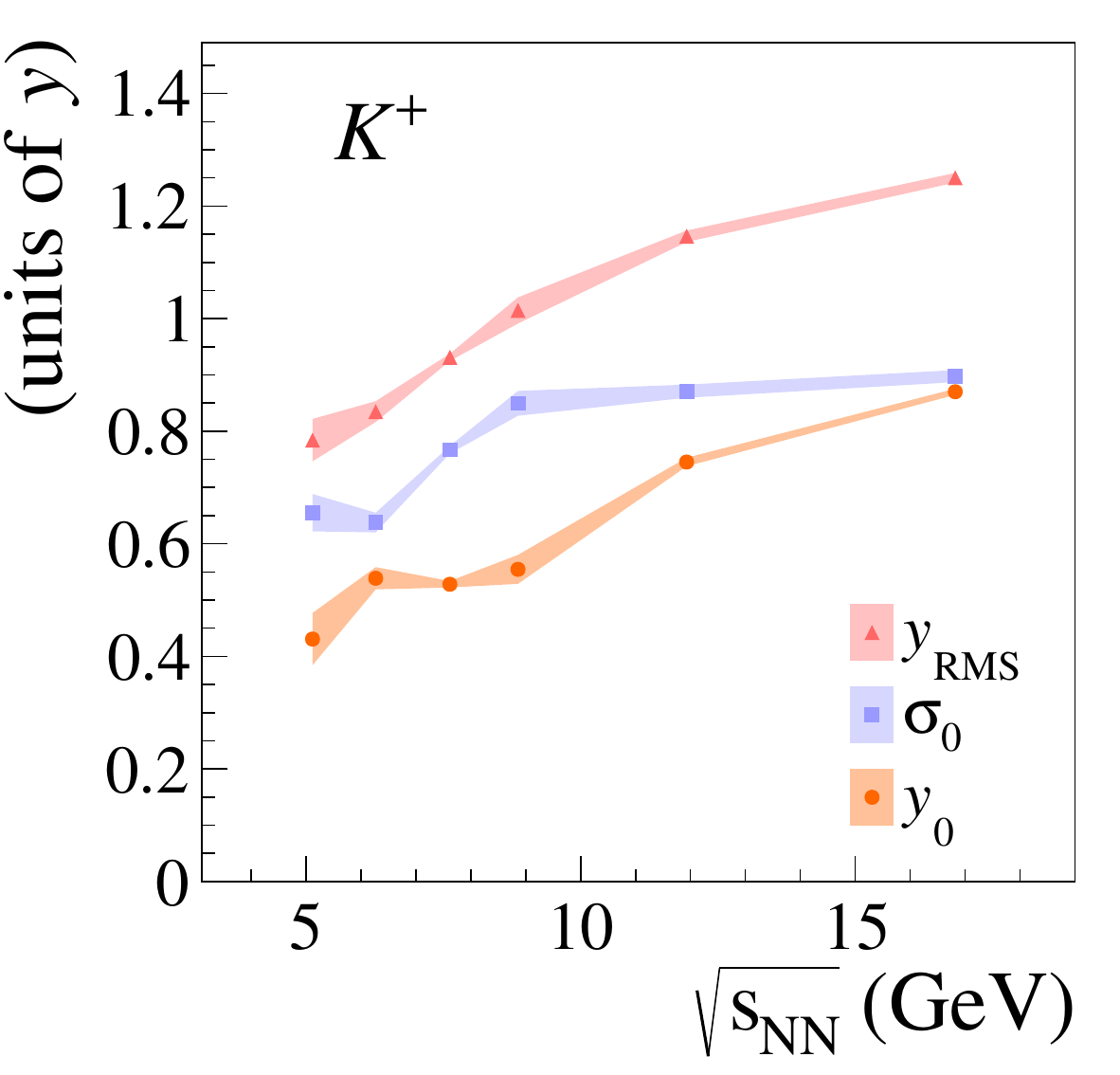}\\
\caption{The parameters of the double-Gaussian fit (Eqs. \ref{eq:pion_yfit} and \ref{eq:rapidiry_rms}) to the charged kaon rapidity spectra in the 10\% most \textit{central} Ar+Sc collisions as a function of beam momentum. All presented quantities ($\sigma_0$, $y_0$, $y_\text{RMS}$) increase with increasing beam momentum. See text for details and exact fitting model formula. The shaded bands illustrate the statistical uncertainties of the fit.
}
\label{fig:yspectra_kaons_fit}
\end{figure}

Figure \ref{fig:yspectra_kaons_fit} shows the properties of the parameters fitted to the rapidity distributions of $K^+$ and $K^-$ at six collision energies.
A smooth evolution of the fit parameters is observed, $\sigma_0$, $y_0$ and $y_\text{RMS}$ increase towards higher beam momenta. Comparison of $y_\text{RMS}$ with other particle species is presented in Fig. \ref{fig:yspectra_y_rms_comparison}.

To obtain the full phase space ($4\pi$) mean kaon multiplicity, the measured $\frac{dn}{dy}$ spectra are supplemented with extrapolated yields by integration of the fitted function (Eq. \ref{eq:pion_yfit}) in the unmeasured region. The mean multiplicities of $K^+$ and $K^-$ calculated at each beam momentum and their respective ratios to mean multiplicities of charged pions are listed in Table \ref{tab:kaon_4pi}.\\

\begin{table}[h]
	\small
	{\centering
		\begin{tabular}{c||c|c|c|c}
		\parbox{1cm}{\,\,$p_{\textrm{beam}}$ (GeV/\textit{c})}& $\langle K^+ \rangle$ & $\langle K^- \rangle$& $\langle K^+ \rangle/\langle \pi^+ \rangle$ & $\langle K^- \rangle/\langle \pi^- \rangle$ \\[0.3cm]
            \hline \hline

13 & 3.67 $\pm$ 0.07 $\pm$ 0.16 & 0.75 $\pm$ 0.01 $\pm$ 0.03 & 0.100 $\pm$ 0.002 $\pm$ 0.019 & 0.020 $\pm$ 0.001 $\pm$ 0.004 \\ 
19 & 5.43 $\pm$ 0.07 $\pm$ 0.26 & 1.45 $\pm$ 0.02 $\pm$ 0.06 & 0.116 $\pm$ 0.002 $\pm$ 0.012 & 0.031 $\pm$ 0.001 $\pm$ 0.003 \\ 
30 & 7.44 $\pm$ 0.04 $\pm$ 0.31 & 2.45 $\pm$ 0.02 $\pm$ 0.08 & 0.124 $\pm$ 0.001 $\pm$ 0.011 & 0.041 $\pm$ 0.001 $\pm$ 0.004 \\ 
40 & 8.76 $\pm$ 0.05 $\pm$ 0.42 & 3.26 $\pm$ 0.02 $\pm$ 0.15 & 0.132 $\pm$ 0.001 $\pm$ 0.012 & 0.049 $\pm$ 0.001 $\pm$ 0.005 \\ 
75 & 11.84 $\pm$ 0.08 $\pm$ 0.39 & 5.33 $\pm$ 0.01 $\pm$ 0.14 & 0.139 $\pm$ 0.001 $\pm$ 0.011 & 0.062 $\pm$ 0.001 $\pm$ 0.005 \\ 
150 & 15.53 $\pm$ 0.09 $\pm$ 0.52 & 8.16 $\pm$ 0.03 $\pm$ 0.25 & 0.145 $\pm$ 0.001 $\pm$ 0.017 & 0.076 $\pm$ 0.001 $\pm$ 0.009 \\ 
		\end{tabular}\\[0.2cm]}
	\normalsize
	\caption{\footnotesize Charged kaon mean multiplicities in 10\% most \textit{central} Ar+Sc collisions, together with $\langle K^+ \rangle/\langle \pi^+ \rangle$ and $\langle K^- \rangle/\langle \pi^- \rangle$ ratios.  The values are provided with statistical ($\sigma_\text{stat}$) and systematic ($\sigma_\text{sys}$) uncertainties in the form: $(\cdots) \pm \sigma_\text{stat} \pm \sigma_\text{sys}$.  The $\pi^+$ and $\pi^-$ mean multiplicities are both taken as the $\pi^-$ yield measured with the $h^-$ method \cite{NA61SHINE:2021nye}, scaled to 10\% most \textit{central} events (see text for details).}
	\label{tab:kaon_4pi}
 \end{table}

\FloatBarrier

\subsection{Protons and antiprotons}
\label{sec:protons}

\subsubsection{Transverse momentum spectra}

\begin{figure}[h]
\centering
\includegraphics[width=0.7\linewidth,trim={0 0 0 15cm},clip]{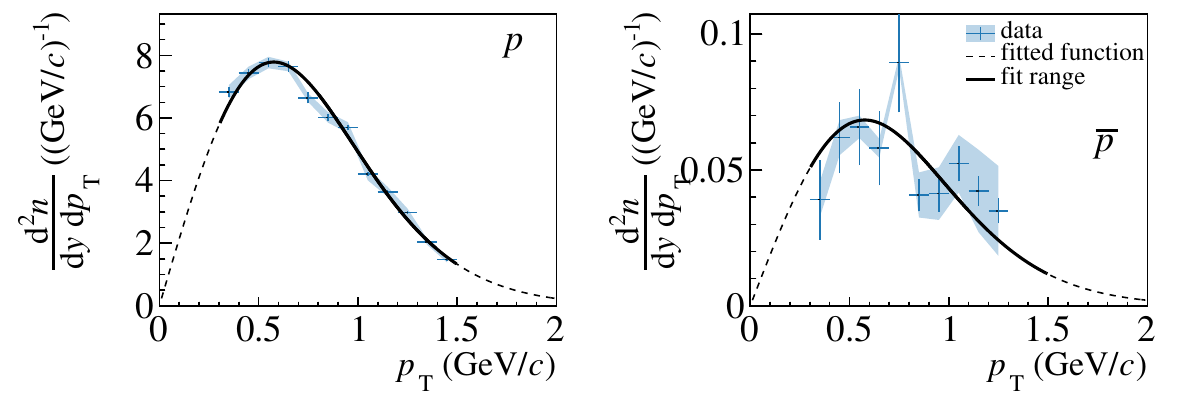}
\includegraphics[width=0.7\linewidth,trim={0 0 0 15cm},clip]{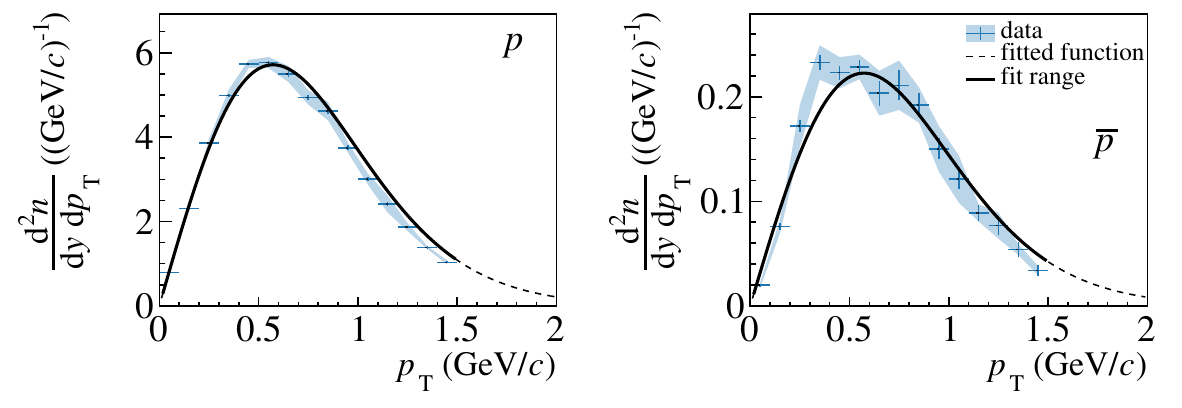}
\caption{Example fits to transverse momentum spectra of protons and anti-protons in 10\% most \textit{central} Ar+Sc collisions (\textit{top}: $p_{\text{beam}}=30$$A$ GeV/$c$ at $-0.2<y<0.0$ and \textit{bottom}: $p_{\text{beam}}=150$$A$ GeV/$c$ at $1.0<y<1.2$). The fitted functions are used to extrapolate the yields beyond $p_T=1.5$ GeV/$c$ and interpolate the yields in case a gap in acceptance appears due to the different coverage of the PID methods. A full set of transverse momentum spectra with corresponding fits is presented in Figures \ref{fig:rap_slices_p_pos} and \ref{fig:rap_slices_p_neg} in Appendix \ref{sec:additional_plots}. The vertical bars represent statistical uncertainties and the shaded bands stand for the systematic uncertainties. The thick solid black line illustrates the region at which the fit was performed and the dashed black line shows the interpolated and extrapolated parts of the spectrum.}
\label{fig:pTfits_protons}
\end{figure}

Similarly as in the case of $K$ mesons, in order to obtain proton and anti-proton $\text{d}n/\text{d}y$ yields, the data is extrapolated in $p_\text{T}$ to account for unmeasured regions of transverse momentum. Due to the effects of radial flow (and possibly other effects), (anti-)proton and spectra cannot be described by an exponential function. Thus, to assure a good description of the measured data, the blastwave model \cite{Schnedermann:1993ws} is fitted to \pt spectra of all identified hadrons ($p,~K^+,~\pi^+$ for the positive charge and $\bar{p},~K^-,~\pi^-$ for the negative charge). Another advantage of using the blastwave model comes from the additional constraints of the fit from other particle species, which allows for obtaining a reliable fit even with limited acceptance. Example fits are shown in Fig. \ref{fig:pTfits_protons}. The $\text{d}n/\text{d}y$ yield of protons and anti-protons at mid-rapidity is obtained in the same way as described for kaons in the previous section. The resulting yields of $p$ and $\bar{p}$, together with statistical and systematic uncertainties, are summarised in Table \ref{tab:midrap_dndy_protons}.

\begin{table}[h]
{\centering
\begin{tabular}{c||c|c}
\parbox{1.2cm}{\,\,$p_{\textrm{beam}}$ (GeV/\textit{c})}& $\left( \frac{\text{d}n}{\text{d}y}\right)_{y\approx0} (p)$ & $\left( \frac{\text{d}n}{\text{d}y}\right)_{y\approx0} (\bar{p})$ \\[0.3cm]
\hline \hline
13$A$  & 10.593 $\pm$ 0.206 $\pm$ 0.632 & - \\
19$A$  & 9.194 $\pm$ 0.115 $\pm$ 0.627 & - \\
30$A$  & 7.652 $\pm$ 0.045 $\pm$ 0.551 & 0.066 $\pm$ 0.003 $\pm$ 0.010 \\
40$A$  & 6.877 $\pm$ 0.013 $\pm$ 0.252 & 0.122 $\pm$ 0.002 $\pm$ 0.020 \\
75$A$  & 5.256 $\pm$ 0.011 $\pm$ 0.180 & 0.248 $\pm$ 0.002 $\pm$ 0.022 \\
150$A$ & 4.481 $\pm$ 0.014 $\pm$ 0.200 & 0.499 $\pm$ 0.004 $\pm$ 0.053 \\
\end{tabular}\\[0.2cm]}
\normalsize
\caption{Numerical values of proton and anti-proton yields ($\text{d}n/\text{d}y$) at mid-rapidity in 10\% most \textit{central} Ar+Sc collisions. The values are provided with statistical ($\sigma_\text{stat}$) and systematic ($\sigma_\text{sys}$) uncertainties in the form: $\text{d}n/\text{d}y \pm \sigma_\text{stat} \pm \sigma_\text{sys}$.}
\label{tab:midrap_dndy_protons}
\end{table}

\FloatBarrier

\subsubsection{Rapidity spectra}
\label{sec:proton_yspectra}
\begin{figure}
	\centering
	\includegraphics[width=0.33\linewidth]{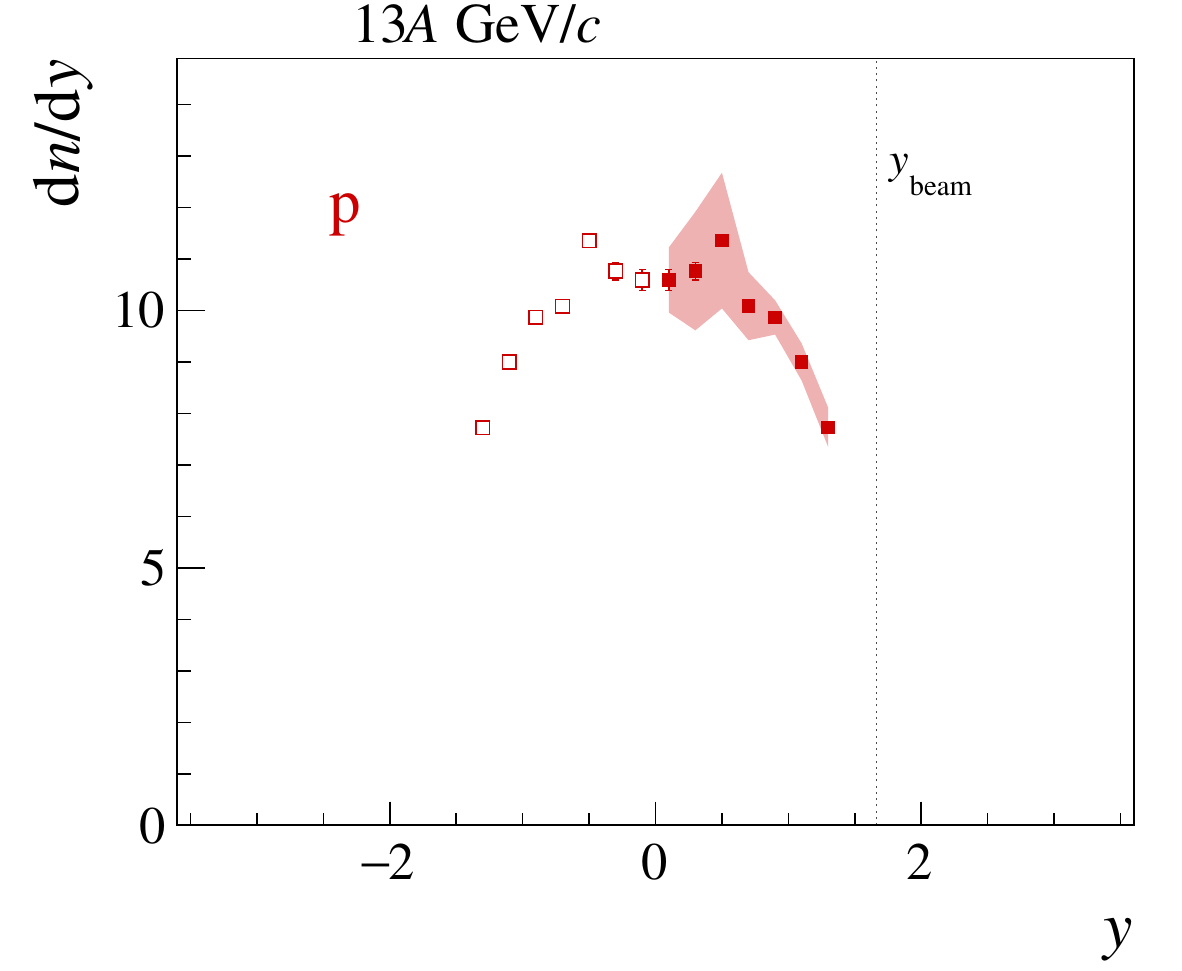}~
	\includegraphics[width=0.33\linewidth]{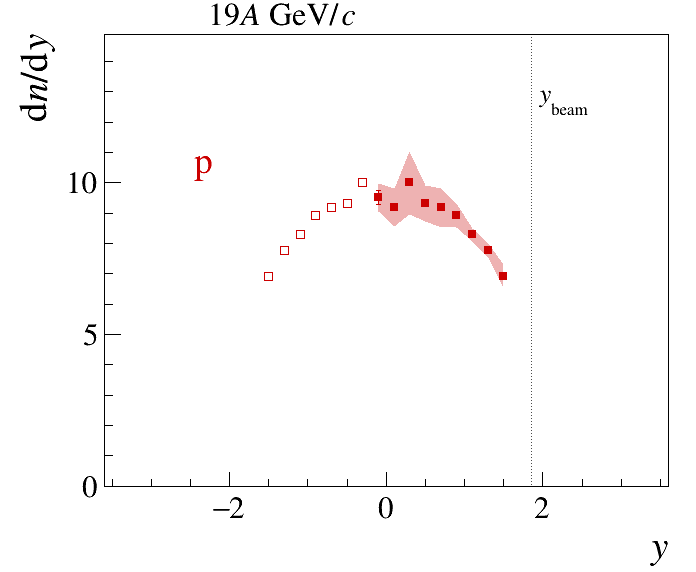}~
	\includegraphics[width=0.33\linewidth]{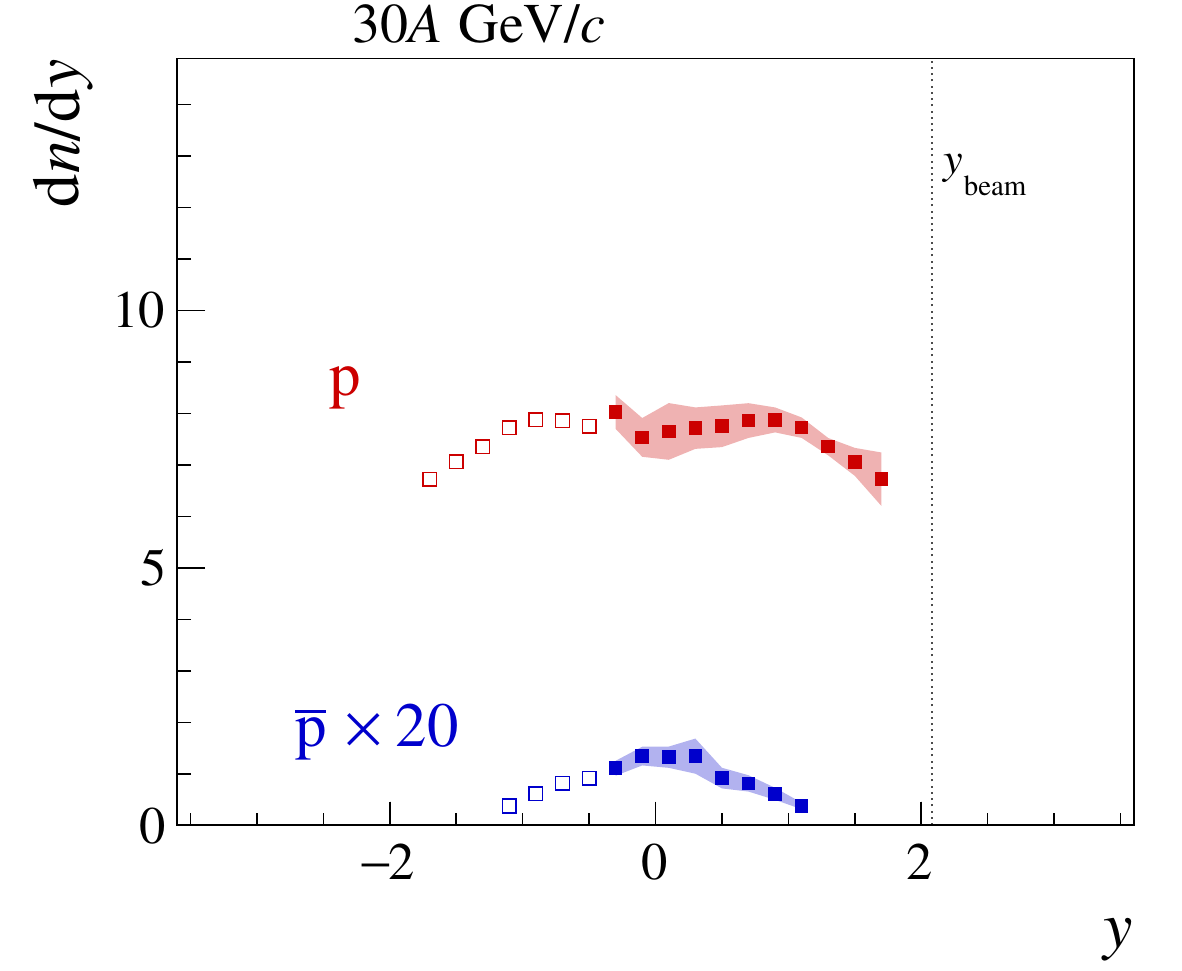}\\
	\includegraphics[width=0.33\linewidth]{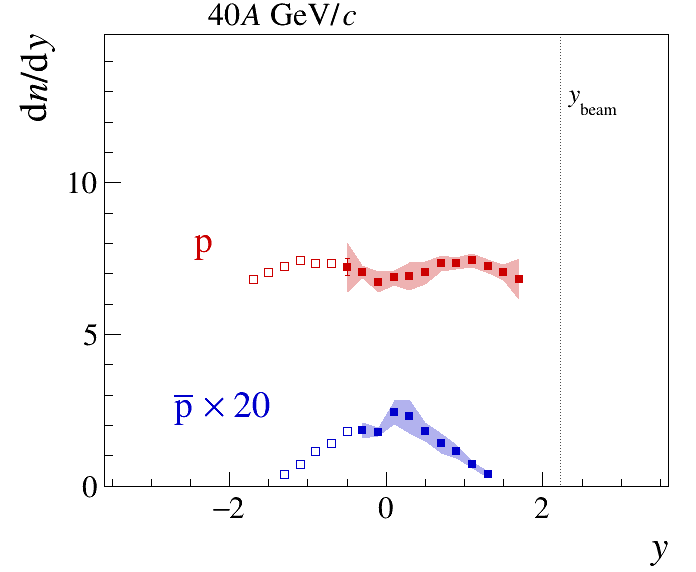}~
	\includegraphics[width=0.33\linewidth]{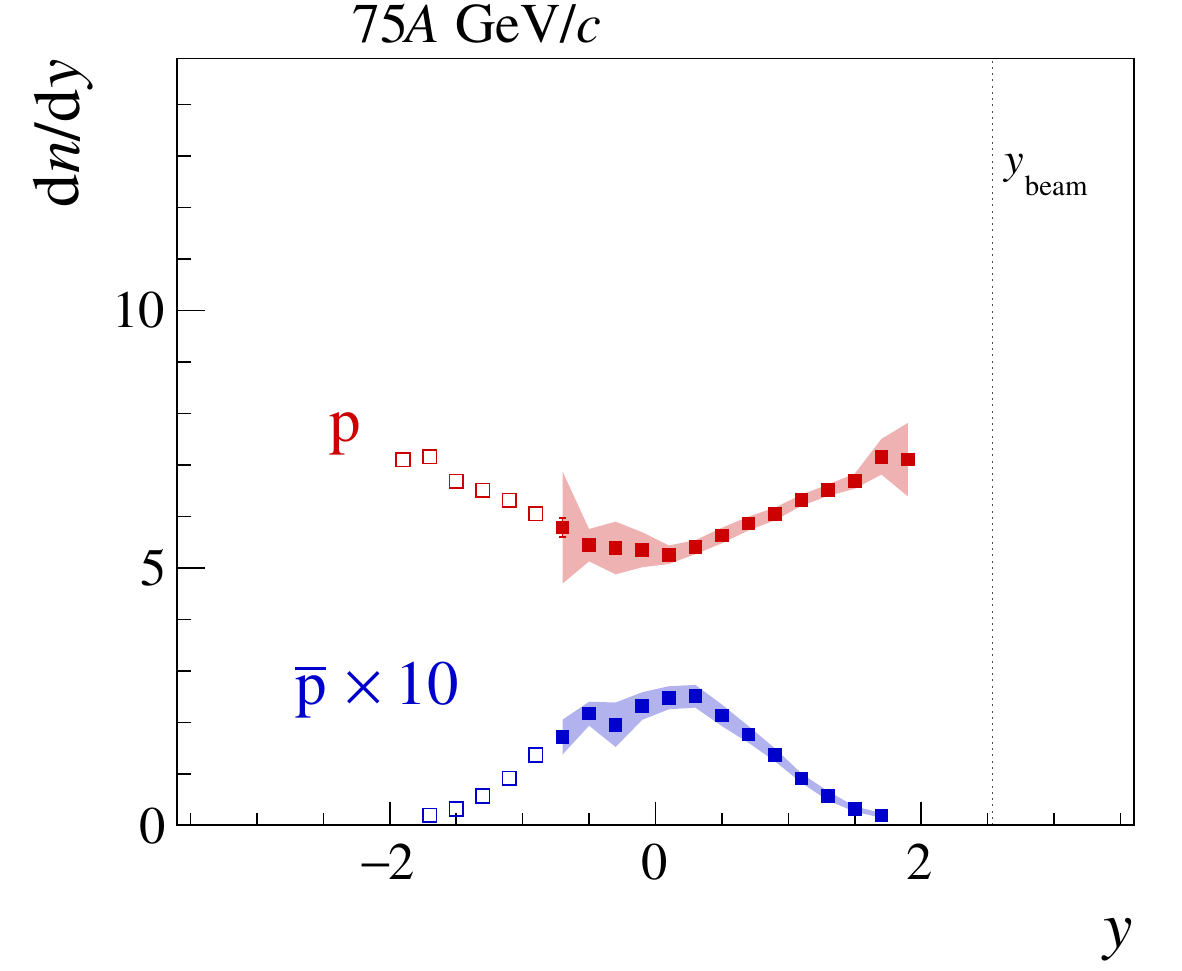}~
	\includegraphics[width=0.33\linewidth]{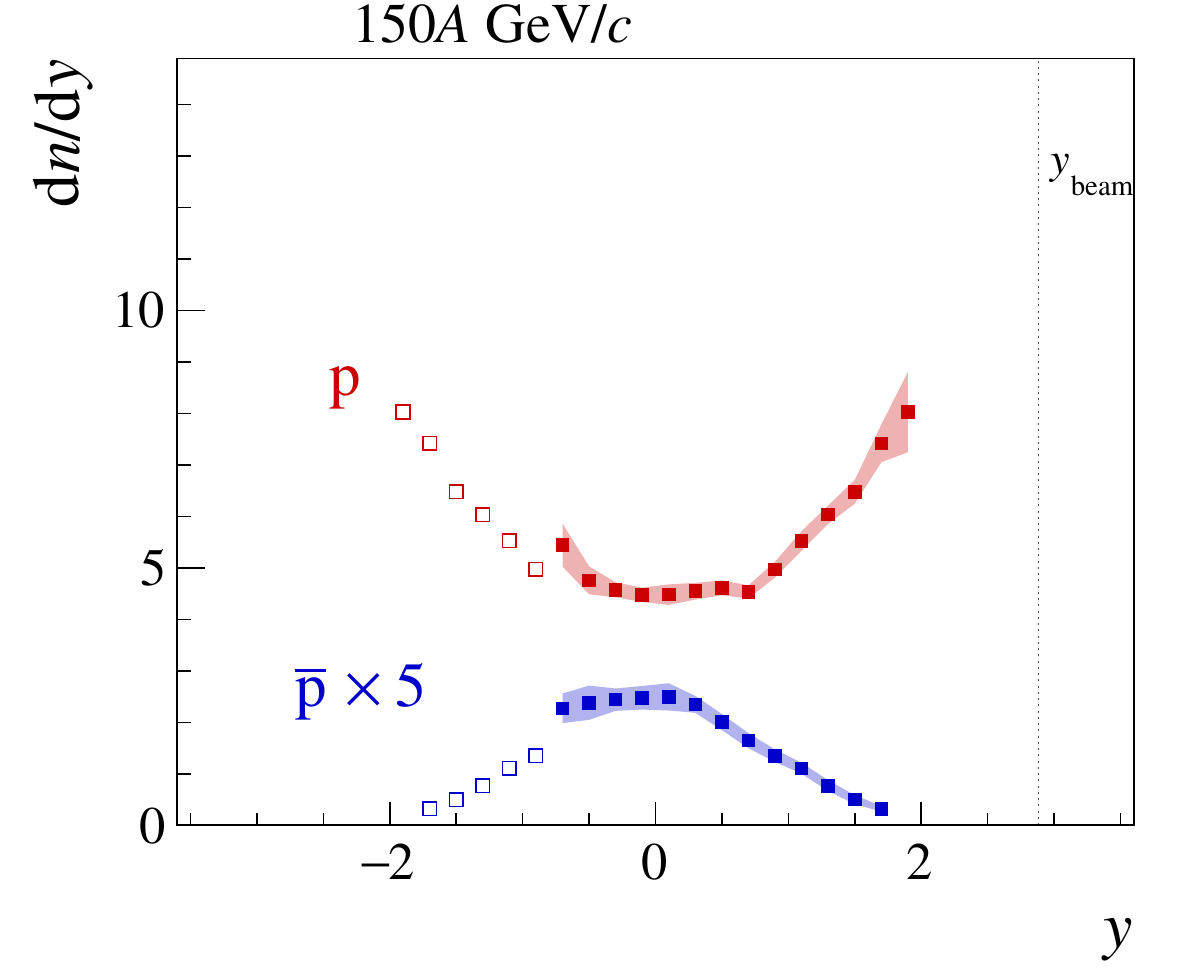}\\	
	\caption{Rapidity spectra of protons and antiprotons produced in 10\% most \textit{central} Ar+Sc collisions at 13\textit{A} - 150\textit\AGeVc. No data was plotted for anti-protons produced in collisions at $p_\text{beam}=13A$ and $19A$ GeV/$c$ due to large statistical and systematic uncertainties. Error bars show statistical uncertainties, while shaded bands correspond to systematic uncertainties. The open markers represent the data reflected wrt. $y=0$.}
	\label{fig:proton_yspectra}
\end{figure}

Figure \ref{fig:proton_yspectra} shows experimental results on the proton and antiproton rapidity distributions.
Contrarily to the well-understood two-Gaussian extrapolation of charged pions and kaons rapidity spectra, a similar procedure cannot be employed for protons. The shape of proton rapidity distributions changes significantly with beam momentum and the unmeasured regions, containing the maxima, constitute a significant portion of total multiplicity. Thus, the calculation of mean multiplicity was not attempted for protons.

\subsubsection{Antiproton mean multiplicities}
\begin{figure}[h]
\centering
\includegraphics[width=0.4\linewidth, page=1]{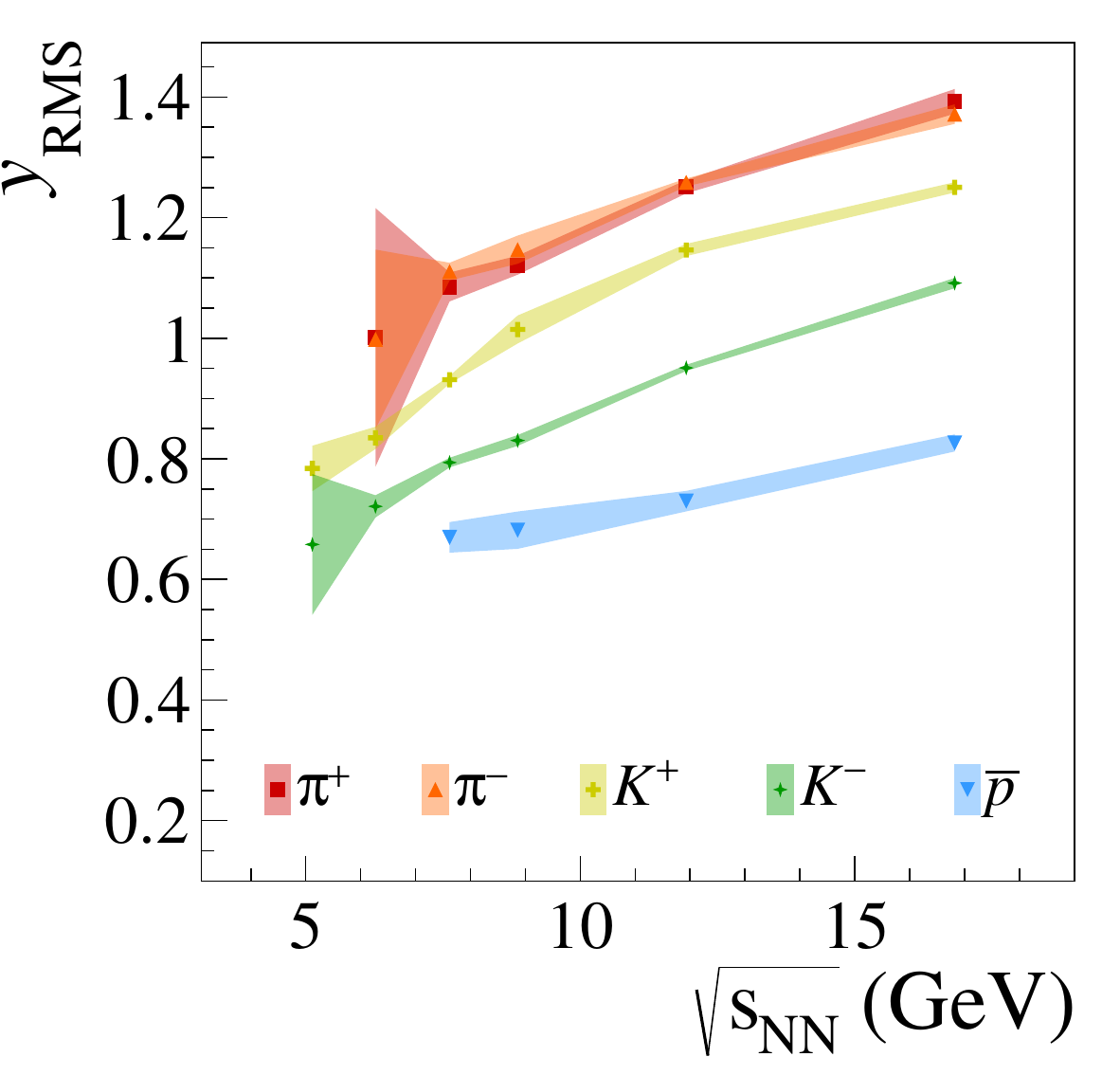}~
\includegraphics[width=0.4\linewidth, page=2]{figures/newresults/rapidity_fit_parameters}\\
\caption{The comparison of RMS widths of rapidity distributions $y_\text{RMS}$ (\textit{left)} and $y_\text{RMS}/y_\text{beam}$ (\textit{right}) (Eq. \ref{eq:rapidiry_rms}) of  $\pi^{-}$, $\pi^{+}$, $K^{-}$, $K^{+}$ and $\bar{p}$ produced in the 10\% most \textit{central} Ar+Sc collisions. The shaded bands illustrate the statistical uncertainties of the fit.
}
\label{fig:yspectra_y_rms_comparison}
\end{figure}

\begin{figure}
\centering
\includegraphics[width=0.5\linewidth]{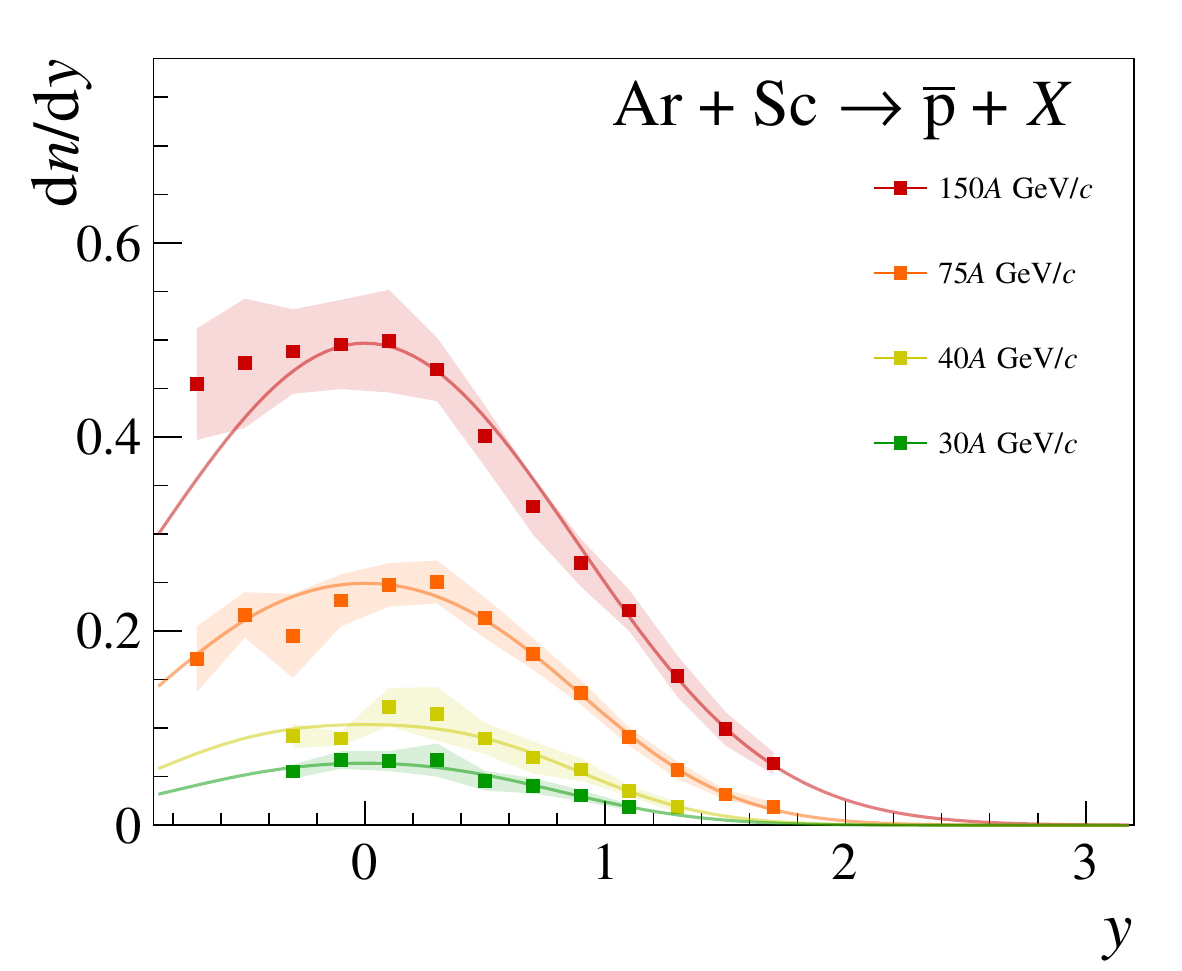}\\
\caption{Rapidity distributions of anti-protons measured in 10\% most \textit{central} Ar+Sc collisions at 30\textit{A}--150\textit{A} GeV/\textit{c} beam momenta. Data points are plotted along with double-Gaussian fits (Eq. \ref{eq:pion_yfit}). Error bars show statistical uncertainties, while shaded bands correspond to systematic uncertainties.}
\label{fig:antiproton_yspectra}
\end{figure}

Mean antiproton multiplicities ($4\pi$) were calculated in a similar way as in the case of charged pions and kaons. The measured $\text{d}n/\text{d}y$ spectra were supplemented for the missing acceptance with the integral of the symmetric double Gaussian (Eq. \ref{eq:pion_yfit}) fitted to the data. The assumption of symmetry was necessary due to large uncertainties and limited acceptance of the data in the backward hemisphere. The comparison with models in Sec. \ref{sec:models_protons} validates this assumption. Figure \ref{fig:antiproton_yspectra} illustrates both the measured data and the fitted model. The comparison of $y_\text{RMS}$ and $y_\text{RMS}/y_\text{beam}$ (Eq. \ref{eq:rapidiry_rms}) for the particle species for which the rapidity spectra were fitted with Eq. \ref{eq:pion_yfit} is presented in Figure \ref{fig:yspectra_y_rms_comparison}. Scaling of RMS widths of rapidity distributions with $y_\text{beam}$ was observed. 

The $\bar{p}$ mean multiplicities calculated at each beam momentum are given in Table \ref{tab:barp_4pi}.
\begin{table}[h]
\centering
\begin{tabular}{c||c}
	\parbox{1.2cm}{\,\,$p_{\textrm{beam}}$ (GeV/\textit{c})}& $\langle \bar{p} \rangle$ \\[0.3cm]
        \hline \hline
30 & 0.116 $\pm$ 0.002 $\pm$ 0.014 \\ 
40 & 0.198 $\pm$ 0.003 $\pm$ 0.025 \\ 
75 & 0.479 $\pm$ 0.003 $\pm$ 0.050 \\ 
150 & 1.080 $\pm$ 0.005 $\pm$ 0.092 \\ 
\end{tabular}\\
\caption{Mean multiplicities of antiprotons produced in 10\% most \textit{central} Ar+Sc collisions at 30$A$-150\AGeVc beam momenta. The values are provided with statistical ($\sigma_\text{stat}$) and systematic ($\sigma_\text{sys}$) uncertainties in the form: $\langle \bar{p} \rangle \pm \sigma_\text{stat} \pm \sigma_\text{sys}$.}
\label{tab:barp_4pi}
\end{table}

Similarly as discussed in Sec. \ref{sec:kaon_rapidity}, the employed extrapolation procedure results in a systematic uncertainty, that increases with shrinking acceptance towards lower energies. The fraction of measured multiplicity contribution to the total multiplicity decreases from over 50\% at top SPS collision energy to only $\approx$10\% at 13\textit{A} GeV/c. This effect, combined with large uncertainties of antiproton $dn/dy$ data prevented the calculation of mean multiplicities at 13\textit{A} and 19\textit{A} GeV/\textit{c}.

\FloatBarrier
\section{Collision energy and system size dependence}
\label{sec:discussion}

In this section, the selected results on identified hadron production in 10\% most \textit{central} Ar+Sc collisions at 13$A$--150\AGeVc beam momenta are reviewed in the context of signatures of the onset of deconfinement. The energy and system size dependence of hadron production properties are compared with available data on inelastic \pp~\cite{Abgrall:2013pp_pim,Aduszkiewicz:2017sei,Aduszkiewicz:2019zsv}, \textit{central} Be+Be~\cite{NA61SHINE:2020ggt,NA61SHINE:2020czq} and \textit{central} Pb+Pb~\cite{Afanasiev:2002mx,Alt:2007aa} collisions.
The remaining world data originate from various experiments at the AGS, SPS, RHIC, and LHC accelerators extracted from Refs. \cite{Gazdzicki:1991ih,NA49:1999uyd,LEBC-EHS:1991mrc,Rossi:1974if,Golokhvastov:2001ei,Becattini:1997rv,PHOBOS:2003nrr,Blobel:1975ka,Aguilar-Benitez:1991hzq,Melissinos:1962zz,Alexander:1967zz,Mueck:1972qz,PhysRevD.3.645,Antinucci:1972ib,French-Soviet:1976tin} and references therein. They concern mostly Pb+Pb and Au+Au collisions.

\subsection{$K^+$ and $K^-$ inverse slope parameter $T$ dependence on collision energy}

\begin{figure}[h]
\centering
\includegraphics[width=0.5\textwidth]{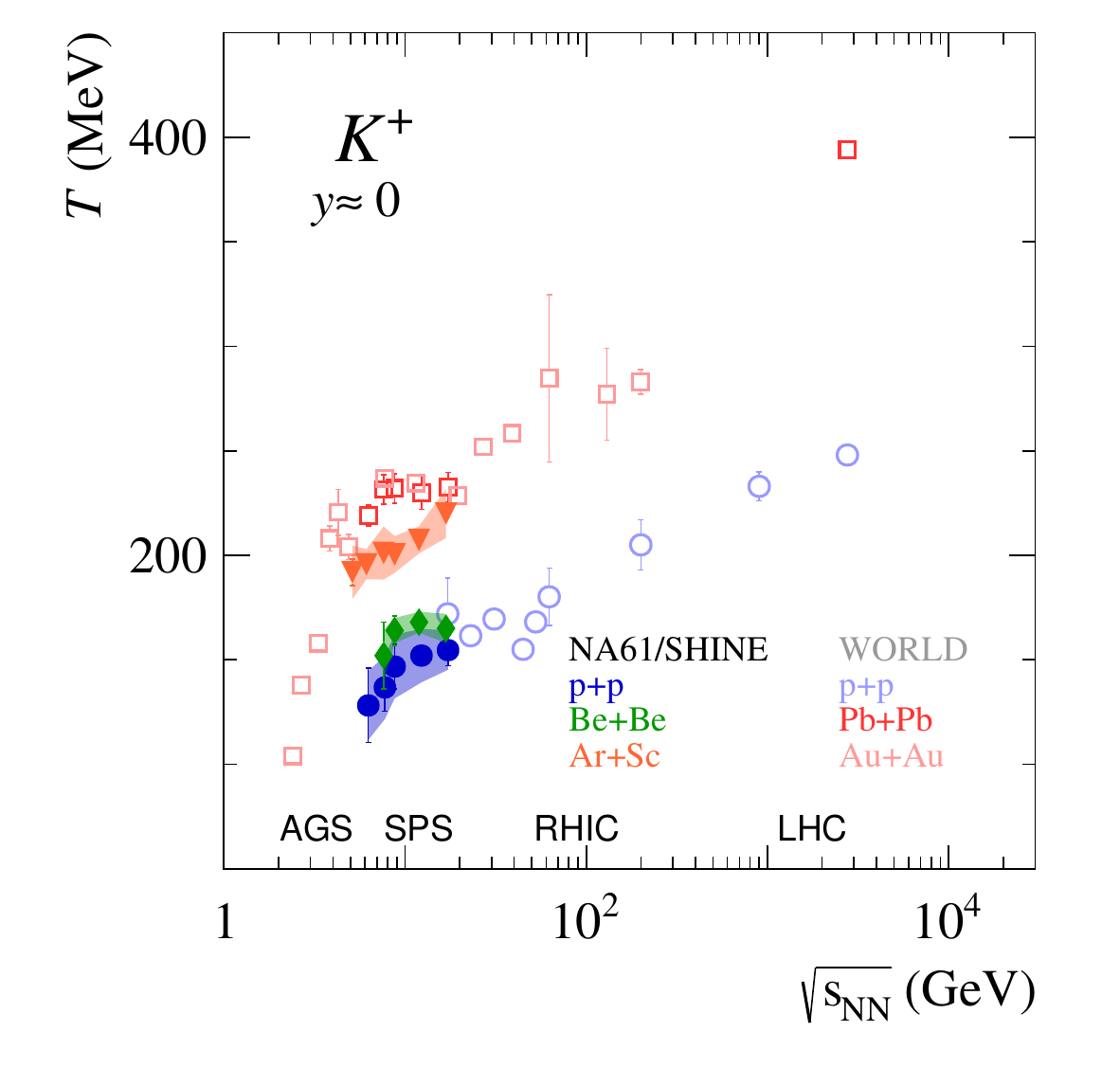}~
\includegraphics[width=0.5\textwidth]{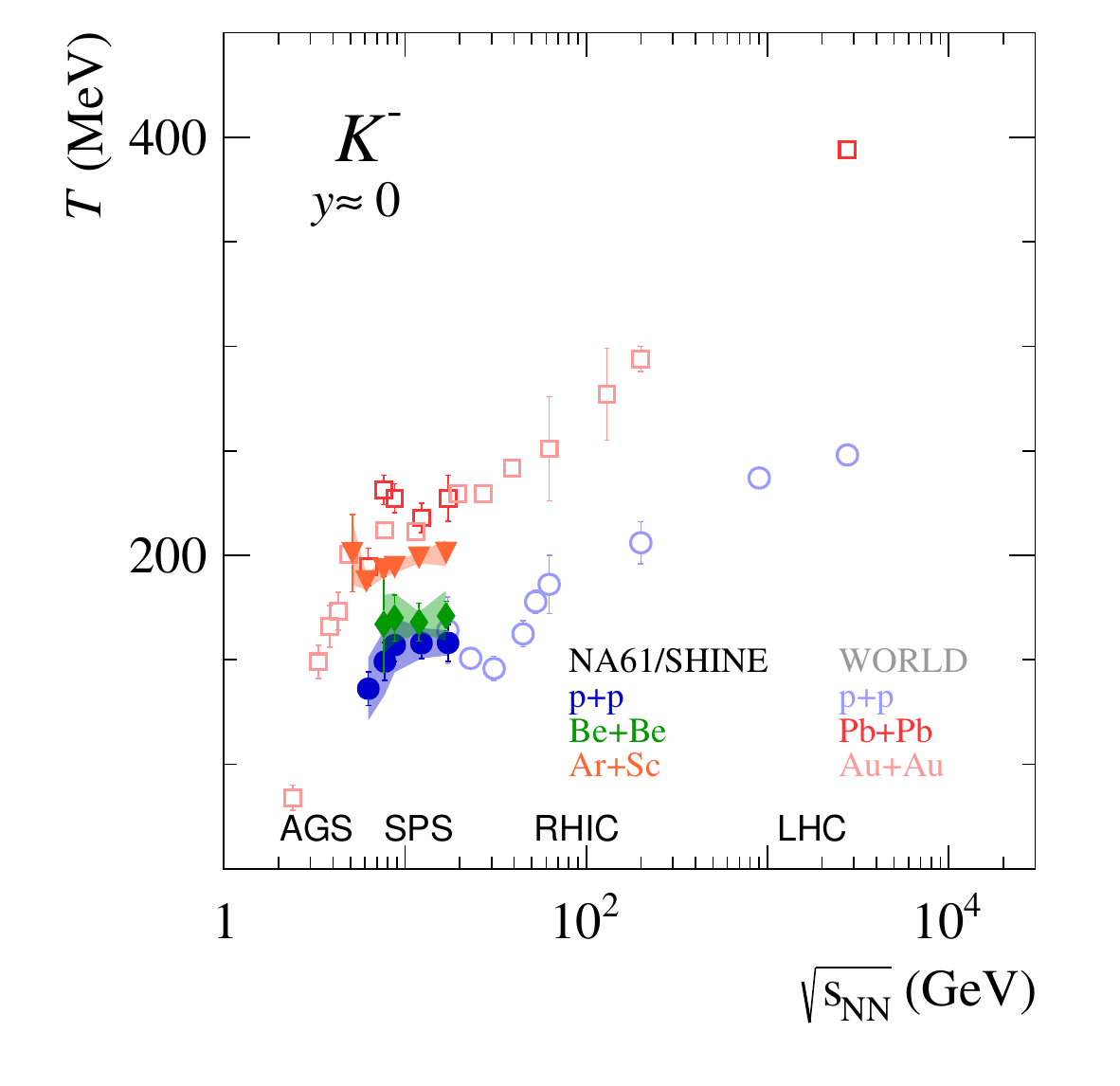}
\caption{The energy dependence of the inverse slope parameter $T$ of \pt spectra at mid-rapidity of positively (\textit{left}) and negatively (\textit{right}) charged $K$ mesons for central Ar+Sc, Be+Be (\hspace{1sp}\cite{NA61SHINE:2021nye}), Pb+Pb (\hspace{1sp}\cite{Alt:2007aa, Anticic:2012ay, Afanasiev:2002mx, ALICE:2012ovd, ALICE:2014juv}) and Au+Au (\hspace{1sp}\cite{Adamczyk:2017iwn,Adams:2003xp,Adcox:2003nr, Adler:2003cb, BRAHMS:2005gow,E-802:1999ewk, Schuldes:2017arj}) collisions as well as inelastic \pp (\hspace{1sp}\cite{Aduszkiewicz:2017sei, STAR:2008med, Kliemant:2003sa, ALICE:2011gmo, ALICE:2014juv}) interactions. Both statistical (vertical bars) and systematic (shaded bands) uncertainties are shown.}
\label{fig:slope}
\end{figure}

The simple exponential parametrization of the kaon transverse momentum spectra (Eq. \ref{eq:pion_yfit}) fits the data well and yields values for the inverse slope parameter $T$, summarized in Table \ref{tab:midrap_inv_slope_mean_pt}. The $T$ values obtained for \textit{central} Ar+Sc collisions at six beam momenta from the CERN SPS energy range as a function of the collision energy ($\sqrt{s_{NN}}$) for positively and negatively charged kaons are presented in Fig. \ref{fig:slope}. The Ar+Sc values of the $T$ parameter are slightly below Pb+Pb, yet still significantly higher than Be+Be.
The value of the inverse slope parameter within hydrodynamical models is interpreted as a kinetic freeze-out temperature with modifications from transverse flow. In this context, the results presented here may indicate that the kinetic freeze-out temperature and transverse flow in Ar+Sc are closer to Pb+Pb (large system) than Be+Be and \pp (small systems).

\FloatBarrier

\subsection{$K/\pi$ ratio dependence on collision energy}
\begin{figure}
\centering
\includegraphics[width=0.5\textwidth]{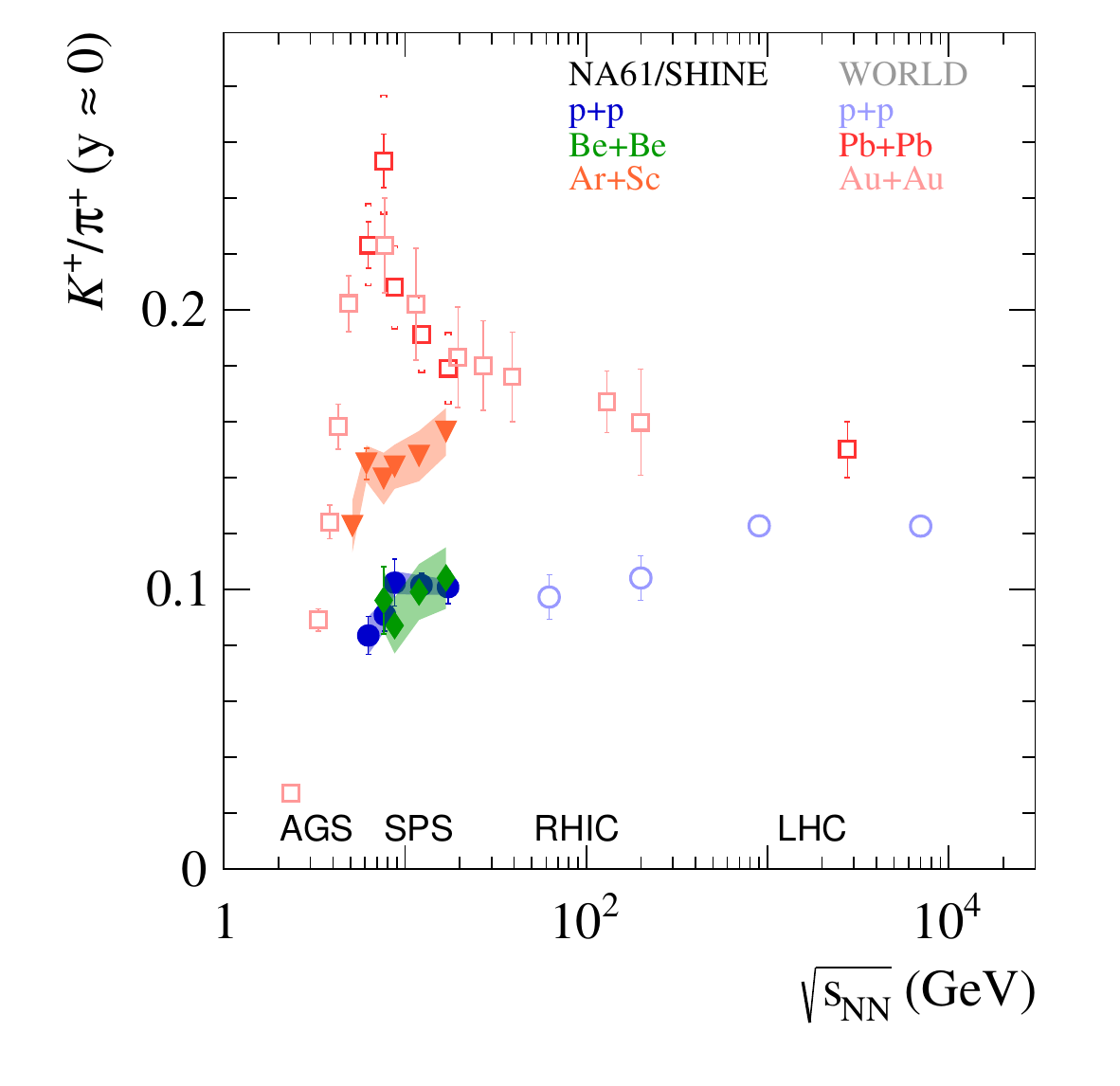}~
\includegraphics[width=0.5\textwidth]{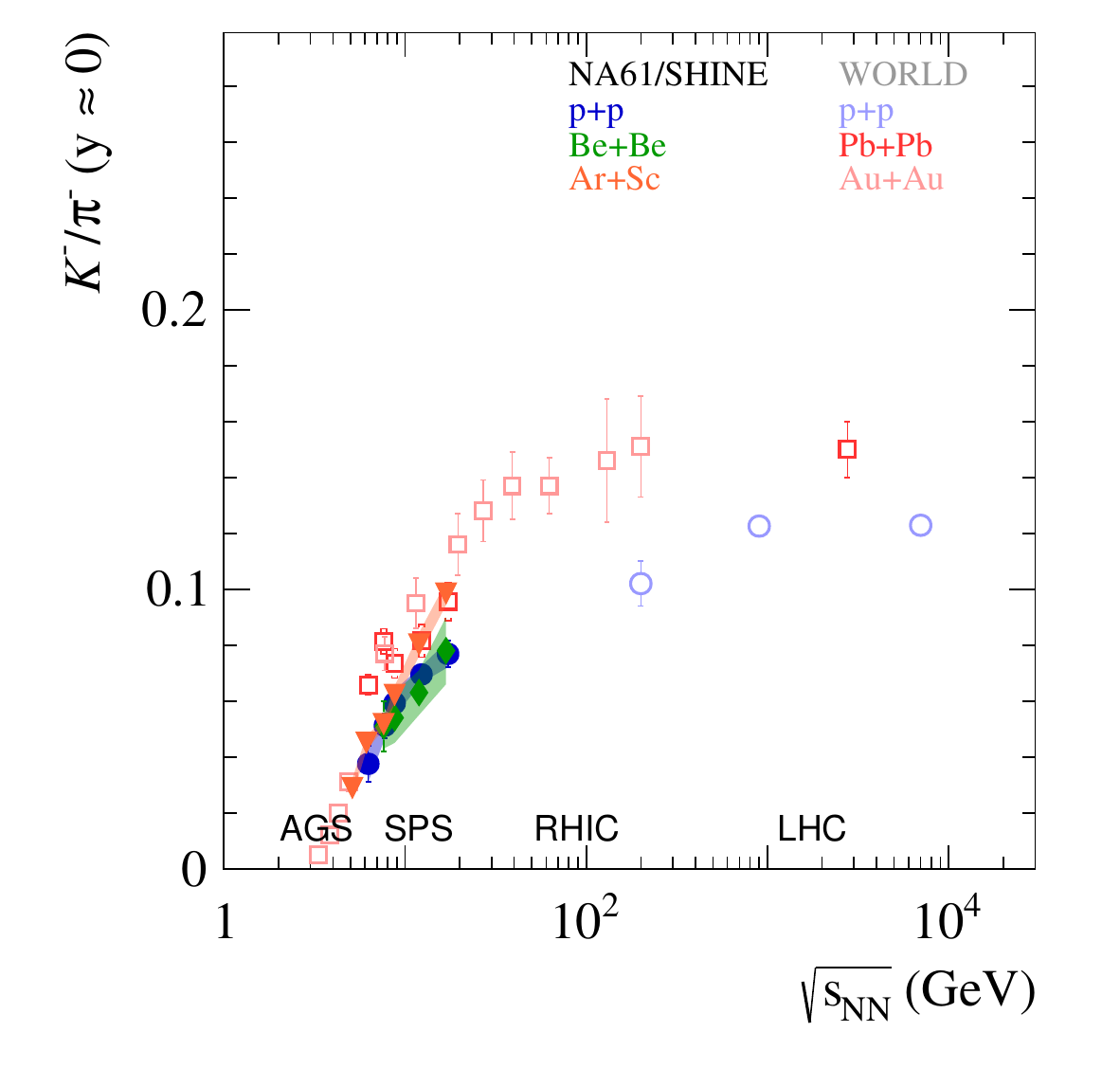}
\caption{The energy dependence of the $K/\pi$ ratio at mid-rapidity of positively (\textit{left}) and negatively (\textit{right}) charged particles for central Ar+Sc, Be+Be (\hspace{1sp}\cite{NA61SHINE:2021nye}), Pb+Pb (\hspace{1sp}\cite{Alt:2007aa, Anticic:2012ay, Afanasiev:2002mx, ALICE:2012ovd, ALICE:2014juv}) and Au+Au (\hspace{1sp}\cite{Adamczyk:2017iwn,Adams:2003xp,Adcox:2003nr, Adler:2003cb, BRAHMS:2005gow,E-802:1999ewk, Schuldes:2017arj}) collisions as well as inelastic \pp (\hspace{1sp}\cite{Aduszkiewicz:2017sei, STAR:2008med, Kliemant:2003sa, ALICE:2011gmo, ALICE:2014juv}) interactions. Both statistical (vertical bars) and systematic (shaded bands) uncertainties are shown.}
\label{fig:horn_midrap}
\end{figure}

\begin{figure}[h]
	\centering
	\includegraphics[width=0.5\textwidth]{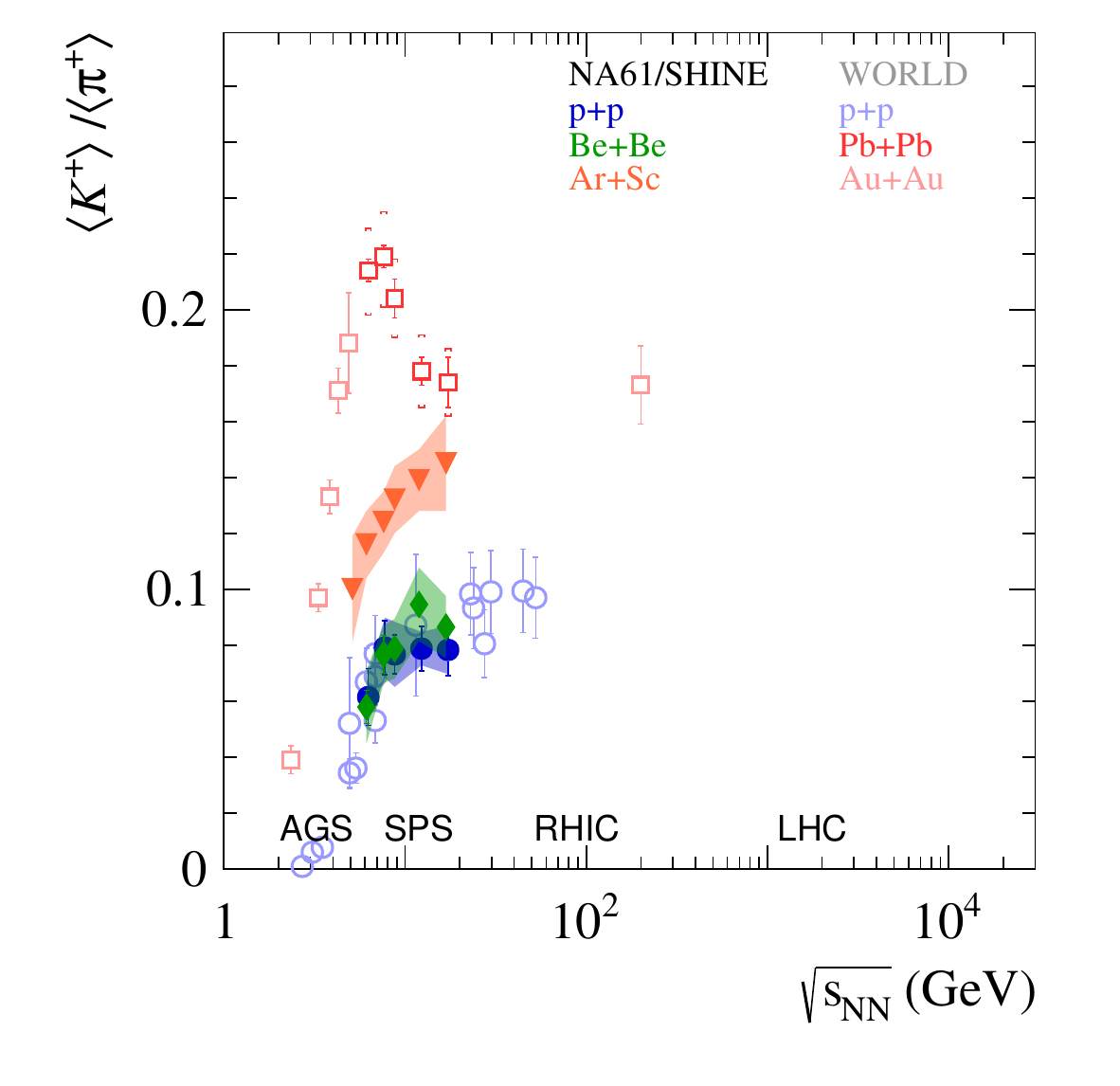}~
\includegraphics[width=0.5\textwidth]{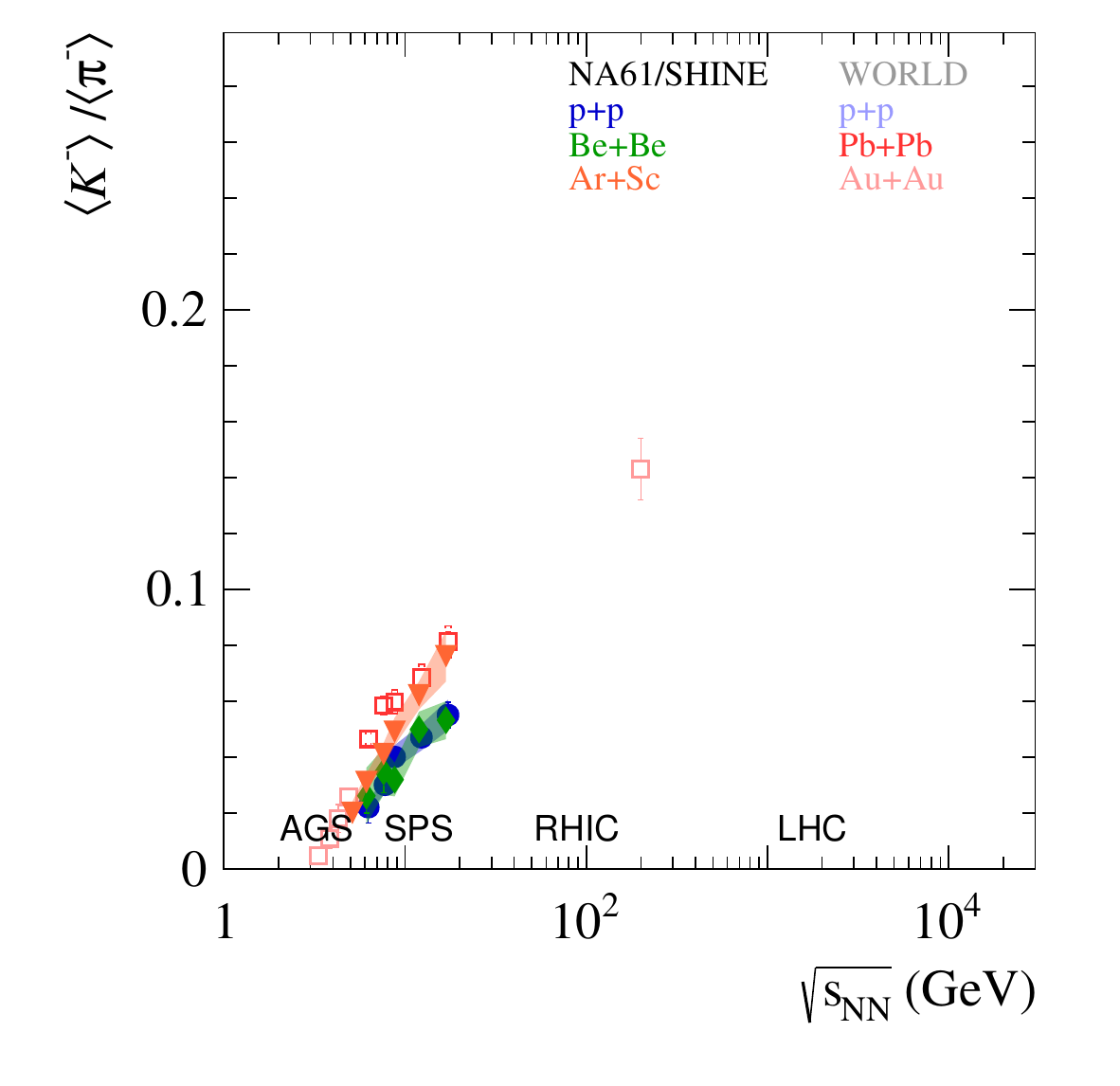}
\caption{The energy dependence of the $\langle K\rangle/\langle\pi\rangle$ mean multiplicity ratio of positively (\textit{left}) and negatively (\textit{right}) charged particles for central Ar+Sc, Be+Be (\hspace{1sp}\cite{NA61SHINE:2021nye}), Pb+Pb (\hspace{1sp}\cite{Alt:2007aa, Anticic:2012ay, Afanasiev:2002mx, ALICE:2012ovd, ALICE:2014juv}) and Au+Au (\hspace{1sp}\cite{Adamczyk:2017iwn,Adams:2003xp,Adcox:2003nr, Adler:2003cb, BRAHMS:2005gow,E-802:1999ewk, Schuldes:2017arj}) collisions as well as inelastic \pp (\hspace{1sp}\cite{Aduszkiewicz:2017sei, STAR:2008med, Kliemant:2003sa, ALICE:2011gmo, ALICE:2014juv}) interactions. Both statistical (vertical bars) and systematic (shaded bands) uncertainties are shown.}
	\label{fig:horn}
\end{figure}

The characteristic, non-monotonic behavior of the $K^+$ over $\pi^+$ ratio observed in central heavy-ion collisions (see Pb+Pb and Au+Au in Figs. \ref{fig:horn_midrap} and \ref{fig:horn}) agrees qualitatively with predictions of \Smes \cite{Gazdzicki:1998vd}, in which quarks and gluons are the relevant degrees of freedom in the early stage of the collision at high energies. Within SMES, the sharp peak in the $K^+/\pi^+$ ratio (\textit{the horn}) seen at $\sqrt{s_{NN}}\approx8$ GeV/\textit{c} is interpreted as an indication of the onset of deconfinement -- a beginning of the creation of quark-gluon plasma in the early stage of collision.
In the case of intermediate-size systems, however, no such structure is visible, neither at mid-rapidity (Fig. \ref{fig:horn_midrap}) nor in full phase-space (Fig. \ref{fig:horn}). However, a clear distinction between the two data subsets is visible---\textit{p}+\textit{p} and Be+Be results show similar values and collision energy dependence, while Pb+Pb, Au+Au, and Ar+Sc collisions show much higher $K^+/\pi^+$ ratios. Moreover, although Ar+Sc is clearly separated from small systems, its energy dependence does not show the horn seen in Pb+Pb and Au+Au reactions.
No available theoretical description agrees with this behavior -- neither the models of statistical hadron production in thermal equilibrium nor the microscopic transport models (see discussion in Sec. \ref{sec:models}).

Figures \ref{fig:horn_midrap} and \ref{fig:horn} display also the energy dependence of the $K^-/\pi^-$ ratio. While the number of $s$ and $\bar{s}$ quarks produced in a collision is equal, their distribution among strange hadrons is heavily affected by the large net-baryon density, characteristic for nucleus-nucleus collisions at SPS energies. A lot of the s quarks will therefore be distributed in $\Lambda$ baryons, while in the case of $\bar{s}$ quarks, the production of $\bar{\Lambda}$ is heavily suppressed. Thus, the vast majority of $\bar{s}$ quarks are carried out of the collision by $K^+$ and $K^0$ mesons, which are expected to be produced in similar quantities. Consequently, the $K^+$ yields are a more sensitive measure of the strangeness content than the $K^-$ yields.
Therefore, a characteristic maximum close to $\sqrt{s_{NN}}=8$~\GeV, a \textit{horn}, is not expected for the $K^-/\pi^-$ ratio and all systems studied at SPS energies display an approximately monotonous rise towards higher collision energies.
Similarly to observations made in the previous paragraph, the yields obtained for Ar+Sc interactions closely resemble Pb+Pb data at high collision energies (75\textit{A}, 150\textit{A} GeV/\textit{c}) and at low energies ($<$40\AGeVc) they are more similar to small systems.

\FloatBarrier

\subsection{Collision energy and system size dependence of proton rapidity spectra}
\begin{figure}[h]
	\centering
	\includegraphics[width=\linewidth]{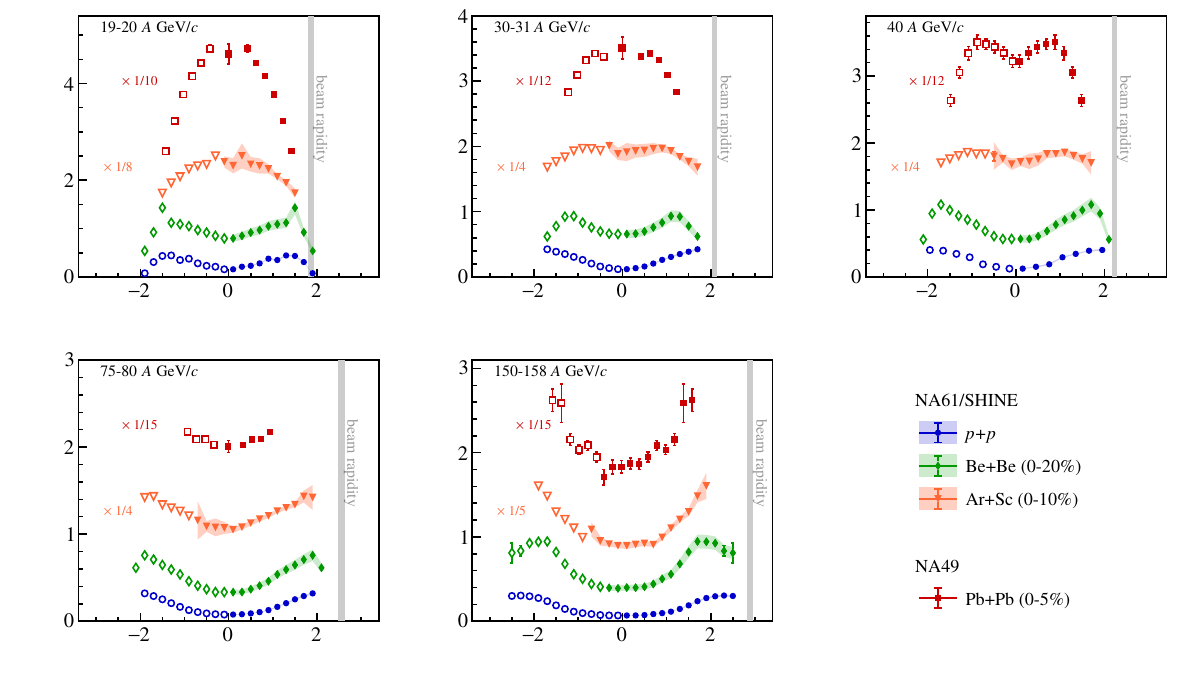}
	\caption{Proton rapidity spectra in 10\% most \textit{central} Ar+Sc collisions in comparison with other systems: \pp, \textit{central} Be+Be and \textit{central} Pb+Pb at five corresponding collision energies. Full markers denote the experimental points, while open ones plot their reflections with respect to $y=0$, where there is no corresponding data point. Error bars display statistical uncertainties and shaded bands stand for systematic uncertainties. Vertical lines represent beam rapidities. Spectra were scaled by numbers indicated in the plots for better display.}
	\label{fig:protonssystems}
\end{figure}

Figure \ref{fig:protonssystems} displays proton rapidity spectra in the 10\% most \textit{central} Ar+Sc collisions in comparison with \pp, Be+Be and Pb+Pb interactions at matching collision energies. 
At the highest collision energies (75\textit{A} and 150\AGeVc) all reactions show approximately similar shapes of the proton rapidity distribution. At 40\AGeVc an inflection of rapidity spectra is observed in Ar+Sc interactions, which is not the case for small systems.
At beam momenta of 150$A$-158\AGeVc, the spectral shape of Ar+Sc data is similar to results on Pb+Pb collisions, while at 40\AGeVc the Ar+Sc data clearly fit in the overall trend of baryon stopping \cite{Busza:1983rj} increasing as a function of system size between Be+Be and Pb+Pb reactions (see Fig.~\ref{fig:protonssystems}). A more extensive discussion on the system size dependence of proton rapidity spectra is presented in Ref. \cite{Panova:2022lnd}.
The discussion of proton rapidity spectra in view of phenomenological models is continued in Sec.~\ref{sec:models_protons}.

\FloatBarrier

\subsection{Comparison with models}
\label{sec:models}

This subsection compares experimental results expected to be sensitive to the onset of deconfinement with corresponding model predictions. The 
Energy conserving quantum mechanical multiple scattering approach, based on Partons (parton ladders) Off-shell remnants, and Splitting of parton ladders 
(\Epos) 1.99~\cite{Werner:2008zza}, Parton-Hadron-String Dynamics (\Phsd)~4.1~\cite{PhysRevC.78.034919,CASSING2009215} and Simulating Many Accelerated Strongly-interacting Hadrons (\Smash)~2.1.4~\cite{Mohs:2019iee,PhysRevC.94.054905} models were chosen for this study. In \Epos, the reaction proceeds from the excitation of strings according to Gribov-Regge's theory to string fragmentation into hadrons. 
\Phsd is a microscopic off-shell transport approach that describes the evolution of a relativistic nucleus-nucleus collision from the initial hard scatterings and string formation through the dynamical deconfinement phase transition to the quark-gluon plasma as well as hadronization and the subsequent interactions in the hadronic phase.
\Smash uses the hadronic transport approach where the free parameters of the string excitation and decay are tuned to match the experimental measurements in inelastic \pp collisions. 
The selection of events in all model calculations follows the procedure for \textit{central} collisions to which experimental results correspond, see Sec.~\ref{sec:centrality}. This is particularly important when comparisons of yields with measurements are to be performed.

\subsubsection{Charged pion spectra and multiplicities}
\begin{figure}
\centering
\includegraphics[width=0.33\linewidth]{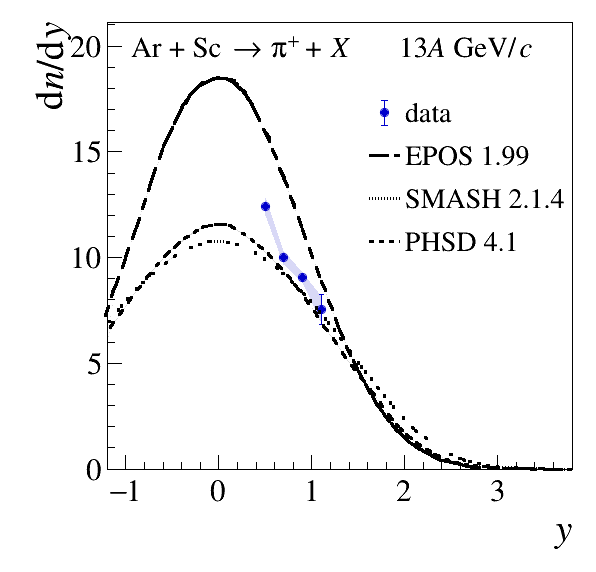}~
\includegraphics[width=0.33\linewidth]{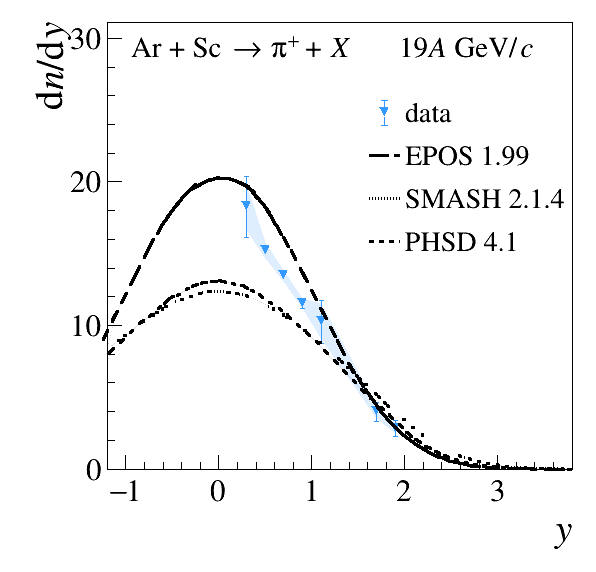}~
\includegraphics[width=0.33\linewidth]{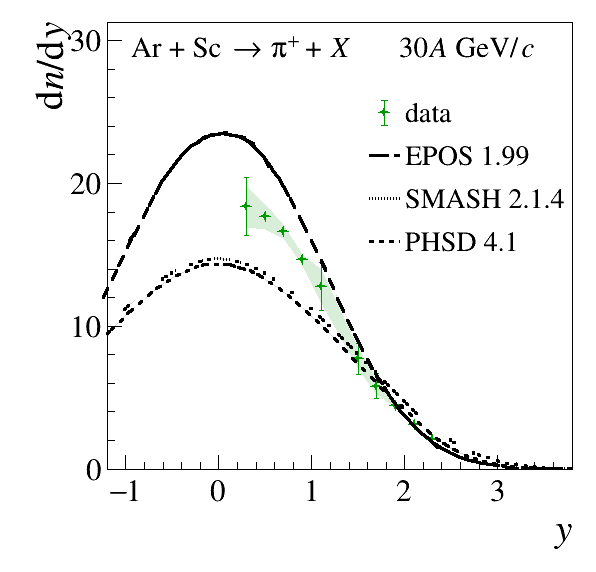}\\
\includegraphics[width=0.33\linewidth]{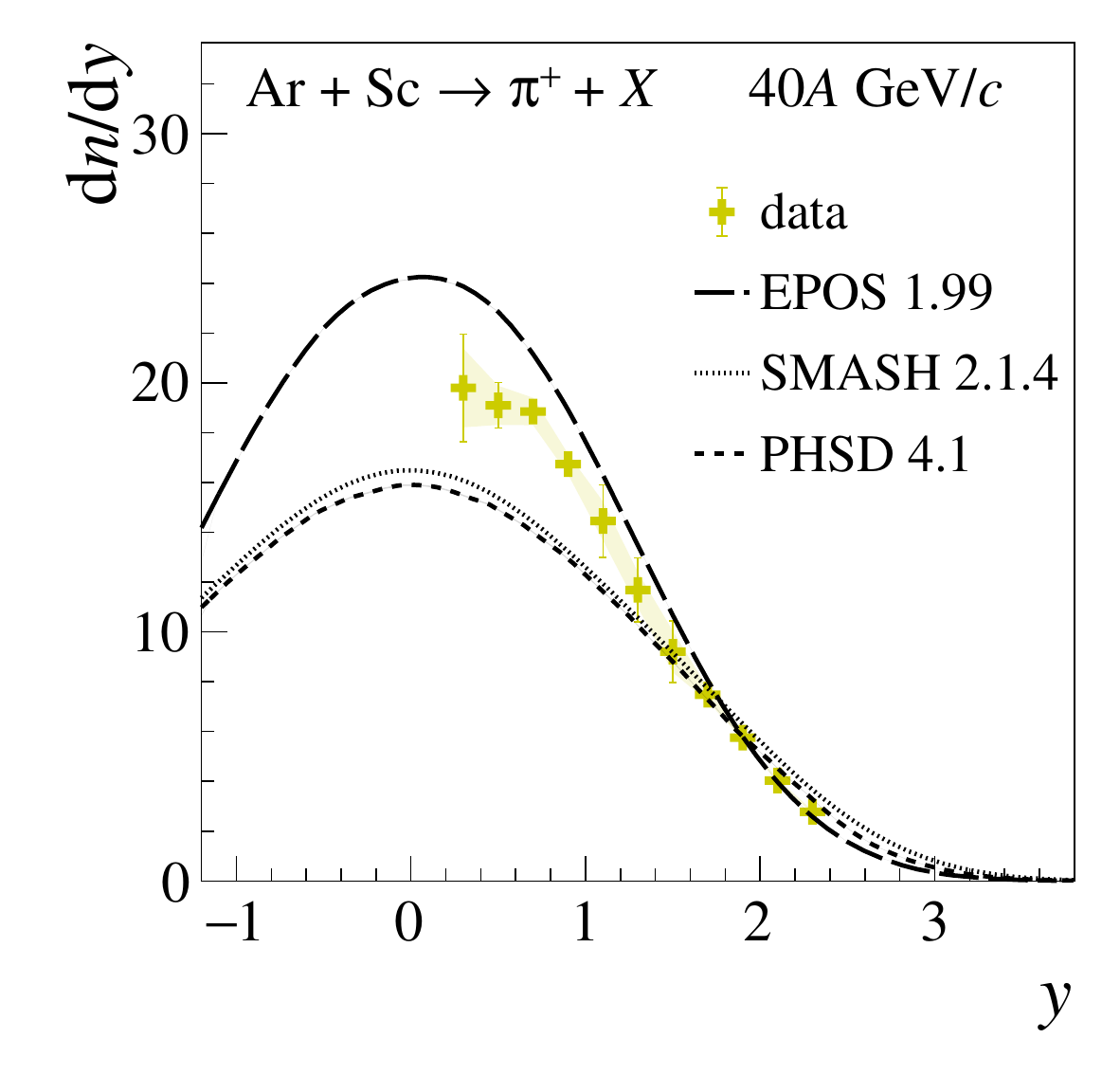}~
\includegraphics[width=0.33\linewidth]{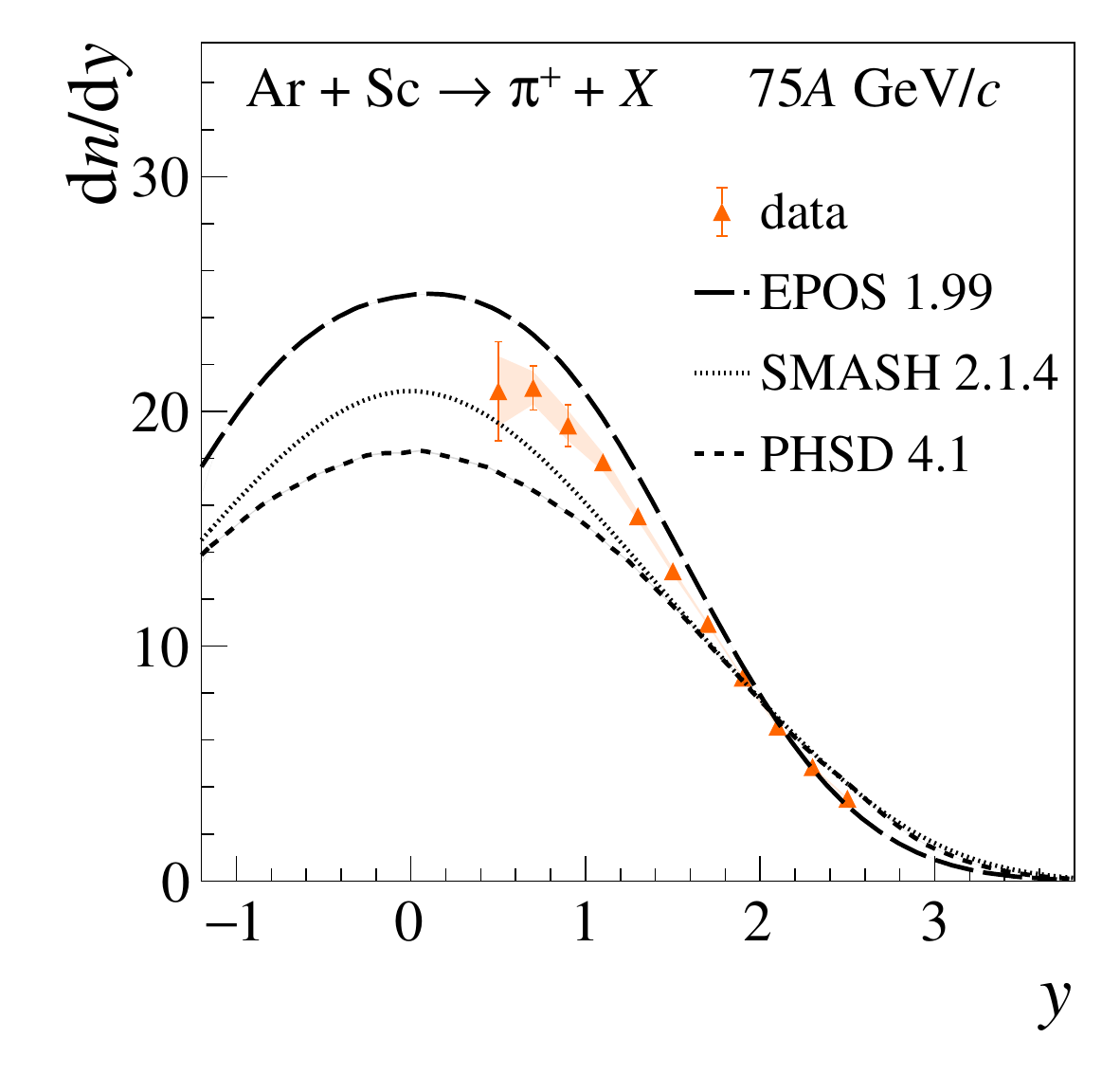}~
\includegraphics[width=0.33\linewidth]{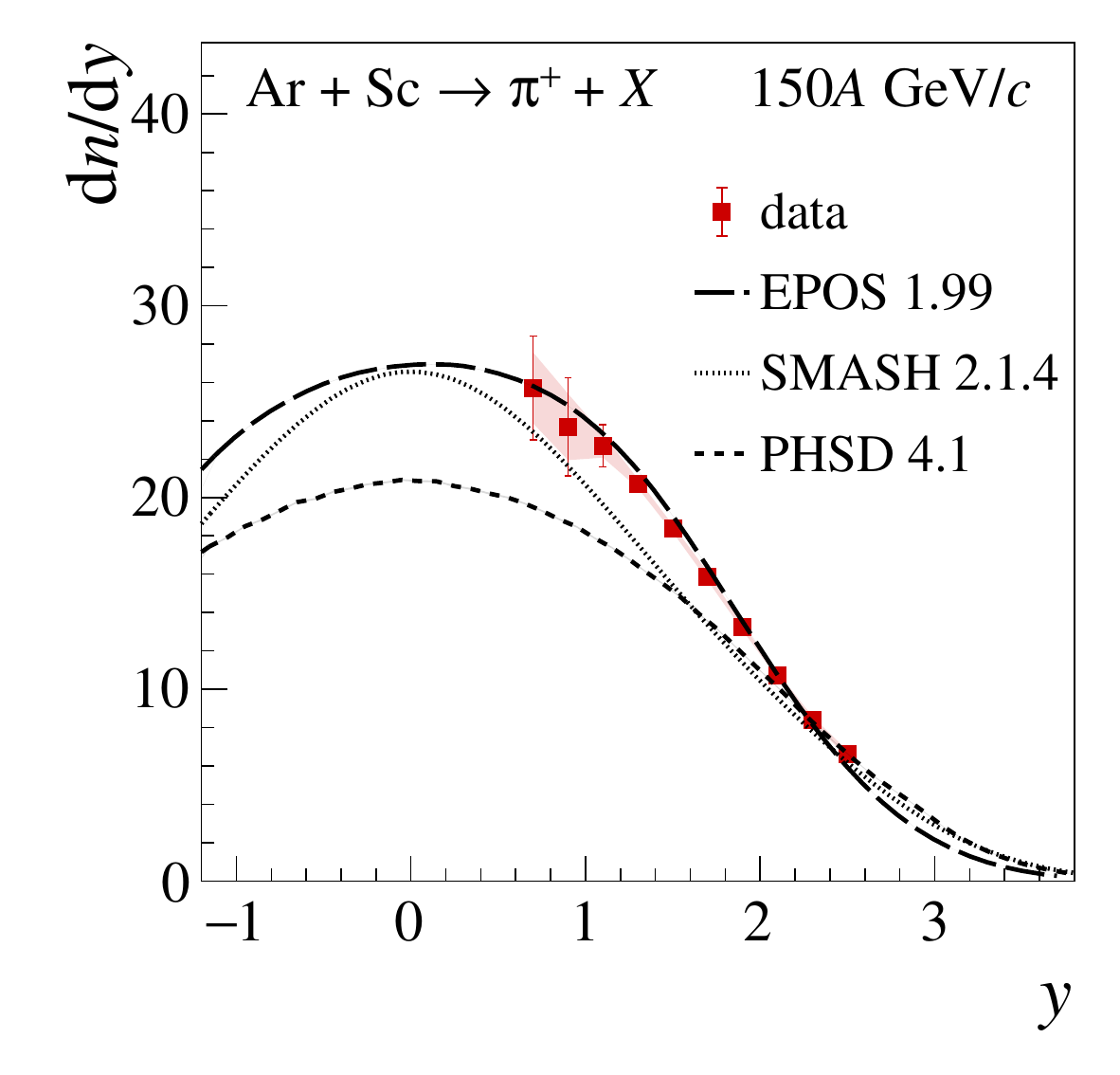}\\
\caption{Rapidity spectra of $\pi^+$ mesons produced in 10\% most \textit{central} Ar+Sc interactions at 13$A$, 19$A$, 30$A$, 40$A$, 75$A$ and 150\AGeVc in comparison with models: \Epos 1.99~\cite{Werner:2008zza}, \Phsd~4.1~\cite{PhysRevC.78.034919,CASSING2009215} and \Smash~2.1.4~\cite{Mohs:2019iee,PhysRevC.94.054905}.}
\label{fig:model_pipos_yspectra}
\end{figure}

\begin{figure}
\centering
\includegraphics[width=0.33\linewidth]{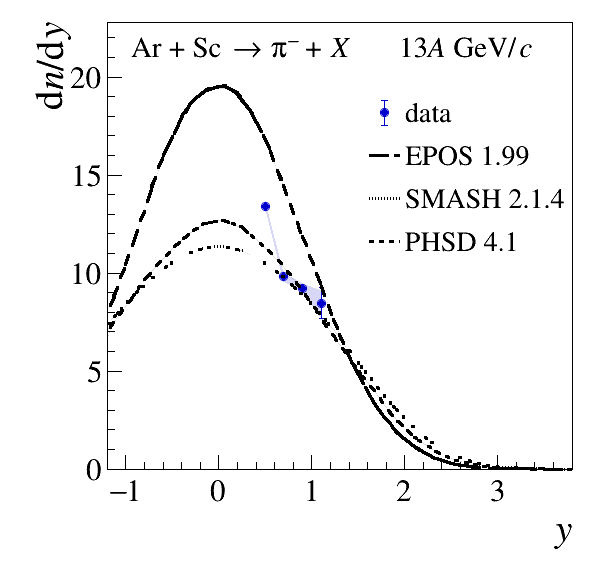}~
\includegraphics[width=0.33\linewidth]{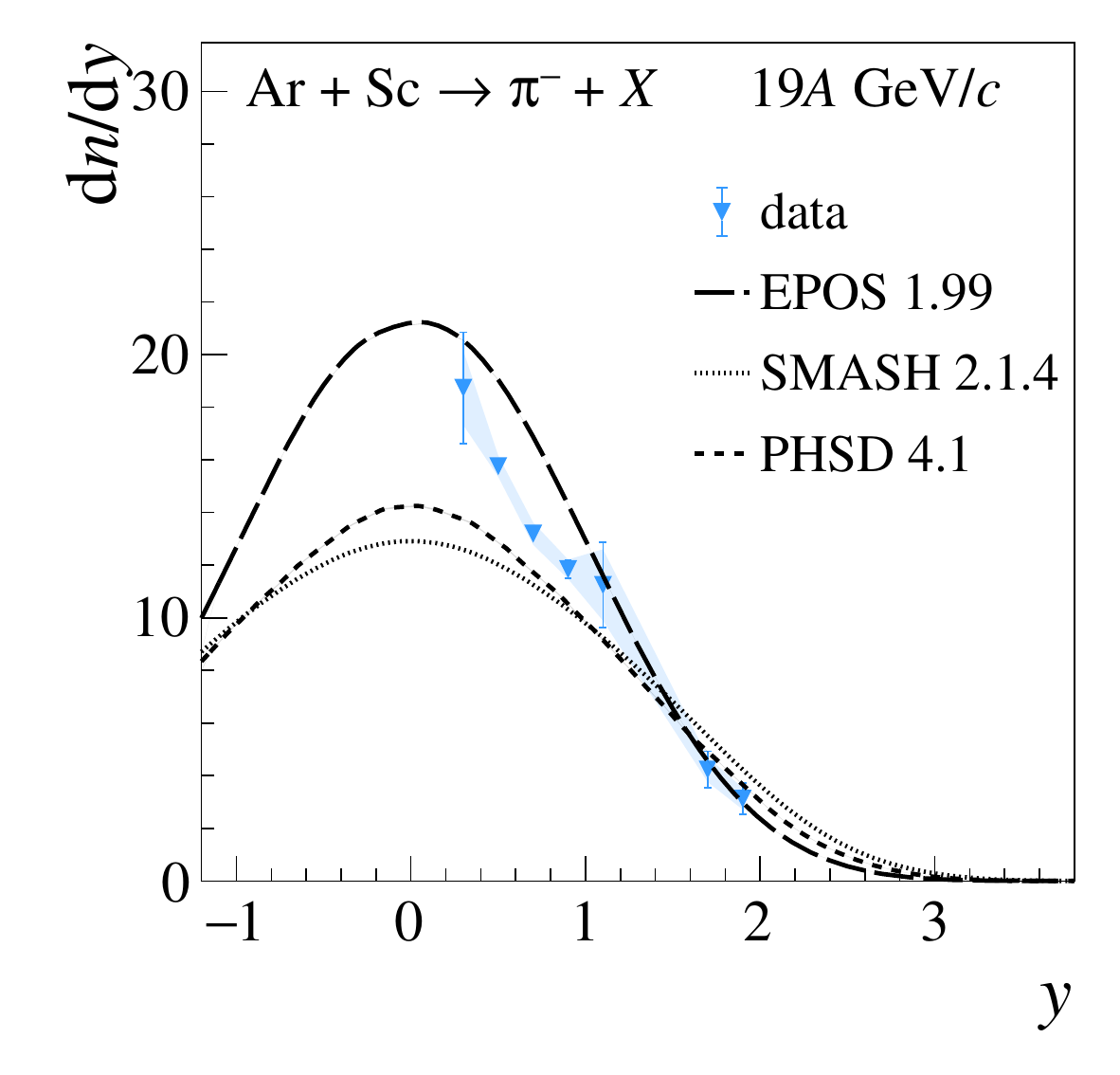}~
\includegraphics[width=0.33\linewidth]{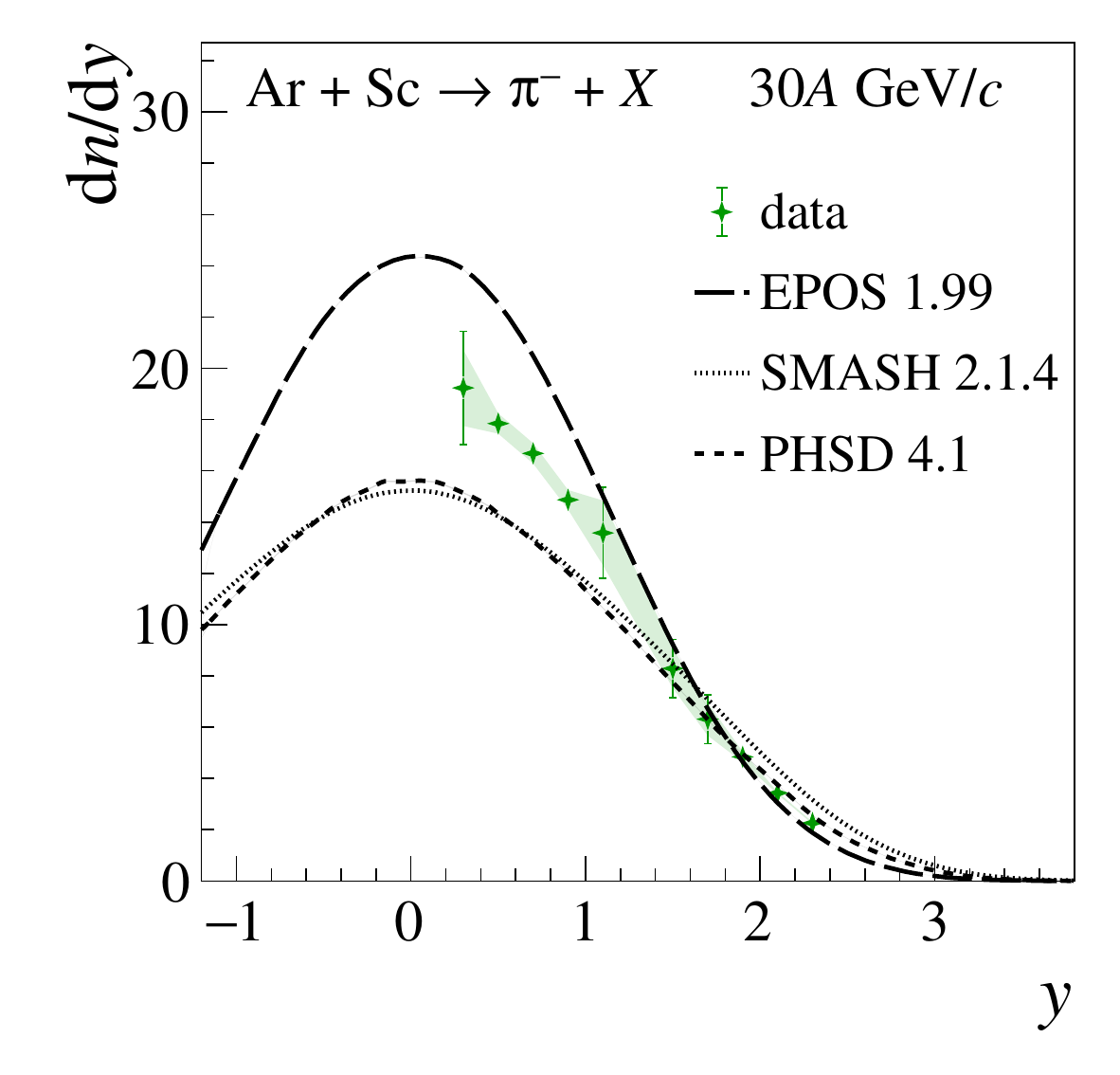}\\
\includegraphics[width=0.33\linewidth]{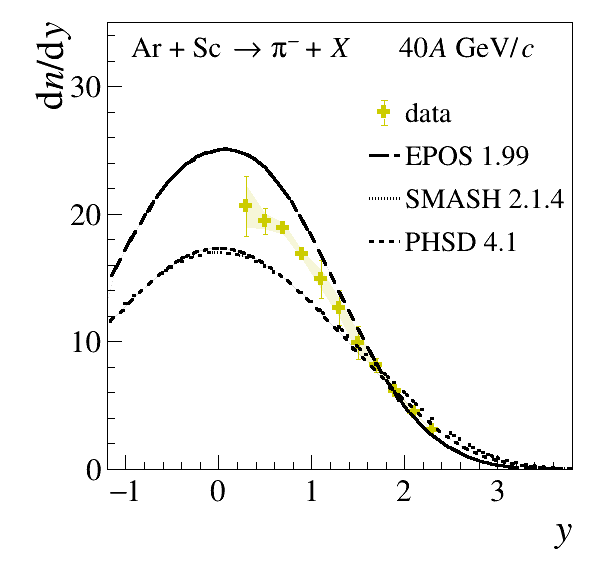}~
\includegraphics[width=0.33\linewidth]{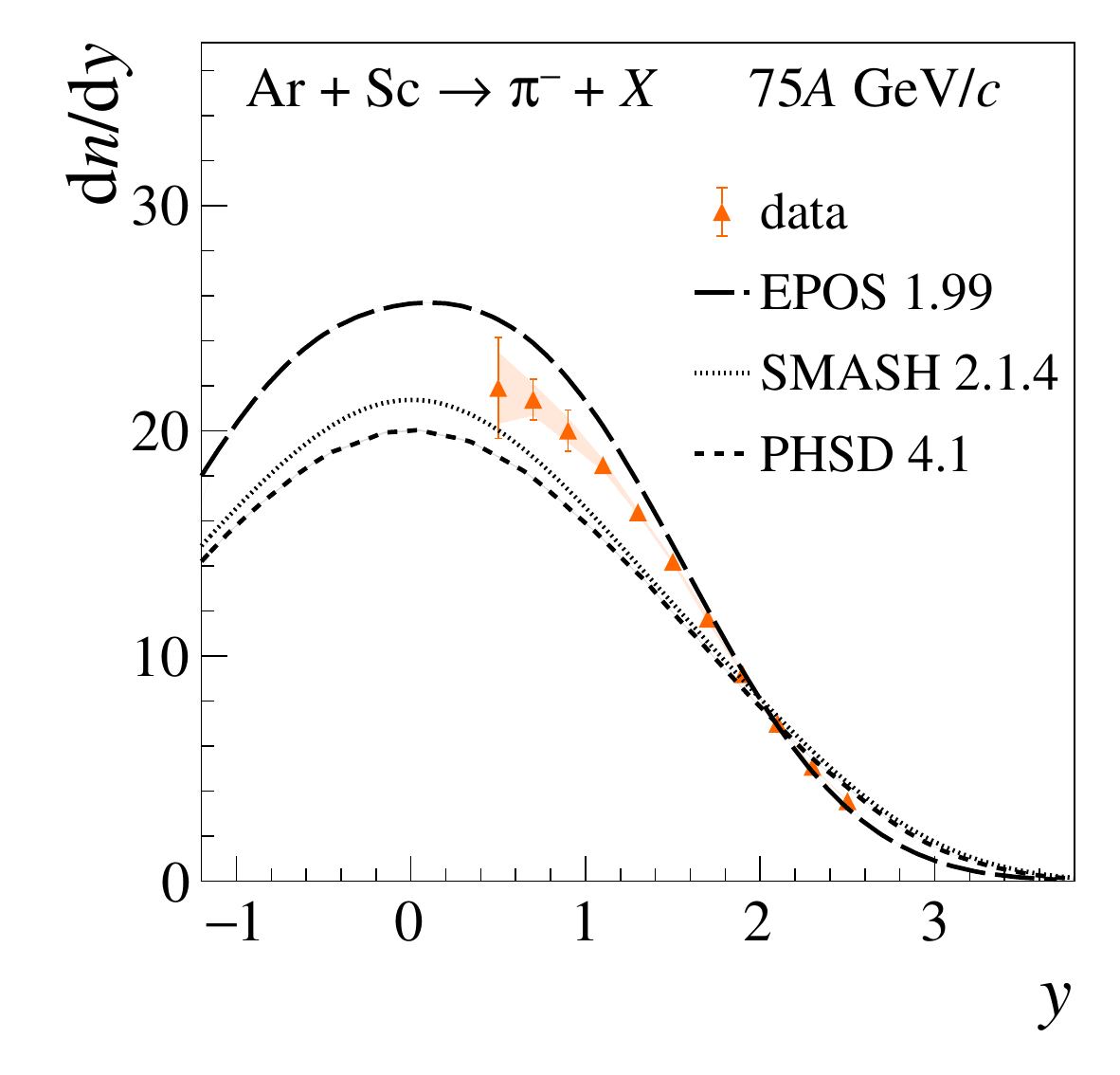}~
\includegraphics[width=0.33\linewidth]{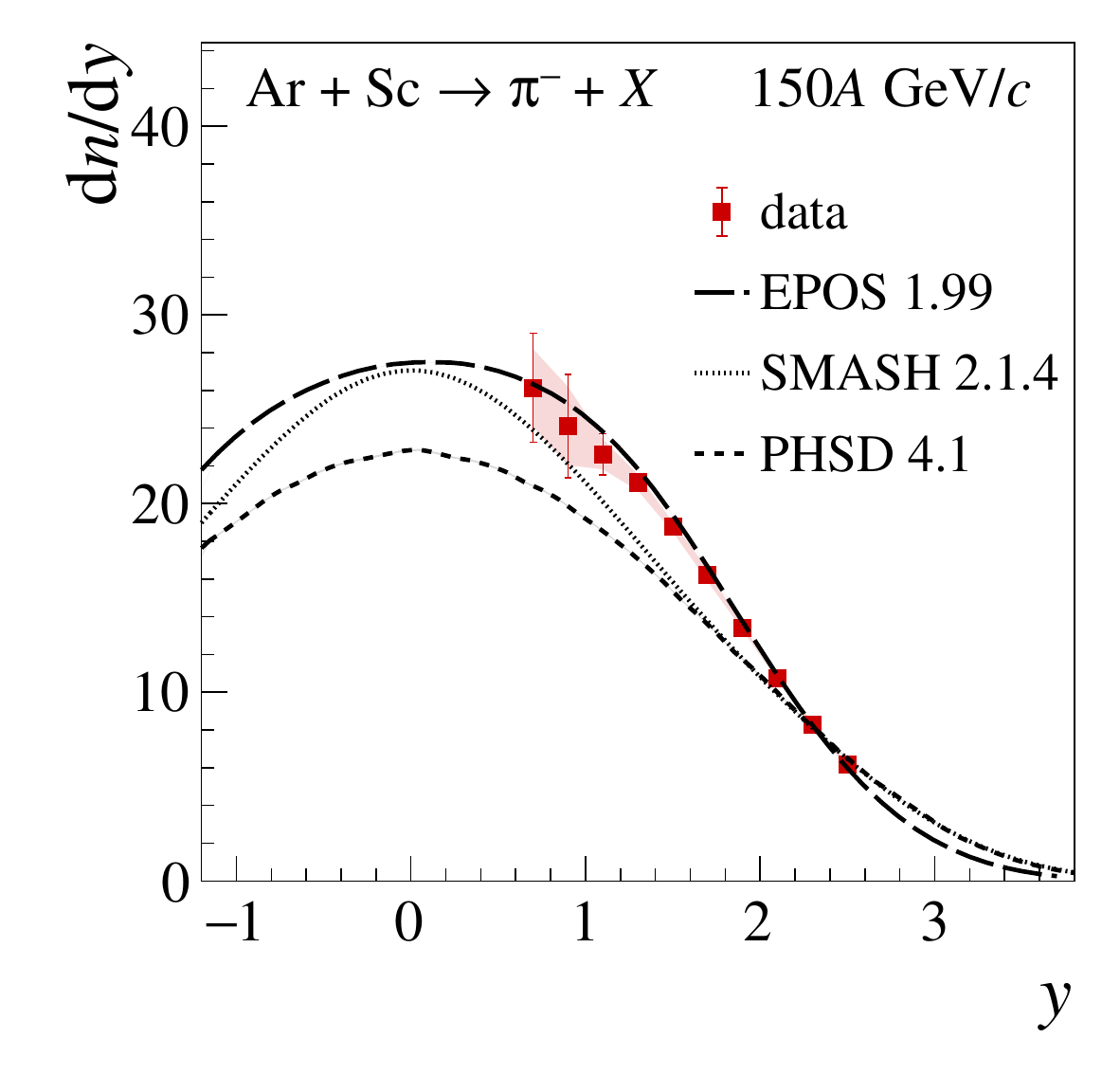}\\
\caption{Rapidity spectra of $\pi^-$ mesons produced in 10\% most \textit{central} Ar+Sc interactions at 13$A$, 19$A$, 30$A$, 40$A$, 75$A$ and 150\AGeVc in comparison with models: \Epos 1.99~\cite{Werner:2008zza}, \Phsd~4.1~\cite{PhysRevC.78.034919,CASSING2009215} and \Smash~2.1.4~\cite{Mohs:2019iee,PhysRevC.94.054905}.}
\label{fig:model_pineg_yspectra}
\end{figure}

Both the shape and the magnitude of $\pi^+$ and $\pi^-$ rapidity spectra in \textit{central} Ar+Sc collisions (Figs. \ref{fig:model_pipos_yspectra} and \ref{fig:model_pineg_yspectra}) are relatively well-described by all analyzed models at higher collision energies. The \Epos model overestimates the charged pion yield at $p_\text{beam} \leq 40$\AGeVc and the $\pi^\pm$ rapidity spectrum in the \Smash model is narrower than observed in data.

\begin{figure}
\centering
\includegraphics[width=0.45\linewidth]{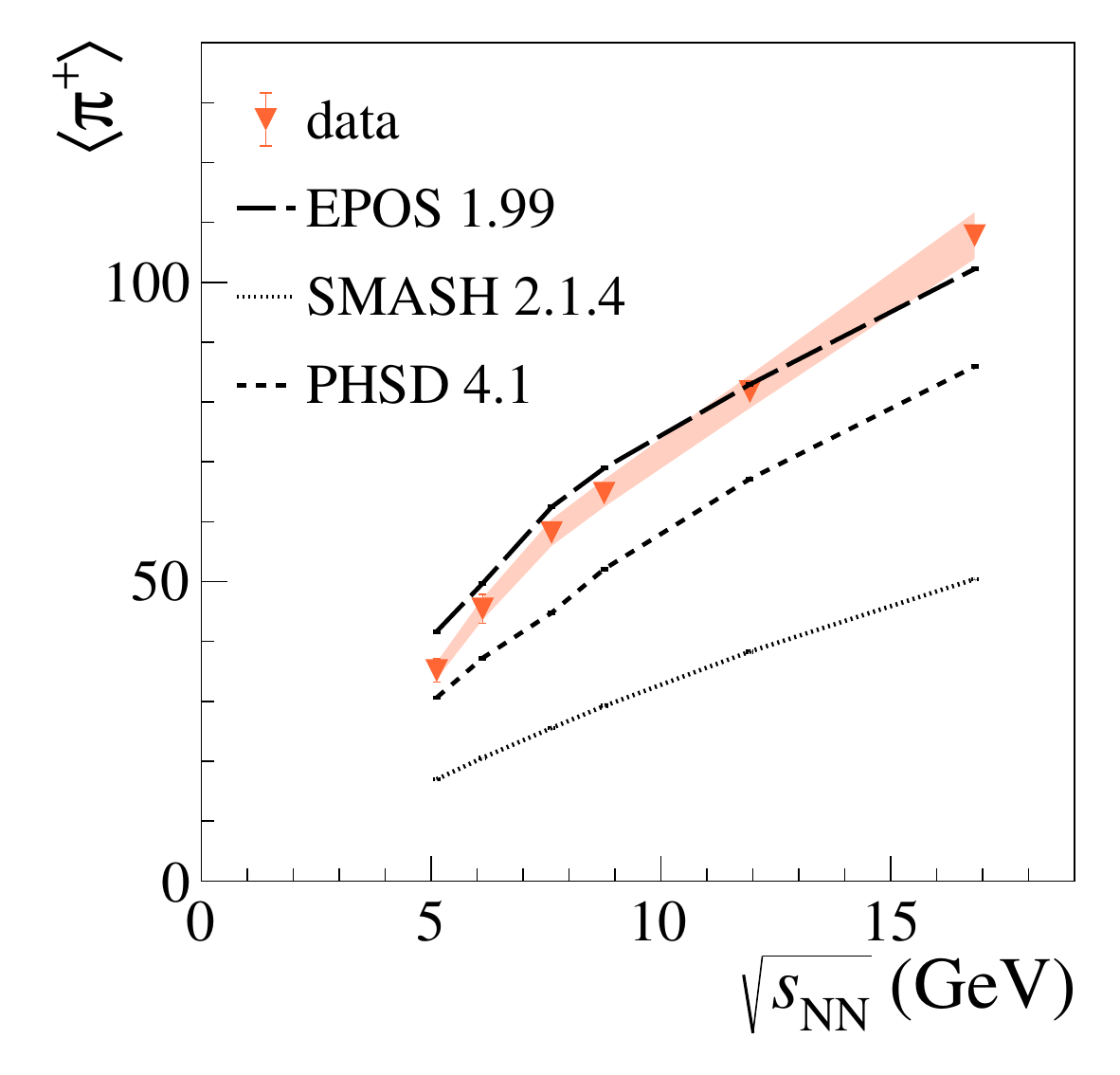}~
\includegraphics[width=0.45\linewidth]{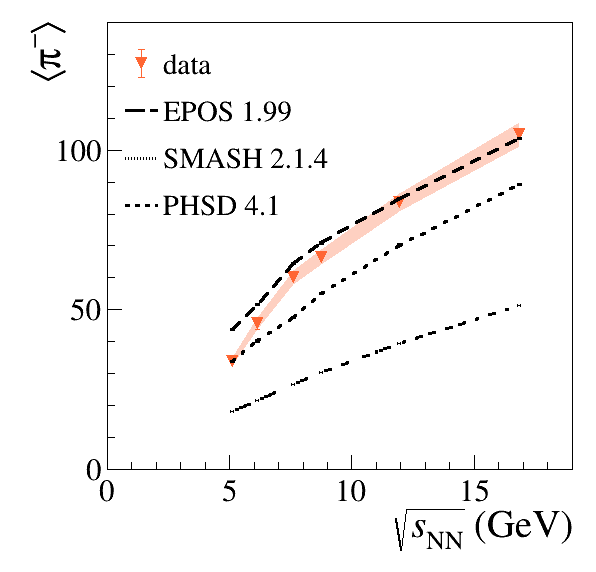}
\caption{Mean multiplicities of $\pi^+$ and $\pi^-$ mesons produced in 10\% most \textit{central} Ar+Sc interactions at 13$A$, 19$A$, 30$A$, 40$A$, 75$A$ and 150\AGeVc in comparison with models: \Epos 1.99~\cite{Werner:2008zza}, \Phsd~4.1~\cite{PhysRevC.78.034919,CASSING2009215} and \Smash~2.1.4~\cite{Mohs:2019iee,PhysRevC.94.054905}.}
\label{fig:model_pimult}
\end{figure}

Figure \ref{fig:model_pimult} also displays the mean multiplicities confronted with the model predictions, which are typically within a 5--10\% to measured data. The yields calculated with the \Smash model are systematically lower than the experimental points at each collision energy.

A detailed analysis of charged pion production in \textit{central} Ar+Sc collisions in the context of the onset of deconfinement is published in Ref. \cite{NA61SHINE:2021nye}. The main observation of the study includes the apparent similarity between the Ar+Sc results and those from the Pb+Pb system in the measurement of $\langle\pi\rangle/\langle W \rangle$ ratio at the top SPS collision energy. However, at low beam momenta (13$A$, 19$A$ GeV/$c$) the ratio measured in Ar+Sc falls closer to the results from $N$+$N$ interactions. At these collision energies, the suppressed pion yield per wounded nucleon observed in central Pb+Pb collisions is attributed to pion absorption in the evolving fireball. This effect is not found for the intermediate-size Ar+Sc system. The referenced study also shows that the $\pi^-$ transverse spectra exhibit features related to the collective flow, in particular, an enhancement of yields towards high \pt values.

\subsubsection{Charged kaon spectra and multiplicities}

\begin{figure}
\centering
    \includegraphics[page=1, width=0.32\textwidth]{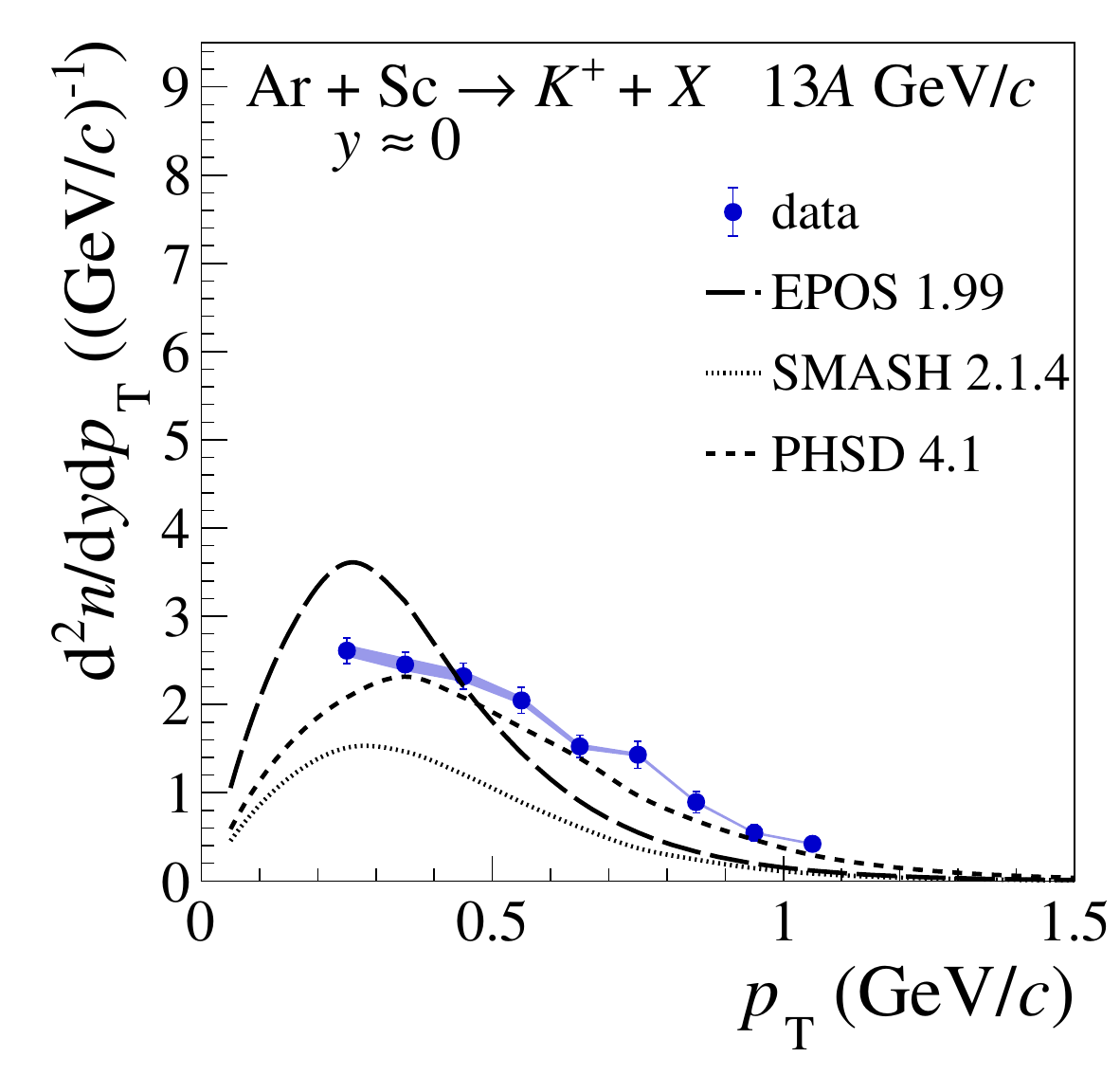}
    \includegraphics[page=2, width=0.32\textwidth]{figures/newresults/model_comp_kpos_pT.pdf}
    \includegraphics[page=3, width=0.32\textwidth]{figures/newresults/model_comp_kpos_pT.pdf}\\
    \includegraphics[page=4, width=0.32\textwidth]{figures/newresults/model_comp_kpos_pT.pdf}
    \includegraphics[page=5, width=0.32\textwidth]{figures/newresults/model_comp_kpos_pT.pdf}
    \includegraphics[page=6, width=0.32\textwidth]{figures/newresults/model_comp_kpos_pT.pdf}
    \caption{Mid-rapidity transverse momentum spectra of $K^+$ mesons produced in 10\% most \textit{central} Ar+Sc interactions at 13$A$, 19$A$, 30$A$, 40$A$, 75$A$ and 150\AGeVc in comparison with models: \Epos 1.99~\cite{Werner:2008zza}, \Phsd~4.1~\cite{PhysRevC.78.034919,CASSING2009215} and \Smash~2.1.4~\cite{Mohs:2019iee,PhysRevC.94.054905}.}
\label{fig:modelcompkpospt}
\end{figure}

\begin{figure}
\centering
    \includegraphics[page=1, width=0.32\textwidth]{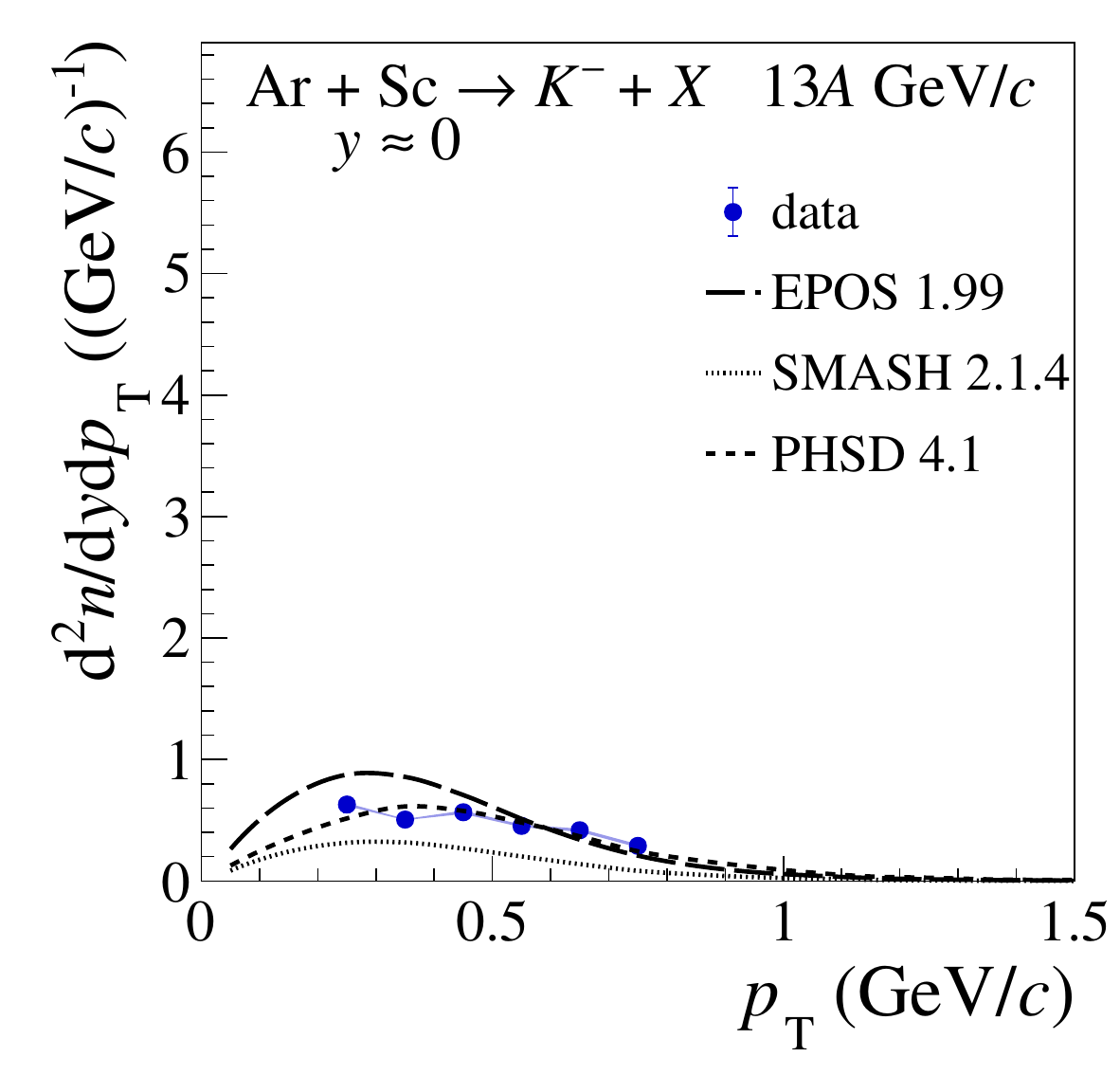}
    \includegraphics[page=2, width=0.32\textwidth]{figures/newresults/model_comp_kneg_pT.pdf}
    \includegraphics[page=3, width=0.32\textwidth]{figures/newresults/model_comp_kneg_pT.pdf}\\
    \includegraphics[page=4, width=0.32\textwidth]{figures/newresults/model_comp_kneg_pT.pdf}
    \includegraphics[page=5, width=0.32\textwidth]{figures/newresults/model_comp_kneg_pT.pdf}
    \includegraphics[page=6, width=0.32\textwidth]{figures/newresults/model_comp_kneg_pT.pdf}
\caption{Mid-rapidity transverse momentum spectra of $K^+$ mesons produced in 10\% most \textit{central} Ar+Sc interactions at 13$A$, 19$A$, 30$A$, 40$A$, 75$A$ and 150\AGeVc in comparison with models: \Epos 1.99~\cite{Werner:2008zza}, \Phsd~4.1~\cite{PhysRevC.78.034919,CASSING2009215} and \Smash~2.1.4~\cite{Mohs:2019iee,PhysRevC.94.054905}.}
\label{fig:modelcompknegpt}
\end{figure}

\begin{figure}
\centering
\includegraphics[width=0.45\linewidth]{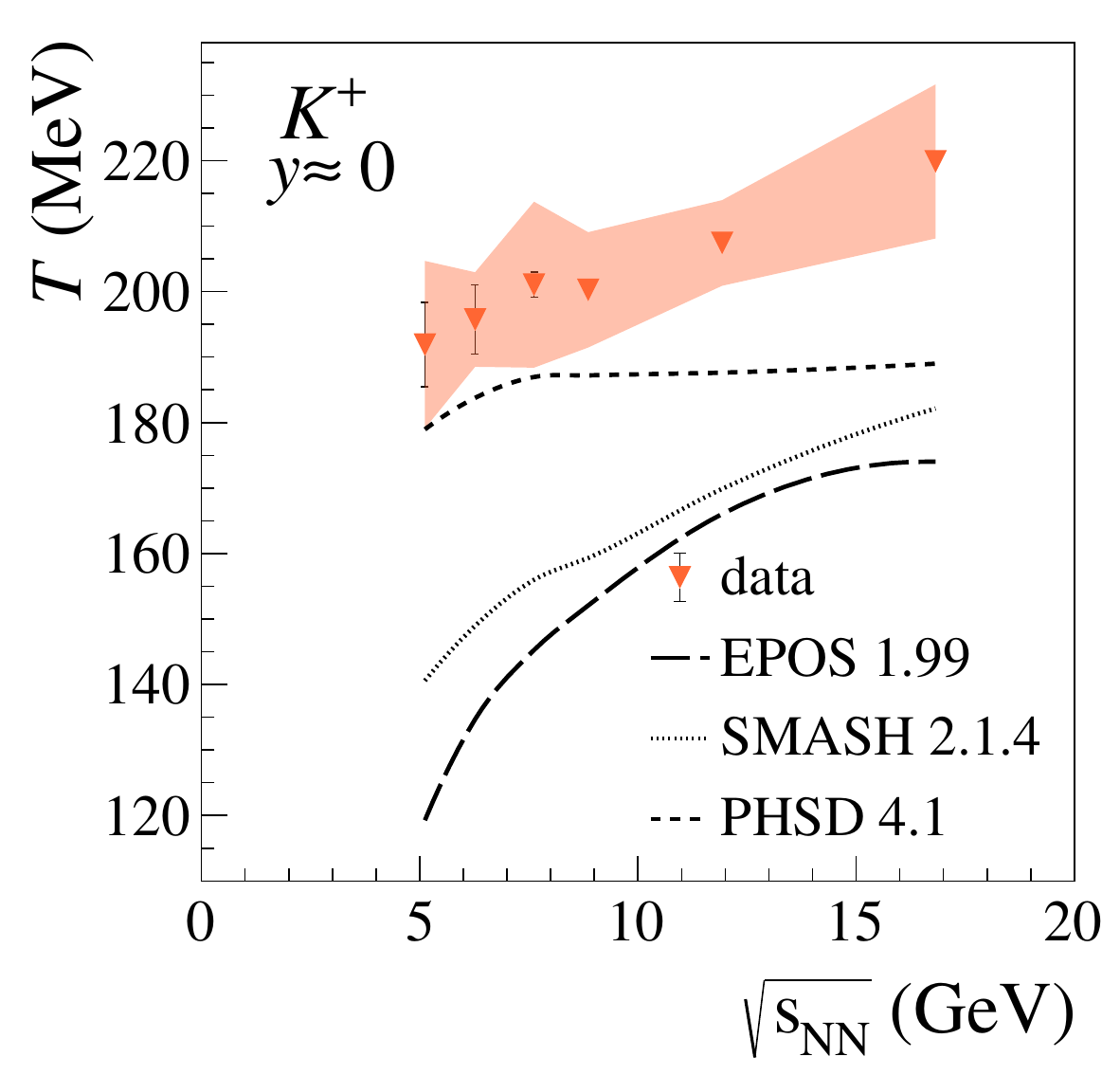}
\includegraphics[width=0.45\linewidth]{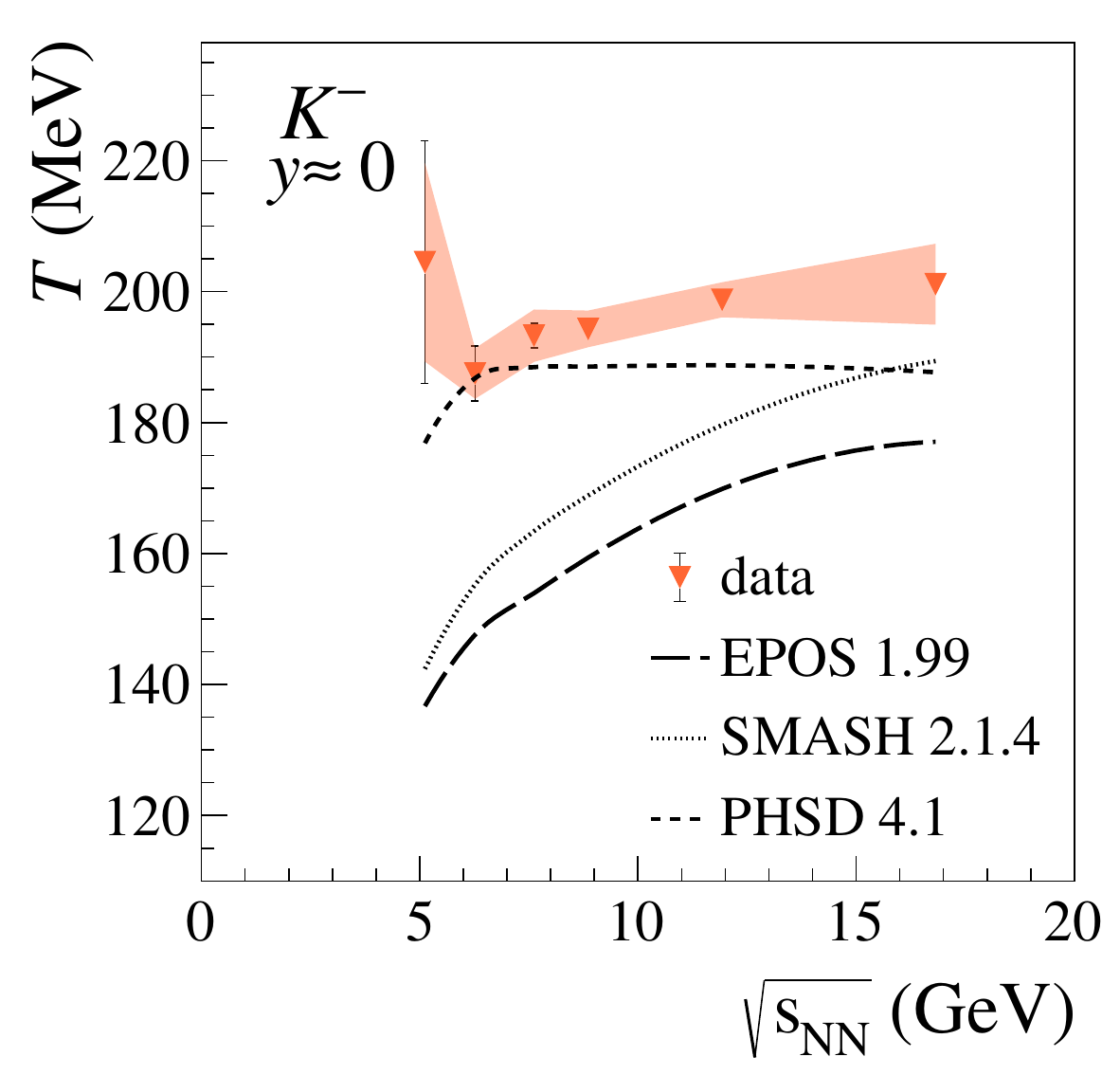}
\caption{Inverse slope parameter of mid-rapidity transverse momentum spectra of $K^+$ and $K^-$ produced in 10\% most \textit{central} Ar+Sc interactions at 13$A$, 19$A$, 30$A$, 40$A$, 75$A$ and 150\AGeVc in comparison with models: \Epos 1.99~\cite{Werner:2008zza}, \Phsd~4.1~\cite{PhysRevC.78.034919,CASSING2009215} and \Smash~2.1.4~\cite{Mohs:2019iee,PhysRevC.94.054905}.}
\label{fig:modelcompkt}
\end{figure}

Comparisons of the \pt spectra at mid-rapidity of $K^+$, $K^-$ mesons in 10\% most \textit{central} Ar+Sc collisions at all analyzed beam momenta are shown in Figs.~\ref{fig:modelcompkpospt} and \ref{fig:modelcompknegpt}. The peaks in the transverse momentum spectra are at lower \pt values in the \Epos and \Smash models relative to the data, which is also reflected in the underestimated inverse slope parameter $T$ displayed in Fig.~\ref{fig:modelcompkt}. The charged kaon transverse momentum spectra are relatively well described with the \Phsd model. 

\begin{figure}
\centering
    \includegraphics[width=0.49\textwidth]{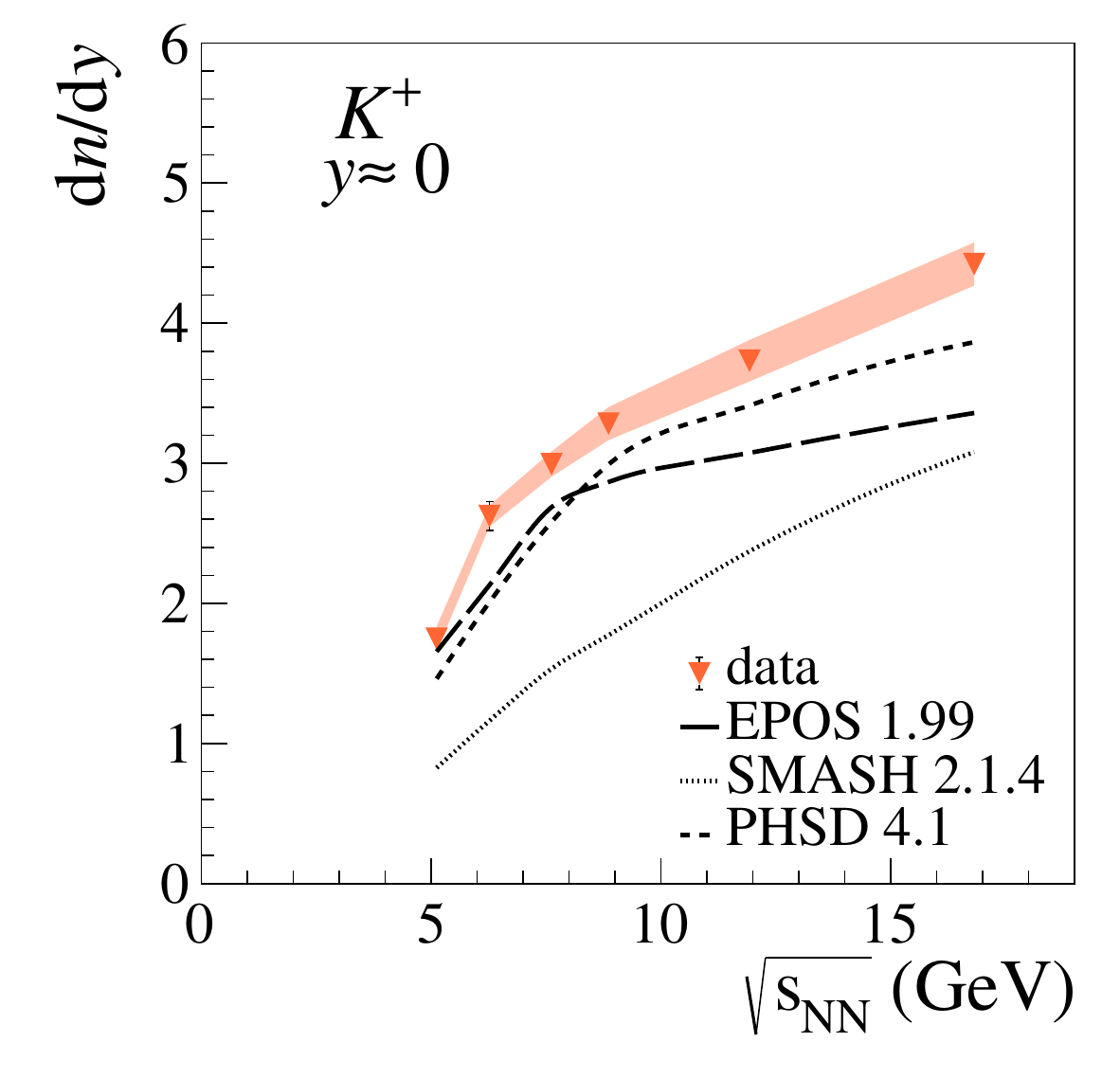}
    \includegraphics[width=0.49\textwidth]{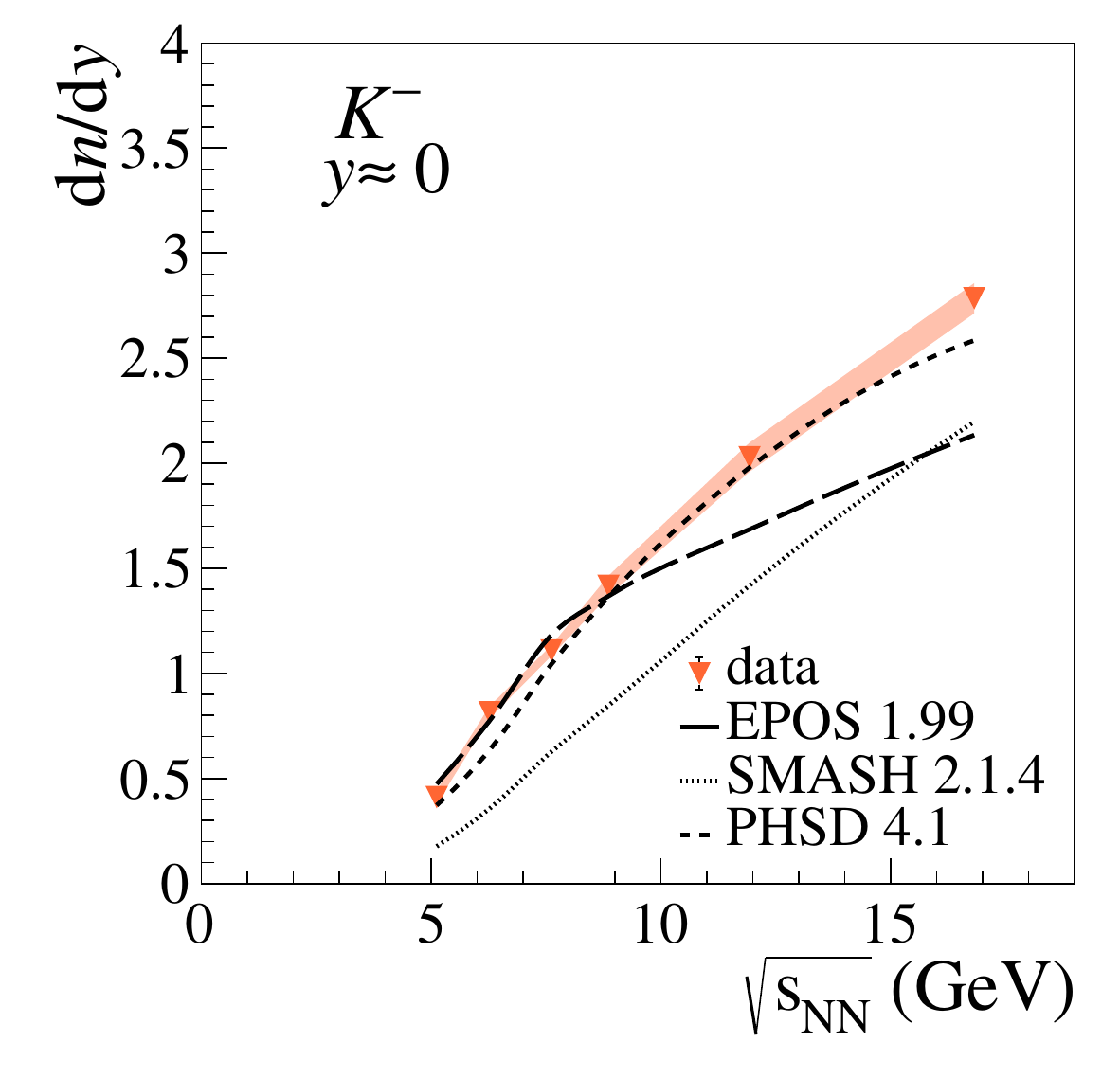}
    \caption{Mid-rapidity $\text{d}n/\text{d}y$ yields of $K^+$ and $K^-$ mesons produced in 10\% most \textit{central} Ar+Sc interactions at 13$A$, 19$A$, 30$A$, 40$A$, 75$A$ and 150\AGeVc in comparison with models: \Epos 1.99~\cite{Werner:2008zza}, \Phsd~4.1~\cite{PhysRevC.78.034919,CASSING2009215} and \Smash~2.1.4~\cite{Mohs:2019iee,PhysRevC.94.054905}.}
\label{fig:modelcompkdndy}
\end{figure}

Figure \ref{fig:modelcompkdndy} displays the charged kaon $\text{d}n/\text{d}y$ yields measured at mid-rapidity. None of the models reproduces accurately the collision energy dependence in the whole studied range, however, a good agreement is observed between the measured data and the \Phsd model in the case of $K^+$ mesons. The \Epos model agrees well with the data on $K^-$ yields only at $p_\text{beam}\leq40$\AGeVc. The \Smash model underestimates charged hadron yields at all studied collision energies.

\begin{figure}
\centering
\includegraphics[width=0.33\linewidth]{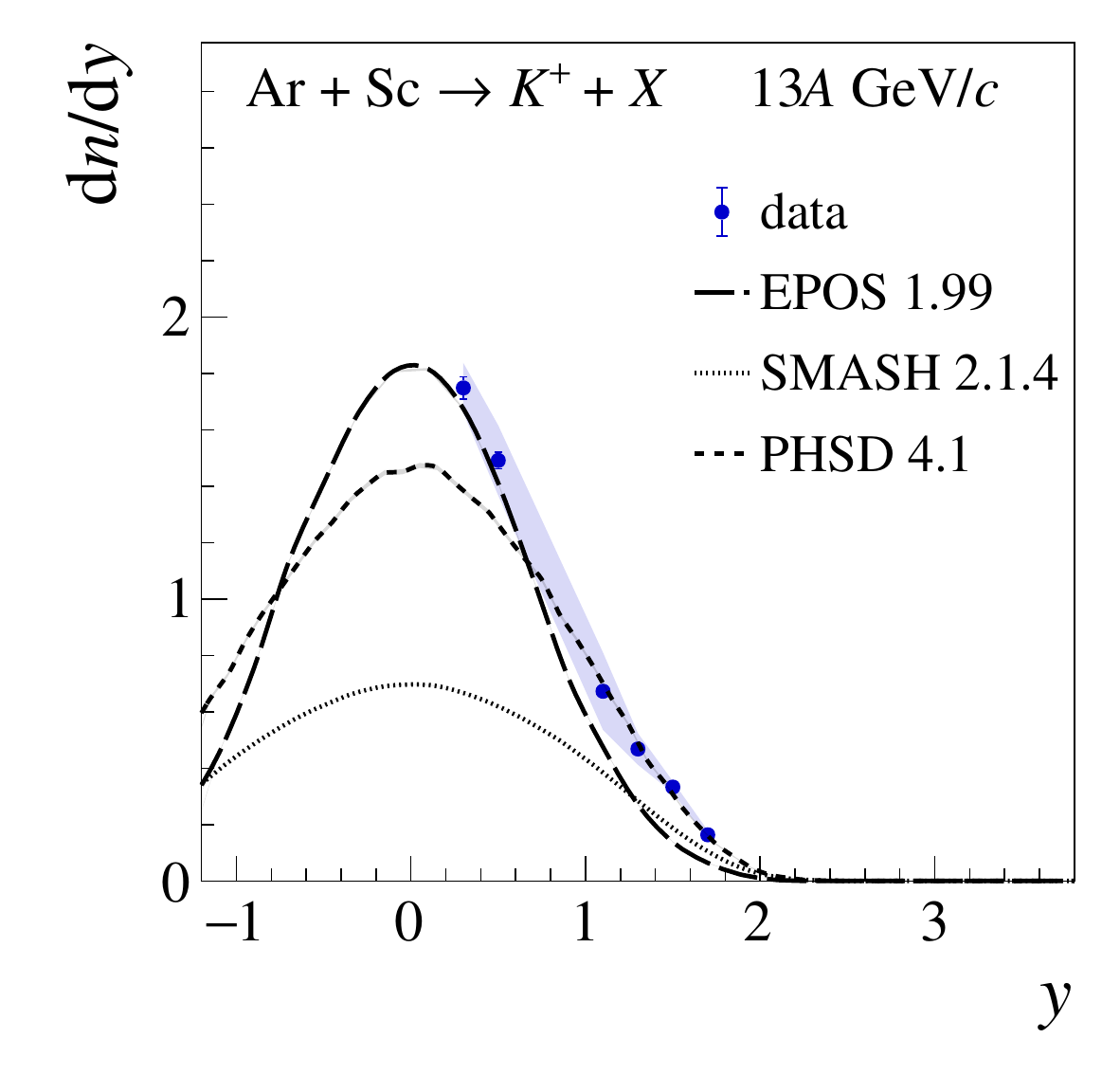}~
\includegraphics[width=0.33\linewidth]{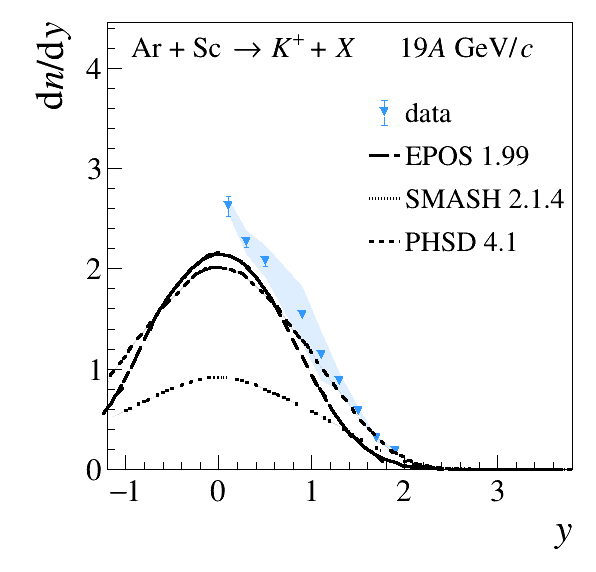}~
\includegraphics[width=0.33\linewidth]{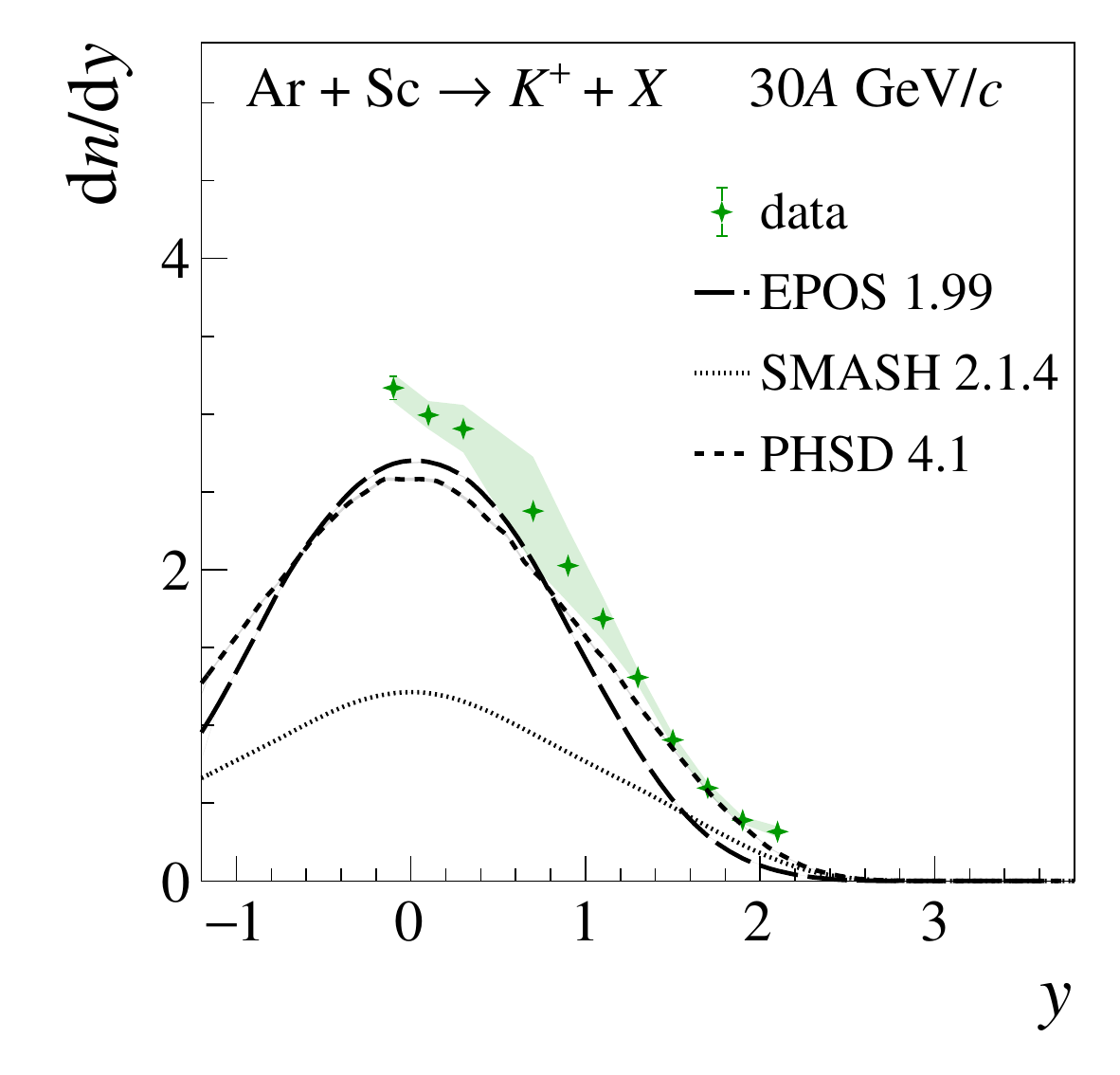}\\
\includegraphics[width=0.33\linewidth]{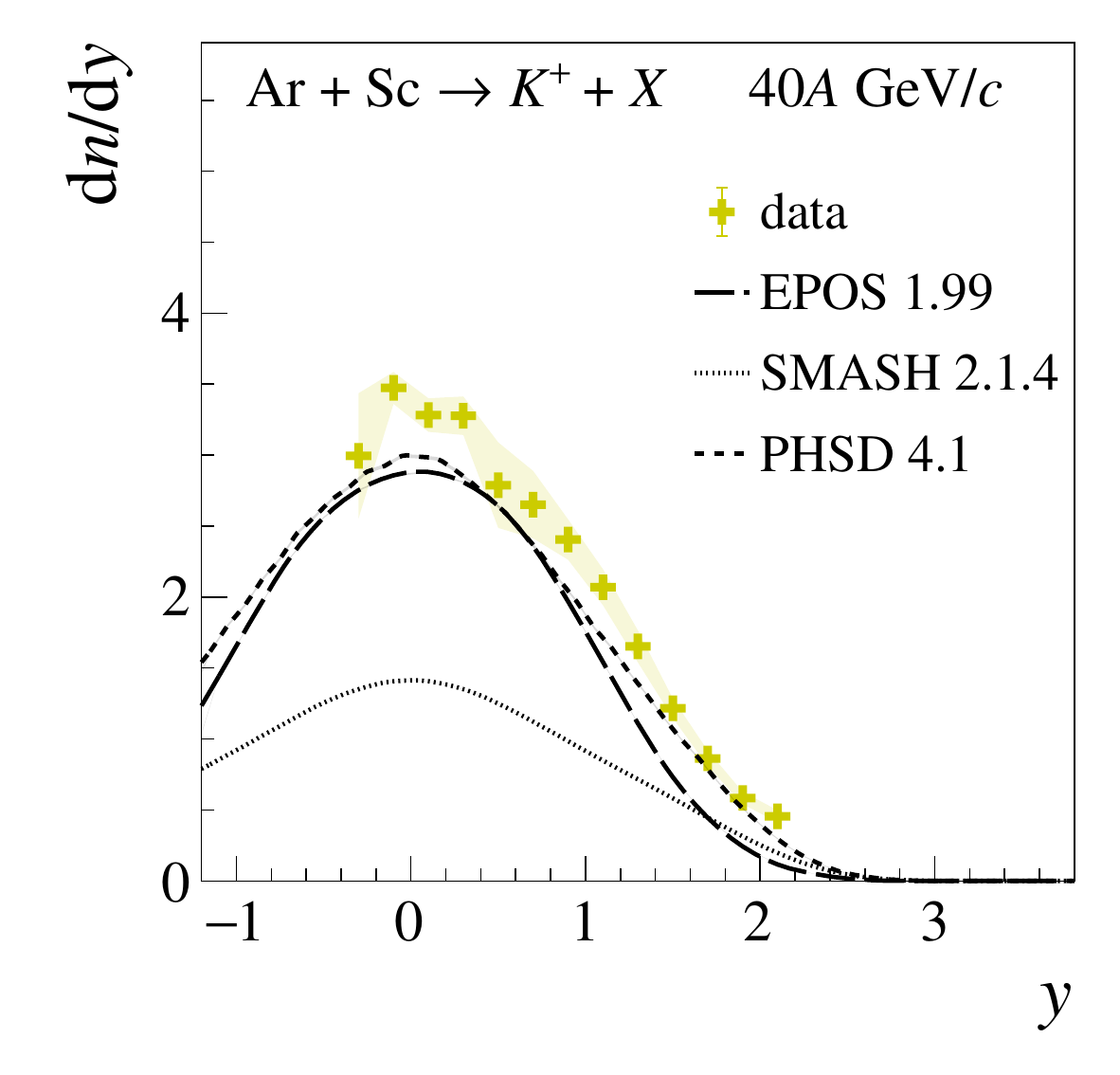}~
\includegraphics[width=0.33\linewidth]{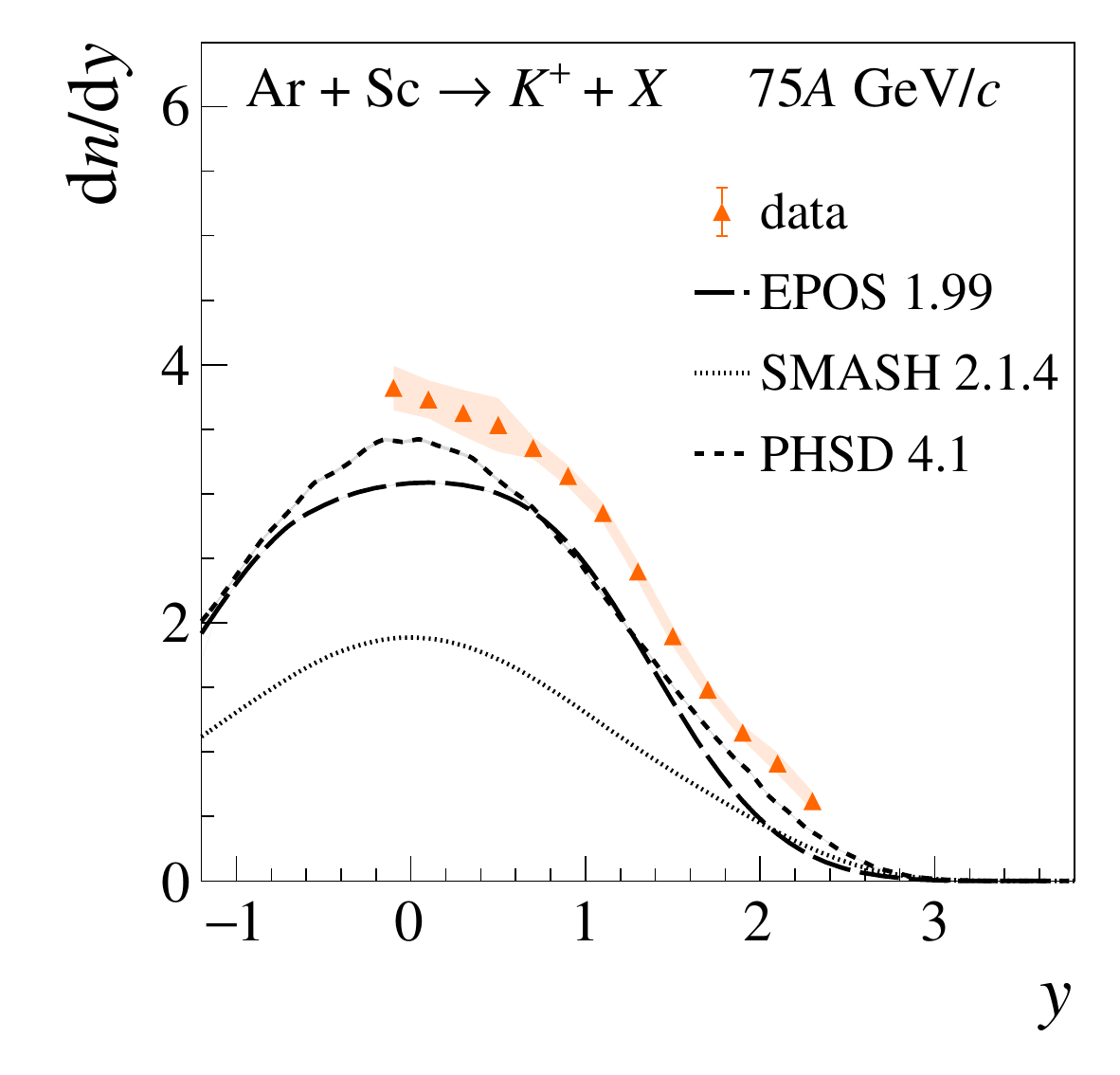}~
\includegraphics[width=0.33\linewidth]{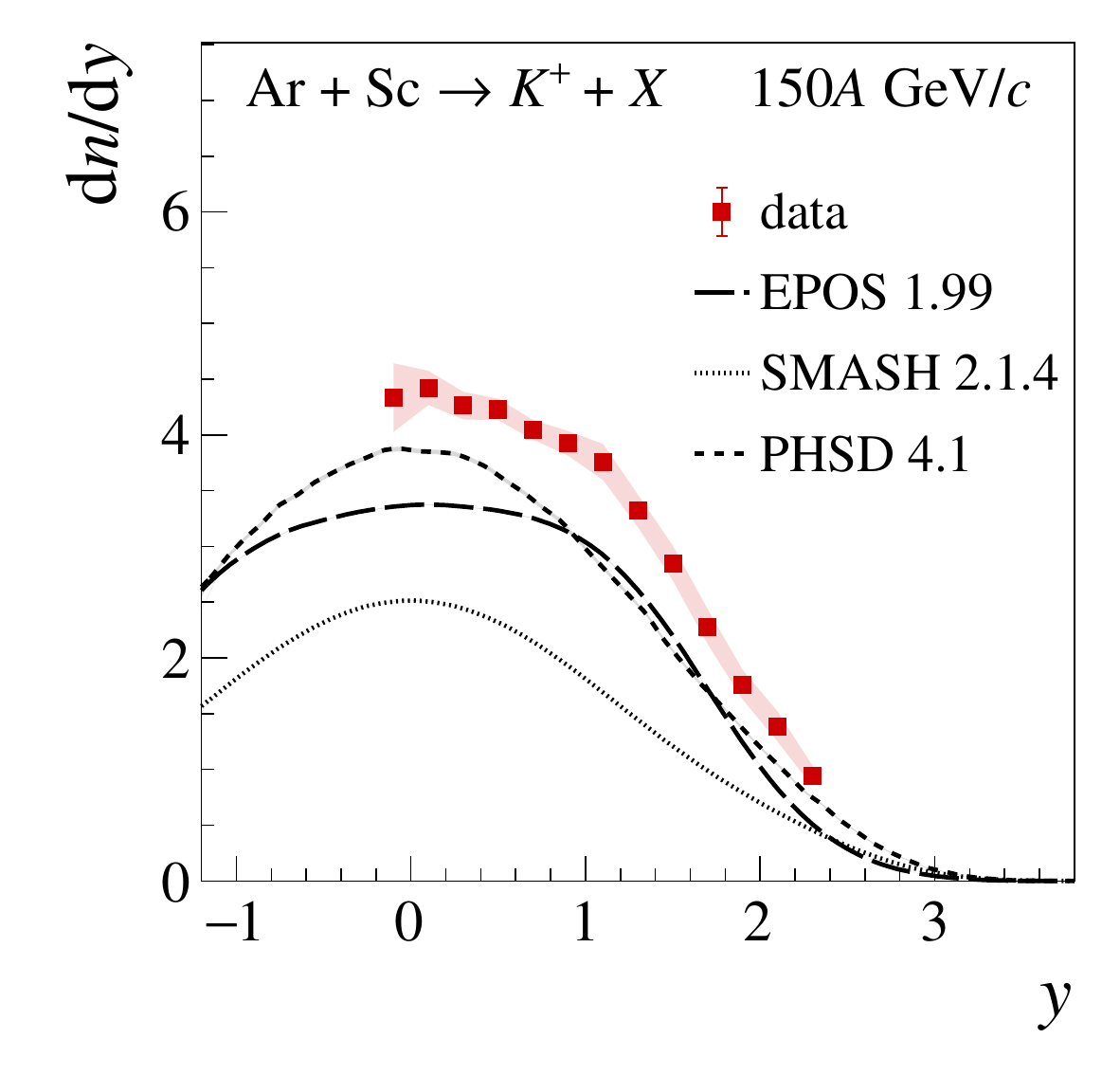}\\
\caption{Rapidity spectra of $K^+$ mesons produced in 10\% most \textit{central} Ar+Sc interactions at 13$A$, 19$A$, 30$A$, 40$A$, 75$A$ and 150\AGeVc in comparison with models: \Epos 1.99~\cite{Werner:2008zza}, \Phsd~4.1~\cite{PhysRevC.78.034919,CASSING2009215} and \Smash~2.1.4~\cite{Mohs:2019iee,PhysRevC.94.054905}.}
\label{fig:model_kpos_yspectra}
\end{figure}

\begin{figure}
\centering
\includegraphics[width=0.33\linewidth]{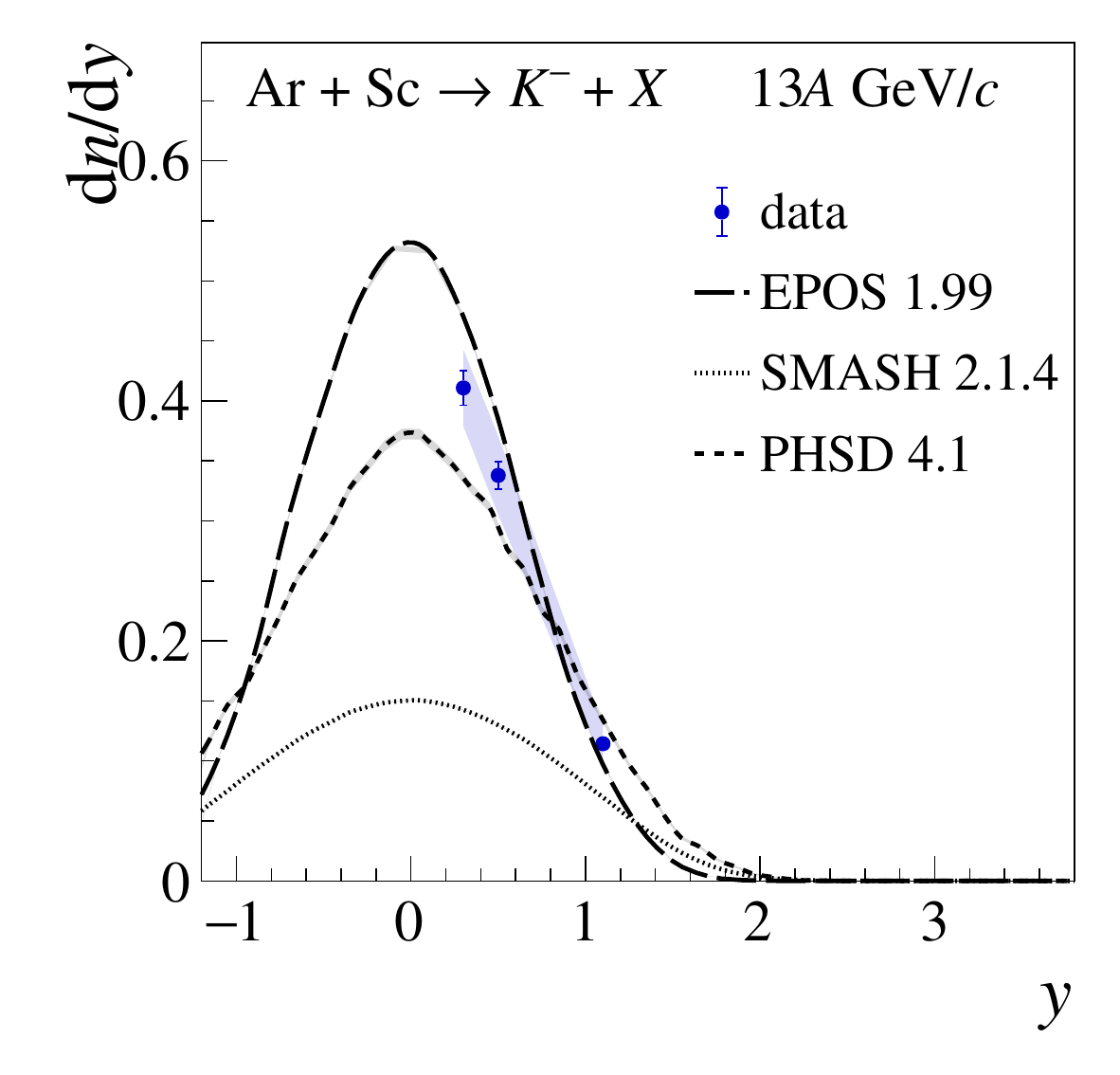}~
\includegraphics[width=0.33\linewidth]{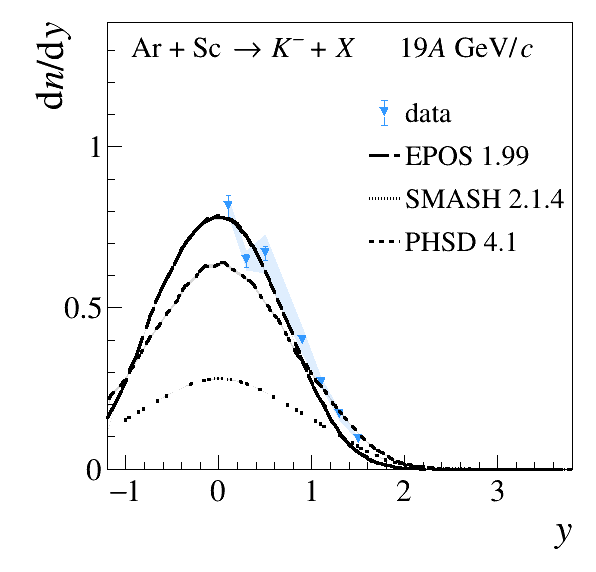}~
\includegraphics[width=0.33\linewidth]{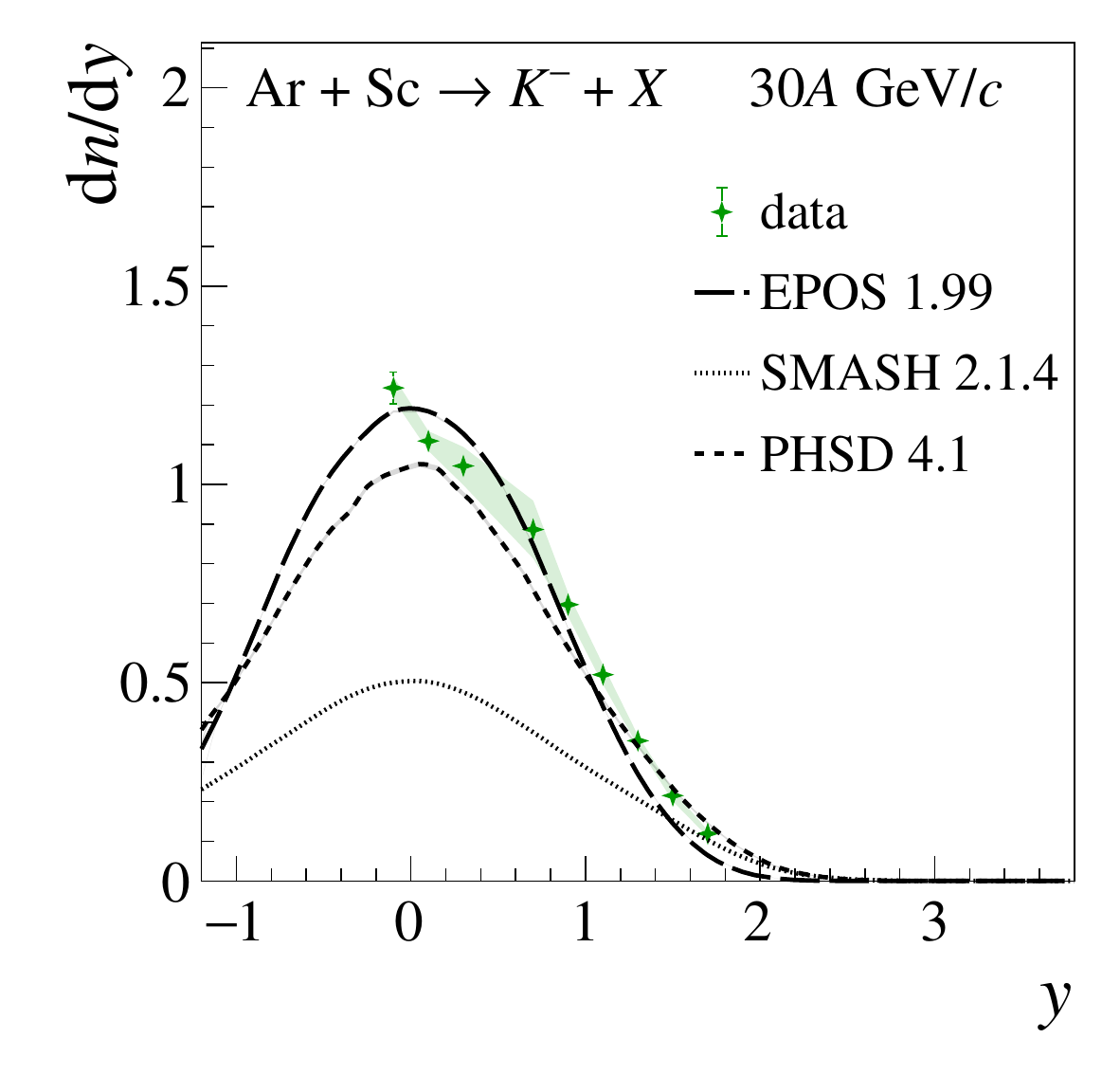}\\
\includegraphics[width=0.33\linewidth]{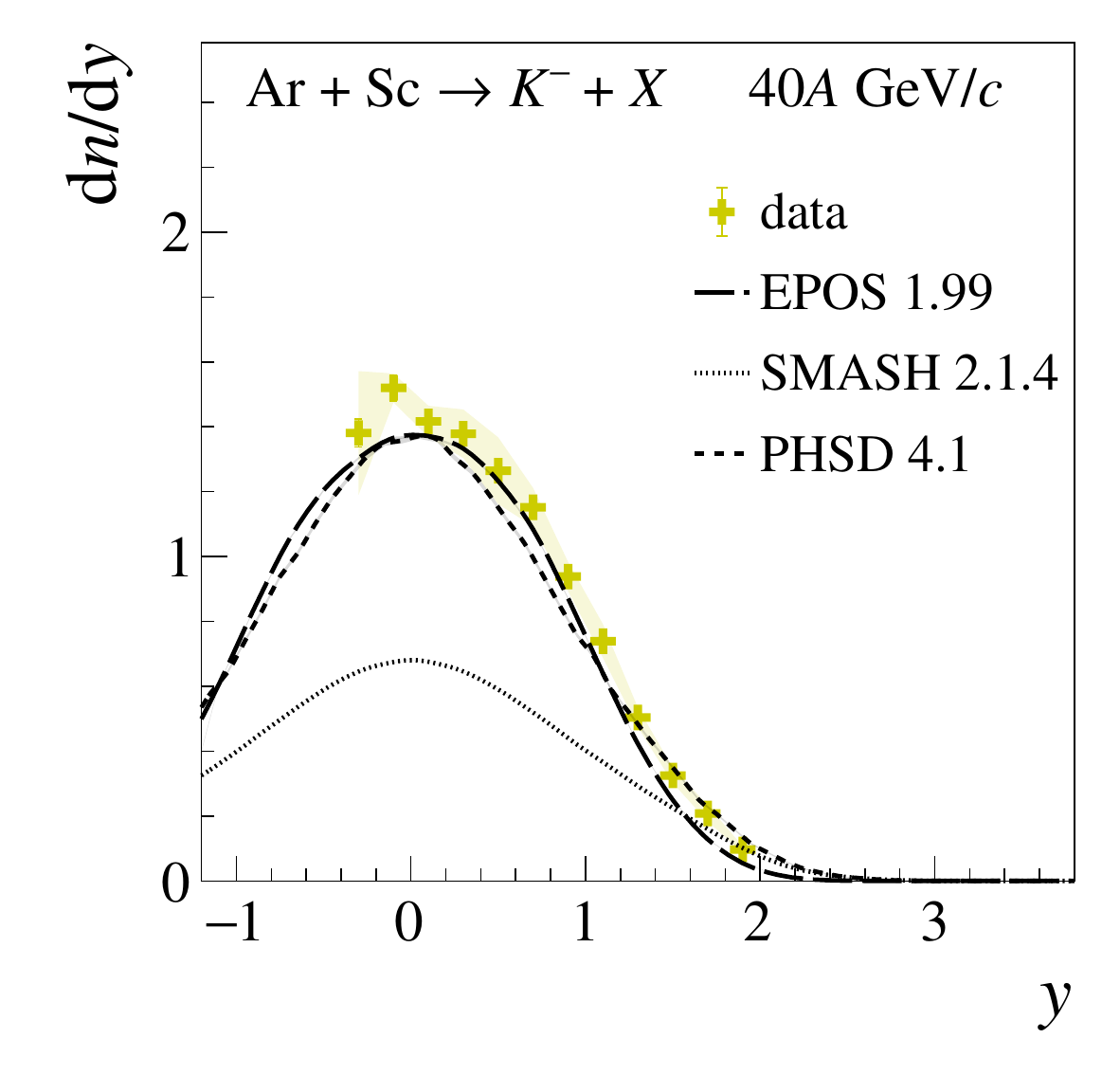}~
\includegraphics[width=0.33\linewidth]{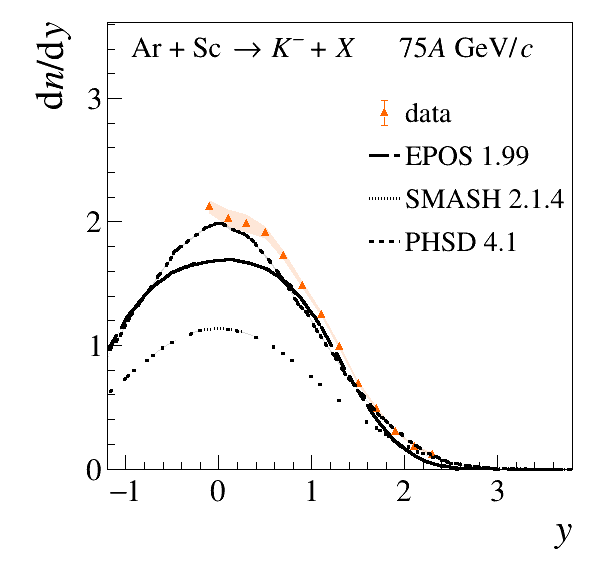}~
\includegraphics[width=0.33\linewidth]{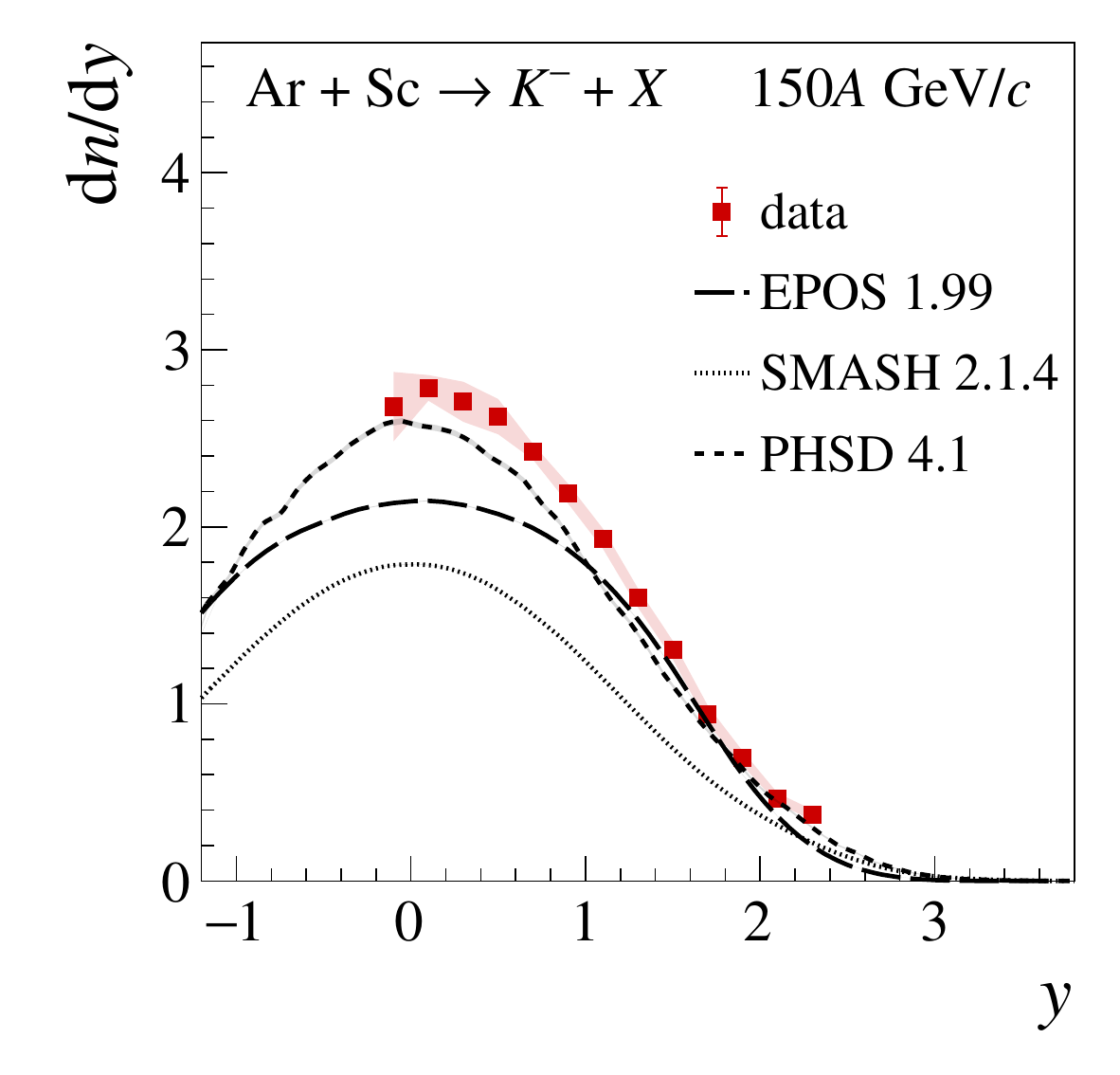}\\
\caption{Rapidity spectra of $K^-$ mesons produced in 10\% most \textit{central} Ar+Sc interactions at 13$A$, 19$A$, 30$A$, 40$A$, 75$A$ and 150\AGeVc in comparison with models: \Epos 1.99~\cite{Werner:2008zza}, \Phsd~4.1~\cite{PhysRevC.78.034919,CASSING2009215} and \Smash~2.1.4~\cite{Mohs:2019iee,PhysRevC.94.054905}.}
\label{fig:model_kneg_yspectra}
\end{figure}

\begin{figure}
\centering
\includegraphics[width=0.45\linewidth]{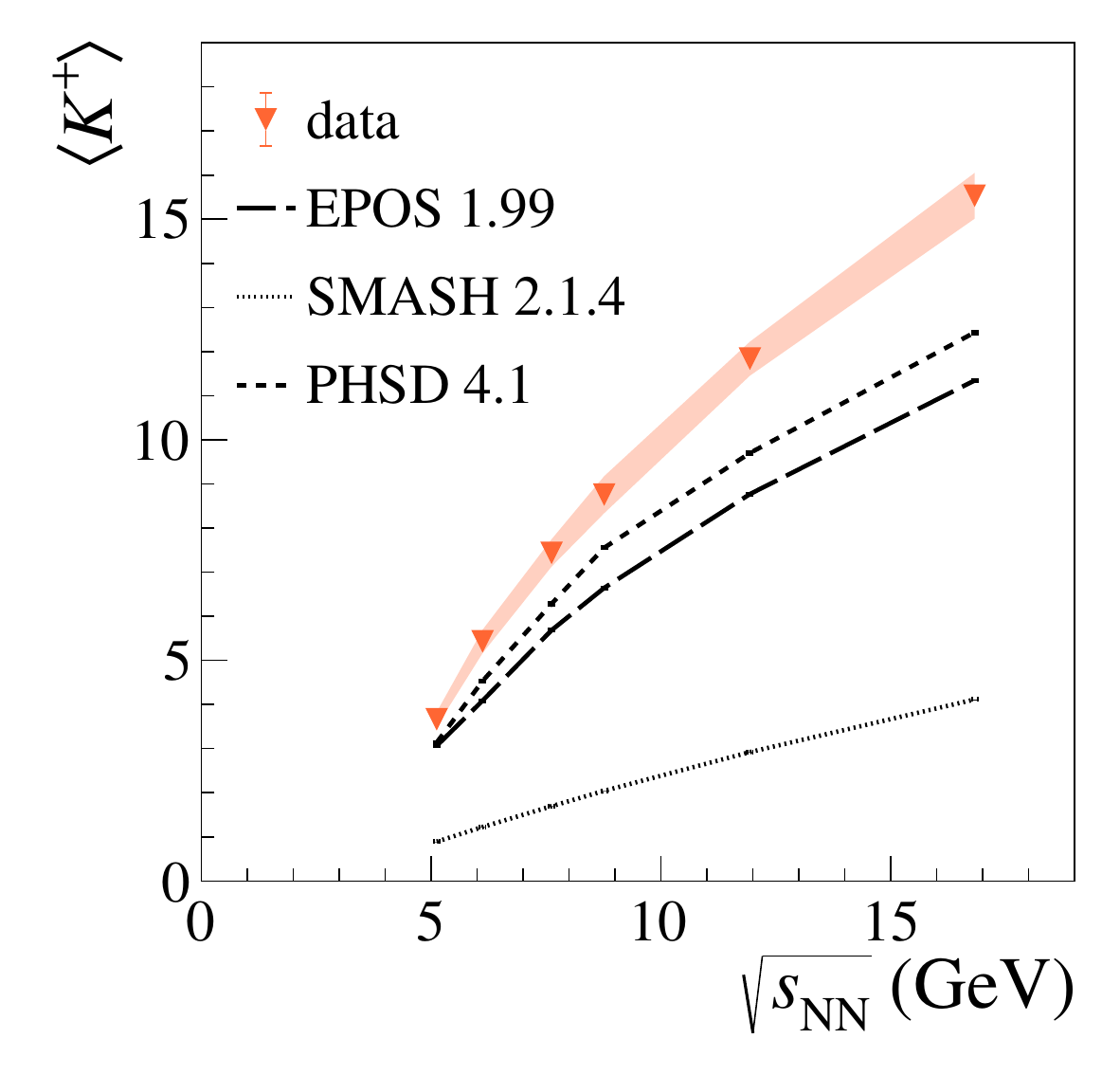}~
\includegraphics[width=0.45\linewidth]{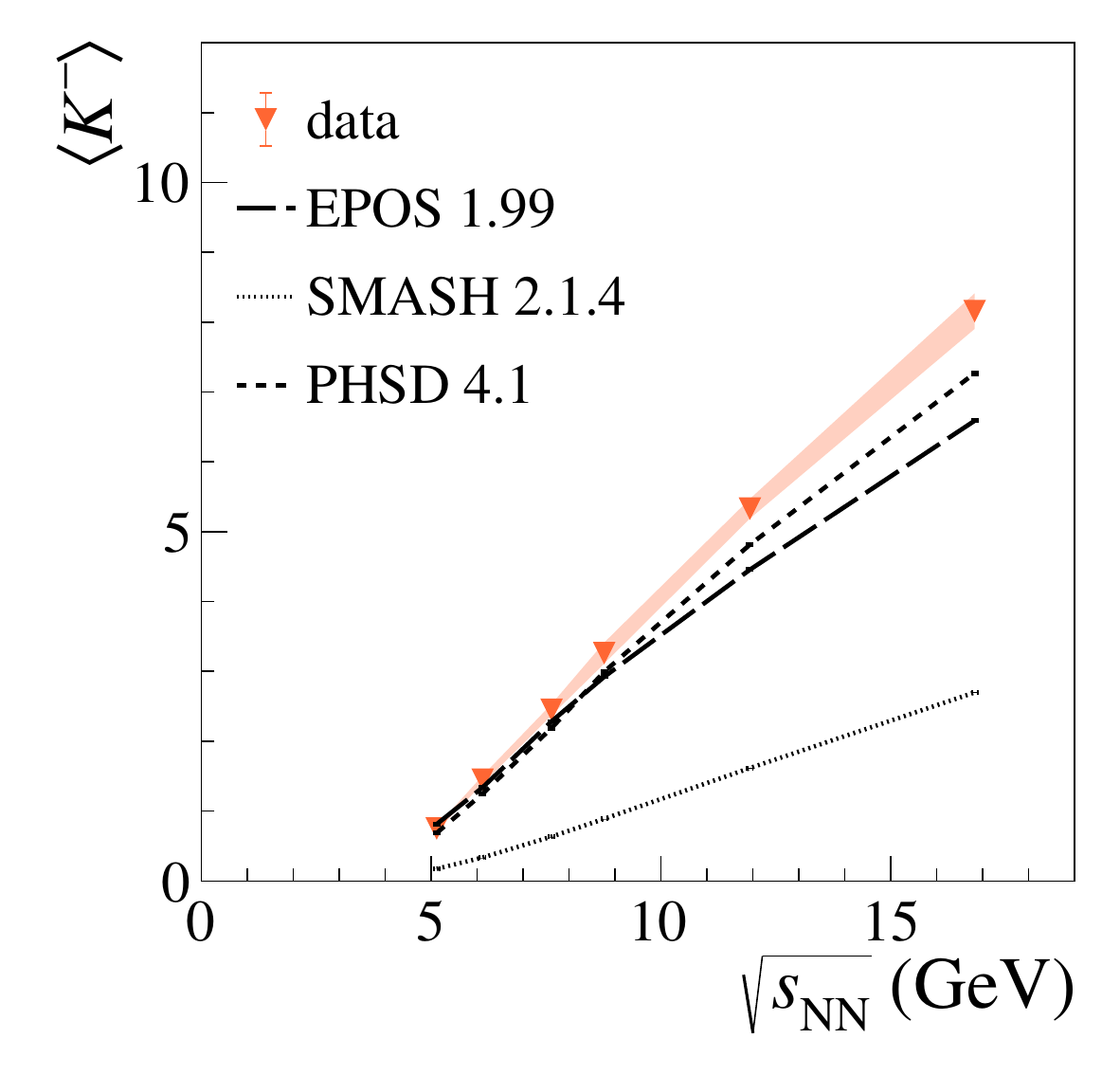}
\caption{Mean multiplicities of charged kaons ($\langle K^+\rangle$ and $\langle K^-\rangle$) produced in 10\% most \textit{central} Ar+Sc interactions at 13$A$, 19$A$, 30$A$, 40$A$, 75$A$ and 150\AGeVc plotted in dependence on collision energy compared with models: \Epos 1.99~\cite{Werner:2008zza}, \Phsd~4.1~\cite{PhysRevC.78.034919,CASSING2009215} and \Smash~2.1.4~\cite{Mohs:2019iee,PhysRevC.94.054905}.}
\label{fig:model_kmult}
\end{figure}

The model predictions of the rapidity spectra are shown in Figs.~\ref{fig:model_kpos_yspectra} and \ref{fig:model_kneg_yspectra}. Similar to previous observations, the \Smash model gives largely lower yields than both data and other investigated models. A relatively good agreement with the measured data is seen for \Epos and \Phsd at lower collision energies, however, the discrepancies become substantial at 75$A$ and 150$A$ GeV/$c$, exceeding 15\%. Described trends propagate into the collision energy dependence of mean multiplicities, which is displayed in Fig. \ref{fig:model_kmult}.

\begin{figure}
\centering
    \includegraphics[width=0.49\textwidth]{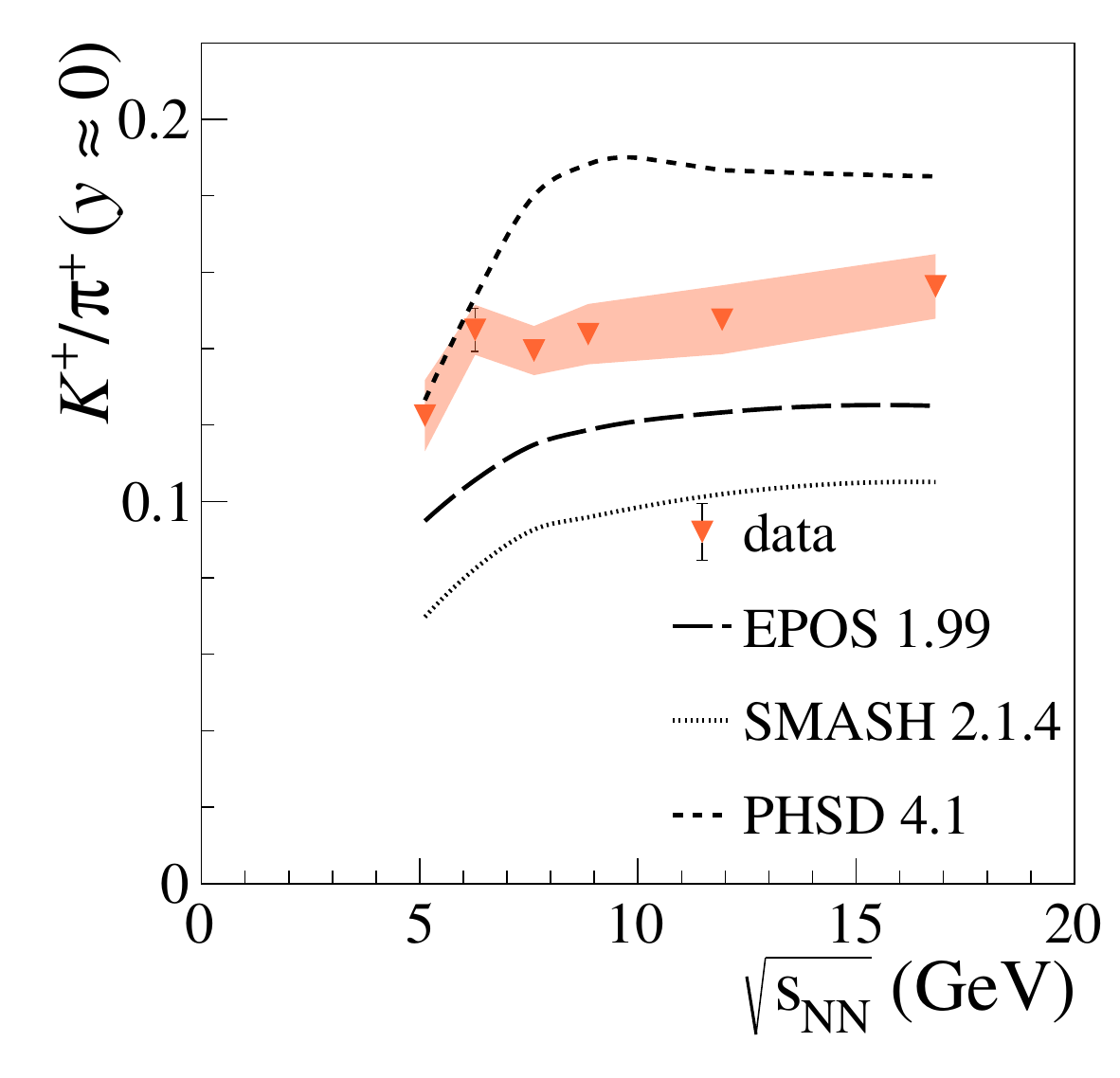}
    \includegraphics[width=0.49\textwidth]{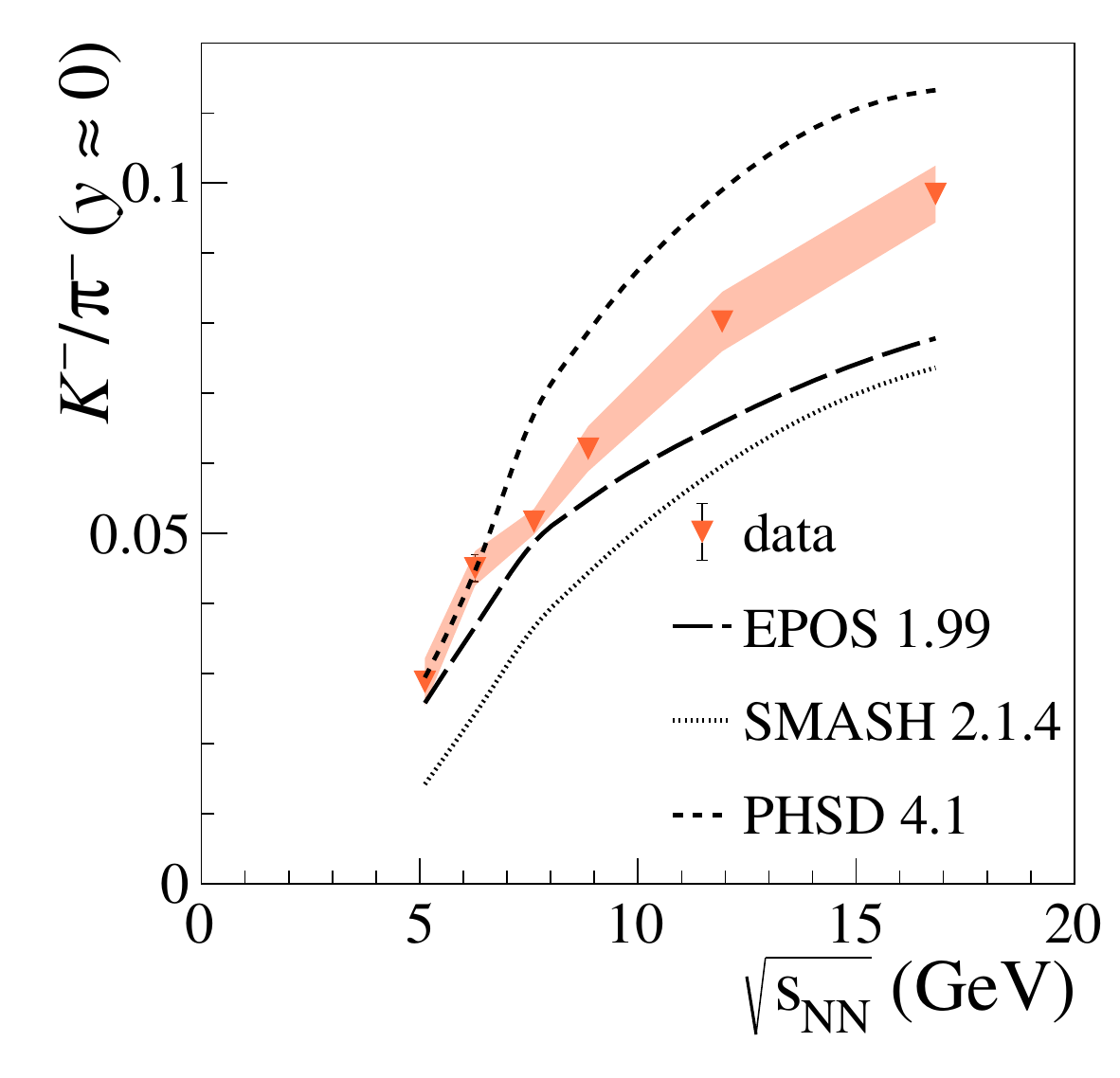}
\caption{The ratios of $\text{d}n/\text{d}y$ yields at mid-rapidity $K^+/\pi^+$ and $K^-/\pi^-$  produced in 10\% most \textit{central} Ar+Sc interactions at 13$A$, 19$A$, 30$A$, 40$A$, 75$A$ and 150\AGeVc plotted in dependence on collision energy compared with models: \Epos 1.99~\cite{Werner:2008zza}, \Phsd~4.1~\cite{PhysRevC.78.034919,CASSING2009215} and \Smash~2.1.4~\cite{Mohs:2019iee,PhysRevC.94.054905}.
}
\label{fig:modelcompktopi}
\end{figure}

\begin{figure}
\centering
\includegraphics[width=0.49\linewidth]{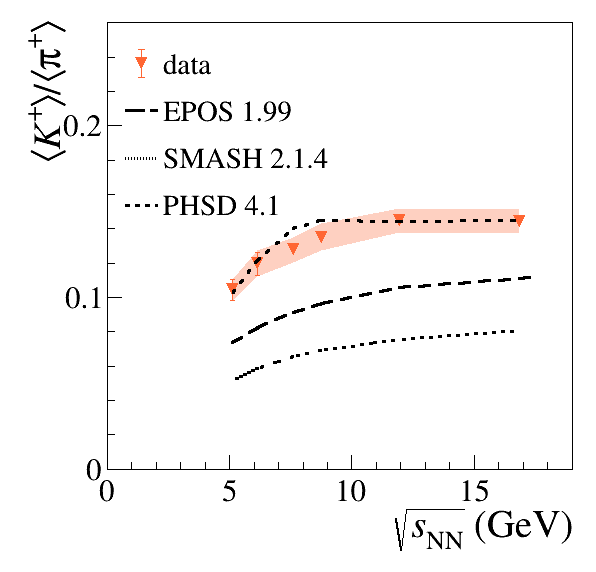}~
\includegraphics[width=0.49\linewidth]{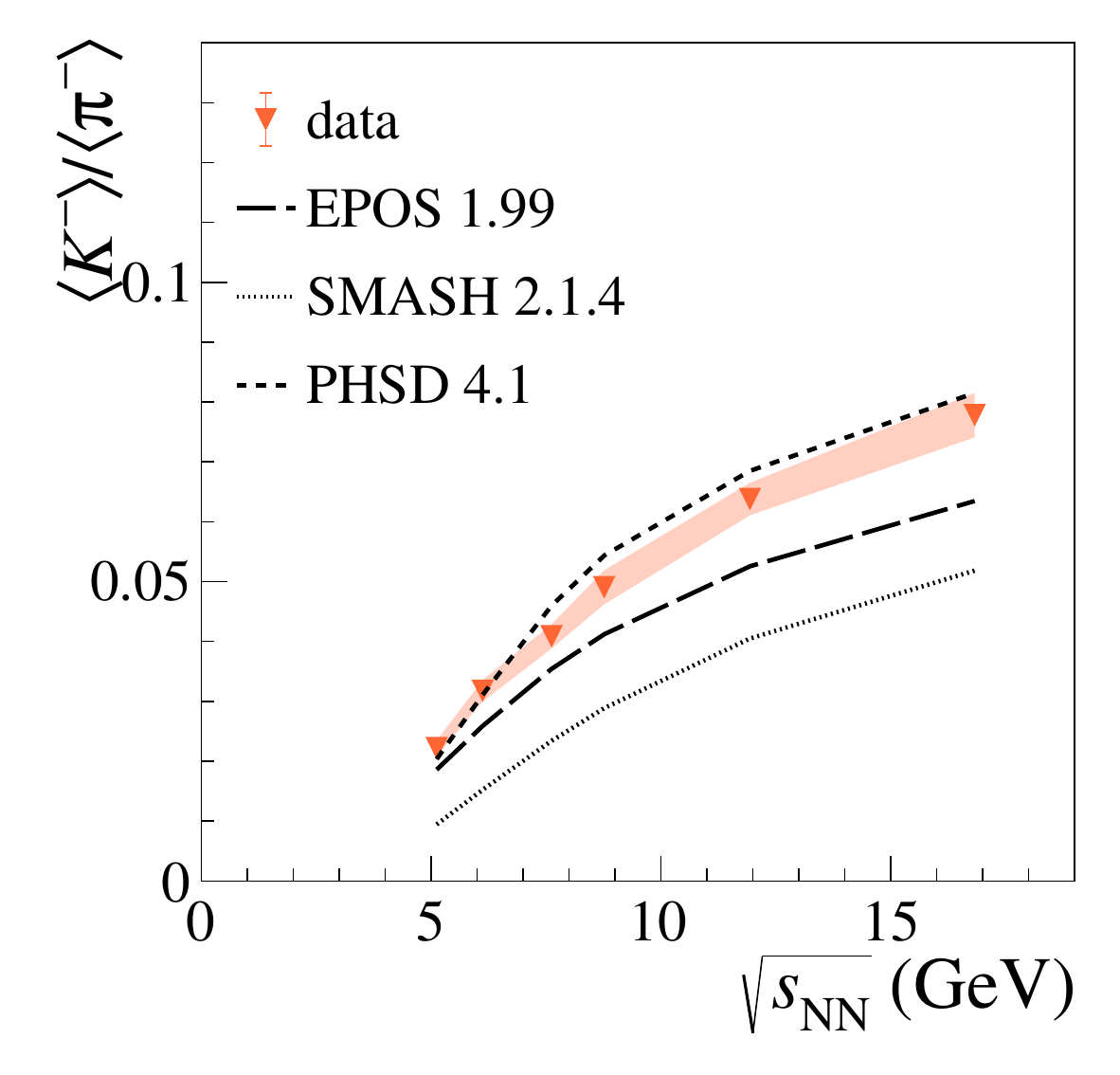}~
\caption{The ratios of mean multiplicity
$\langle K^+ \rangle $/$\langle \pi^+ \rangle $ and $\langle K^- \rangle $/$\langle \pi^- \rangle $ produced in 10\% most \textit{central} Ar+Sc interactions at 13$A$, 19$A$, 30$A$, 40$A$, 75$A$ and 150\AGeVc in dependence on collision energy compared with models: \Epos 1.99~\cite{Werner:2008zza}, \Phsd~4.1~\cite{PhysRevC.78.034919,CASSING2009215} and \Smash~2.1.4~\cite{Mohs:2019iee,PhysRevC.94.054905}.}
\label{fig:model_kratios}
\end{figure}

Finally, the energy dependence of the ratio of kaon and pion yields is compared to model predictions. Figure~\ref{fig:modelcompktopi} shows the mid-rapidity results for $K^+$/$\pi^+$ and $K^-$/$\pi^-$, and  Fig.~\ref{fig:model_kratios} displays the corresponding results obtained for the full phase space. Unlike particle yields, particle ratios are not sensitive to the details of the event selection assuming that the shapes of the spectra do not change significantly in the studied centrality range.

The collision energy dependence of the $K^+$/$\pi^+$, $K^-$/$\pi^-$ ratios is generally well reproduced by the analyzed models. However, the PHSD predicts a non-monotonic behavior of the $K^+$/$\pi^+$ ratio, while it is not observed in the experimental results, or for other models. All predictions diverge significantly from the measured data in terms of absolute values.

\FloatBarrier

\subsubsection{Proton and antiproton spectra}
\label{sec:models_protons}

\begin{figure}
\centering
    \includegraphics[page=1, width=0.32\textwidth]{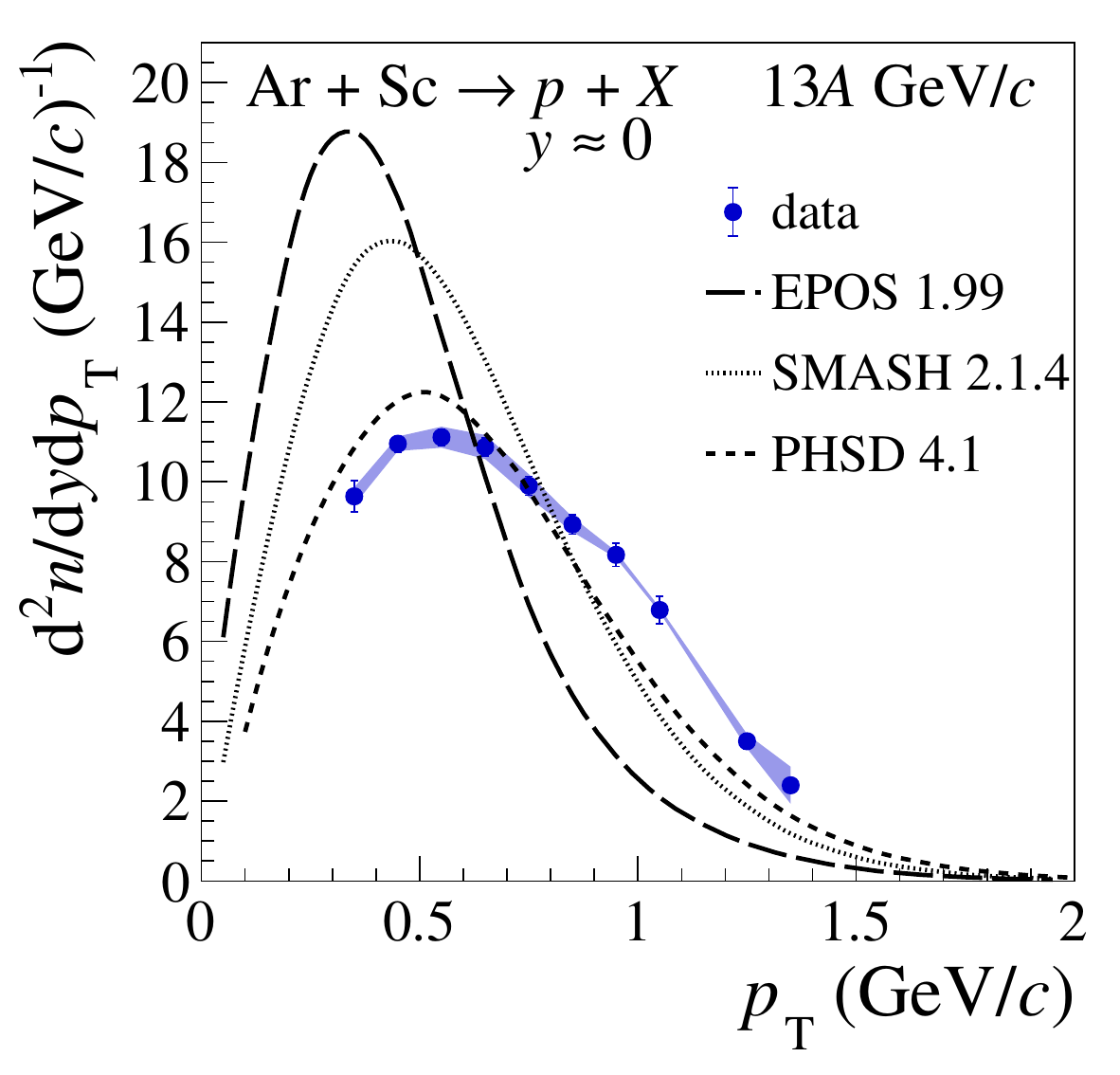}
    \includegraphics[page=2, width=0.32\textwidth]{figures/newresults/model_comp_ppos_pT.pdf}
    \includegraphics[page=3, width=0.32\textwidth]{figures/newresults/model_comp_ppos_pT.pdf}\\
    \includegraphics[page=4, width=0.32\textwidth]{figures/newresults/model_comp_ppos_pT.pdf}
    \includegraphics[page=5, width=0.32\textwidth]{figures/newresults/model_comp_ppos_pT.pdf}
    \includegraphics[page=6, width=0.32\textwidth]{figures/newresults/model_comp_ppos_pT.pdf}
\caption{Transverse momentum distributions of protons at mid-rapidity produced in 10\% most \textit{central} Ar+Sc interactions at 13$A$, 19$A$, 30$A$, 40$A$, 75$A$ and 150\AGeVc compared with models: \Epos 1.99~\cite{Werner:2008zza}, \Phsd~4.1~\cite{PhysRevC.78.034919,CASSING2009215} and \Smash~2.1.4~\cite{Mohs:2019iee,PhysRevC.94.054905}.}
\label{fig:modelcompppt}
\end{figure}

\begin{figure}
\centering
    \includegraphics[page=3, width=0.32\textwidth]{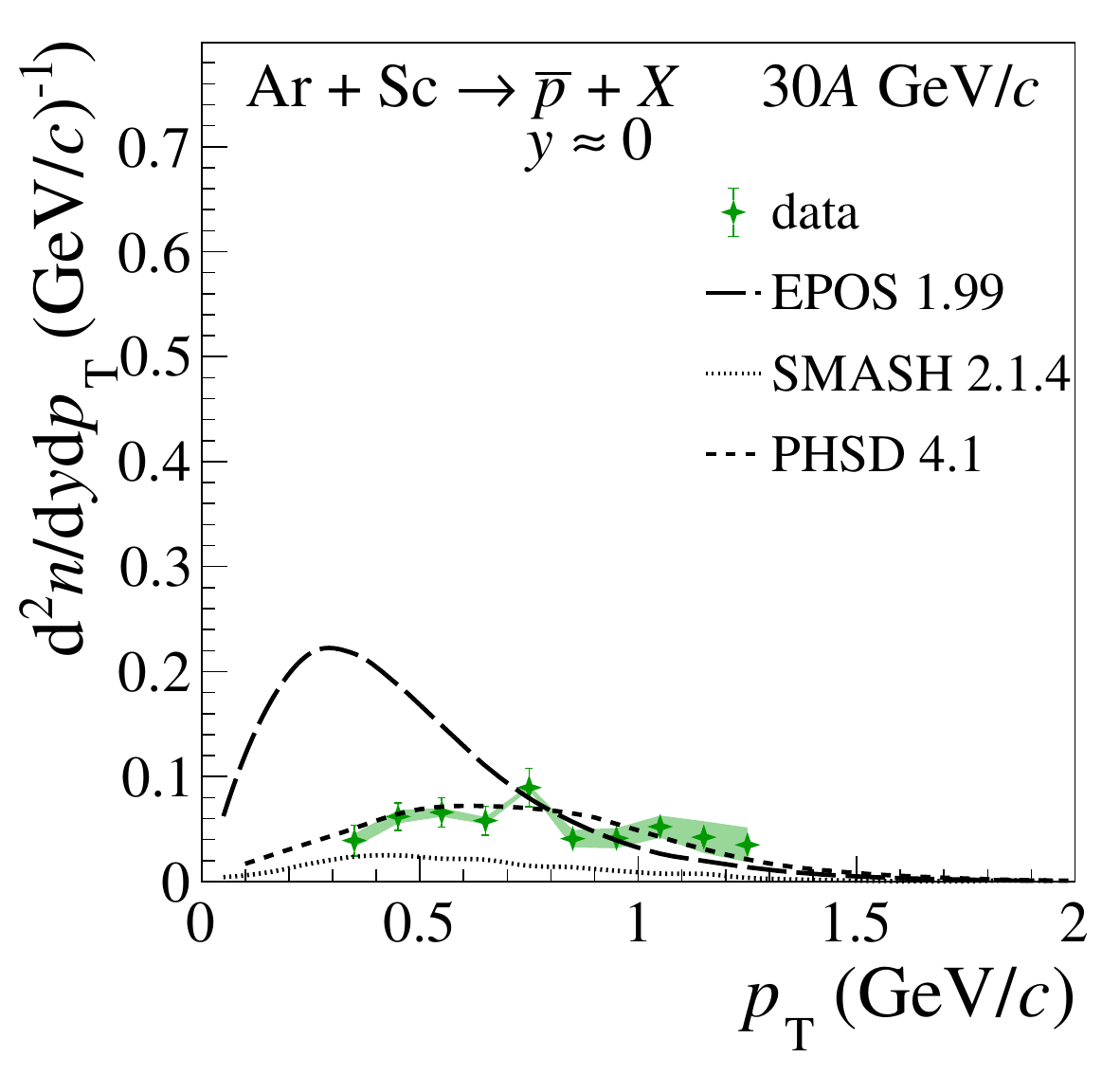}
    \includegraphics[page=4, width=0.32\textwidth]{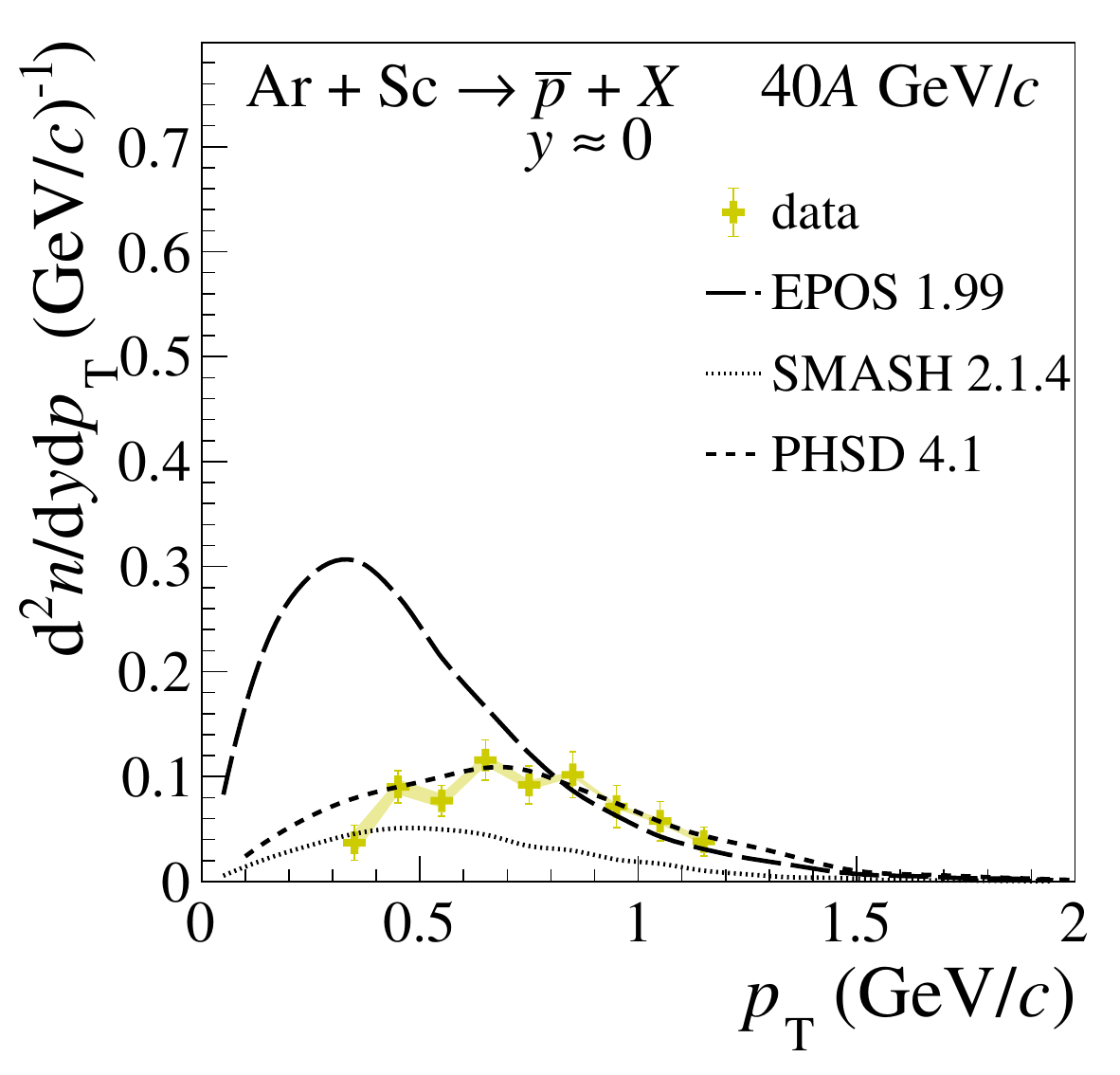}\\
    \includegraphics[page=5, width=0.32\textwidth]{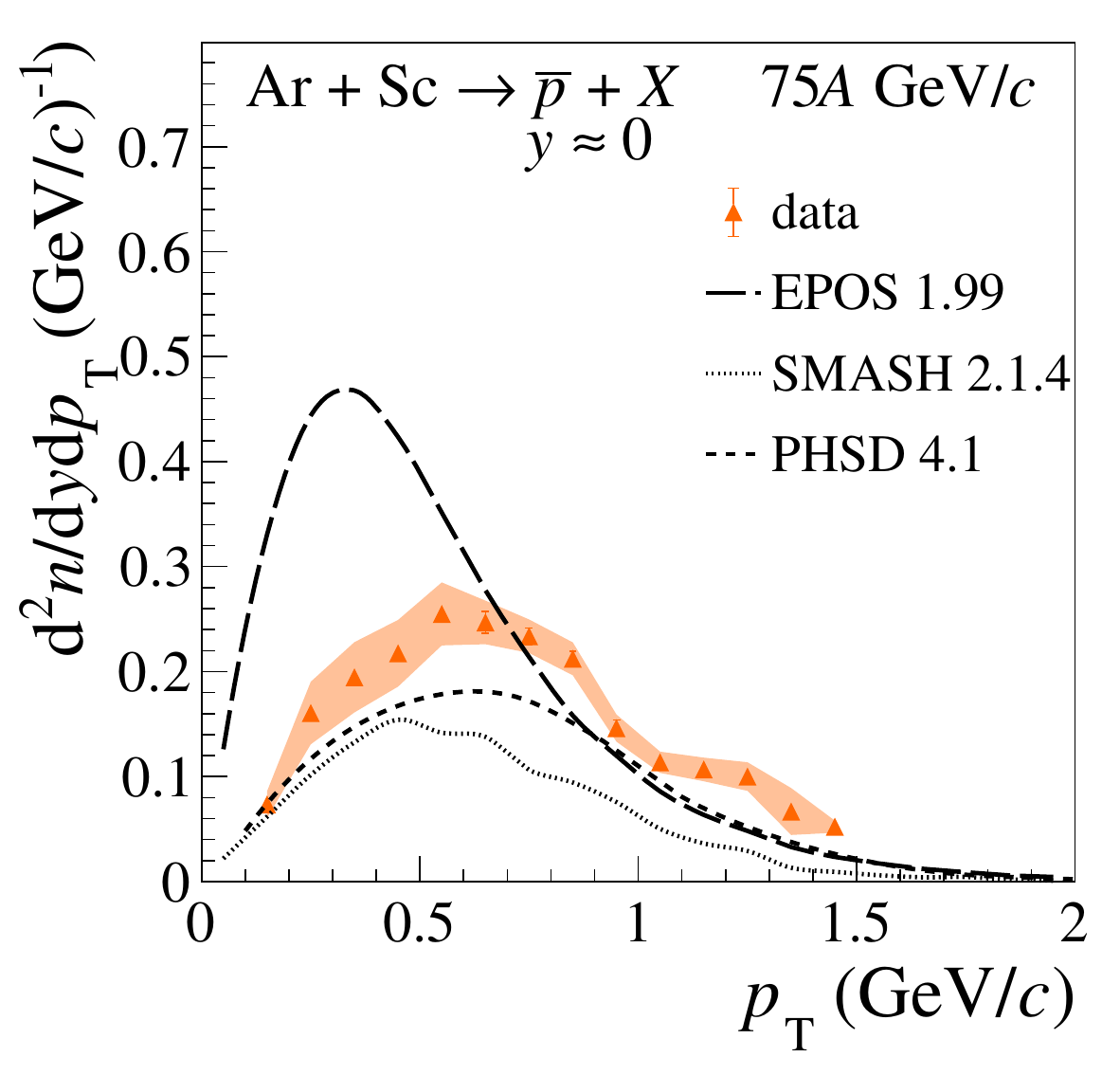}
    \includegraphics[page=6, width=0.32\textwidth]{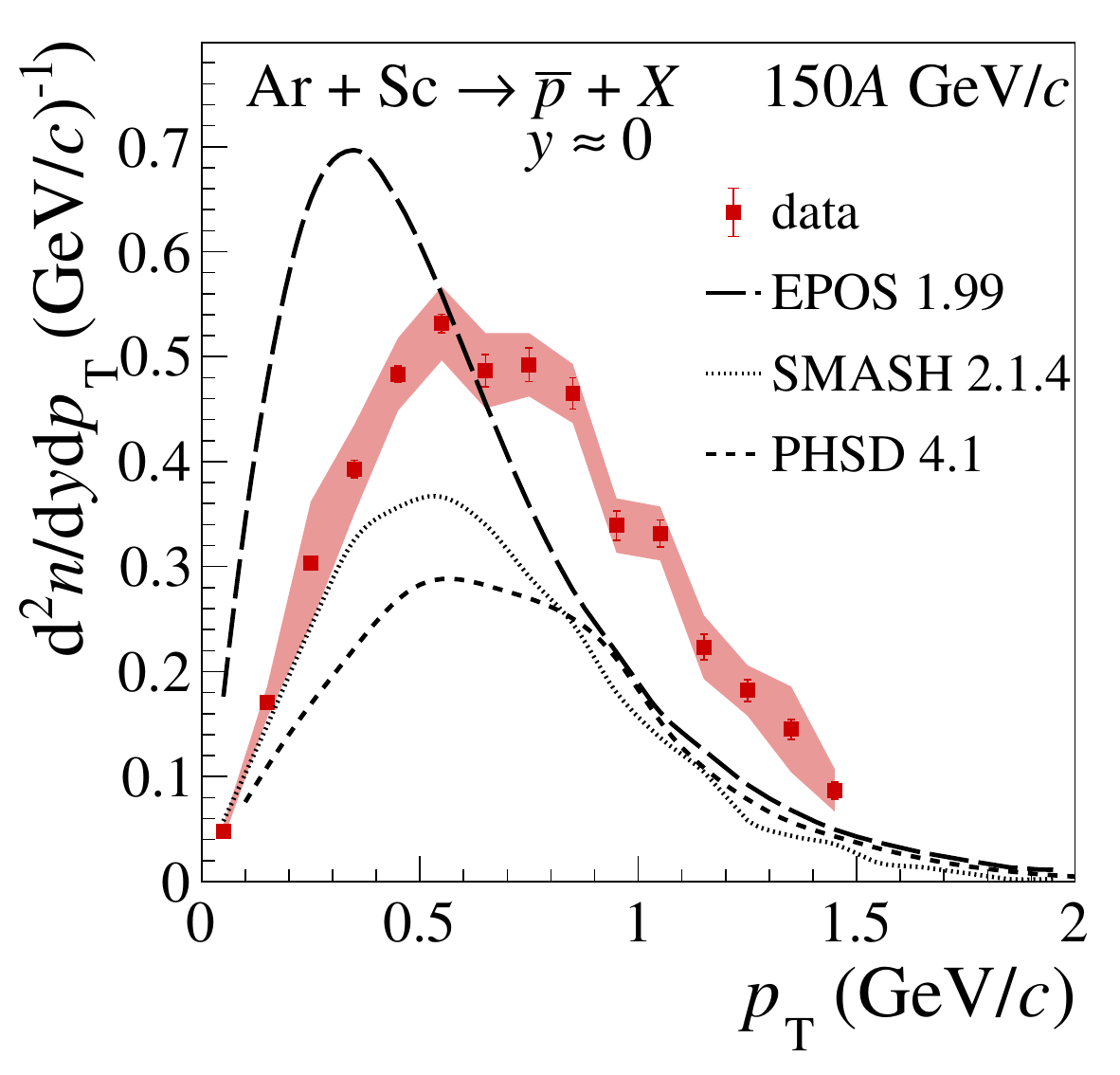}
\caption{Transverse momentum distributions of antiprotons at mid-rapidity produced in 10\% most \textit{central} Ar+Sc interactions at 30$A$, 40$A$, 75$A$ and 150\AGeVc compared with models: \Epos 1.99~\cite{Werner:2008zza}, \Phsd~4.1~\cite{PhysRevC.78.034919,CASSING2009215} and \Smash~2.1.4~\cite{Mohs:2019iee,PhysRevC.94.054905}.}
\label{fig:modelcompbarppt}
\end{figure}

Figures \ref{fig:modelcompppt} and \ref{fig:modelcompbarppt} display a comparison with model predictions of $p$ and $\bar{p}$ transverse momentum distributions at mid-rapidity. The general trend observed for all beam momenta is a maximum at higher \pt values than expected from \Epos and \Smash models, while \Phsd describes the measured spectra shape much better.

\begin{figure}
\centering
\includegraphics[width=0.33\linewidth]{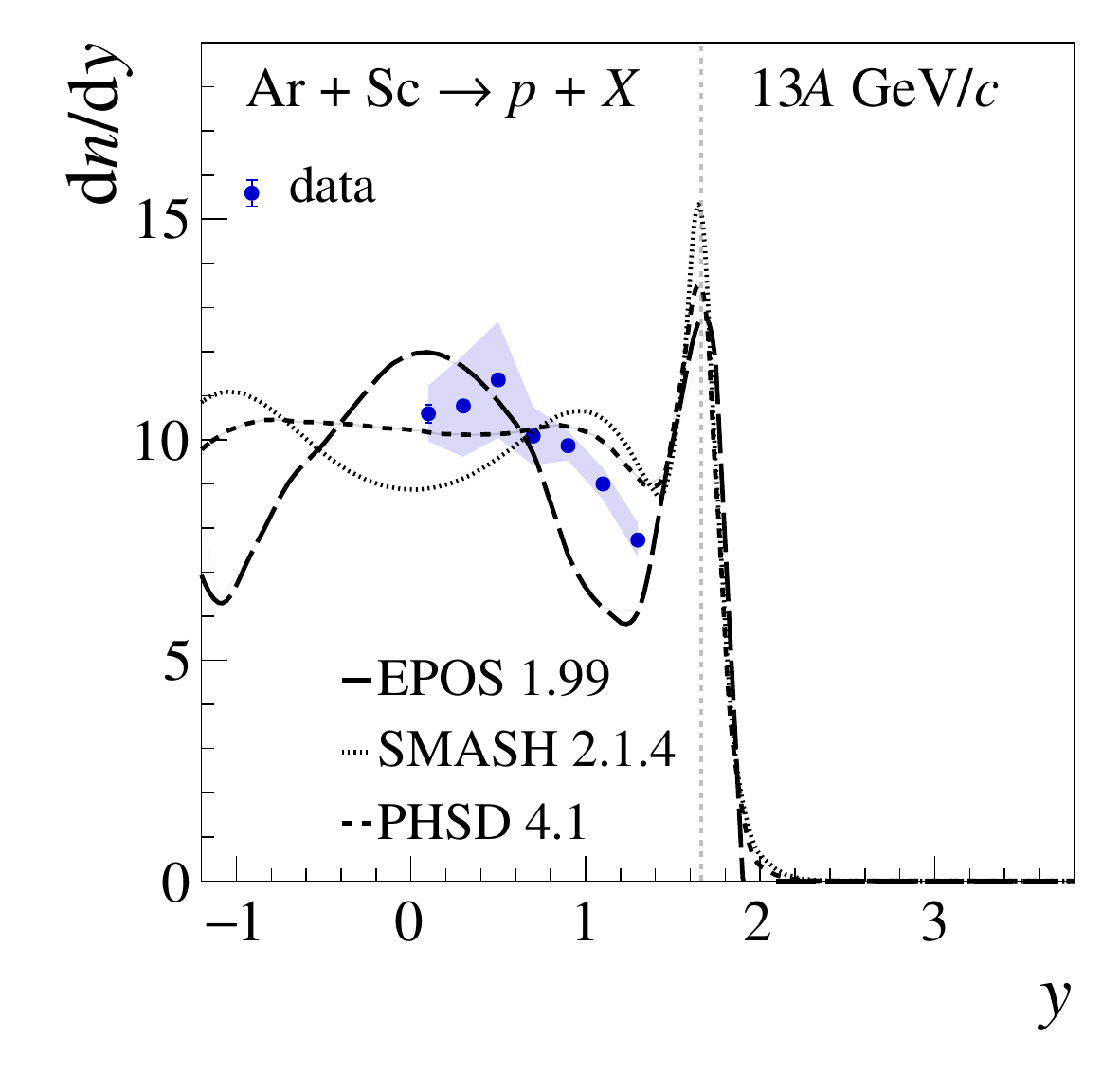}~
\includegraphics[width=0.33\linewidth]{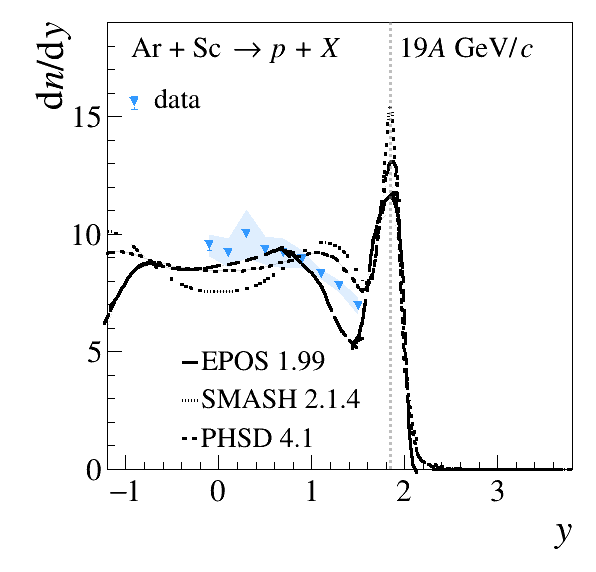}~
\includegraphics[width=0.33\linewidth]{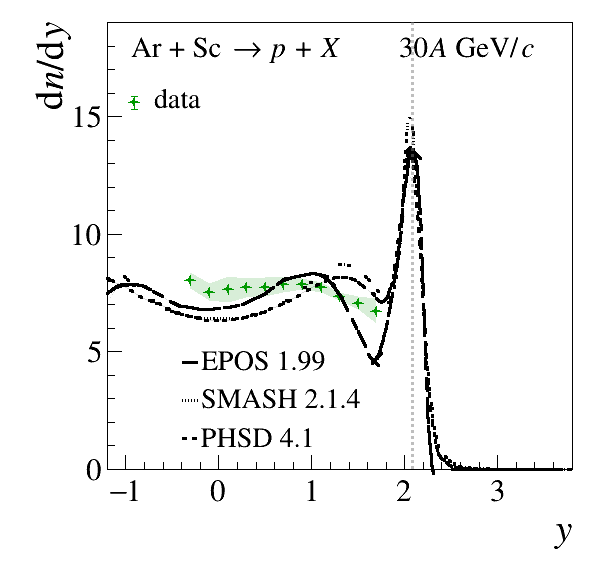}\\
\includegraphics[width=0.33\linewidth]{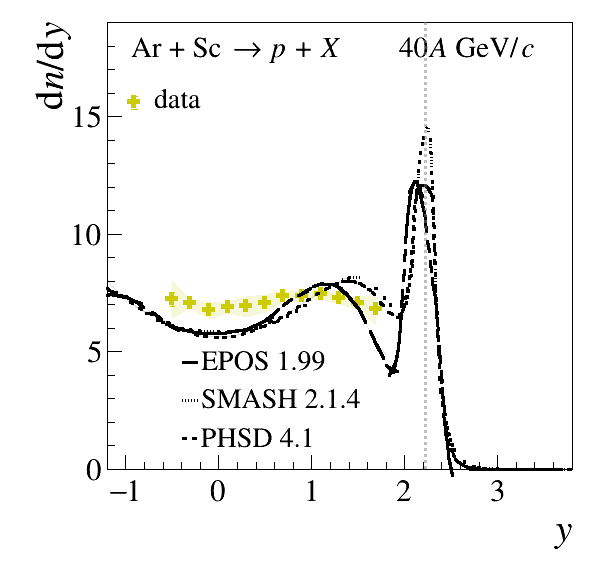}~
\includegraphics[width=0.33\linewidth]{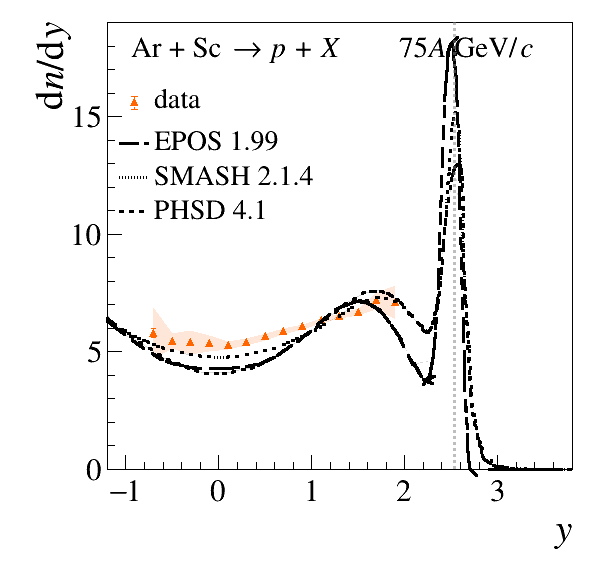}~
\includegraphics[width=0.33\linewidth]{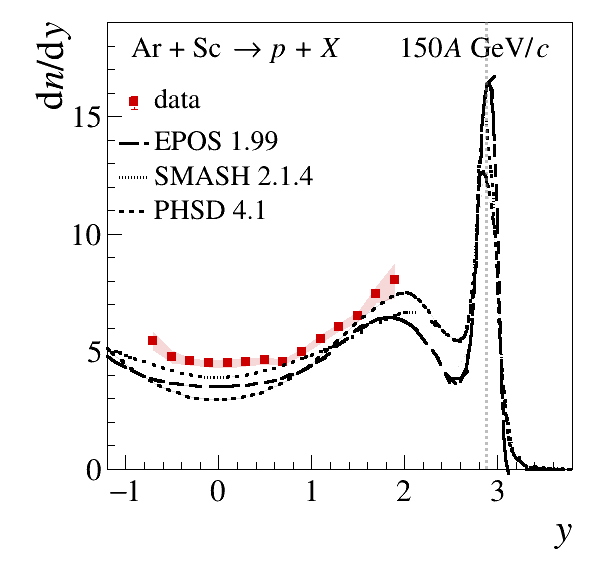}\\
\caption{Rapidity spectra of protons produced in 10\% most \textit{central} Ar+Sc interactions at 13$A$, 19$A$, 30$A$, 40$A$, 75$A$ and 150\AGeVc in comparison with models: \Epos 1.99~\cite{Werner:2008zza}, \Phsd~4.1~\cite{PhysRevC.78.034919,CASSING2009215} and \Smash~2.1.4~\cite{Mohs:2019iee,PhysRevC.94.054905}.}
\label{fig:model_ppos_yspectra}
\end{figure}

\begin{figure}
\centering
\includegraphics[width=0.33\linewidth]{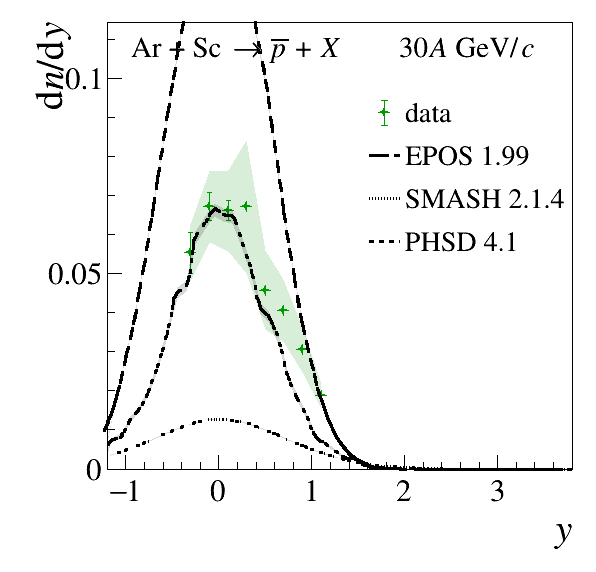}~
\includegraphics[width=0.33\linewidth]{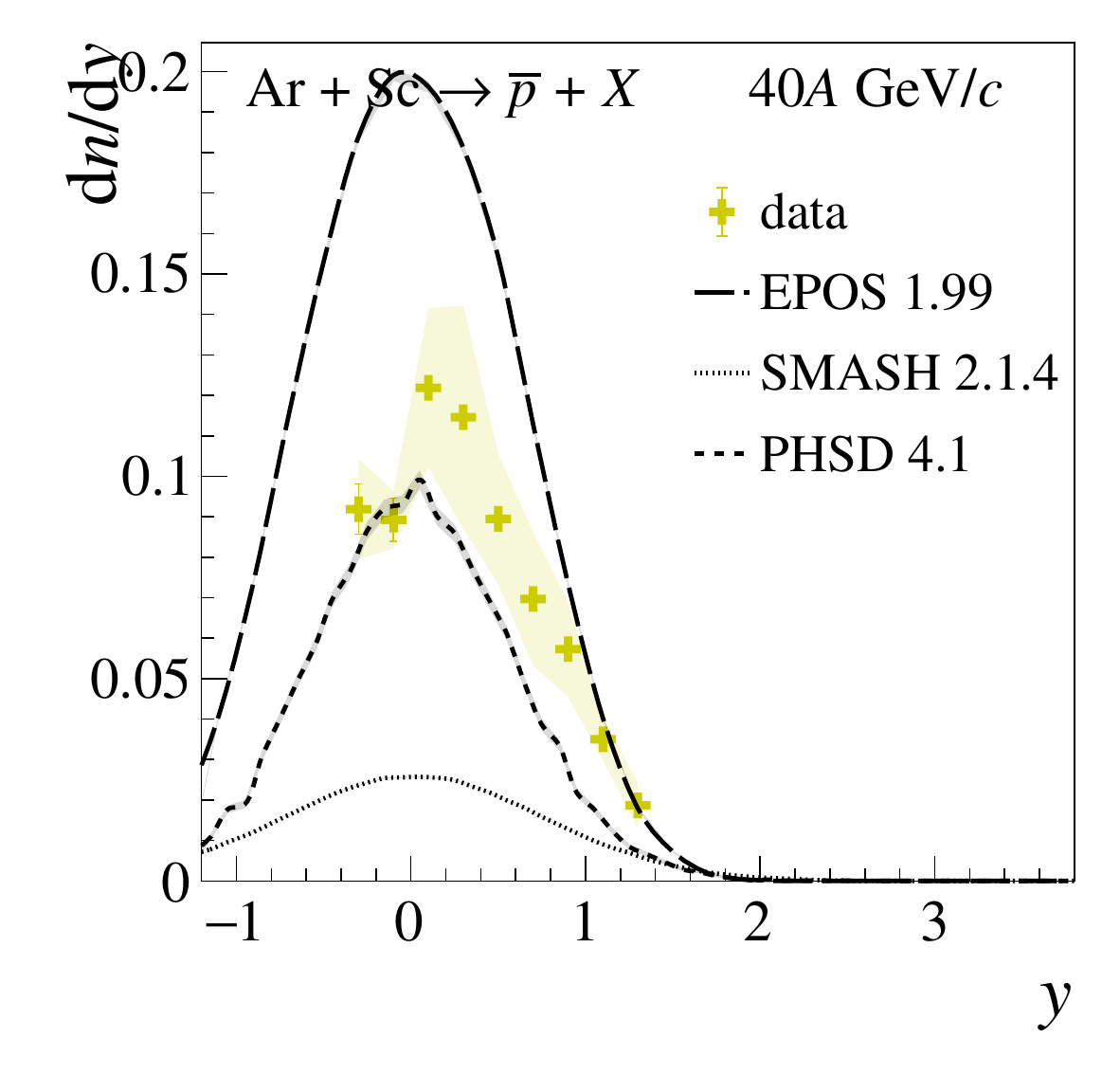}\\
\includegraphics[width=0.33\linewidth]{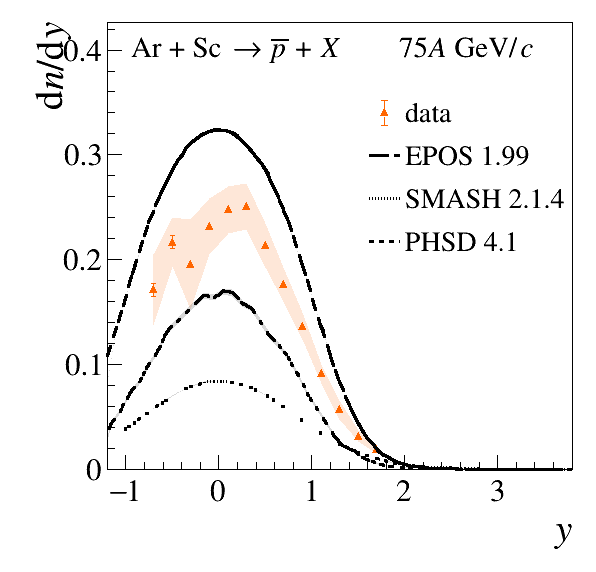}~
\includegraphics[width=0.33\linewidth]{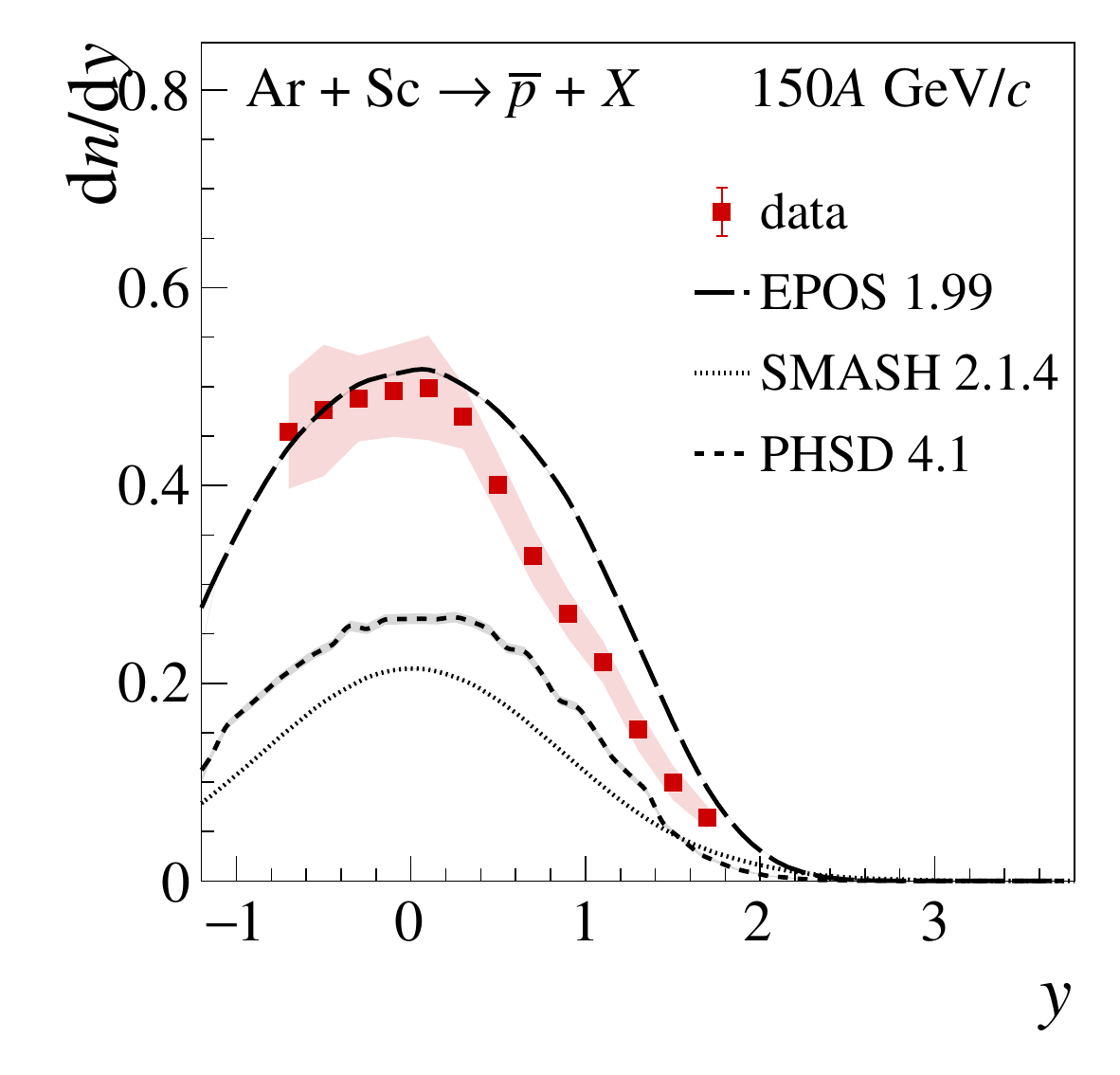}\\
\caption{Rapidity spectra of antiprotons produced in 10\% most \textit{central} Ar+Sc interactions at 30$A$, 40$A$, 75$A$ and 150\AGeVc in comparison with models: \Epos 1.99~\cite{Werner:2008zza}, \Phsd~4.1~\cite{PhysRevC.78.034919,CASSING2009215} and \Smash~2.1.4~\cite{Mohs:2019iee,PhysRevC.94.054905}.}
\label{fig:model_pneg_yspectra}
\end{figure}

Figures \ref{fig:model_ppos_yspectra} and \ref{fig:model_pneg_yspectra} show $p$ and $\bar{p}$ rapidity distributions compared with model predictions. Note that the discussed measurements of proton spectra do not cover the so-called ``diffractive region'', which features a characteristic peak structure close to the beam rapidity, observed experimentally in \pp reactions \cite{NA49:2009brx}. The \Phsd model provides the best description of proton and antiproton rapidity spectra, in particular at lower collision energies. At 75$A$ and 150\AGeVc, the antiproton yield is significantly underestimated and the proton rapidity spectrum aligns with data only close to mid-rapidity. The \Epos model generally underestimates the yields of protons and overestimates the yield of antiprotons, while the \Smash model's relation to measured data is exactly the opposite. 

The properties of proton rapidity distributions may be sensitive to the changes in the equation of state. In particular, a scenario of first-order phase transition could feature a characteristic concave-convex interchange (``peaks'' and ``dips'') in the central part of the proton rapidity spectrum in dependence on collision energy \cite{Ivanov:2016xev}.

With the data presented here, we observe that at the beam momenta of 13$A$ and 19\AGeVc the proton rapidity spectrum features a global maximum at mid-rapidity, while starting from 30$A$--40\AGeVc a local minimum appears at $y=0$. Such observations are not consistent with either the hadronic or double-phase equation of state within the framework presented in Ref. \cite{Ivanov:2016xev}. 

Notably, the \Epos and \Phsd models describe well the concave shape of the spectra at 13$A$, the flattening at 19$A$ and 30\AGeVc, as well as the convex characteristic of the distributions at higher beam momenta.

\subsubsection{System size dependence of the $K^+/\pi^+$ ratio}

\begin{figure}
\centering
\includegraphics[width=0.5\linewidth]{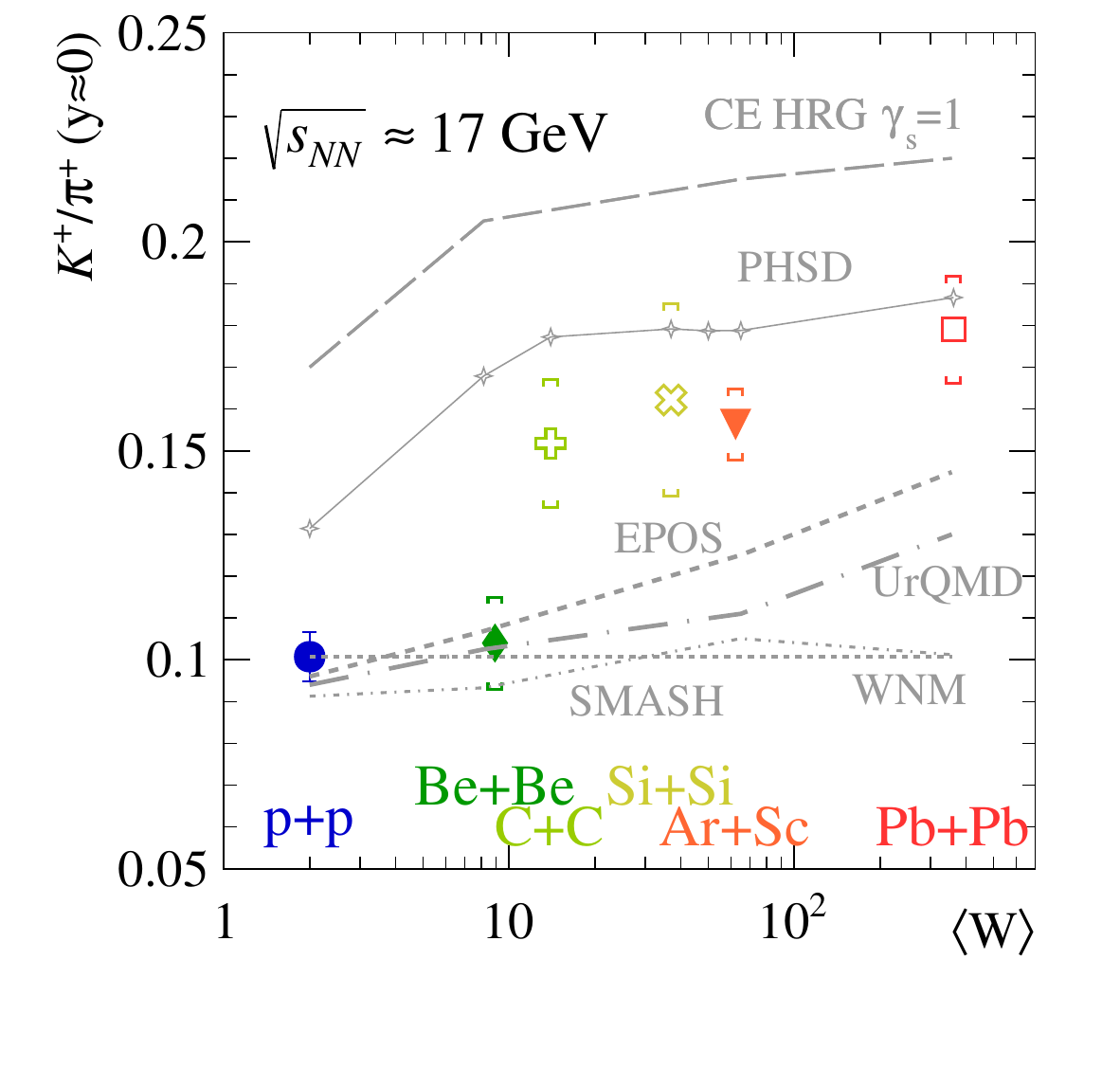}~
\caption{System size dependence of the $K^+/\pi^+$ ratio (at $y\approx 0$) in \textit{central} nucleus-nucleus and inelastic \pp interactions obtained at beam momenta of $\approx$150\AGeVc ($\snn~\approx$~17~\GeV) compared with dynamical (\Epos 1.99~\cite{Werner:2008zza}, \Phsd~4.1~\cite{PhysRevC.78.034919,CASSING2009215} and \Urqmd~\cite{Bass:1998ca,Bleicher:1999xi}) and statistical (HRG \cite{Motornenko:2018gdc}) models. Additionally, WNM \cite{Bialas:1976ed} prediction is shown as a horizontal dashed line.}
\label{fig:modelsyssizedep}
\end{figure}

Figure \ref{fig:modelsyssizedep} presents the $K^+/\pi^+$ multiplicity ratio as a function of the system size for the highest SPS energy ($\snn~\approx$~17~\GeV, 150\AGeVc beam momentum). System size is quantified by the mean number of wounded nucleons in collisions $\langle W \rangle$. Dynamical models, \Epos~\cite{Werner:2008zza}, \Urqmd~\cite{Bass:1998ca,Bleicher:1999xi} and 
 Smash~\cite{Mohs:2019iee,PhysRevC.94.054905}, successfully describe the $K^+/\pi^+$ ratio for light systems ($p$+$p$ and Be+Be) but fail for heavier ones (Ar+Sc, Pb+Pb). On the other hand, PHSD, the model with phase transition, reproduces the data for heavy systems but overestimates the $K^+/\pi^+$ ratio for lighter ones. The statistical hadron resonance gas model (HRG, \cite{Motornenko:2018gdc}) significantly overestimates the ratio for light systems. The overestimation for heavier systems is still present but is less pronounced.

\subsubsection{Summary on model predictions}
\label{sec:model_summary}

The comparison of measured data on Ar+Sc collisions at $p_{beam}=13A-150\AGeVc$ with the dynamical models of nucleus-nucleus collisions (\Epos, \Smash, \Phsd) provides interesting insights into the challenges of understanding the dynamics of intermediate-size system collisions. The charged-pion rapidity spectra are relatively well described by all analyzed models at high beam momenta (75$A$, 150\AGeVc), where overall deviations do not exceed 10\%. Larger discrepancies appear at lower collision energies. The d$n/$d$y$ yields of charged kaons are, in general, poorly described by the models. However, a good description of the $K^+$ yields is observed in the case of \Phsd. Particularly interesting are the rapidity spectra of protons, which feature an inflection point at mid-rapidity---from convex to concave shape---with increasing collision energy. Both \Epos and \Phsd display similar behavior, while the spectra shapes in \Smash are qualitatively different at 13$A$ and 19\AGeVc. The \Phsd model predictions concerning charged kaon, proton, and antiproton transverse momentum spectra are in good agreement with measured data. It is in contrast to the \Epos and \Smash models, which feature a shift of transverse momentum distribution peaks towards lower values of \pt, which may be attributed to unimplemented effects of radial flow. The collision energy dependence of the $K^+/\pi^+$ ratio is not well reproduced by any of the analyzed models and a similar statement is true in the case of the $K^-/\pi^-$ ratio.

The most interesting observable, in the context of this article, is the system size dependence of the $K^+/\pi^+$ ratio. For the comparison with the models, data on \pp, Be+Be, Ar+Sc and Pb+Pb collisions at $\snn~\approx$~17~\GeV are selected. Neither the dynamical nor statistical models describe a rapid change of the $K^+/\pi^+$ ratio between \textit{central} Be+Be and Ar+Sc collisions at the highest SPS energy, as it can be seen in Fig. \ref{fig:modelsyssizedep}.

It is apparent that the influence of the system size on particle production and strangeness production, in particular, is not well understood and requires more theoretical and phenomenological studies. The models analyzed in this work implement different particle creation mechanisms. The widely used approach is modeling collisions via the formation and fragmentation of parton-strings. Such a mechanism is utilized in all three discussed dynamical models, however, at low collision energies, the applicability of this approach is questionable --- in both \Phsd and \Smash, the string approach is replaced with the creation of resonances and their decay. Additionally, \Phsd also features a chiral symmetry restoration with increasing energy followed by the creation of the QGP. The best overall agreement with the data is observed in the case of \Phsd.

\FloatBarrier
\section{Summary and conclusions}
\label{sec:summary}

\color{black}
This paper reports measurements by the \NASixtyOne experiment at the CERN SPS of spectra 
and mean multiplicities of $\pi^\pm, K^\pm, p$ and $\bar{p}$  produced in 
the 10\% most central Ar+Sc collisions at beam momenta of 13$A$, 19$A$, 30$A$, 40$A$, 75$A$ and 150\AGeVc.  This is an intermediate-size nucleus-nucleus system investigated
in the system size scan of \NASixtyOne. In this program, data were also recorded 
for \pp, Be+Be, Xe+La and Pb+Pb collisions. While the analysis for the two largest systems is still ongoing, an emerging system size dependence already shows interesting features.

Ar+Sc system appears to be the smallest for which a significant enhancement of the $K^+/\pi^+$ ratio with respect to \pp collisions is observed.
Similarly, the measured charged-kaon transverse momentum spectra are characterized by significantly larger values of the inverse slope parameter ($T$) than in the case of small systems (\pp, Be+Be).
The similarity between intermediate and heavy systems becomes more evident towards higher collision energies. At the same time, the measurements presented in this article show no indications of a horn structure at SPS energies for intermediate-size collision systems in contrast to the results from central Pb+Pb interactions.

The third property that distincts the Ar+Sc system from the small systems is the qualitatively different shape of the proton rapidity spectra. Similarly as in the collisions of Pb+Pb, the spectra measured in the Ar+Sc reaction at 19$A$ \GeVc are characterized by a convex shape near midrapidity, in contrast to the concave structure present for $p$+$p$ and Be+Be systems.

The experimental results were compared with predictions of the models: \Epos 1.99, \Phsd, and \Smash. None of the models reproduces all features of the presented results. We also highlight the failure of analyzed models in the description of the $K^+/\pi^+$ ratio in terms of both collision energy and system size dependence.

\clearpage
\section*{Acknowledgments}
We would like to thank the CERN EP, BE, HSE and EN Departments for the
strong support of NA61/SHINE.

This work was supported by
the Hungarian Scientific Research Fund (grant NKFIH 138136\slash138152),
the Polish Ministry of Science and Higher Education
(DIR\slash WK\slash\-2016\slash 2017\slash\-10-1, WUT ID-UB), the National Science Centre Poland (grants
2014\slash 14\slash E\slash ST2\slash 00018, 
2016\slash 21\slash D\slash ST2\slash 01983, 
2017\slash 25\slash N\slash ST2\slash 02575, 
2017\slash 27\slash N\slash ST2\slash 00778, 
2018\slash 29\slash N\slash ST2\slash 02595, 
2018\slash 30\slash A\slash ST2\slash 00226, 
2018\slash 31\slash G\slash ST2\slash 03910, 
2019\slash 33\slash B\slash ST9\slash 03059, 
2020\slash 39\slash O\slash ST2\slash 00277), 
the Norwegian Financial Mechanism 2014--2021 (grant 2019\slash 34\slash H\slash ST2\slash 00585),
the Polish Minister of Education and Science (contract No. 2021\slash WK\slash 10),
the European Union's Horizon 2020 research and innovation program under grant agreement No. 871072,
the Ministry of Education, Culture, Sports,
Science and Tech\-no\-lo\-gy, Japan, Grant-in-Aid for Sci\-en\-ti\-fic
Research (grants 18071005, 19034011, 19740162, 20740160 and 20039012),
the German Research Foundation DFG (grants GA\,1480\slash8-1 and project 426579465),
the Bulgarian Ministry of Education and Science within the National
Roadmap for Research Infrastructures 2020--2027, contract No. D01-374/18.12.2020,
Ministry of Education
and Science of the Republic of Serbia (grant OI171002), Swiss
Nationalfonds Foundation (grant 200020\-117913/1), ETH Research Grant
TH-01\,07-3 and the Fermi National Accelerator Laboratory (Fermilab), a U.S. Department of Energy, Office of Science, HEP User Facility managed by Fermi Research Alliance, LLC (FRA), acting under Contract No. DE-AC02-07CH11359 and the IN2P3-CNRS (France).\\

The data used in this paper were collected before February 2022.

\clearpage
\appendix
\section{Additional plots}
\label{sec:additional_plots}
\begin{figure}[h]
    \centering
    \includegraphics[width=0.32\textwidth, page=1, trim = 0 10 0 30, clip]{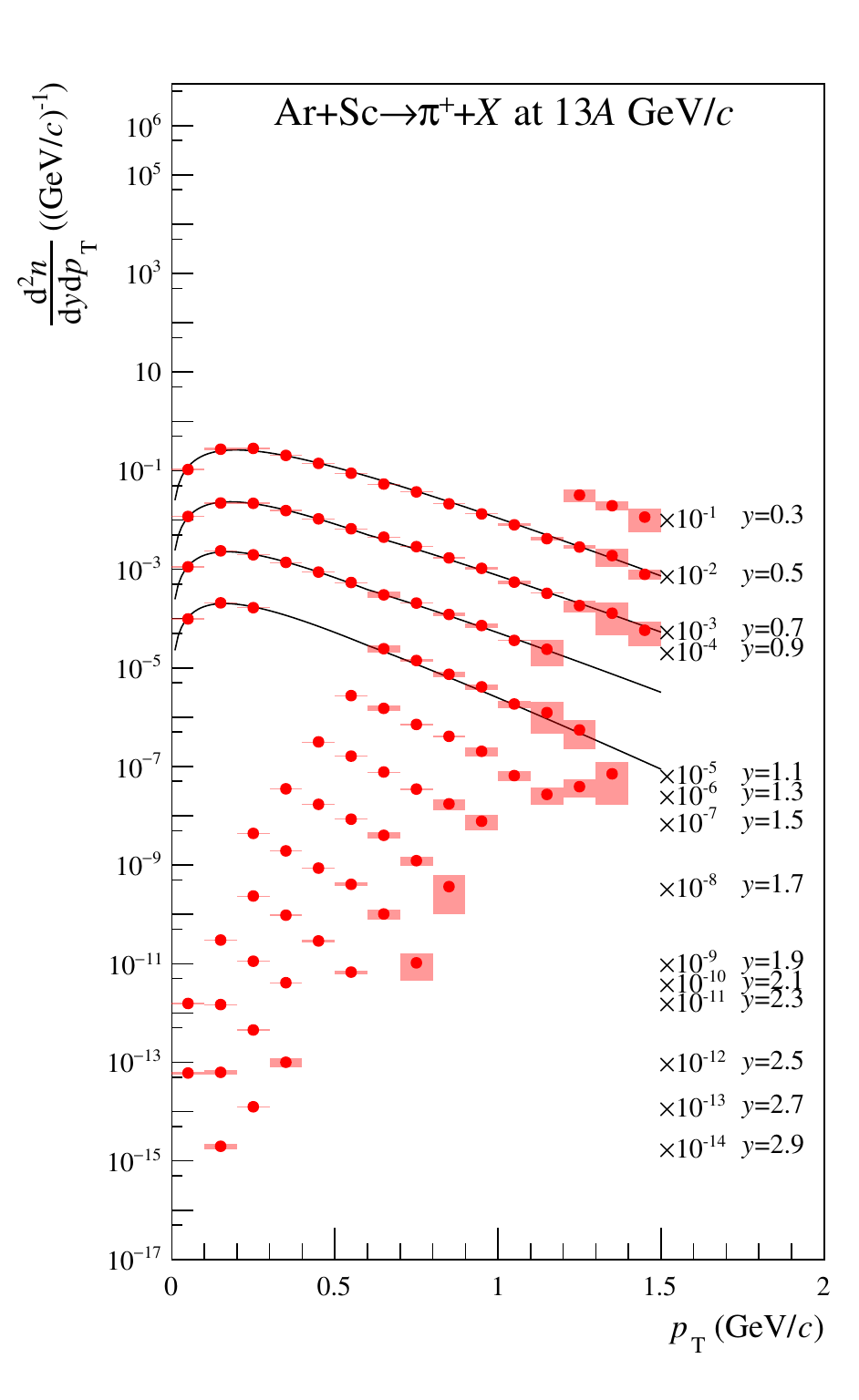}
    \includegraphics[width=0.32\textwidth, page=2, trim = 0 10 0 30, clip]{figures/rapidity_slices/rap_slices_pi_pos.pdf}
    \includegraphics[width=0.32\textwidth, page=3, trim = 0 10 0 30, clip]{figures/rapidity_slices/rap_slices_pi_pos.pdf}\\
    \includegraphics[width=0.32\textwidth, page=4, trim = 0 10 0 30, clip]{figures/rapidity_slices/rap_slices_pi_pos.pdf}
    \includegraphics[width=0.32\textwidth, page=5, trim = 0 10 0 30, clip]{figures/rapidity_slices/rap_slices_pi_pos.pdf}
    \includegraphics[width=0.32\textwidth, page=6, trim = 0 10 0 30, clip]{figures/rapidity_slices/rap_slices_pi_pos.pdf}
    \caption{ Transverse momentum spectra in rapidity slices of $\pi^+$ produced in 10\% most \textit{central} Ar+Sc collisions at 13$A$ to 150\AGeVc. For graphical presentations, the spectra are multiplied by factors given in the plots. Rapidity values given in the legends correspond to the middle of the corresponding interval. Lines correspond to fitted exponential functions (Eq. \ref{eq:pion_pT}). Error bars show statistical uncertainties, while shaded boxes correspond to systematic uncertainties.}
    \label{fig:rap_slices_pi_pos}
\end{figure}

\begin{figure}[h]
    \centering
    \includegraphics[width=0.32\textwidth, page=1, trim = 0 10 0 30, clip]{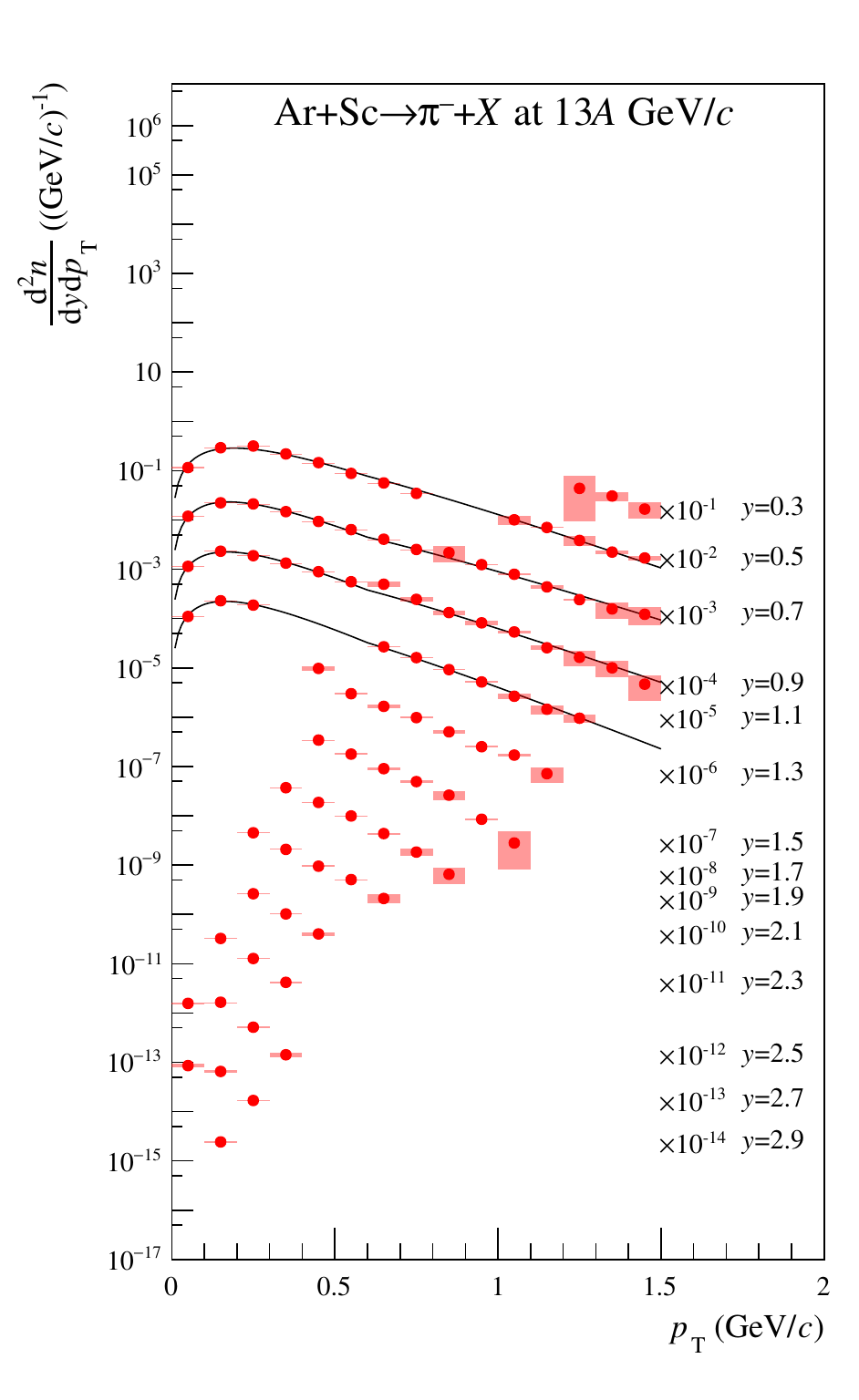}
    \includegraphics[width=0.32\textwidth, page=2, trim = 0 10 0 30, clip]{figures/rapidity_slices/rap_slices_pi_neg.pdf}
    \includegraphics[width=0.32\textwidth, page=3, trim = 0 10 0 30, clip]{figures/rapidity_slices/rap_slices_pi_neg.pdf}\\
    \includegraphics[width=0.32\textwidth, page=4, trim = 0 10 0 30, clip]{figures/rapidity_slices/rap_slices_pi_neg.pdf}
    \includegraphics[width=0.32\textwidth, page=5, trim = 0 10 0 30, clip]{figures/rapidity_slices/rap_slices_pi_neg.pdf}
    \includegraphics[width=0.32\textwidth, page=6, trim = 0 10 0 30, clip]{figures/rapidity_slices/rap_slices_pi_neg.pdf}
    \caption{ Transverse momentum spectra in rapidity slices of $\pi^-$ produced in 10\% most \textit{central} Ar+Sc collisions at 13$A$ to 150\AGeVc. For graphical presentations, the spectra are multiplied by factors given in the plots. Rapidity values given in the legends correspond to the middle of the corresponding interval. Lines correspond to fitted exponential functions (Eq. \ref{eq:pion_pT}). Error bars show statistical uncertainties, while shaded boxes correspond to systematic uncertainties.}
    \label{fig:rap_slices_pi_neg}
\end{figure}

\begin{figure}[h]
    \centering
    \includegraphics[width=0.32\textwidth, page=1, trim = 0 10 0 30, clip]{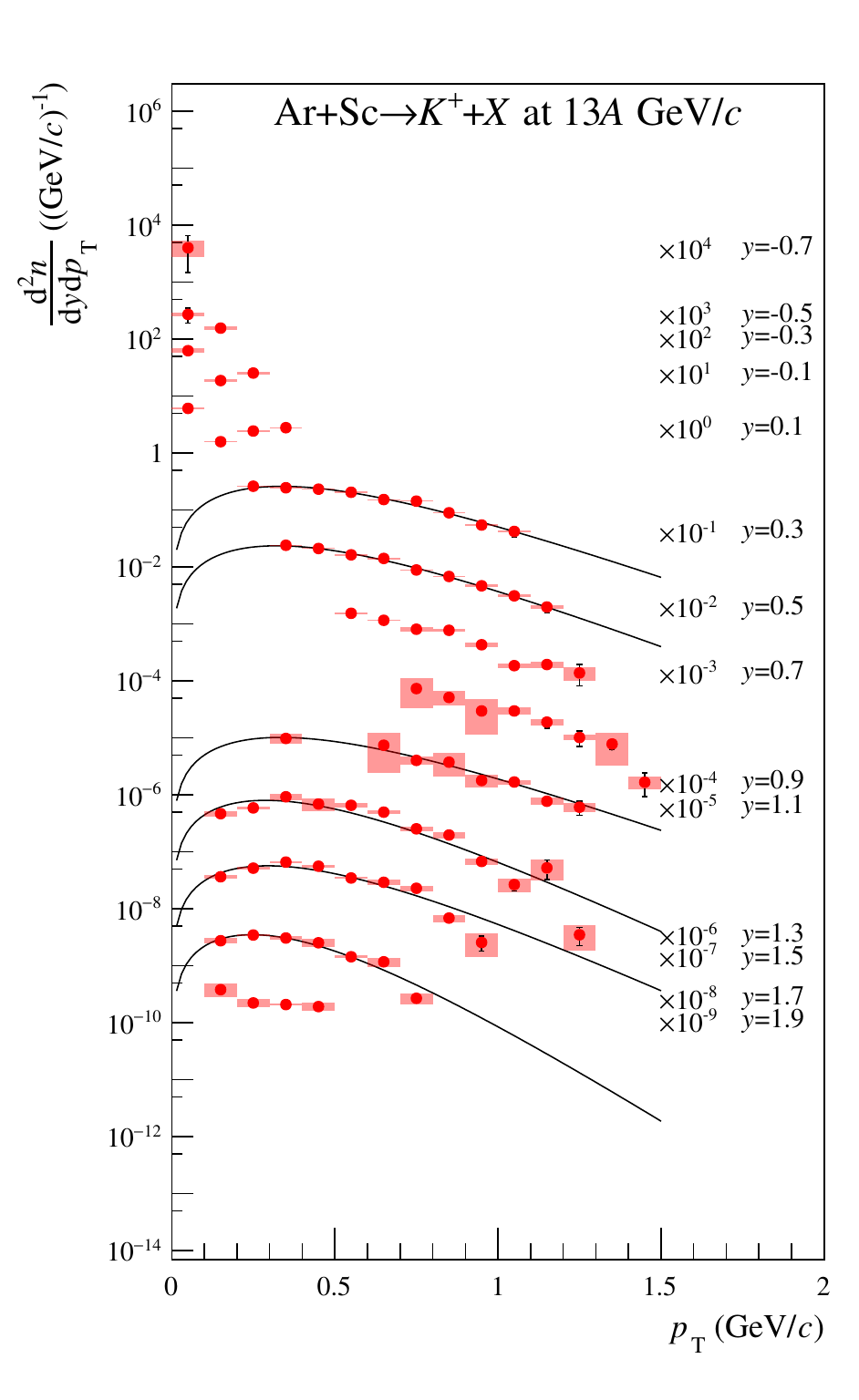}
    \includegraphics[width=0.32\textwidth, page=2, trim = 0 10 0 30, clip]{figures/rapidity_slices/rap_slices_k_pos.pdf}
    \includegraphics[width=0.32\textwidth, page=3, trim = 0 10 0 30, clip]{figures/rapidity_slices/rap_slices_k_pos.pdf}\\
    \includegraphics[width=0.32\textwidth, page=4, trim = 0 10 0 30, clip]{figures/rapidity_slices/rap_slices_k_pos.pdf}
    \includegraphics[width=0.32\textwidth, page=5, trim = 0 10 0 30, clip]{figures/rapidity_slices/rap_slices_k_pos.pdf}
    \includegraphics[width=0.32\textwidth, page=6, trim = 0 10 0 30, clip]{figures/rapidity_slices/rap_slices_k_pos.pdf}
    \caption{ Transverse momentum spectra in rapidity slices of $K^+$ produced in 10\% most \textit{central} Ar+Sc collisions at 13$A$ to 150\AGeVc. For graphical presentations, the spectra are multiplied by factors given in the plots. Lines correspond to fitted exponential functions (Eq. \ref{eq:pion_pT}). Rapidity values given in the legends correspond to the middle of the corresponding interval. Error bars show statistical uncertainties, while shaded boxes correspond to systematic uncertainties.}
    \label{fig:rap_slices_k_pos}
\end{figure}

\begin{figure}[h]
    \centering
    \includegraphics[width=0.32\textwidth, page=1, trim = 0 10 0 30, clip]{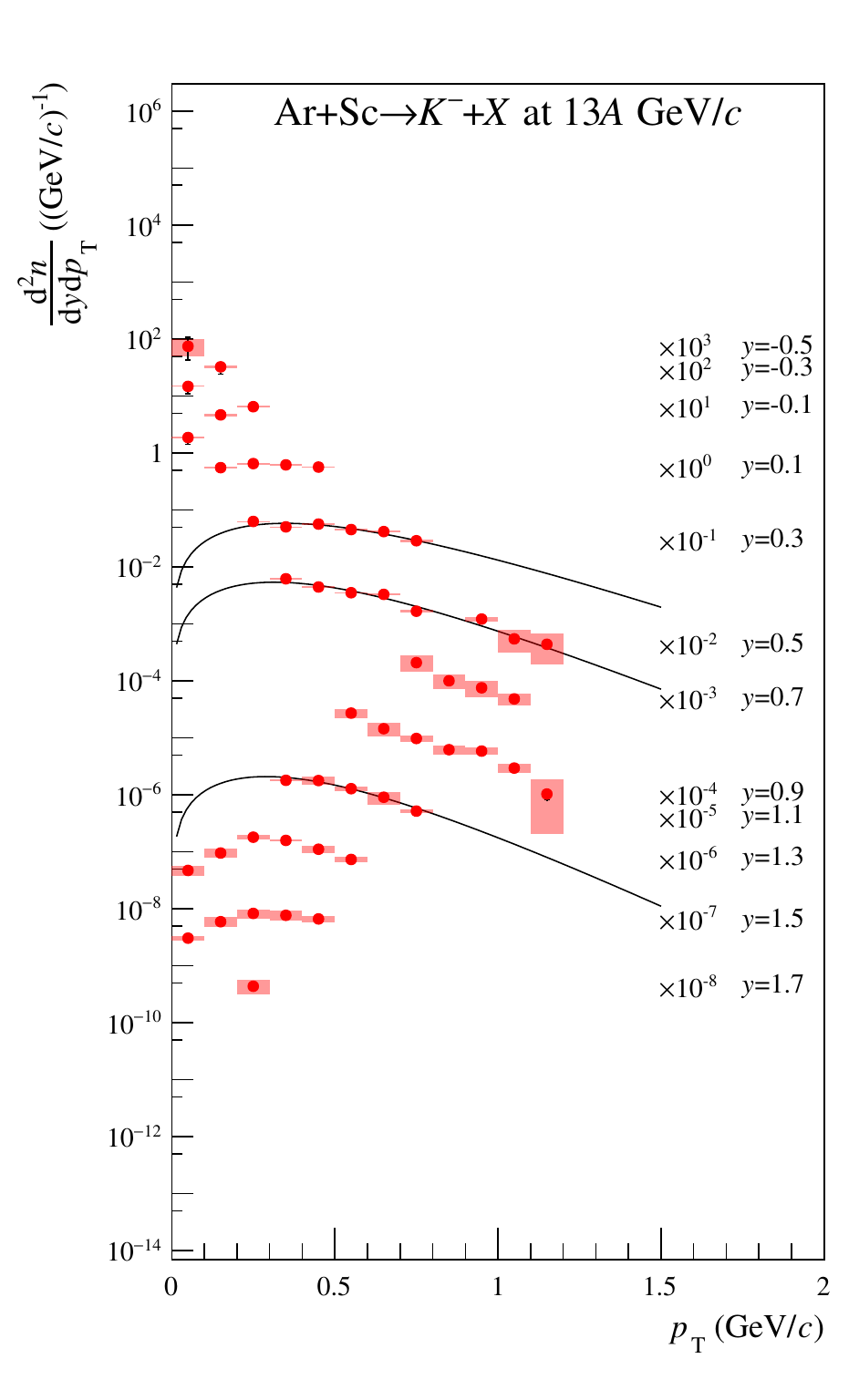}
    \includegraphics[width=0.32\textwidth, page=2, trim = 0 10 0 30, clip]{figures/rapidity_slices/rap_slices_k_neg.pdf}
    \includegraphics[width=0.32\textwidth, page=3, trim = 0 10 0 30, clip]{figures/rapidity_slices/rap_slices_k_neg.pdf}\\
    \includegraphics[width=0.32\textwidth, page=4, trim = 0 10 0 30, clip]{figures/rapidity_slices/rap_slices_k_neg.pdf}
    \includegraphics[width=0.32\textwidth, page=5, trim = 0 10 0 30, clip]{figures/rapidity_slices/rap_slices_k_neg.pdf}
    \includegraphics[width=0.32\textwidth, page=6, trim = 0 10 0 30, clip]{figures/rapidity_slices/rap_slices_k_neg.pdf}
    \caption{ Transverse momentum spectra in rapidity slices of $K^-$ produced in 10\% most \textit{central} Ar+Sc collisions at 13$A$ to 150\AGeVc. For graphical presentations, the spectra are multiplied by factors given in the plots. Lines correspond to fitted exponential functions (Eq. \ref{eq:pion_pT}). Rapidity values given in the legends correspond to the middle of the corresponding interval. Error bars show statistical uncertainties, while shaded boxes correspond to systematic uncertainties.}
    \label{fig:rap_slices_k_neg}
\end{figure}

\begin{figure}[h]
    \centering
    \includegraphics[width=0.32\textwidth, page=1, trim = 0 10 0 30, clip]{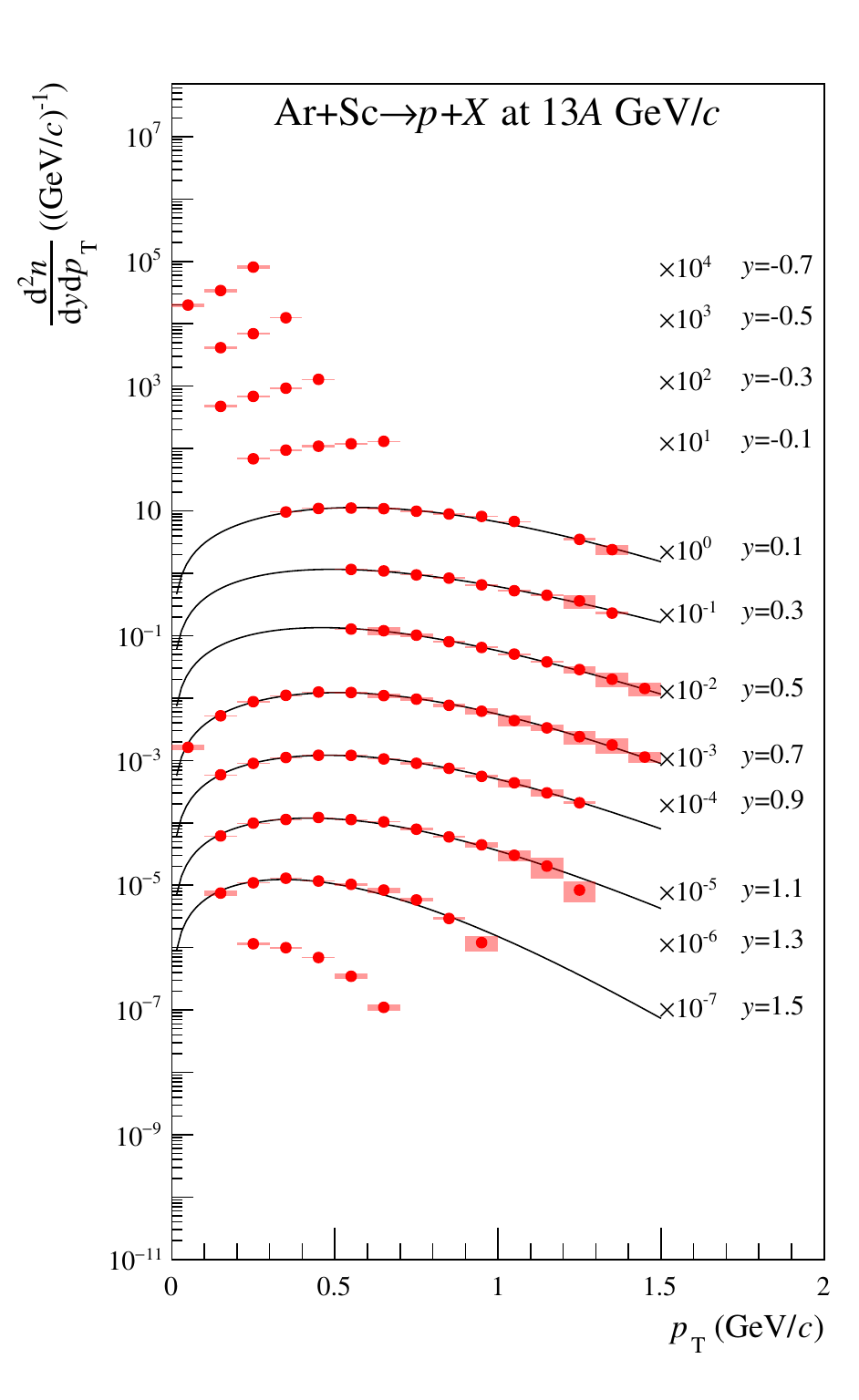}
    \includegraphics[width=0.32\textwidth, page=2, trim = 0 10 0 30, clip]{figures/rapidity_slices/rap_slices_p_pos.pdf}
    \includegraphics[width=0.32\textwidth, page=3, trim = 0 10 0 30, clip]{figures/rapidity_slices/rap_slices_p_pos.pdf}\\
    \includegraphics[width=0.32\textwidth, page=4, trim = 0 10 0 30, clip]{figures/rapidity_slices/rap_slices_p_pos.pdf}
    \includegraphics[width=0.32\textwidth, page=5, trim = 0 10 0 30, clip]{figures/rapidity_slices/rap_slices_p_pos.pdf}
    \includegraphics[width=0.32\textwidth, page=6, trim = 0 10 0 30, clip]{figures/rapidity_slices/rap_slices_p_pos.pdf}
    \caption{ Transverse momentum spectra in rapidity slices of $p$ produced in 10\% most \textit{central} Ar+Sc collisions at 13$A$ to 150\AGeVc. For graphical presentations, the spectra are multiplied by factors given in the plots. Lines correspond to fitted exponential functions (Eq. \ref{eq:pion_pT}). Rapidity values given in the legends correspond to the middle of the corresponding interval. Error bars show statistical uncertainties, while shaded boxes correspond to systematic uncertainties.}
    \label{fig:rap_slices_p_pos}
\end{figure}

\begin{figure}[h]
    \centering
    \includegraphics[width=0.32\textwidth, page=1, trim = 0 10 0 30, clip]{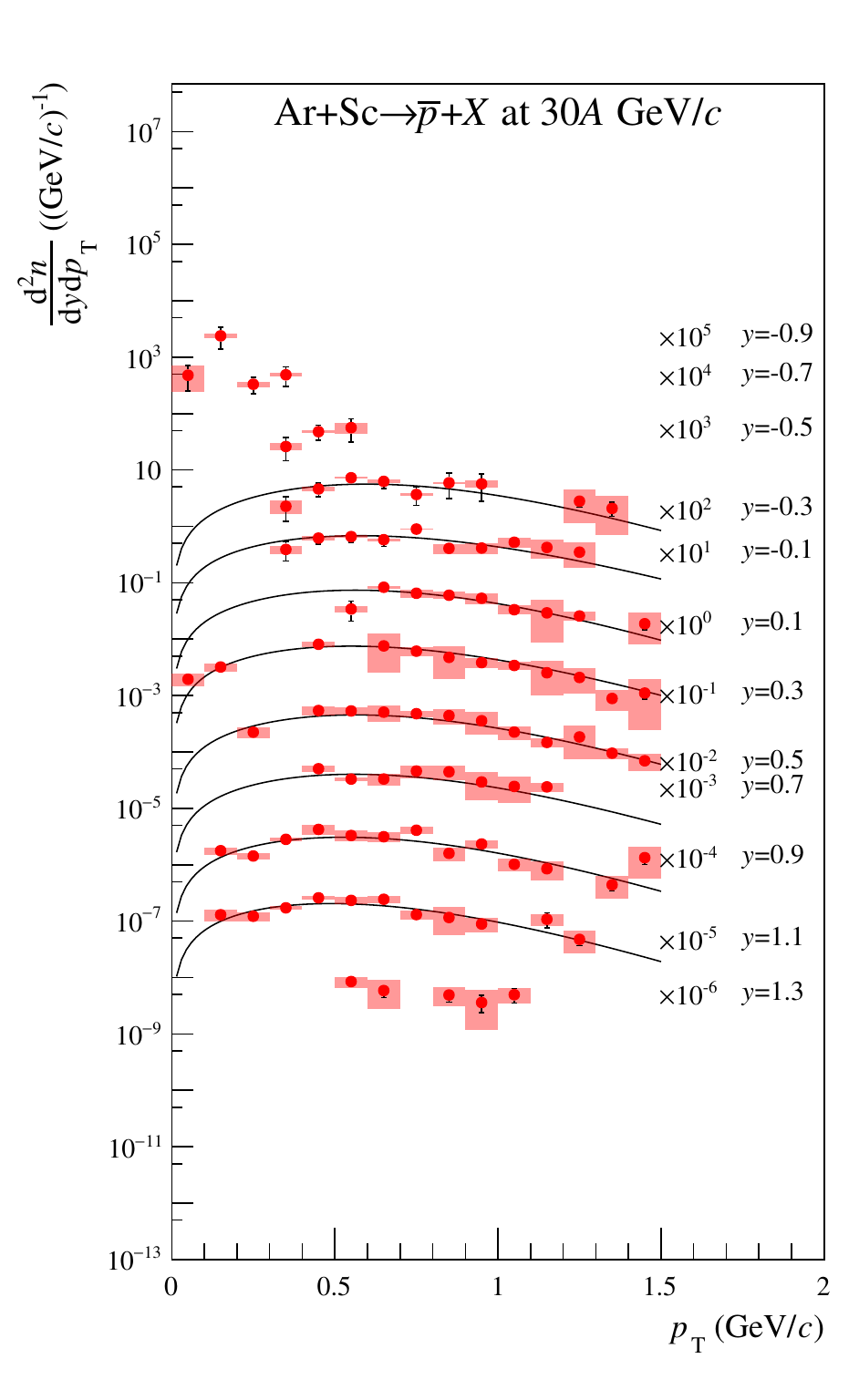}
    \includegraphics[width=0.32\textwidth, page=2, trim = 0 10 0 30, clip]{figures/rapidity_slices/rap_slices_p_neg.pdf}\\
    \includegraphics[width=0.32\textwidth, page=3, trim = 0 10 0 30, clip]{figures/rapidity_slices/rap_slices_p_neg.pdf}
    \includegraphics[width=0.32\textwidth, page=4, trim = 0 10 0 30, clip]{figures/rapidity_slices/rap_slices_p_neg.pdf}
    \caption{ Transverse momentum spectra in rapidity slices of $\bar{p}$ produced in 10\% most \textit{central} Ar+Sc collisions at 30$A$ to 150\AGeVc. For graphical presentations, the spectra are multiplied by factors given in the plots. Rapidity values given in the legends correspond to the middle of the corresponding interval. Lines correspond to fitted exponential functions (Eq. \ref{eq:pion_pT}). Error bars show statistical uncertainties, while shaded boxes correspond to systematic uncertainties.}
    \label{fig:rap_slices_p_neg}
\end{figure}

\begin{figure}[h]
    \centering
    \includegraphics[width=0.32\textwidth, page=1]{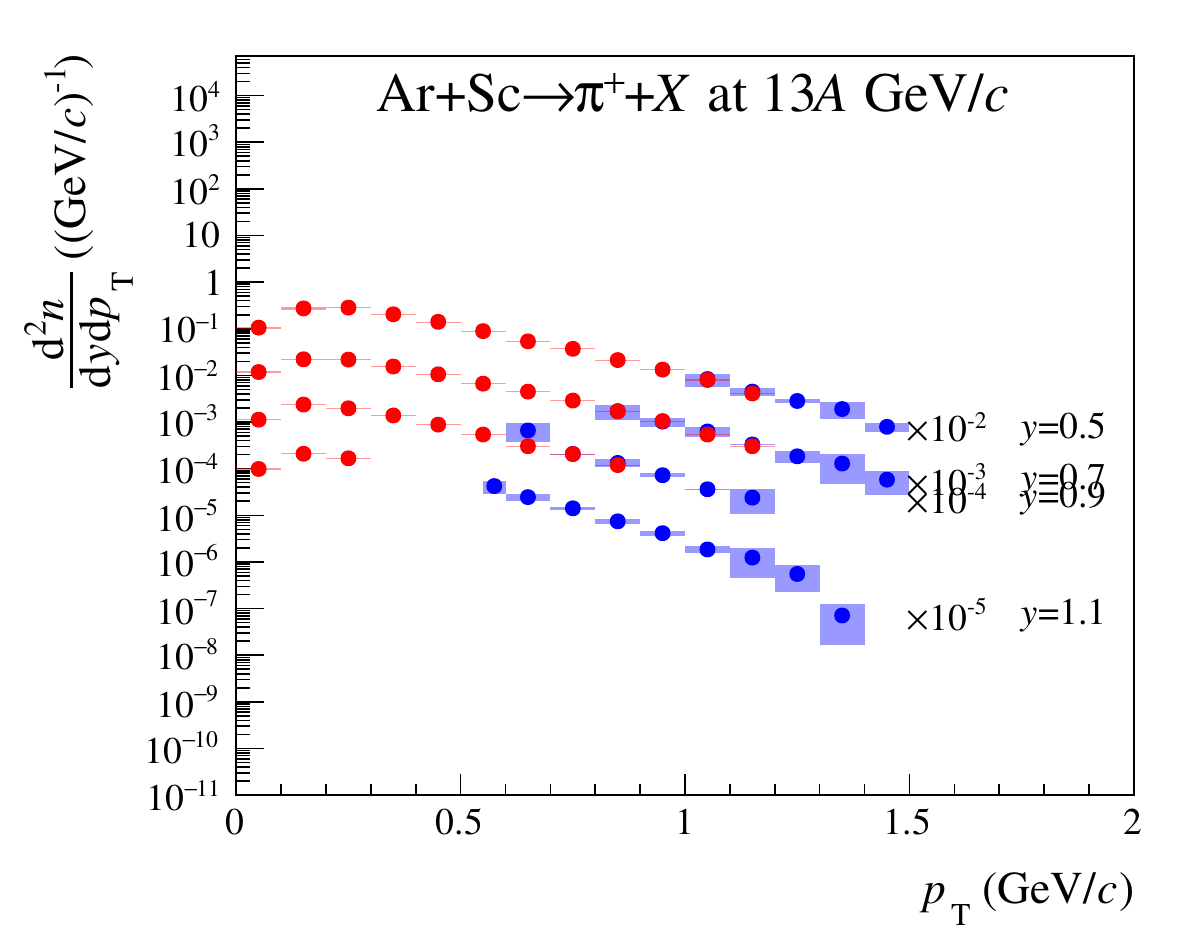}
    \includegraphics[width=0.32\textwidth, page=2]{figures/rapidity_slices/rap_comp_pi_pos.pdf}
    \includegraphics[width=0.32\textwidth, page=3]{figures/rapidity_slices/rap_comp_pi_pos.pdf}\\
    \includegraphics[width=0.32\textwidth, page=4]{figures/rapidity_slices/rap_comp_pi_pos.pdf}
    \includegraphics[width=0.32\textwidth, page=5]{figures/rapidity_slices/rap_comp_pi_pos.pdf}
    \includegraphics[width=0.32\textwidth, page=6]{figures/rapidity_slices/rap_comp_pi_pos.pdf}
    \caption{ Comparison of transverse momentum spectra in rapidity slices of $\pi^+$ obtained with \dedx (blue) and \tofdedx (red) methods, produced in 10\% most \textit{central} Ar+Sc collisions at 13$A$ to 150\AGeVc. For graphical presentations, the spectra are multiplied by factors given in the plots. Rapidity values given in the legends correspond to the middle of the corresponding interval. Error bars show statistical uncertainties, while shaded boxes correspond to systematic uncertainties.}
    \label{fig:rap_comp_pi_pos}
\end{figure}

\begin{figure}[h]
    \centering
    \includegraphics[width=0.32\textwidth, page=1]{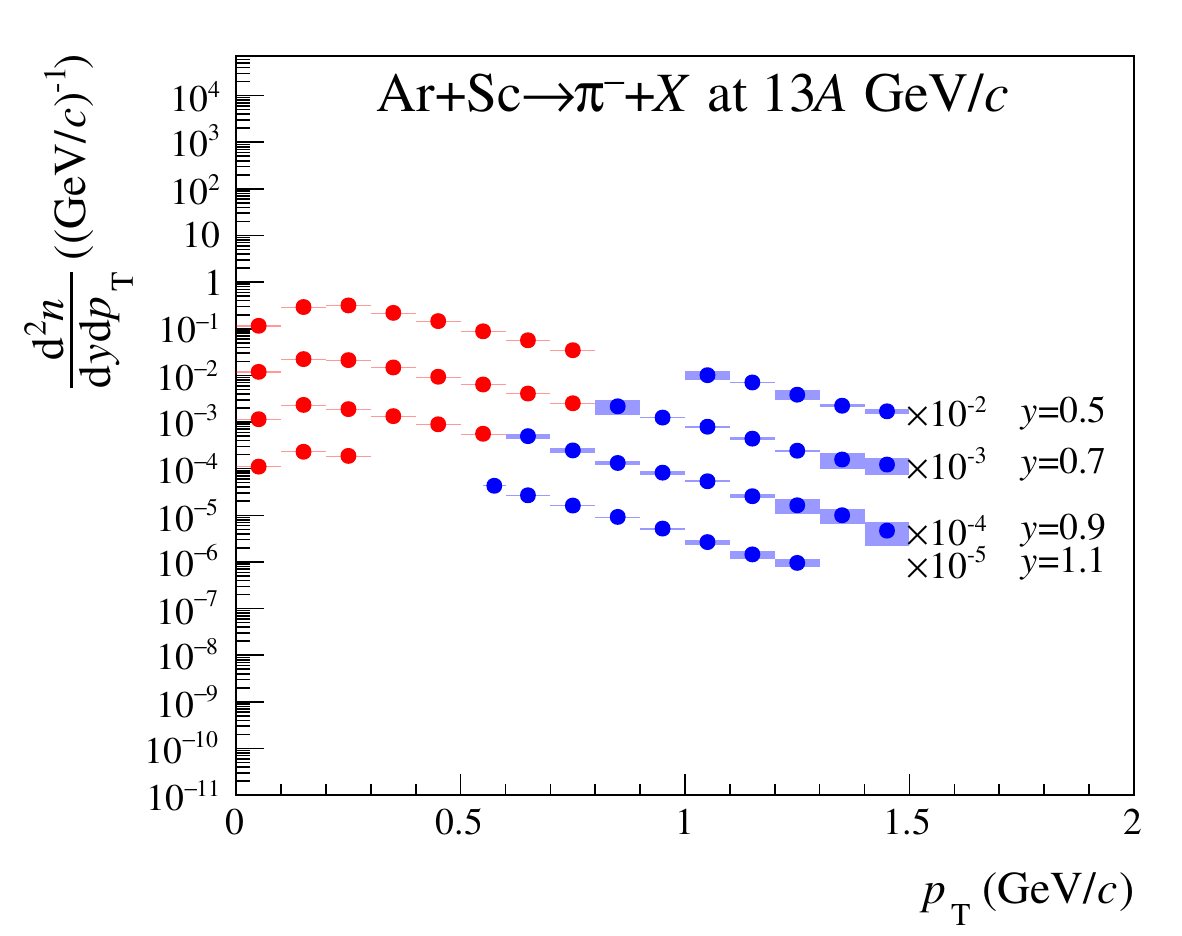}
    \includegraphics[width=0.32\textwidth, page=2]{figures/rapidity_slices/rap_comp_pi_neg.pdf}
    \includegraphics[width=0.32\textwidth, page=3]{figures/rapidity_slices/rap_comp_pi_neg.pdf}\\
    \includegraphics[width=0.32\textwidth, page=4]{figures/rapidity_slices/rap_comp_pi_neg.pdf}
    \includegraphics[width=0.32\textwidth, page=5]{figures/rapidity_slices/rap_comp_pi_neg.pdf}
    \includegraphics[width=0.32\textwidth, page=6]{figures/rapidity_slices/rap_comp_pi_neg.pdf}
    \caption{ Comparison of transverse momentum spectra in rapidity slices of $\pi^-$ obtained with \dedx (blue) and \tofdedx (red) methods, produced in 10\% most \textit{central} Ar+Sc collisions at 13$A$ to 150\AGeVc. For graphical presentations, the spectra are multiplied by factors given in the plots. Rapidity values given in the legends correspond to the middle of the corresponding interval. Error bars show statistical uncertainties, while shaded boxes correspond to systematic uncertainties.}
    \label{fig:rap_comp_pi_neg}
\end{figure}

\begin{figure}[h]
    \centering
    \includegraphics[width=0.32\textwidth, page=1]{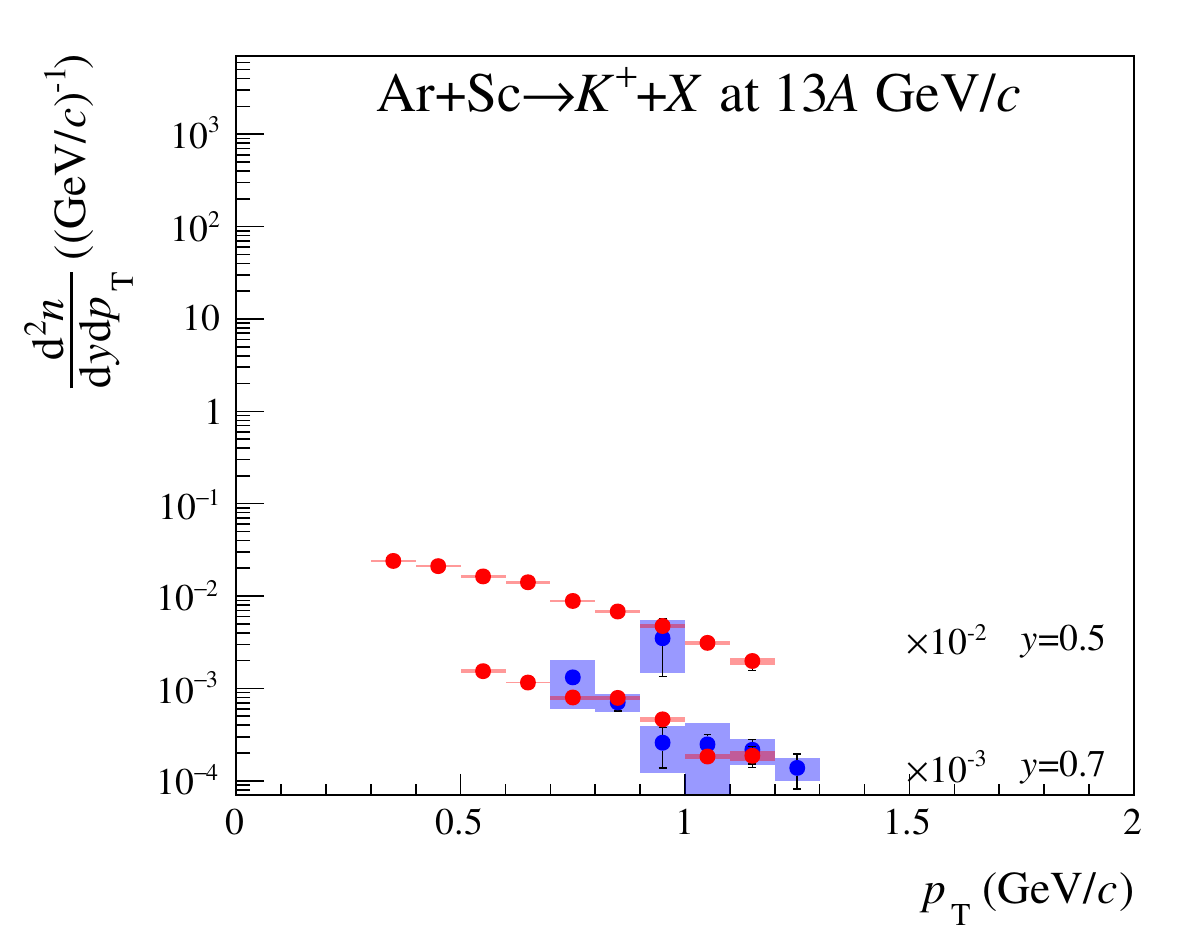}
    \includegraphics[width=0.32\textwidth, page=2]{figures/rapidity_slices/rap_comp_k_pos.pdf}
    \includegraphics[width=0.32\textwidth, page=3]{figures/rapidity_slices/rap_comp_k_pos.pdf}\\
    \includegraphics[width=0.32\textwidth, page=4]{figures/rapidity_slices/rap_comp_k_pos.pdf}
    \includegraphics[width=0.32\textwidth, page=5]{figures/rapidity_slices/rap_comp_k_pos.pdf}
    \includegraphics[width=0.32\textwidth, page=6]{figures/rapidity_slices/rap_comp_k_pos.pdf}
    \caption{ Comparison of transverse momentum spectra in rapidity slices of $K^+$ obtained with \dedx (blue) and \tofdedx (red) methods, produced in 10\% most \textit{central} Ar+Sc collisions at 13$A$ to 150\AGeVc. For graphical presentations, the spectra are multiplied by factors given in the plots. Rapidity values given in the legends correspond to the middle of the corresponding interval. Error bars show statistical uncertainties, while shaded boxes correspond to systematic uncertainties.}
    \label{fig:rap_comp_k_pos}
\end{figure}

\begin{figure}[h]
    \centering
    \includegraphics[width=0.32\textwidth, page=1]{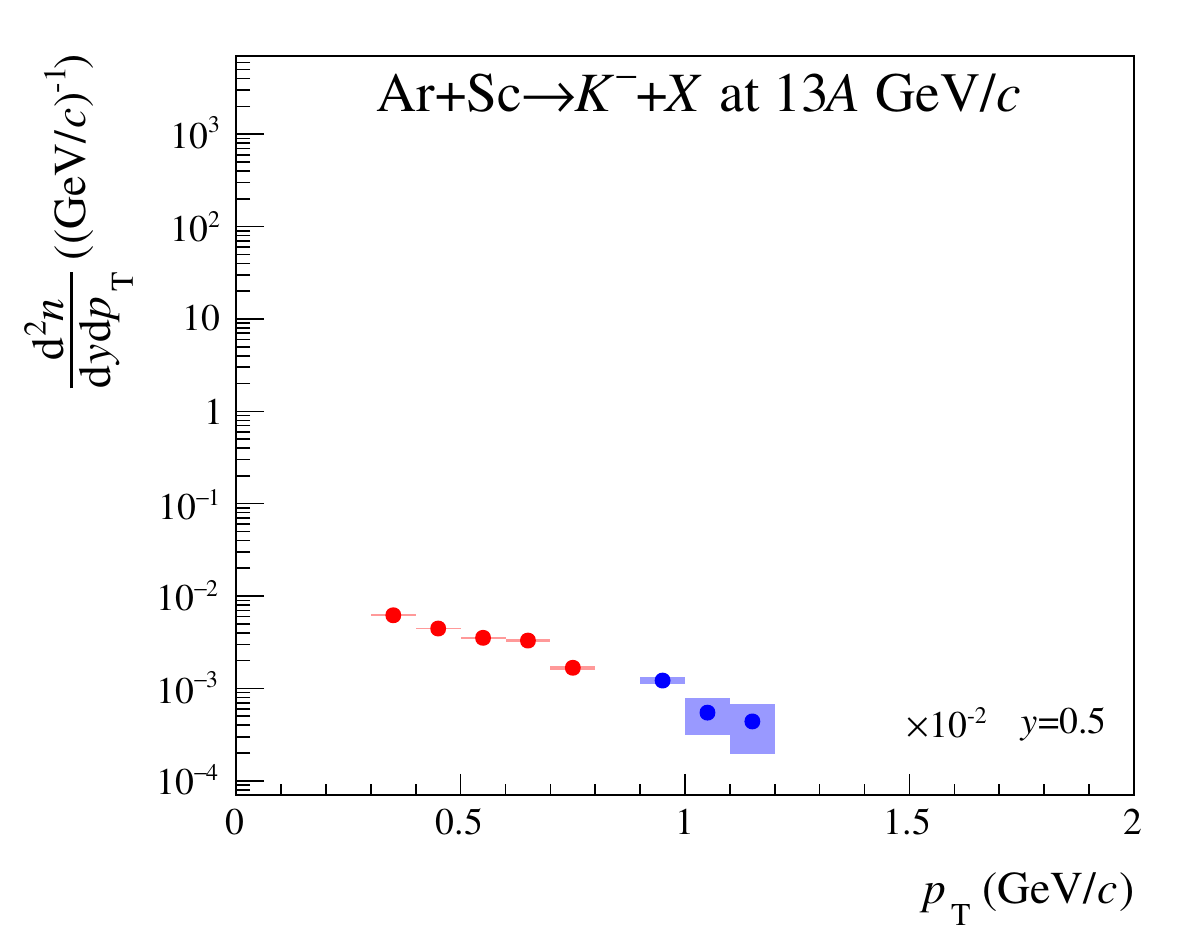}
    \includegraphics[width=0.32\textwidth, page=2]{figures/rapidity_slices/rap_comp_k_neg.pdf}
    \includegraphics[width=0.32\textwidth, page=3]{figures/rapidity_slices/rap_comp_k_neg.pdf}\\
    \includegraphics[width=0.32\textwidth, page=4]{figures/rapidity_slices/rap_comp_k_neg.pdf}
    \includegraphics[width=0.32\textwidth, page=5]{figures/rapidity_slices/rap_comp_k_neg.pdf}
    \includegraphics[width=0.32\textwidth, page=6]{figures/rapidity_slices/rap_comp_k_neg.pdf}
    \caption{ Comparison of transverse momentum spectra in rapidity slices of $\pi^-$ obtained with \dedx (blue) and \tofdedx (red) methods, produced in 10\% most \textit{central} Ar+Sc collisions at 13$A$ to 150\AGeVc. For graphical presentations, the spectra are multiplied by factors given in the plots. Rapidity values given in the legends correspond to the middle of the corresponding interval. Error bars show statistical uncertainties, while shaded boxes correspond to systematic uncertainties.}
    \label{fig:rap_comp_k_neg}
\end{figure}

\begin{figure}[h]
    \centering
    \includegraphics[width=0.32\textwidth, page=1]{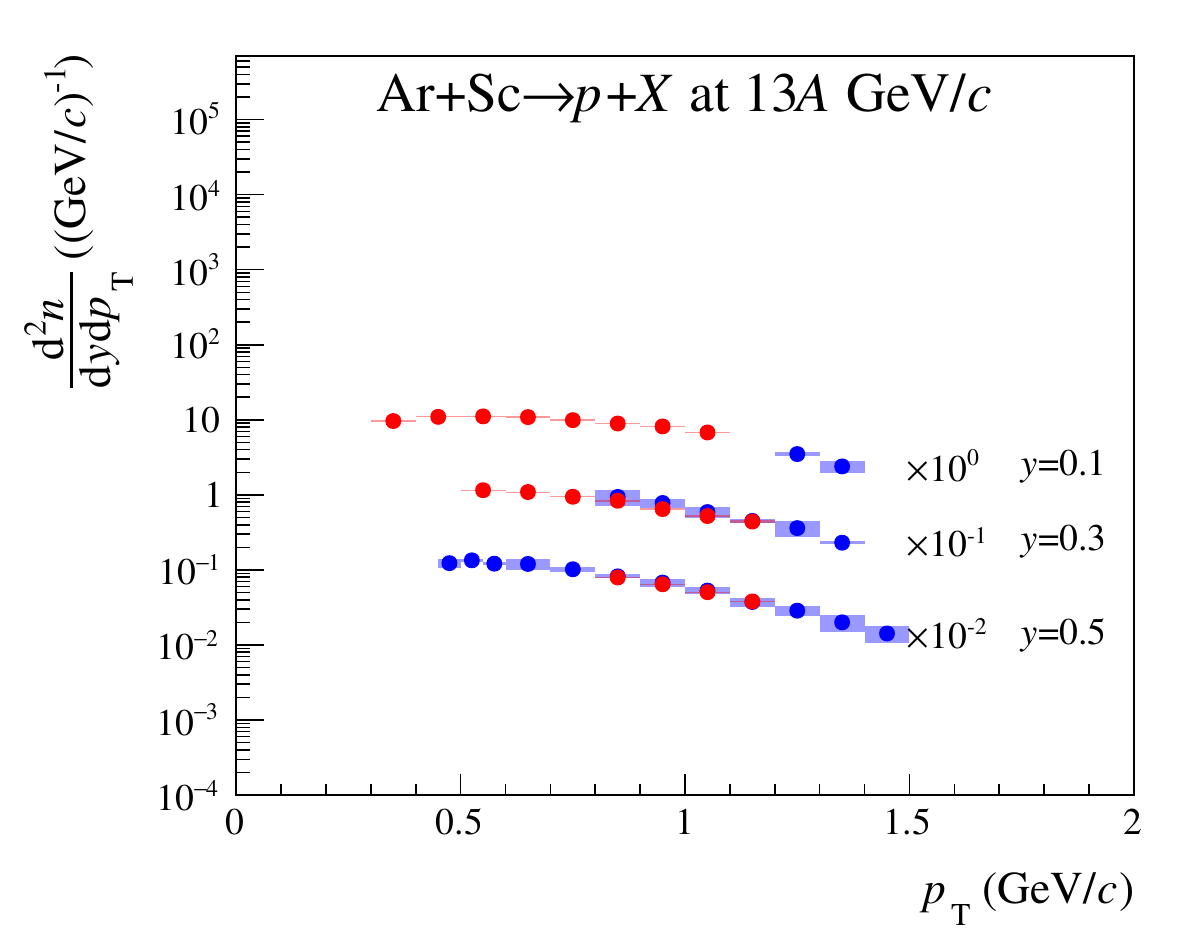}
    \includegraphics[width=0.32\textwidth, page=2]{figures/rapidity_slices/rap_comp_p_pos.pdf}
    \includegraphics[width=0.32\textwidth, page=3]{figures/rapidity_slices/rap_comp_p_pos.pdf}\\
    \includegraphics[width=0.32\textwidth, page=4]{figures/rapidity_slices/rap_comp_p_pos.pdf}
    \includegraphics[width=0.32\textwidth, page=5]{figures/rapidity_slices/rap_comp_p_pos.pdf}
    \includegraphics[width=0.32\textwidth, page=6]{figures/rapidity_slices/rap_comp_p_pos.pdf}
    \caption{ Comparison of transverse momentum spectra in rapidity slices of $p$ obtained with \dedx (blue) and \tofdedx (red) methods, produced in 10\% most \textit{central} Ar+Sc collisions at 13$A$ to 150\AGeVc. For graphical presentations, the spectra are multiplied by factors given in the plots. Rapidity values given in the legends correspond to the middle of the corresponding interval. Error bars show statistical uncertainties, while shaded boxes correspond to systematic uncertainties.}
    \label{fig:rap_comp_p_pos}
\end{figure}

\begin{figure}[h]
    \centering
    \includegraphics[width=0.32\textwidth, page=1]{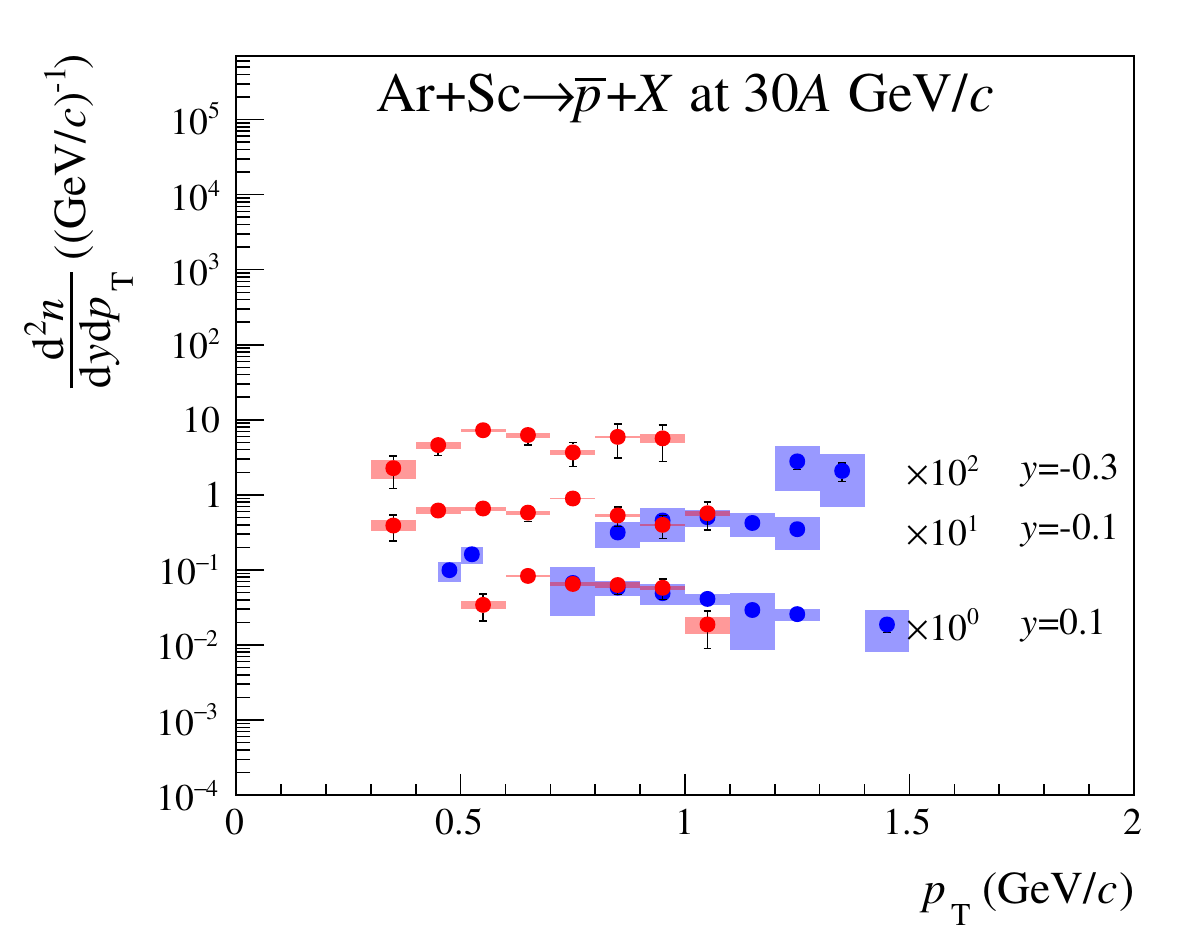}
    \includegraphics[width=0.32\textwidth, page=2]{figures/rapidity_slices/rap_comp_p_neg.pdf}\\
    \includegraphics[width=0.32\textwidth, page=3]{figures/rapidity_slices/rap_comp_p_neg.pdf}
    \includegraphics[width=0.32\textwidth, page=4]{figures/rapidity_slices/rap_comp_p_neg.pdf}
    \caption{ Comparison of transverse momentum spectra in rapidity slices of $\bar{p}$ obtained with \dedx (blue) and \tofdedx (red) methods, produced in 10\% most \textit{central} Ar+Sc collisions at 30$A$ to 150\AGeVc. For graphical presentations, the spectra are multiplied by factors given in the plots. Rapidity values given in the legends correspond to the middle of the corresponding interval. Error bars show statistical uncertainties, while shaded boxes correspond to systematic uncertainties.}
    \label{fig:rap_comp_p_neg}
\end{figure}

This Appendix contains a supplementary set of plots showing transverse momentum spectra in rapidity slices of $\pi^\pm$, $K^\pm$, $p$ and $\bar{p}$ (Figs. \ref{fig:rap_slices_pi_pos}, \ref{fig:rap_slices_pi_neg}, \ref{fig:rap_slices_k_pos}, \ref{fig:rap_slices_k_neg}, \ref{fig:rap_slices_p_pos}, \ref{fig:rap_slices_p_neg}), together with transverse momentum spectra showing the comparison between \dedx and \tofdedx identification methods on the acceptance overlaps (Figs. \ref{fig:rap_comp_pi_pos}, \ref{fig:rap_comp_pi_neg}, \ref{fig:rap_comp_k_pos}, \ref{fig:rap_comp_k_neg}, \ref{fig:rap_comp_p_pos}, \ref{fig:rap_comp_p_neg}).

\clearpage

\bibliographystyle{include/na61Utphys}
{\footnotesize\raggedright
\bibliography{include/na61References}
}

\end{document}